\documentclass[aps,prx,a4paper,reprint,twocolumn,floatfix,superscriptaddress,notitlepage,usenames,dvipsnames,svgnames,x11names,table,nofootinbib,longbibliography,amsmath,amssymb]{revtex4-2}
\usepackage[english]{babel}
\usepackage[T1]{fontenc}
\usepackage{float}
\usepackage{graphicx}
\usepackage{natbib}
\usepackage{bm}
\usepackage{tabularx}
\usepackage{braket}
\usepackage[svgnames]{xcolor}
\usepackage{amsfonts}
\usepackage{dsfont}
\usepackage{stmaryrd}
\usepackage{comment}
\usepackage{dirtytalk}
\usepackage{times}
\usepackage[pdfusetitle]{hyperref}
\hypersetup{pdflang={English},colorlinks=true,linkcolor=RoyalBlue,citecolor=RoyalBlue,urlcolor=RoyalBlue,breaklinks=true}
\usepackage[nameinlink,capitalize,capitalise]{cleveref}

\usepackage[autostyle,english=american]{csquotes}
\MakeOuterQuote{"}

\usepackage[linesnumbered, ruled, vlined]{algorithm2e}
\usepackage[normalem]{ulem}
\usepackage{url}
\usepackage{bbold}
\usepackage{realboxes}

\usepackage{enumitem}
\usepackage{makecell} 
\usepackage{multirow}
\usepackage{tikz}
\usepackage{siunitx}
\usepackage{booktabs}
\usepackage{algpseudocode}
\usepackage[page]{appendix}
\usepackage{ragged2e}
\usepackage[font=scriptsize,labelfont=bf,justification=justified,format=plain]{subcaption}
\usepackage[font=scriptsize,labelfont=bf,justification=justified,format=plain]{caption}
\DeclareCaptionJustification{reallyjustified}{\justifying}
\usepackage{placeins}

\usepackage{listings}

\definecolor{bkgd}{RGB}{240,242,246}
\definecolor{ceruleanblue}{rgb}{0.16, 0.32, 0.75}
\definecolor{orange-red}{rgb}{1.0, 0.27, 0.0}
\definecolor{anotherblue}{RGB}{37,92,243}
\definecolor{blackblue}{RGB}{46,60,85}
\definecolor{goldyellow}{RGB}{199,146,12}
\definecolor{purple}{RGB}{157,0,255}
\lstdefinestyle{altstyle2}{
    backgroundcolor=\color{bkgd},
    basicstyle=\ttfamily\footnotesize\color{blackblue},
    breakatwhitespace=false,
    breaklines=true,
    captionpos=b,
    commentstyle=\color{goldyellow},
    keepspaces=true,
    keywordstyle=\color{orange-red},
    language=Python,
    numbersep=5pt,
    numberstyle=\tiny\color{ceruleanblue},
    showspaces=false,
    showstringspaces=false,
    showtabs=false,
    stringstyle=\color{anotherblue},
    tabsize=2
}
\begin{document}

\title{How to Build a Quantum Supercomputer: \\ Scaling from Hundreds to Millions of Qubits}

\author{Masoud Mohseni}
\email{masoud.mohseni@hpe.com}
\affiliation{HPE Quantum, Emergent Machine Intelligence, HPE Labs, CA, USA}

\author{Artur Scherer}
\affiliation{1QB Information Technologies (1QBit), BC, Canada}

\author{K. Grace Johnson}
\affiliation{HPE Quantum, Emergent Machine Intelligence, HPE Labs, CA, USA}

\author{Oded Wertheim}
\affiliation{Quantum Machines, Israel}

\author{Matthew Otten}
\affiliation{Department of Physics, University of Wisconsin--Madison, WI, USA}

\author{Namit Anand}
\affiliation{HPE Quantum, Emergent Machine Intelligence, HPE Labs, CA, USA}
\affiliation{KBR, Inc., TX, USA}
\affiliation{Quantum Artificial Intelligence Laboratory (QuAIL), NASA Ames Research Center, CA, USA}

\author{\mbox{Navid Anjum Aadit}}
\affiliation{Department of Electrical and Computer Engineering, University of California, Santa Barbara, CA, USA}

\author{Yuri Alexeev}
\affiliation{NVIDIA Corporation, Santa Clara, CA, USA}

\author{Gilad Ben-Shach}
\affiliation{Quantum Machines, Israel}

\author{Kirk M. Bresniker}
\affiliation{Hewlett Packard Labs, CA, USA}

\author{Kerem Y. Camsari}
\affiliation{Department of Electrical and Computer Engineering, University of California, Santa Barbara, CA, USA}

\author{Barbara Chapman}
\affiliation{Hewlett Packard Enterprise, TX, USA}

\author{Soumitra Chatterjee}
\affiliation{Hewlett Packard Enterprise, TX, USA}

\author{Shuvro Chowdhury}
\affiliation{Department of Electrical and Computer Engineering, University of California, Santa Barbara, CA, USA}

\author{Gebremedhin A. Dagnew}
\affiliation{1QB Information Technologies (1QBit), BC, Canada}

\author{Tom Dvir}
\affiliation{Quantum Machines, Israel}

\author{Aniello Esposito}
\affiliation{Hewlett Packard Labs, CA, USA}

\author{Farah Fahim}
\affiliation{Fermi National Accelerator Laboratory, IL, USA}

\author{Michael Ferguson}
\affiliation{HPE Quantum, Emergent Machine Intelligence, HPE Labs, CA, USA}

\author{\mbox{Marco Fiorentino}}
\affiliation{HPE Quantum, Emergent Machine Intelligence, HPE Labs, CA, USA}

\author{Archit Gajjar}
\affiliation{Hewlett Packard Labs, CA, USA}

\author{Katerina Gratsea}
\affiliation{Department of Physics, University of Wisconsin--Madison, WI, USA}

\author{Gaurav Gyawali}
\affiliation{HPE Quantum, Emergent Machine Intelligence, HPE Labs, CA, USA}

\author{Christian Heiter}
\affiliation{HPE Quantum, Emergent Machine Intelligence, HPE Labs, CA, USA}

\author{\mbox{Ali H. Z. Kavaki}}
\affiliation{1QB Information Technologies (1QBit), BC, Canada}

\author{Abdullah Khalid}
\affiliation{1QB Information Technologies (1QBit), BC, Canada}

\author{Xiangzhou Kong}
\affiliation{1QB Information Technologies (1QBit), BC, Canada}

\author{Bohdan Kulchytskyy}
\affiliation{1QB Information Technologies (1QBit), BC, Canada}

\author{Elica Kyoseva}
\affiliation{NVIDIA Corporation, Santa Clara, CA, USA}

\author{Ruoyu Li}
\affiliation{Applied Materials, CA, USA}

\author{P. Aaron Lott}
\affiliation{HPE Quantum, Emergent Machine Intelligence, HPE Labs, CA, USA}
\affiliation{KBR, Inc., TX, USA}
\affiliation{Quantum Artificial Intelligence Laboratory (QuAIL), NASA Ames Research Center, CA, USA}

\author{\mbox{Igor L. Markov}}
\affiliation{Synopsys, CA, USA}

\author{Robert F. McDermott}
\affiliation{Department of Physics, University of Wisconsin--Madison, WI, USA}
\affiliation{Qolab, WI, USA}

\author{Lucas Morais}
\affiliation{Hewlett Packard Labs, CA, USA}

\author{Giacomo Pedretti}
\affiliation{Hewlett Packard Labs, CA, USA}

\author{Pooja Rao}
\affiliation{NVIDIA Corporation, Santa Clara, CA, USA}

\author{Eleanor Rieffel}
\affiliation{Quantum Artificial Intelligence Laboratory (QuAIL), NASA Ames Research Center, CA, USA}

\author{Allyson Silva}
\affiliation{1QB Information Technologies (1QBit), BC, Canada}

\author{John Sorebo}
\affiliation{Synopsys, CA, USA}

\author{Panagiotis Spentzouris}
\affiliation{Fermi National Accelerator Laboratory, IL, USA}

\author{Ziv Steiner}
\affiliation{Quantum Machines, Israel}

\author{Boyan Torosov}
\affiliation{1QB Information Technologies (1QBit), BC, Canada}

\author{Davide Venturelli}
\affiliation{Quantum Artificial Intelligence Laboratory (QuAIL), NASA Ames Research Center, CA, USA}
\affiliation{USRA Research Institute for Advanced Computer Science, CA, USA}

\author{Robert J. Visser}
\affiliation{Applied Materials, CA, USA}

\author{Zak Webb}
\affiliation{1QB Information Technologies (1QBit), BC, Canada}

\author{Xin Zhan}
\affiliation{HPE Quantum, Emergent Machine Intelligence, HPE Labs, CA, USA}

\author{Yonatan Cohen}
\affiliation{Quantum Machines, Israel}

\author{Pooya Ronagh}
\affiliation{1QB Information Technologies (1QBit), BC, Canada}
\affiliation{Institute for Quantum Computing, University of Waterloo, ON, Canada}
\affiliation{Department of Physics \& Astronomy, University of Waterloo, ON, Canada}
\affiliation{Perimeter Institute for Theoretical Physics, ON, Canada}

\author{Alan Ho}
\affiliation{Qolab, WI, USA}

\author{Raymond G. Beausoleil}
\affiliation{HPE Quantum, Emergent Machine Intelligence, HPE Labs, CA, USA}

\author{John M. Martinis}
\email{john@qolab.ai}
\affiliation{Qolab, WI, USA}

\date{\today}


\begin{abstract}
In the span of four decades, quantum computation has evolved from an intellectual curiosity to a potentially realizable technology. Today, small-scale demonstrations have become possible for quantum algorithmic primitives on hundreds of physical qubits. Nevertheless, there are significant outstanding challenges in quantum hardware, fabrication, software architecture, and algorithms on the path towards a full-stack scalable quantum computing technology.
Here, we provide a comprehensive review of these scaling challenges. We show how to facilitate scaling by adopting existing semiconductor technology to build much higher-quality qubits, employing systems engineering approaches, and performing distributed heterogeneous quantum-classical computing. We provide a detailed resource and sensitivity analysis for quantum applications on surface-code error-corrected quantum computers given current, target, and desired hardware specifications based on superconducting qubits, accounting for a realistic distribution of errors. We provide comprehensive resource estimates for several utility-scale applications including quantum chemistry calculations, catalyst design, NMR spectroscopy, and Fermi-Hubbard simulation. We show that orders of magnitude enhancement in performance could be obtained by a combination of hardware improvements and tight quantum-HPC integration. Furthermore, we introduce high-performance architectures for quantum-probabilistic computing with custom-designed accelerators to tackle today's industry-scale classical optimization, machine learning, and quantum simulation tasks in a cost-effective manner.
\end{abstract}

\clearpage
\maketitle
\tableofcontents

\section{Introduction: Past, present, and future challenges of building quantum computers}
\label{sec:challenges-intro}

Quantum computing has seen remarkable progress over the past few decades despite facing tremendous conceptual, theoretical, and technical challenges. There have been a few significant breakthroughs, such as Shor's algorithm for integer factoring \cite{shor_factoring_1994} to solve a seemingly exponentially hard classical problem, or the invention of quantum error-correction (QEC)~\cite{PhysRevA.52.R2493,gottesman1997stabilizercodesquantumerror} to tackle the fundamental problem of  \textit{decoherence}, without which Shor's algorithm would remain merely a mathematical curiosity. Starting from small experiments that manipulate single- or few-qubit systems, research groups using a variety of technologies can now make and operate quantum processors with on the order of 100 physical qubits. Under certain conditions state-of-the-art classical simulation techniques such as tensor networks or quantum Monte Carlo could struggle to simulate quantum dynamics on $\sim 100$ noisy physical qubits, thereby providing a natural benchmark for \textit{beyond classical} computation with quantum hardware. Some proof-of-principle speedups over conventional (classical) supercomputers called ``quantum supremacy'' \cite{arute2019quantum} or ``quantum advantage'' \cite{quantum-advantage} have been demonstrated, but only for carefully crafted problems. The next step is to scale up quantum processors to demonstrate significant speedup for a practical problem in a cost effective manner, thereby achieving {\em quantum utility} \cite{utility-scale}. 

Recently, the high-performance computing (HPC) community has shown significant interest in employing quantum computation as a complementary paradigm beyond exascale supercomputing \cite{alexeev_quantum-centric_2024,ibmPositionPaper,humbleQuantumClassical}. Rather than replacing classical computers as general-purpose processors, quantum computers can be better understood as accelerators or coprocessors that can efficiently carry out specialized tasks within an HPC framework. Hybrid quantum--classical frameworks will be crucial not only in the near term---the noisy, intermediate-scale quantum  (NISQ)  era~\cite{preskill2018quantum}---but also for future fault-tolerant quantum computation (FTQC), as error-correction schemes will rely heavily on classical HPC and the number of logical qubits will be fairly small for the foreseeable future. To achieve true utility-scale quantum computing, successful integration with existing heterogeneous HPC infrastructures and the development of a hybrid quantum--classical full computing stack are necessary. 

Practical quantum--HPC integration runs into a number of challenges. At the hardware-architecture and system-design levels, there are significant differences between the quantum and classical components in physical scale, hardware reliability, control electronics, communication bandwidth, and time scales of operation. At the algorithmic level, the challenges lie within memory access, data sharing and movement, and information extraction. To build a quantum-centric supercomputer \cite{alexeev_quantum-centric_2024,ibmPositionPaper}  at scale, much better quantum components must be built at different layers that ultimately rely on much higher quality qubits. While basic research remains critical, a comprehensive engineering approach targeting the full stack must be taken in parallel, with the aim of steadily increasing the technology readiness level (TRL) of the components at various levels of the full stack. This mindset is needed for NISQ devices, logical intermediate-scale quantum (LISQ) processors or early-FTQC and for large-scale FTQC.

Utility-scale quantum computing ultimately involves deep quantum circuits requiring logical error rates far below those experimentally feasible with physical qubits and gate operations. Quantum error correction~\cite{PhysRevA.52.R2493,gottesman1997stabilizercodesquantumerror} is necessary, as even small errors rapidly accumulate, resulting in significant errors. Similar to classical error correction, QEC utilizes redundancy in the encoding of information to detect and correct errors. However, unlike in the case of 
classical telecommunication, copying quantum information is prohibited due to the no-cloning theorem~\cite{Wootters1982Single}. In addition, quantum measurements irreversibly collapse the wavefunction. Therefore, the key mechanism for error correction without destroying logical information is through exploiting quantum entanglement with ancillary subsystems. Quantum error-correction codes (QECC) protect a smaller logical state of computation within a much larger highly entangled quantum state involving many noisy physical qubits. Therefore, errors below a certain weight can be detected and corrected. Multi-qubit stabilizer measurements produce a set of syndromes that detect whether some of the physical qubits have been corrupted. A QEC decoder can then infer the most likely errors for the detected syndromes. With this knowledge, corrective operations can be either physically applied (active error correction) or kept track of in software.

While QEC is necessary, it is by no means sufficient for reliable large-scale quantum computation. When performing QEC, the operations on the physical qubits necessary to implement the QECC can eventually themselves add more errors than they are able to correct. Thus, achieving FTQC requires implementation schemes in which QEC succeeds in suppressing errors faster than it causes them. According to the threshold theorem, once the physical errors of the quantum system are below a certain threshold, the overhead of these schemes scales poly-logarithmically with the precision of computation~\cite{aharonov1997fault-tolerant,knill1998resilient,Kitaev_2003}. However, these schemes introduce substantial overhead in physical resources, whose estimation and validation is a critical step toward practical, cost-efficient realization of FTQC at utility scale.

At the lowest level of the full stack---the physical components for storing and processing quantum information---we focus on superconducting qubits, in the context of digital (gate-based) fault-tolerant quantum computing. Our discussion of scaling challenges and opportunities does not necessarily apply to quantum annealers based on superconducting qubits \cite{Johnson2011annealing}. Historically, the success of superconducting qubits has come from a sustained focus on improving qubit coherence using different approaches. At present, although there are a variety of ideas on how to make further advances, there is sufficient justification for using advanced semiconductor processing and tools.  This approach will need to be guided by and integrated with more complex device architectures and structures, such as bump-bonding and advanced packaging, which are designed around specific decoherence mechanisms such as large two-level state loss from amorphous insulators.   

These advances were also shaped by an understanding that qubit architecture would be improved by designs that turned off the qubit--qubit interaction, which does not occur naturally using simple coupling capacitors. The adjustable coupler was pioneered by UCSB and Google and is an example of careful systems engineering that was initially assessed skeptically by a majority of the superconducting qubit community. Although the adjustable coupler was significantly more complex, as it required using a qubit to turn interactions on and off, the reduction in crosstalk enabled Google to achieve the quantum supremacy milestone \cite{arute2019quantum}. Today, the adjustable coupler has been adopted and integrated into superconducting systems from IBM, USTC, Rigetti, and IQM. Alternative superconducting qubit architectures such as the fluxonium are only being realized \cite{fluxonium} thanks to the adjustable couplers technology. Further innovations which, like adjustable couplers, trade simplicity for performance are needed to push errors in large qubit systems to the $10^{-4}$ range.  

Despite the large size of superconducting qubits (roughly millimeters), the numbers of qubits have been scaled up from single- and few-qubit systems. 
Recent advances enabled large enough quantum computers to test system performance and demonstrate the execution of simple quantum algorithms. Experiments have demonstrated physical qubit performance that are projected to reach logical error rates at the $10^{-10}$ level in the surface code, provided steps are taken to mitigate correlated errors from high-energy cosmic ray and gamma ray impacts in the chip \cite{Wilen2021,mcewen2024cosmicray,martinis2021saving,Iaia2022}. Despite the large footprint of superconducting qubits, in this position paper we describe how modern semiconductor processing facilitates their scaling.

While we introduce a definitive architecture for coupling transmon qubits, alternative designs might also improve performance. Our expectation is that advanced fabrication targeted to improve both coherence and scaling could be used for a variety of approaches. We believe that these ideas will greatly enhance the likelihood of making useful quantum computers. 

\subsection{Systems engineering for quantum hardware}
\label{sec:system-eng}

Systems engineering simultaneously optimizes many system parameters.  In contrast, prior research often treats quantum computation as a custom-configured quantum physics experiment focusing on isolated physical phenomena to fully understand each of them.  For example, reported metrics often emphasize the best-case quality of a given approach, whereas systems engineering is typically sensitive to average and worst-case quality. 

Each technology is sensitive to
a different set of systems engineering parameters.  Here, we focus on the four most important parameters that serve as a concise but powerful way to compare quantum hardware approaches. 

\textbf{Quality.}  Qubits are different from classical bits in that they are fundamentally prone to errors.  This is in part due to their analog-control nature, but also due to their quantum properties, such as decoherence (i.e., entanglement of the qubits with their environment).  It is important to note that the scalability of an approach, expressed by the number of qubits and the length of circuits successfully executed, is now mostly bottlenecked by qubit errors.  For example, a 50-qubit system with 1\% errors will allow only about two layers of gates (at 50 qubits per layer) before there is an error in the execution of the circuit: this clearly limits the system's utility.  Errors in the range 0.01--0.1\% are believed necessary for both NISQ and FTQC algorithms, as is described in more detail in \Cref{sec:fab}.  

\textbf{Quantity.}  This is the most intuitively understood metric. As discussed above, at first we need to optimize for low error rates. However, as errors reach the 0.01--0.1\% range, the ability to scale up the number of qubits becomes more important.  This is, in part, because the number of qubits scales logarithmically with the error-suppression rate of QECCs such as the surface code. So, once the error rate is sufficiently low, the strategy should change to prioritizing qubit counts.  Scaling to thousands or millions of qubits is required in the long term, and thus careful deliberation is needed on how to achieve that with any particular technology.

\textbf{Speed}.  The clock speed of qubits is often not being highlighted in public roadmaps, which put emphasis instead on harnessing quantum advantage from ``exponential quantum parallelism''.  Although this Hilbert space size advantage can be argued for theoretically, we show in our resource estimations (\Cref{sec:resource-estimates-pbenzyne}) that speed is crucially important for practical utility. This is because the speed of various qubit technologies differs greatly, by up to four orders of magnitude.  For example, superconducting qubits have typical single- and two-qubit gate times of 50 ns or less, whereas typical clock speeds of atomic systems are $\sim$$\SI{100}{\micro\second}$, often limited by the time scale of mechanical motions of atoms or ions.  In addition, superconducting gates are operated in parallel, whereas ion-trap systems have their primitive gates often processed serially through interaction zones in leading quantum-charge-coupled-device (QCCD) architectures.  For NISQ applications with short circuits, repeated trials are typically necessary to obtain sufficient statistics, and thus end-to-end experiments are slow even for superconducting qubits. Generally, one can understand the need for speed by noting that a 1000$\times$ increase in speed translates to 1000$\times$ more throughput. However, for FTQC, a 1000$\times$ slowdown can render certain applications impractical because, even with the fast clock-speed of superconducting qubits, the execution time of utility-scale algorithms can be on the order of months or years (see \Cref{sec:resource-estimates-pbenzyne} for predicted execution times).

\textbf{Connectivity.}  The number of connections from one physical qubit to another is also an important metric, and has an interesting trade-off with speed.  Neutral-atom and trapped-ion systems often advertise all-to-all connectivity, but this might not scale beyond small systems, e.g., single traps. This all-to-all connectivity also typically requires ample time for shuttling qubits. Superconducting qubits do not need to move but have sparse connectivity; they typically have nearest-neighbor interactions, either to 4 qubits in a regular rectangular lattice or 2.5 qubits on average for the heavy-hex lattice \cite{ibm-heavy-hex}. This more-limited connectivity can be factored into the design of near-term algorithms, and thus it is hard to make a fair performance comparison with respect to connectivity. For a fault-tolerant quantum computer, the present connectivity of superconducting qubits is sufficient to support error-corrected logical qubits such as the surface codes. Additionally, the more-connected rectangular lattice architecture is less likely to suffer from qubit dropouts, an important system constraint. 

\textbf{Tying the four parameters together.} A quantum computer must incorporate a large number of qubits that are well-connected by sufficiently fast high-quality gates. Performing well with respect to all the above metrics is necessary for creating highly entangled quantum states between the qubits. However, there may be trade-offs between these metrics; e.g., with lower connectivity, more gates are needed to entangle qubits, and if these gates are too slow, decoherence may limit the amount of entanglement generated. Given that entanglement is necessary for quantum computational advantage, {\em the size of the entangled states that can be prepared} by the computer is a useful system-level metric that incorporates the four parameters above and their trade-offs~\cite{cao2023entanglement}.

Quantum volume provides a complementary system-level benchmark to the size of the entangled states~\cite{PhysRevA.100.032328}. It probes the largest approximately square random circuits that a device can execute successfully and is therefore sensitive to gate and measurement fidelity, connectivity, and compiler quality \cite{Pelofske2022}. It is useful for comparing generic circuit execution across devices, but it 
does not necessarily provide good predictions for the performance of structured algorithms, dynamic circuits, QEC syndrome-extraction circuits, or long-running FTQC workloads. Moreover, evaluating the heavy-output condition when computing quantum volume requires classically estimating the ideal output probabilities of random circuits, which becomes expensive as the benchmark width grows. This limits the applicability of quantum volume as a scalable and verifiable benchmark. In contrast, the size of a verified multipartite-entangled state probes a different capability: whether many qubits can be coherently connected in a structured experiment. This metric, however, does not by itself capture useful circuit depth or universality. We therefore regard quantum volume, entangled-state size, application-level benchmarks, and logical-QEC performance as complementary rather than interchangeable metrics. 
Together, they provide a method to more holistically benchmark quantum supercomputers than any subset of these methods. Efficient methods to benchmark quantum supercomputers at scale remains an active area of research.

Next, we discuss various quantum hardware, software, and algorithmic challenges that one would face when scaling the system size from 100 physical qubits for NISQ processors to beyond 1 million qubits required for utility-scale applications on fault-tolerant quantum computers. These challenges, in turn, offer significant research and development opportunities.

\subsection{Technical challenges and opportunities  at different scales}
\label{sec:challenges-scaling}

The success of quantum computing at large scale will require overcoming major obstacles. Notably, much of the current practical know-how is for NISQ computers, complemented only by theoretical developments for larger scales. 
As quantum processors increase in size, with physical qubits from $\sim$100 for NISQ to $\sim$$10^7$ for utility-scale FTQC (corresponding to $\sim$$10^4$ logical qubits, given QEC overheads), challenges of  different natures emerge at each scale. Such challenges can be mitigated with {\em ad hoc} approaches at small scales \cite{maksymov2023mitigation}, but require fundamentally new solutions for true scalability. To benchmark a quantum computer consisting of 1--10 million physical qubits, innovations are required at intermediate scales. Thus, a comprehensive multi-scale roadmap for superconducting qubits is needed that tackles these challenges and outlines appropriate testing, validation, and benchmarking at each scale. This roadmap must address quantum device design, fabrication, control electronics, calibrations, and interconnects. Various classes of noise sources such as $T_1$, $T_2$, single- and two-qubits errors, crosstalk, 1/f noise, two-level systems (TLS) defects, fat-tail of error distributions, cosmic background radiation, and low-fidelity interconnects must be characterized and dealt with at their relevant scales. 
 
Ultimately, efforts to scale the number of physical and logical qubits in quantum processors must rely on QECCs. We foresee that different scales and different applications may favor different types and sizes of error-correcting codes. Even in the case of surface codes, different variants (e.g., the XZZX or XY codes~\cite{Bonilla_Ataides_2021,PhysRevX.14.031003}) may be preferable for different hardware noise profiles. These codes must be supported by architectural provisions for fast classical decoding and control based on syndrome measurement results. Additional support is required at the operating and compilation level. 

Although many of the scaling challenges are rooted in device and architecture research, we acknowledge the lack of industry-scale impactful applications of quantum computing in its current state. Utilizing quantum computers to perform useful tasks such as quantum simulation reveals an additional set of scaling challenges, including problem identification and data management, e.g., loading, pre- and post-processing, co-processing, and scheduling.

Here, we highlight some important challenges at four different scales characterized by the number of physical qubits. This multi-scale approach not only categorizes known challenges, but reveals untold or overlooked challenges that could present significant stumbling blocks to building useful and cost-effective quantum computers.  For each scale we describe the challenges in a bottom-up order, from qubit fabrication, to hardware control, calibration, error correction, hybrid quantum--classical coprocessing, micro- and instruction-set architectures, and finally algorithms and applications. 

\subsubsection{Challenges at 100--1000 physical qubits}
\label{sec:100-1000_qubits}

At the intermediate scale, key challenges involve individual qubits and gates operating within the system, as well as fabrication and basic operation \cite{markov2014limits}. At this scale, the execution of quantum algorithms is used primarily to demonstrate and characterize hardware capabilities.

\textbf{Fat-tail distribution of errors.} The subtlety of decoherence mechanisms for superconducting qubits is underappreciated because researchers often report the coherence times of their best qubits. Measuring the median is clearly better, but reporting the worst 1\% would be a more faithful reflection of system performance at scale. Indeed, for published Google and IBM data, the worst 10\% of the $T_1$ data drops significantly (30--100$\times$) away from a Gaussian distribution (see \Cref{sec:impact_of_fat-tails} for further analysis of these effects). 

\textbf{Qubit fabrication.}
CMOS manufacturing is proven to scale to trillions of transistors per die, therefore adapting known scalable fabrication techniques is desirable. However, superconducting qubits create several constraints. Commonly used amorphous insulators (i.e. SiOx) cannot be used because it causes decoherence. Metals must also be superconducting, which eliminates many commonly used metals in CMOS manufacturing and introduces new metals that are rarely used. Cryogenic requirements around heat load and space creates significant constraints on wiring and components. Working around these constraints sometimes requires significant cost and modification at a semiconductor tool level.

We have recently experimentally determined (\Cref{sec:hardware_perf}) that better fabrication can improve $T_1$ tails so that a smaller fraction of qubits show degradation, and the drop in $T_1$ is smaller. This data points to the fact that better benchmarking and process control is needed for superconducting fabrication. This is especially important for cryogenic quantum devices, as low-temperature testing is much more difficult than wafer probing of semiconductor devices at room temperature.  We discuss process control and component-level testing in \Cref{sec:fab} for qubit fabrication.

\textbf{Recalibration.} Another overlooked technological risk is that coherent TLS defects fluctuate in time, requiring recalibration of the quantum computer. Today, with systems consisting of 100 qubits, full recalibration is needed approximately once per day and can take up to two hours, even though leading methods for QPU calibration involve representation as a directed acyclic graph~\cite{kitaev2003fault}, which is amenable to GPU-accelerated and reinforcement learning-based approaches~\cite{fluxonium}. Because the rate of emergence of outlier qubits with low coherence is proportional to the number of qubits, a 1000-qubit computer becomes effectively unusable because it requires constant recalibration. We discuss how to reduce the TLS defects to improve coherence, two-qubit error rates, and outlier emergence in \Cref{sec:fab}. 

\textbf{Catastrophic error bursts.} A technical challenge recently revealed by a Google experiment on error correction is the impact of cosmic rays on qubit error rates \cite{googlequantumai2023suppressing,mcewen2024cosmicray}. Although this may impose a lower bound on the error rate of logical superconducting qubits at the $\sim$$10^{-10}$ range with the help of gap engineering, additional mitigation strategies \cite{jmquasiparticle} are described in \Cref{sec:fab}.

\textbf{Early FTQC demonstrations and real-time decoding.} As a critical step towards enabling a faster path to future FTQC architectures at larger scales, various small-scale experimental efforts using systems consisting of $\sim $$100$ qubits are currently underway to demonstrate the break-even point at which a quantum processor is capable of executing logical quantum circuits on error-corrected logical qubits with a higher outcome fidelity than equivalent circuits run on uncorrected physical qubits. Such proof-of-principle practical demonstrations of small-scale fault-tolerant logical circuits 
with better-than-physical error rates 
typically rely on validation strategies that combine error-correction and error-detection schemes with \textit{post-selection}. Important examples of such demonstrations include recent small-scale experiments performed on trapped-ion~\cite{Hong_2024,Daguerre_2025, paetznick2024demonstrationlogicalqubitsrepeated}, neutral-atom~\cite{Sales_Rodriguez_2025,Zhou_2025}, and  superconducting~\cite{googlequantumai2023suppressing,zhang2025demonstrating,wang2025demonstrationlowoverheadquantumerror}  quantum processors. However, it should be possible to create logical qubits and benchmark real-time error correction at this scale.

The challenge of performing real-time error correction for superconducting qubits is the speed at which the qubits operate. Today, state-of-the-art decoders for superconducting qubits take $\sim$$\SI{60}{\micro\second}$ to decode $d=7$ surface codes \cite{Acharya2024}. Smaller fast-feedback experiments have demonstrated a decoding response times of $\SI{9.6}{\micro\second}$~\cite{caune2024decoding}.
However, at $\sim$0.5-$\SI{}{\micro\second}$-long stabilization rounds, the total latency inclusive of decoding and redirecting the waveform in an FPGA needs to be within \mbox{$\sim$5--20 $\SI{}{\micro\second}$} for code distances obtained in our resource estimation studies (see \Cref{tab:resource_estimates_physical_trotter}) to avoid compilation bottlenecks \cite{litinski2019game}. Even faster decoding is desirable to eliminate more sources of coherent and incoherent errors. Minimizing the practical decoding latency additionally requires a low-latency connection between qubit control systems and the significant classical compute resources (FGPA or GPU) performing the decoding. See \Cref{sec:other-FTQC-protocols} for further discussion on the effects of decoder delay and \Cref{sec:decoder} for details on an approach to fast and tightly integrated real-time decoding with petaflop/s processing speed, as well as \Cref{sec:dgx}.

{\bf Circuit knitting overhead.} In the past few years, circuit knitting methods have been introduced to allow running quantum circuits that require more qubits than are available on a single processor at the current scale \cite{peng2020simulating, piveteau_quasiprobability_2022, piveteau2023circuit}. Formally, these methods incur an exponential classical post processing overhead for exact reconstruction of a quantum observable. To enable distributed quantum circuit execution with classical communication at scale, innovative techniques must be devised for reducing this exponential overhead. While the challenge of quantum workload distribution emerges at the scale of 100--1000 physical qubits, it will be present at all later scales. Indeed, optimal circuit partitioning and patching will be essential at the logical layer above FTQC compiling stack to reduce industry-size problems into maximum logical circuits that can be embedded on any finite size QPU irrespective of modality. In \Cref{sec:hpc-qc-workload-ack} we provide a detailed discussion on this topic as well as present a family of adaptive circuit knitting methods.  In \Cref{sec:adaptive-circuit-knitting-experiments}, we show a particular implementation of this adaptive circuit knitting approach that could significantly reduce overheads for quantum simulation of quantum spin glasses via an approximate tensor-network contraction over distributed quantum circuits.

{\bf NISQ computing.} Systems consisting of 100--1000 physical qubits create unique challenges for NISQ algorithms. First, the larger number of qubits requires a higher shot count. This is due to the fact that many variational algorithms produce information spread across multiple qubits and the output quantum states are not localized to a small number of qubits. The second challenge is that for potentially useful applications, e.g., simulating quantum dynamics, typically one needs more than 100+ qubits at depths larger than what can be achieved by a 10\textsuperscript{-3} two-qubit error rate.  We discuss how to address these challenges in \Cref{sec:hpc-qc} on high-performance quantum--classical coprocessors.

Due to the short coherence times of the qubits at this scale, combined with the lack of practical error correction, algorithms will continue to focus on iterative strategies that limit circuit depth such as variational quantum algorithms~\cite{Cerezo_2021}. Current QPUs are predominantly available through a cloud access model with very limited classical compute. However, using HPC resources for this classical compute runs into the problem that, without adjustments, the cloud access model has prohibitive latencies, observed as hours-to-days in one study \cite{BECK202411}. Tighter integration between HPC and quantum systems addresses this problem. However, achieving such integration requires addressing additional challenges which are discussed in \Cref{sec:hpc-qc-challenges}, including coordinated resource allocation and scheduling.

More recent developments in NISQ algorithms have utilized the notion of adaptive circuits, where mid-circuit measurements and feed-forward information are used to reduce circuit depth~\cite{smith2023deterministic,smith2024constant,foss2023experimental}. Making use of such constructions will require the ability to make rapid measurements and, within the coherence time of the qubit, perform additional operations based on the measured results.
For certain applications of quantum computing such as calculating the ground-state energy of a chemical system, extensive preprocessing is necessary to formulate the problem in a way amenable to quantum computers, even for small but challenging systems. For example, identifying the proper active space for the iron-molybdenum cofactor (FeMoco), which has long been hailed as a premier application of quantum computing~\cite{reiher2017elucidating}, is itself a complicated computational task~\cite{li2019the}. Integrating the quantum computer in an HPC environment can help mitigate these issues (see \Cref{sec:hpc-qc}). Such tight quantum and HPC integration will be also critical for early-FTQC and full FTQC as many error mitigation strategies and variational or adaptive quantum algorithms will involve built-in feedback and feedforward mechanisms, thus could demand short latencies and high-performance compute.

We also note that, with qubit counts below 1000 in the near future, automated testing techniques at the system and component level are necessary. We discuss these procedures in \Cref{sec:fab} on qubit fabrication for component-level testing.

\subsubsection{Challenges at 1000--10k physical qubits}
\label{sec:1000-10k_qubits}

At the large scale, system integration and orchestration challenges become more prominent, including those related to high power consumption and costs as well as availability of established fabrication technologies \cite{markov2014limits}. At this scale, algorithmic benchmarking becomes necessary to assess and optimize performance.

\textbf{Wiring and packaging.} Beyond 1000 qubits, a new under-appreciated systems challenge emerges, that of how to compactly address wiring, control, and circulation within today's dilution refrigerators. A secondary aspect is the opportunity to drastically reduce the cost. For example, a cryostat for a 150-qubit processor with coaxial wires is \$5M, with \$4M devoted to wiring alone. Without circulators, 10--100$\times$ more qubits can fit into a single dilution refrigerator. This will allow the packing of 20k qubits on a single 14$\times$14-cm die. However, with new packaging of 1000--10k wires, crosstalk will likely be a dominant hardware error, requiring new designs based on electromagnetic simulations. Regrettably, state-of-the-art electromagnetic simulations have been validated only on the order of six qubits.  We discuss these issues and mitigation strategies in \Cref{sec:wafer-scale} (wafer-scale integration), including scaling up crosstalk simulations to thousands of qubits. 

\textbf{Control electronics.} The ability to control several thousands of qubits is necessary, but it would drive up both the cost of the electronics and the total thermal budget required for the control electronics, which in turn would increase cooling costs. We discuss opportunities to reduce both costs and power consumption for classical CMOS control in \Cref{sec:control-hw}. We also discuss the need for advanced qubit calibration, which will be necessary even with improved qubit fabrication.

The largest risk of this phase is the cost of development of these processes, which could be mitigated by leveraging the semiconductor industry. In \Cref{sec:wafer-scale} on wafer-scale integration, we also discuss how to leverage the existing semiconductor industry to drastically reduce costs. Because of the high costs of developing, building, and operating fault-tolerant quantum computers, there must be a strong emphasis on understanding the impact of exact hardware noise profiles on the choice of error correcting codes, as well as the resulting resource estimates for useful applications with utility-scale value. In \Cref{sec:hardware-noise-modeling} we explain how we use hardware noise profiles at this scale to inform FTQC compilation and assembly at the utility scale.

{\bf Cryogenic thermal and space budgets.} A key systems engineering challenge is to design the components in a way that they fits within the thermal and space budgets afforded by the cryogenic system. In general, the lower the temperature stage within the cryostat is, the less cooling power and space there is available for the various components. The proposed approach carefully systems engineers the various components accounting for this constraint, balancing qubit quality, complexity of the components, and complexity of fabrication. These constraints become especially acute when scaling from 1000 to 10k qubits. 
A complete stage-by-stage thermal, crosstalk, yield, and reliability budget remains to be established. The wafer-scale capacities discussed below are design targets, not validated system capacities. The FTQC estimates are conditional on the physical error rates and operation times in Table~I; they do not establish that the proposed hardware meets those specifications.

{\bf Verification, testing, and debugging.} As QEC circuits become more sophisticated and undergo optimization to reduce overhead, the possibilities for introducing design errors during these optimizations increase. Verification aims to catch design bugs as soon as they are introduced \cite{ViamontesMH07checking}, testing looks for problematic behaviors in a physical quantum computer (by running specific circuits) \cite{PatelHM04fault,MaksymovNCNM22detecting}, and debugging attempts to diagnose and correct problems \cite{neilson2024testing}. Historically, each of these steps became a bottleneck to scaling of classical semiconductor circuits, and required the development of new algorithmic technologies and hardware solutions to sustain scaling~\cite{Lavagno2016EDA}.
These tasks are much more complicated for quantum circuits than for classical ones. For example, a quantum counterpart to the conventional equivalence-checking technique \cite{Yamashita2010} must tackle unitary operators acting on exponentially large Hilbert spaces. Similarly, testing must be heavily optimized to handle the large number of trials required in view of the non-deterministic nature of quantum measurements and the frequent need of QPUs for recalibration \cite{Maksymov2022}. More-sophisticated approximate testing \cite{ROMANIK199779} techniques must also be developed to take the error tolerance of quantum computation into account. Finally, it is much harder to diagnose and eliminate errors \cite{metwalli2023testing}; therefore, more-scalable debugging techniques are needed which can benefit from HPC hardware support.

\textbf{Near-term applications.} Algorithmically, the problem of data input and output starts to become challenging at the scale of 1000--10k physical qubits. Target problems at this scale could require a large amount of classical data to either be loaded onto the quantum computer or to be read from the quantum computer. Both the classical processing of this data and the quantum resources (circuit depth or measurements) can grow quickly. Without quantum error correction, quantum computers at this scale will not be able to execute standard fault-tolerant quantum algorithms such as quantum phase estimation or Shor's algorithm. However, they will be capable of executing relatively deep circuits that are well beyond anything classically simulable, even with approximations. Therefore, there is an opportunity for discovering heuristic quantum algorithms that could provide potential utility. Rigorously benchmarking such algorithms against the classical and HPC-accelerated state of the art will be necessary to convincingly demonstrate their accuracy and effectiveness. As such, good, hardware-agnostic benchmarks in various application domains (like chemistry, materials science, and optimization) are necessary to enable testing newly discovered heuristic quantum algorithms. 

\subsubsection{Challenges at 10k--100k physical qubits}
\label{sec:10k-100k_qubits}

 At the very large scale, circuit-level scaling challenges become significant \cite[Table 1]{markov2014limits}, including verification, testing, and debugging. For conventional integrated circuits, the challenge of ``dark silicon'' arises, where a significant fraction of the chip performs various service roles \cite{Taylor13landscape}. In quantum computing, FTQC creates a similar overhead.

{\bf FTQC overhead.} A major challenge at this scale is reducing the cross-talk noise and two-qubit gate errors. Unfavorable scaling of accumulated errors can increase the overhead of QECCs needed to compensate for them, further undermining quantum advantage. This issue is the focus of our device-level efforts to mitigate errors (see \Cref{sec:cross-talk} on scaling cross-talk simulation), but it can also be addressed at the architecture level.

At the scale of tens of thousands of physical qubits, many fault-tolerant protocols including full-fledged magic state distillation units can be implemented and validated. Yet, the high space and time overhead of FTQC mean this scale will still fall short of demonstrating quantum utility. This prompts the need for advancements in QEC and FTQC schemes that reduce the overhead of fault tolerance, a goal actively pursued in current research trends for building ``good'' QECCs, that is, those with high encoding rates, such as the quantum LDPC codes \cite{breuckmann2021quantum}.

A promising approach to reducing the FTQC overhead at this scale is adopting partially fault-tolerant compilation of quantum algorithms, for example, the compilation scheme based on error-corrected Clifford gates and space--time-efficient analog rotations~\cite{akahoshi2024partially,toshio2025practicalquantumadvantagepartially,ismail2025transversalstararchitecturemegaquopscale,chung2026partiallyfaulttolerantquantumcomputation}. This approach entirely omits the costly magic state distillation procedures that are typically required for a reliable implementation of non-Clifford gates; it also avoids decompositions of arbitrary-angle rotation gates into the Clifford$+T$ gate set via the Solovay--Kitaev algorithm or alternative techniques (see \Cref{sec:resource-estimates-pbenzyne} and \Cref{sec:error-analysis}), which typically result in a large number of $T$ gates. Instead, arbitrary-angle rotation gates are directly executed through analog rotation and ancilla state injection followed by error detection and post-selection in a repeat-until-success fashion. These Space--Time-efficient Analog Rotation (STAR) architectures~\cite{akahoshi2024partially,toshio2025practicalquantumadvantagepartially,ismail2025transversalstararchitecturemegaquopscale,chung2026partiallyfaulttolerantquantumcomputation}, have recently gained attention as a promising approach for bridging the gap between the NISQ and FTQC eras. Partial FTQC alternatives could be candidates for demonstrating a practical quantum advantage on early-FTQC hardware platforms, such as \lq\lq{}megaquop machines\rq\rq{}~\cite{preskill2025beyondNISQ} capable of reliably executing quantum circuits with a million quantum operations. Among various other approaches to reduce QEC overheads are device-level hardware-efficient QEC schemes exploiting a natural hardware-specific hierarchy of physical and logical errors in their design. A promising example for such schemes is the dual-rail encoding implemented with superconducting cavities~\cite{teoh2023dualrail}, which, along with using only one additional dispersively coupled transmon ancilla per dual-rail qubit, can detect the dominant first-order photon-loss errors in the cavities as well as dephasing errors in the transmon ancilla, converting them into easier-to-correct erasure errors, leaving the system with background Pauli errors and leakage, both of which are orders of magnitude smaller than erasure errors. 

Moreover, the successful realization of QEC requires low-latency integration of QPUs with significant classical compute resources. Keeping the decoding latency low is critical for system performance. One strategy is to use GPUs, which are effective at executing a large number of identical, relatively shallow computations in parallel on different input data. Such advantages are relevant to real-time decoding, as discussed in \Cref{sec:decoder}. Another approach to relaxing the requirements of decoders is construction of better codes with faster decoding algorithms. It has been speculated that there may exist QECCs whose decoding time is independent of the code distance~\cite{Gottesman2022Opportunities}.

\subsubsection{Challenges at 100k--1M physical qubits and beyond}
\label{sec:100k-1M_qubits}

Computation at extreme scales faces system-level and complexity-theoretical challenges as well as those related to physical embedding and distributed computation as quantum interconnects become a bottleneck \cite[Table 1]{markov2014limits}. For quantum computers dominated by QEC, these challenges take on specific forms, as described below. Additionally, finding {\em killer apps} for quantum computers and validating their performance remains challenging.

{\bf Distributed FTQC.} To achieve utility scale, tens to thousands of logical qubits are needed (\Cref{tab:resource_estimates_logical_trotter}). Even at a 10\textsuperscript{-4} two-qubit error rate, this translates to 1 million (or more) physical qubits, which is about an order of magnitude more than what can practically be placed in today's dilution refrigerators (DR). Significantly larger DRs will be costly. Therefore, performing large-scale FTQC will likely require quantum interconnects between multiple DRs, which brings its own set of challenges \cite{caleffi2024distributed}. Optical interconnects have been proposed for providing such quantum links between distinct DRs~\cite{lee2020high,youssefi2021cryogenic,lecocq2021control,awschalom2021development,delaney2022superconducting}. We discuss the first analysis of the effects of noisy optical interconnects for QECCs on superconducting computers in \Cref{sec:distributed-ftqc} and discuss the compilation of FTQC algorithms on a multi-DR architecture in \Cref{sec:distributed-ftqc}.

Recent proposals for scalable networking of neutral-atom fault-tolerant architectures~\cite{sunami2024scalablenetworkingneutralatomqubits,sinclair2024faulttolerantopticalinterconnectsneutralatom} suggest promising prospects for distributed quantum computing. However, while the creation of atom--photon entanglement and heralded atom--atom entanglement generation based on atom--photon interface have been demonstrated in numerous experiments, deploying entanglement among remote superconducting qubits through an interface with microwave photons is a less mature but promising direction for distributed superconducting quantum computing~\cite{storz2023loopholefreebell}.

Since quantum networking technologies are still in their infancy, in \Cref{sec:hpc-qc-workload-ack}, we discuss how strategies like adaptive circuit knitting could delay the need for optical interconnects via high-performance distributed quantum evaluation of sub-circuits and classical post-processing to merge the solutions. With large-scale integration of a quantum accelerator with a classical supercomputer, dynamically dispatching workloads to and from the quantum computer will become a complex challenge.

{\bf Scalable compilation and decoding for FTQC.} Compiling a quantum circuit for execution on a fully FTQC system brings significant challenges. Arbitrary-angle rotation gates need to be decomposed into Clifford+T gate set. Then, the quantum circuit needs to be mapped onto error correction constructs such as lattice surgery and magic-state distillation, which are discussed in \Cref{sec:ftqc-compile}. The challenges of FTQC compilation are compounded by the need to scale to hundreds of thousands of physical qubits and the need to compile for the limited classical compute capabilities of the QPU control systems.
When scaling to large circuit sizes with $\sim$100 or more qubits, current declarative frameworks that describe complete quantum circuits without classical control flow struggle to handle compilation bottlenecks. Circuit representations that include control flow can substantially reduce the time and space required for compilation, as demonstrated for Shor's algorithm in a recent study \cite{ittah2025constanttimehybridcompilationshors}. While promising, several challenges remain unsolved: these techniques do not necessarily apply to arbitrary hybrid-classical programs; the techniques need to be extended to FTQC compilation; and lastly the techniques may require more classical compute than is available on the QPU control systems. An alternative strategy is scalable compilation on tightly-integrated HPC resources.

Realtime decoding at this scale will also require significant classical compute resources that are tightly integrated with QPU control systems to minimize latency and maximize throughput, since the decoding latency impacts the error corrected cycle time and overall system performance. We discuss realtime decoding in \Cref{sec:decoder}.

{\bf Microarchitecture standardization.} Efficient compilation and execution of large FTQC programs demands optimized and modular instruction-set architectures that are drastically different from that of conventional computers. A challenge at this scale is how to converge to an effective and versatile microarchitecture. Such microarchitectures will face additional challenges of dependability; they must be designed to be reproducible, resilient, and secure \cite{giusto2024dependable}. Recent studies \cite{litinski2019game, beverland2022assessing, silva2024optimizing, silva2024lattice} suggest that efficient instruction pipelines for FTQC using solid-state qubits comprise (potentially multi-level) magic state factories for producing high-quality resource states for consumption in a memory unit (which we call the core processor; see \Cref{sec:ftqc-compile}). Therefore, unlike the conventional von Neumann architecture wherein data is taken from memory blocks to the gates, in the FTQC pipelines, high-quality gates are brought to the memory block. Thus, FTQC micro-architectures may better resemble that of in-memory computing technologies \cite{verma2019inmemory}. This suggests a rethinking of conventional memory hierarchies (e.g., cache, L1, and L2) for quantum memory blocks. Notably, quantum random access memory (QRAM) \cite{giovannetti2008qram} is used in quantum computing literature as means for accessing and manipulating classical or quantum data in superposition. However, monumental resources are required to ensure its fault tolerance \cite{matteo2020fault-tolerant}. Quantum LDPC codes may provide a path forward for realization of feasible quantum memory blocks.

{\bf High-performance quantum operating system.} The combination of long runtimes for FTQC algorithms (see \Cref{sec:resource-estimates-pbenzyne}) operating within a frequently changing noise environment lead to the need for a software component that manages QPU calibration as well as noise-aware quantum circuit compilation. Here, we refer to this software component as the quantum operating system (OS).

To address the need for frequent QPU recalibration, the quantum OS must dispatch quantum characterization, verification, and validation (QCVV) protocols on the QPUs with an appropriate cadence to inform circuit compilation. This reveals a fundamental difference between FTQC compilers and the conventional compilers for classical computers: classical compilers do not require knowledge of the noise profile of the hardware, whereas the choice of QEC codes, their sizes, and ancillary modules (e.g., magic state factories) all require detailed information about the hardware noise characteristics. We describe preliminary steps toward addressing these challenges for FTQC compilers in \Cref{sec:ftqc-compile}.

An OS for a fault-tolerant computer must therefore perform offline and real-time QPU and decoder characterization, modelling, and performance analysis, and incorporate this information into the compilation pipeline for FTQC execution.
The compute-heavy real-time processes of the quantum OS suggest that it will ultimately be hosted on a well-integrated HPC platform communicating with the decoder and controllers in real time and adjusting the compilation/assembly of the quantum program given real-time QPU noise profiling data as depicted in \Cref{figure1_Arc}.

{\bf Quantum algorithm discovery.} The absence of efficient quantum memory blocks and methods for reading from and writing into them is one of the reasons for the lack of useful quantum algorithms for conventional enterprise problems involving classical big data, for which classical AI has provided major breakthroughs at an increasing pace. Efficient quantum memory block usually refers to "quantum read-only memory" (QROM) or more broadly "quantum random access memory" (QRAM), which serve as key subroutines in several quantum machine learning algorithms. Broadly speaking, the presence of efficient quantum memory blocks refers to the fast reading and writing of quantum states in a coherent way. This read--write bottleneck eliminates the exponential speedups promised by many algorithms, for example, quantum machine learning (QML) applied to classical data. Such algorithms include quantum linear system solvers \cite{hhl2009quantum,childs2017quantum}, quantum clustering \cite{Lloyd2013-cj,PhysRevLett.88.018702}, quantum principal component analysis \cite{Lloyd_2014}, and quantum support vector machines \cite{Rebentrost_2014}.
However, the promise for exponential quantum advantage still holds for quantum data \cite{Lloyd_2014,huang_power_2021,Huang_learningexperiments_2022} and substantial progress has been made toward overcoming the known trainability limits of quantum neural networks \cite{skolik2021layerwise,larocca2024review}. 

Moreover, processing classical data, for example, via coherent arithmetic operations, is costly for quantum computers \cite{haener2018optimizingquantumcircuitsarithmetic,haener2018quantumcircuitsfloatingpointarithmetic}. Indeed, the qubitized electronic-structure quantum simulation for which we provide resource estimates in \Cref{sec:resource-estimates-pbenzyne} uses look-up tables (coined as quantum read-only memory, or QROM) to avoid calculating trigonometric functions \cite{babbush2018encoding}.  
In general, applications with classical inputs and outputs are fundamentally limited because quantum computers do not provide a universal advantage. To wit, there is provably no quantum advantage for standard comparison-based sorting---no quantum algorithm can solve the task asymptotically faster than conventional algorithms do \cite{Hoyer2001quantum}.

Even if a given subroutine in an important application is accelerated by a quantum oracle, it must be a single distinctive bottleneck, otherwise the impact of quantum speedup will be capped by Amdahl's law \cite{10.1145/327070.327215}. Shor's algorithm for integer factorization and its variants offer a rare combination of likely exponential quantum speedup with a practical need for running the algorithm on many different inputs. In general, quantum optimization could offer a quadratic speedup \cite{szegedy2004quantum} with many implicit assumptions such as lack of explicit structures in the problem instances. However, the opportunity for such quadratic speedups in practice might be slim and most likely eventually washed away at scale by the huge overhead of QEC~\cite{Babbush_PRXquantum_2021}. Engineering at scale requires developing novel quantum heuristic algorithms that work in concert with their classical counterparts, see \Cref{sec:pbits} for a discussion on the possibility of accelerating classical probabilistic sampling by quantum-inspired subroutines or quantum fluctuations. Recently, there has been tremendous interest in the family of quantum approximate optimization algorithms (QAOA) \cite{farhi2014quantum,blekos2024review}, including numerical or theoretical studies that claim potential scaling advantages might be possible~\cite{Shaydulin24, Boulebnane24, augustino2024strategies,montanaro_new_speedup}. However, small-size effects can be misleading, and contrived nature of some benchmarking tasks misrepresent scaling advantage in practical scenarios for which highly-tuned classical heuristics are available.  Historically, it has proven difficult to develop quantum heuristics that remain relevant at large scales. The search for new quantum algorithms on classical inputs remains a major research avenue. 

{\bf Validation of quantum algorithms.} There is little doubt that quantum computation can provide revolutionary advantage for quantum-mechanical problems \cite{Lloyd_2014,Huang_learningexperiments_2022}. But validating such algorithms---ensuring that they produce correct outputs, especially in practice---remains a challenge. For example, quantum computation of electronic spectra of molecules relies on preparation of input states with significant overlap with the ground state (whose energy is to be estimated). However, is it unclear whether the commonly adopted choices (e.g., the Hartree--Fock state) will be sufficient. Indeed, ground state preparation is a QMA-hard problem, and even the FTQC quantum algorithms for such tasks are merely heuristics that may fail in practice. Motivated by these challenges, recent studies have focused on efficient preparation of better initial states for such tasks~\cite{berry2024rapidinitialstatepreparation}. 

\subsubsection{Challenges in integrating high-performance computing and quantum computing}
\label{sec:hpc-qc-challenges}

As we have discussed, integration with high-performance classical computing will be necessary to scale and operate a quantum computer, where tightly integrated HPC resources will enable real-time decoding, QEC orchestration, mid-execution recalibration, faster iterative algorithms, mid-circuit measurements \& feedforward operations, adaptive circuit knitting, and scalable circuit compilation. Additionally, quantum--HPC integration is necessary to employ quantum computing as a complementary paradigm beyond exascale supercomputing as discussed in \cite{alexeev_quantum-centric_2024,ibmPositionPaper,humbleQuantumClassical}. However, there are several challenges that need to be faced in order to successfully integrate HPC and quantum computing into a single system. These challenges must be factored into the practical design and implementation of a hybrid quantum--HPC system.

\textbf{Hardware Challenges.} On the hardware and system design fronts, key differences between quantum and classical components include physical scale, reliability, control electronics, communication bandwidth, and operational time scales. 
Physically co-locating classical and quantum computing resources within the same hardware node would be ideal when classical and quantum components need to exchange data with tight latency or require frequent synchronization. However, such integration within a node is not possible with leading qubit modalities due to the size and temperature requirements of the current devices. Instead, tight integration relies on low-latency networking between QPU control systems and HPC nodes.

\textbf{Algorithmic Challenges.}
Algorithmically, the challenges involve memory access, data sharing and movement, and efficient information extraction. Some quantum algorithms lack clearly defined kernels to be off-loaded to QPUs.  
For certain hybrid quantum--classical algorithms, the data movement overhead associated with offloading portions to a quantum device could diminish or erase performance gains the quantum algorithm could in theory provide. This is especially true for variational quantum algorithms~\cite{Cerezo_2021}, where the quantum kernel is executed multiple times in tight interaction with a classical program. Other examples include iterative hybrid quantum--classical schemes that frequently exchange large amounts of classical inputs and outputs at each iteration, requiring efficient mechanisms both for encoding classical data into quantum states (classical-to-quantum data loading) and for extracting information from quantum registers (quantum-to-classical readout). As a result, such algorithms are subject to the read--write bottleneck discussed in \Cref{sec:100k-1M_qubits}. As a simple illustrative example, consider the computation of scalar products using a quantum subroutine within an iterative hybrid quantum--classical framework. The {\it swap test}~\cite{Buhrman2001,Escartin_2013}, combined with amplitude estimation, enables estimation of the inner product of two $d$-dimensional vectors within error $\epsilon$ with complexity $\mathcal{O}\left(\log d/\epsilon\right)$ \mbox{(see Refs.~\cite{Clader2013,Lloyd2013-cj,Zhang2016,Schuld2016})}. With respect to the scaling in terms of the dimension $d$, this is exponentially faster than computing scalar products classically, which has complexity $\mathcal{O}(d)$. However, the absence of efficient dynamically addressable quantum memory blocks, along with  practical methods for coherently loading classical data into and retrieving data from them, renders 
this exponential speedup useless in realistic implementations. 

\textbf{Programmability.}
Beyond physical integration, it is necessary to ensure the hybrid quantum--HPC system is easily programmable for the end user. There are two distinct audiences for such capabilities, quantum computing programmers and HPC users. Quantum computing programmers will need to develop quantum circuits as well as the classical workflows and subroutines within their algorithms. HPC users will seek to accelerate existing pipelines  with quantum computing and will be looking for programming strategies that view QPUs as accelerators requiring minimal changes to overall HPC program structure. There is usually a language mismatch as well, which adds to the challenge. Quantum computing tools are predominantly available with a Python interface, while HPC applications are predominantly written in compiled languages C, C++, and Fortran. An additional aspect of programmability is portability. Without portability, users of hybrid quantum--HPC systems will need to rewrite their applications to try running them on other systems.

\textbf{Efficient Scheduling.}
Finding a scheduling approach that provides good application performance while minimizing idle resources is a significant challenge. HPC systems already use resource managers such as Slurm and PBS (Portable Batch System) \cite{SLURM, PBS} to schedule work across the system. An initial solution to incorporate quantum circuit execution might be to separately schedule classical work on the HPC system and quantum circuit execution on QPUs. However, that approach will lead to long application runtime for iterative algorithms (such as the Fermi-Hubbard simulation discussed in \Cref{subsec:FTQC_FH2d}), because scheduler delays (potentially taking hours or days \cite{BECK202411}) will impact the application run time. The opposite approach is to reserve HPC and quantum resources simultaneously to avoid scheduler delays, but that approach is likely to lead to idle resources.

\textbf{Minimizing Iterative Latencies.}
As previously mentioned, many quantum algorithms combine classical and quantum steps in an iterative manner (the Fermi-Hubbard simulation discussed in \Cref{subsec:FTQC_FH2d} is one such example). Providing the best performance for these algorithms requires that the system minimize the latency of alternating between classical and quantum portions of the application. In addition to minimizing scheduling delays, it is necessary to minimize circuit compilation time for any circuit compilation steps that occur in an iterative algorithm.

Overall, a natural question for the arrival of utility-scale quantum computing is the estimated timeline for the development cycles. This will be addressed next.

\subsection{Exponential scaling progress: A mirage or reality?}
\begin{figure}
    \centering
    \includegraphics[width=1\linewidth]{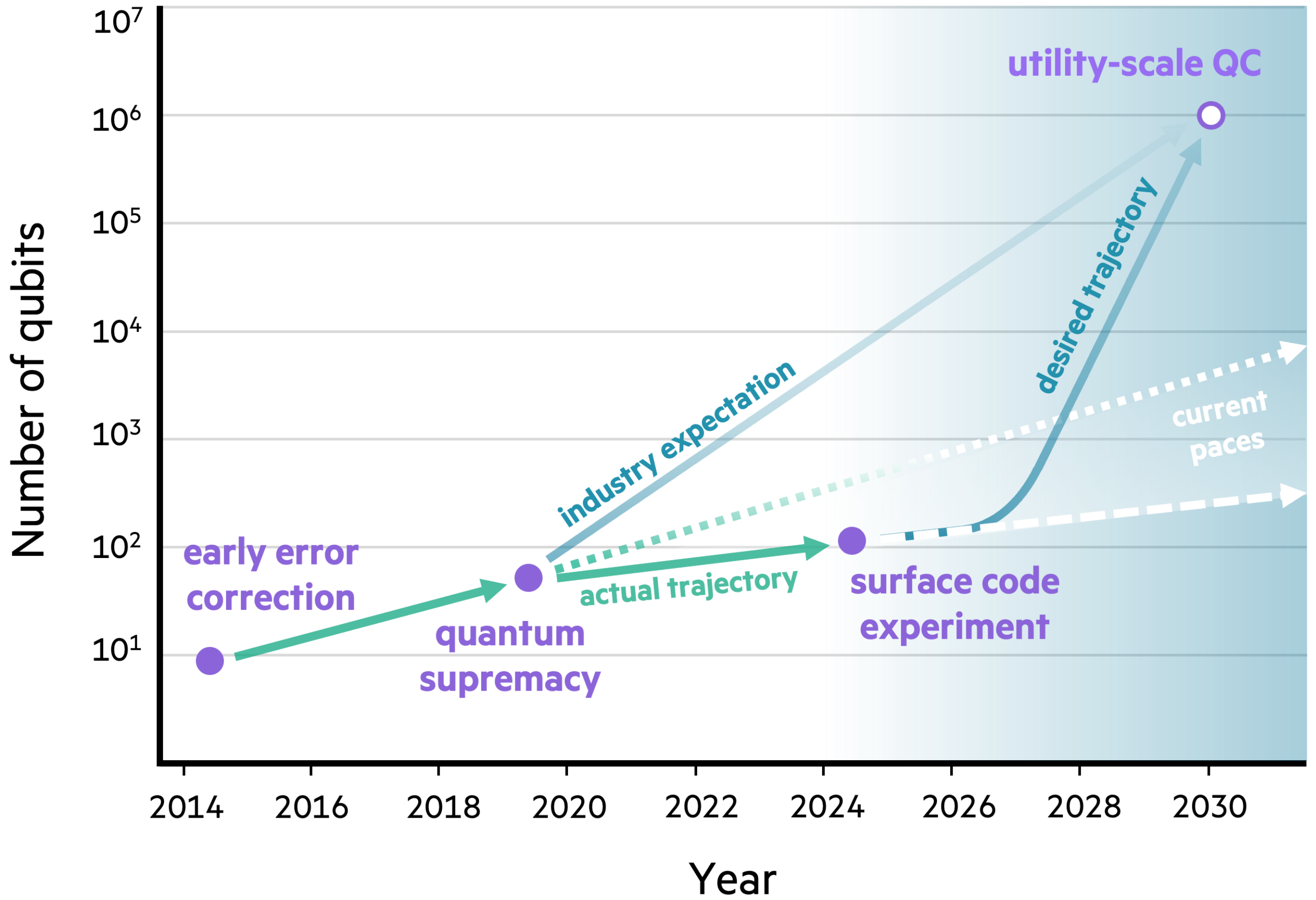}
    \caption{Schematic plot of the number of qubits from three experiments at UCSB and Google over time. After the quantum supremacy experiment in 2019 \cite{arute2019quantum}, a target of one million qubits at the end of the decade was projected by industry leaders.  However, the current pace of hardware progress \cite{Acharya2024} suggests that goal might be postponed by several decades, assuming an optimistic scenario that none of the scaling challenges mentioned here slows or halts the progress. To arrive at utility-scale quantum computers in the 2030--2035 time frame, a major increase in the rate of progress over the next five years is needed. Our thesis is that new fabrication and systems design as well as full-stack HPC integration are required to tackle this challenge. For estimates of future scaling, we suggest using the number of qubits that can be entangled in practice (which assumes sufficient qubit connectivity with multi-qubit gates that are fast and accurate enough)~\cite{cao2023entanglement}. The three historical data points are not used to fit or establish an exponential law.}
    \label{fig:ScalingRate}
\end{figure}

 Since the quantum supremacy milestone \cite{arute2019quantum}, there has been an expectation that scaling the number of qubits will follow a Moore's law (exponential) growth over time from $\sim$50 qubits to a million qubits at the end of this decade (see, for e.g., \cite{google_quantum_ai_roadmap}). This goal would mark the arrival of a practical fault-tolerant quantum computer. We chart this progress in \Cref{fig:ScalingRate}, starting from 2014 when a repetition code experiment was performed at UCSB (nine qubits) \cite{state-prep}, through the Google quantum supremacy on random circuits \cite{arute2019quantum} (53 qubits), on to the most recent surface-code error-correction experiment (105 qubits) \cite{Acharya2024}. Care must be taken in plotting only the qubit quantity since this ignores other important qubit metrics such as quality, speed, and connectivity. Here, these data points represent qubit systems with some degree of consistency in these four key metrics; thus, it is a fairly reasonable timeline that has been plotted.

The trajectory of these three data points appears to follow a nontrivial improvement, but at a much lower rate than is needed to reach one million qubits by 2030. Larger numbers of qubits have been reported for superconducting quantum processors, but even for these more optimistic characterizations of functional qubits, the scaling falls short of the original industry expectation. Will it be possible to greatly increase the slope over the next five years, corresponding to a double exponential growth? We think this is unlikely because the added challenges for scaling beyond 1000 qubits---as detailed here---could hinder even staying on the current growth path. 

\begin{figure*}[htbp]
\centerline{\includegraphics[width=2\columnwidth]{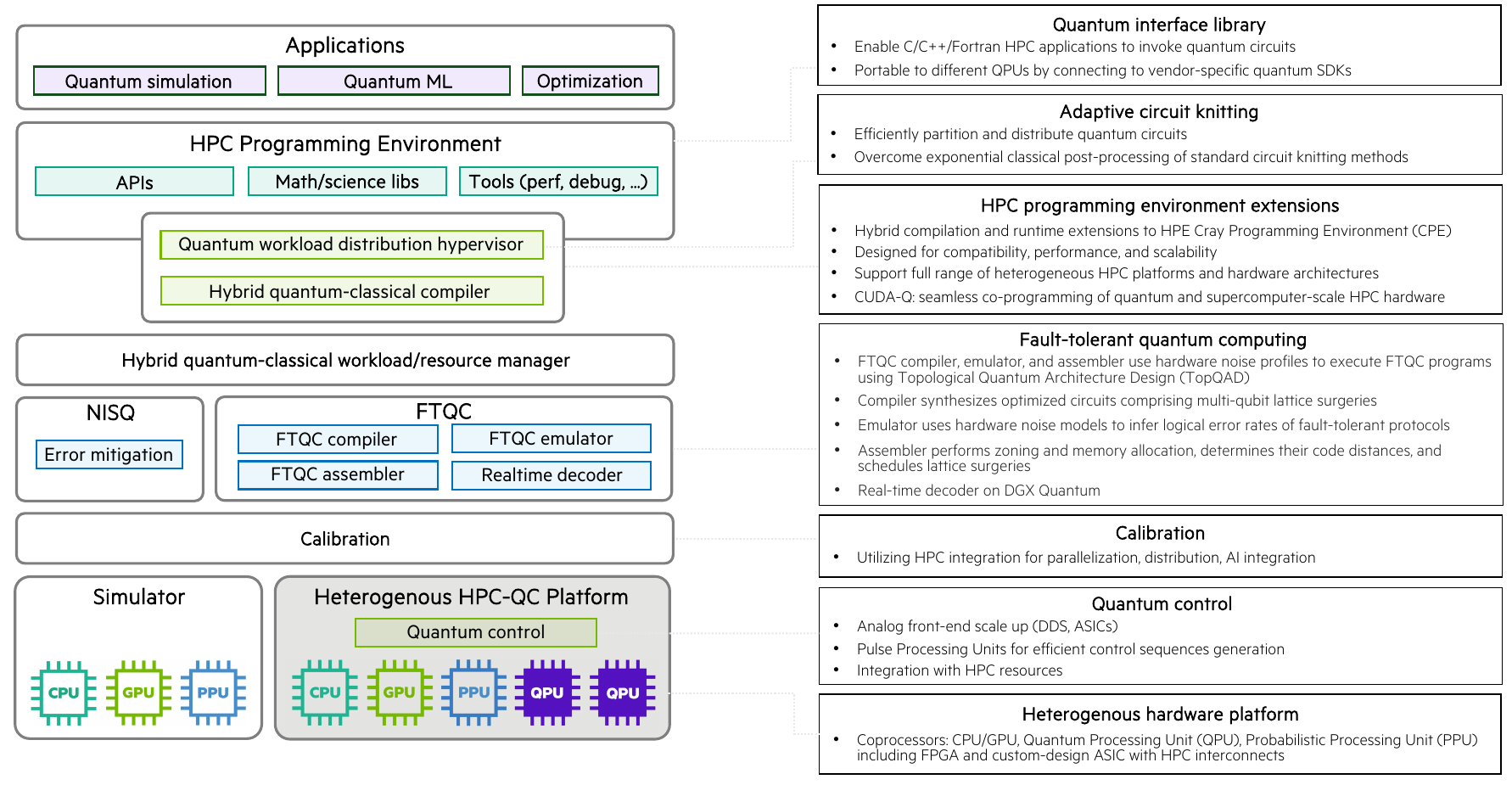}}
\caption{Architecture diagram of a quantum--classical full-stack solution. Extensions within the HPC programming environment include a quantum interface library for seamless invocation of quantum kernels, a hypervisor for efficient quantum workload partitioning and distribution, and quantum compiler and runtime extension for performant quantum circuit compilation. A customized hybrid workload manager ensures maximal quantum resource utilization in a multi-user environment. For fault-tolerant quantum computation, a compiler, emulator, assembler, and real-time decoder work together to use hardware noise profiles to synthesize optimized fault-tolerant circuits. Calibration and control of quantum resources use specialized hardware and are integrated with HPC. At the hardware layer, heterogeneous coprocessors include CPUs/GPUs, quantum processing units (QPU), probabilistic processing units (PPU), FPGA and custom-design ASICs with high-speed low-latency scale-up interconnect.}
\label{figure1_Arc}
\end{figure*}

As we have outlined in this section, to maintain the rate of exponential growth it is important to identify all major technical challenges that are impeding progress and devise mitigation strategies. Such strategies include taking a radically different approach to qubit fabrication and developing a full-stack system integration with heterogeneous high-performance computing infrastructures.

\subsection{A full-stack hardware--software architecture for high-performance quantum computation}
\label{sec:fullstack}

Addressing key technical challenges listed above requires a heterogeneous full-stack quantum--classical hardware and software system architecture \cite{SvoreACCM06layered,humbleQuantumClassical}. To this end, we outline how one could adopt existing semiconductor ecosystems and conventional high-performance infrastructures to build such an architecture, which is schematically illustrated in \Cref{figure1_Arc} (see also \Cref{sec:hpc-qc}). At the highest layer, an HPC programming environment is extended to include a quantum accelerator API consisting of an interface library for seamless invocation of quantum kernels, a hypervisor for efficient quantum workload partitioning and distribution, and a hybrid  quantum--classical compiler. A hybrid quantum--classical workload manager ensures optimal quantum resource utilization in a multi-user environment. To support fault-tolerant quantum computation, a compiler, emulator, assembler, and a real-time decoder together use known hardware noise profiles to synthesize optimized fault-tolerant circuits to solve the problem at hand. Calibration and control of quantum resources use specialized hardware that are integrated with the HPC. Heterogeneous coprocessors---including CPU/GPUs, Quantum Processing Units (QPU), Probabilistic Processing Units (PPU)---allow the system to partition a particular problem into subproblems that can be sent to the appropriate coprocessors.

The inclusion of PPUs in this heterogeneous architecture is not
incidental. Probabilistic computing can be considered as a physics-based classical counterpart of quantum computing: both exploit non-trivial probabilistic emergent (many-body) correlations to accelerate computation over conventional deterministic paradigms. However, quantum systems generalize positive
probabilities to complex amplitudes and classical correlations
to quantum entanglement. The boundary between what each can efficiently
sample can be understood by the sign
problem~\cite{troyer2005computational, chowdhury2020emulating},
making PPUs natural co-processors whose computational niche
directly abuts that of the QPU. This proximity drives
algorithmic co-development. For example, QAOA and its probabilistic counterpart~\cite{weitz2025subuniversal,abdelrahman2025generalized},
quantum annealing and simulated quantum annealing~\cite{kadowaki1998quantum,albash2018adiabatic}, as well as 
neural quantum states~\cite{carleo2017solving} whose sampling
inner loop maps onto the native operation of a PPU \cite{chowdhury2025probabilistic}. The interplay of quantum and probabilistic computing is discussed in detail in \Cref{sec:pbits}.

Figure \ref{Figure_HPCquantum} shows a schematic architecture layout for future accelerated quantum supercomputers. Here, the key approach is a tight integration of QPUs as specialized accelerators within AI supercomputers. This approach combines the capabilities of quantum and classical computation, with AI superchips executing computationally demanding classical tasks and QPUs handling specialized quantum tasks. AI supercomputing can also play a complementary role in enhancing the capabilities of QPUs across hardware-software stack by executing state preparation tasks, calibrations, noise modeling, real-time QEC, mid-circuit measurements and feedforward operations, distributed workload management, and other optimization tasks. A low-latency interconnect between AI superchips and QPUs is a critical component of HPC--quantum integration in executing QEC. In addition, classical interconnects between AI superchips allow scalable parallel processing, while quantum interconnects between QPUs facilitate entanglement distribution and resource sharing for distributed quantum computing. By tightly coupling classical and quantum components through optimized interconnects, this architecture provides a scalable foundation for future quantum supercomputers.

Current approaches to building a quantum computer are vertically integrated and do not leverage either today's semiconductor manufacturing ecosystem or state-of-the-art classical supercomputing infrastructures. One alternative is a more horizontal advanced development approach based on a consortium across supercomputer integrators, HPC platform and EDA tool developers, and semiconductor fabrication specialists, all guided by quantum computing experts. The consortium's combined skill sets could speed the development of a quantum computer that can solve utility-scale problems by enabling the building blocks and their relationships as shown in the schematic. The mandate of the consortium would be benchmarking the five following key components of the hardware--software stack:

\begin{figure*}[htbp]
\centerline{\includegraphics[width=1.8\columnwidth]{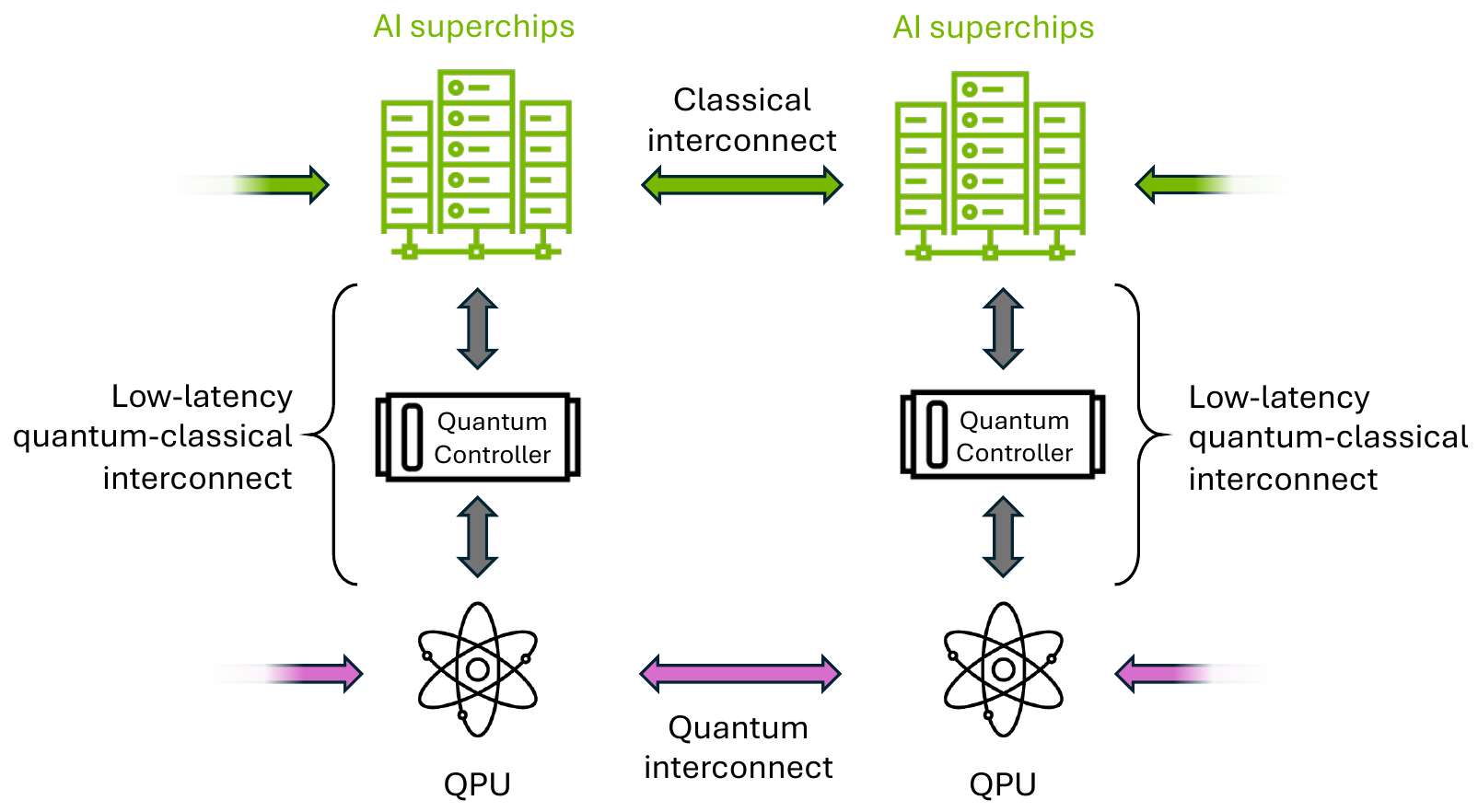}}
\caption{Architecture of two nodes 
 in an accelerated quantum supercomputer. AI superchips are coupled to QPUs via low-latency quantum--classical interconnects and quantum control hardware. AI superchips communicate via classical HPC interconnects, while QPUs are connected by quantum interconnects.
}
\label{Figure_HPCquantum}
\end{figure*}

\begin{enumerate}

    \item \textbf{Qubit fabrication} (\Cref{sec:fab}) can be developed in a new 300-mm prototype foundry using custom state-of-the-art cluster tools that only exist at this scale. The quality and yield of the qubits could be simultaneously improved. For scaling, new metrology tools should be developed that use standard in-line defect tools to benchmark qubit yield.
    
    \item Scaling the quantum computer to 20k qubits/wafer could use \textbf{wafer-scale integration} (\Cref{sec:wafer-scale}), as already demonstrated by the semiconductor industry for 300-mm wafers. Feasibility benchmarking can use established superconducting and micromachining processes, but be concurrently developed at 300 mm. This design allows all electrical connections to be at 3--4 K, making it much easier to scale.
    
    \item \textbf{Control hardware} (\Cref{sec:control-hw}) can be realized and benchmarked by existing room-temperature control electronics for up to a thousand qubits. The 20k-qubit scale could be achieved by a combination of i) wafer-scale integration, ii) high-density cables and interconnects, iii) a moderate level of time and frequency division multiplexing (1:4 or 1:8), and iv) low-power, high-density digital-to-analog front-end development. 
    To reach the 1M-qubit scale, dedicated and integrated CMOS may need to be developed to operate at cryogenic temperatures. In addition, tight integration between  control hardware and compute resources of the HPC system must be developed to allow for efficient calibration workflows that optimize and stabilize fidelities while not limiting uptimes and system utilization.
    
    \item For fault-tolerant quantum computing (\Cref{sec:ftqc,sec:resource-estimates-pbenzyne} and \Cref{sec:app_logical_circuits,sec:resource_estimation,sec:Double-factorized_quantum-chemistry,sec:classical_re,sec:Circuit-Level-Noise-Model}) an \textbf{error-correction decoder} (\Cref{sec:decoder}) can be tested using an HPC-accelerated FTQC emulator (\Cref{sec:hardware-noise-modeling,sec:other-FTQC-protocols} and \Cref{sec:Circuit-Level-Noise-Model}) to synthesize syndrome measurements of QEC codes. This test data could then be directed into the control hardware and used to benchmark the decoding CPU/GPU hardware and software in real-time emulation. For utility-scale FTQC, the decoding software system may require incorporating all ideas of distributed, hierarchical, and moving-window decoding to support 1M+ qubits.
    
    \item The quantum computer can be integrated with conventional HPC infrastructures in six different layers (\Cref{sec:hpc-qc} and \Cref{sec:dgx,sec:cudaq}). From top to bottom, these layers are: quantum workload distribution hypervisors, FTQC compilers, distributed decoders, calibration, control, and \textbf{heterogeneous quantum--classical coprocessors}. A dedicated ultra-low latency, real-time QEC network that includes the control hardware, the decoder hardware, and the logic circuit orchestration hardware could be utilized to execute the FTQC workflow.
    
\end{enumerate}
System integration of the various quantum hardware and software components, particularly as the number of qubits scales, should be tested throughout development based on state-of-the art metrology technologies and protocols currently used by the HPC industry.

\begin{figure*}
    \centerline{\includegraphics[width=1.8\columnwidth]{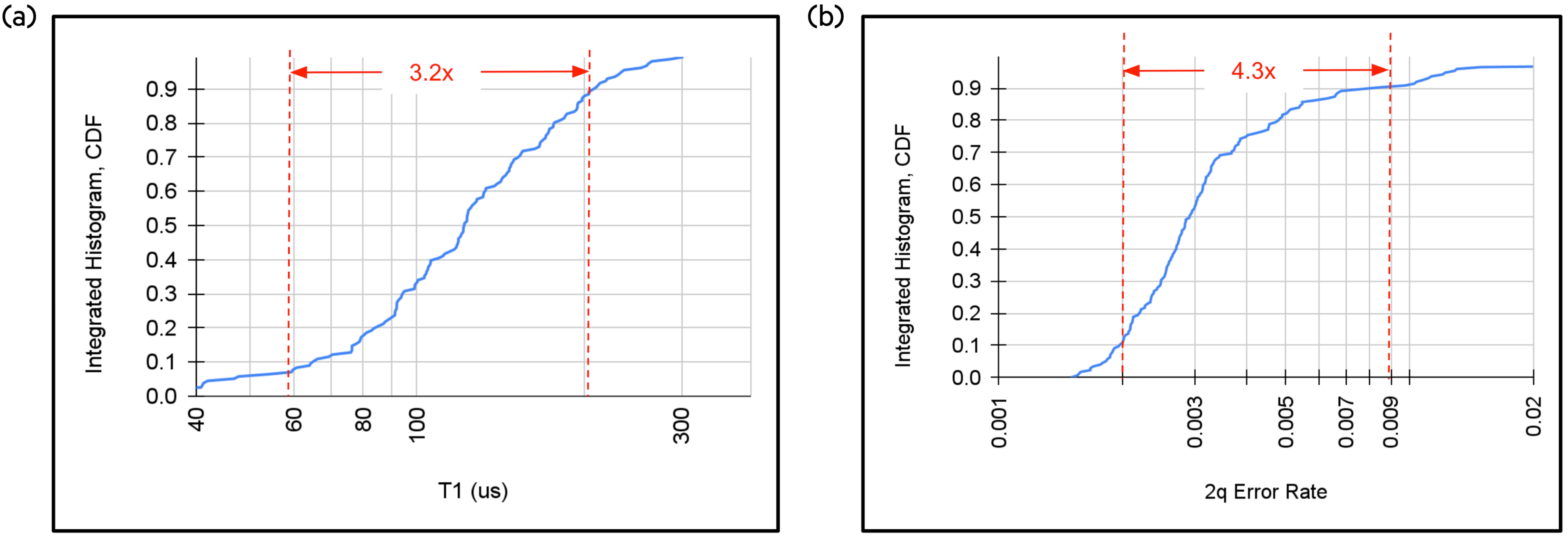}}
    \caption{IBM Heron (Fez) device performance taken on 08/06/2024 from calibration data accessible via IBM's quantum cloud. (a) Distribution of $T_1$ across 155 qubits. (b) Distribution of two-qubit (2q) error rates across 351 qubit pairs.}
    \label{fig:IBM data distribution}
\end{figure*}

\section{Towards high-quality quantum hardware and high-performance control}
\label{sec:hardware}

It is well understood that a limiting factor to realizing a practical quantum computer is construction of high quality qubits with high performance control. In this section we will discuss how to adapt the latest semiconductor manufacturing techniques to superconducting qubits with the additional constraints introduced by decoherence, superconductivity, and cryogenics. Finally, we end with how the microwave control hardware can be engineered in a scalable and economical fashion.

\subsection{Qubit fabrication}
\label{sec:fab}
Researchers often report the performance of their best superconducting qubits, as opposed to more meaningful metrics such as an average, or as more appropriate for systems engineering, the worst device performance.  A current challenge is thus to fairly compare approaches and build reliable models for improving coherence. In \Cref{tab:physical_params}, we derived the target hardware parameters needed to achieve an FTQC error suppression rate for practical quantum advantage. In particular, we introduce a new {\em tailedness} metric that describes the distribution in qubit quality. For more information on how these metrics were obtained, refer to \Cref{sec:resource_estimation}.

\begin{table*}[t]
    \centering
    \begin{tabular}{cccc}
        \hline\hline
       \hspace{0.7cm} 
        \multirow{2}{*}{\shortstack[c]{  \textbf{Hardware Parameter}}}\hspace{1cm} 
       & 
        \multirow{2}{*}{\shortstack[c]{\textbf{Baseline}  }}  
       \hspace{1cm}  
       & 
        \multirow{2}{*}{\shortstack[c]{\textbf{Target}  }}  
       \hspace{1cm} 
       &
       \multirow{2}{*}{\shortstack[c]{\textbf{Desired}  }}  
       \hspace{1cm}\\
        &&&\\ 
        \hline
        \hspace{0.5cm} $T_1$, $T_2$ times\hspace{1cm} & \SI{100}{\micro\second}\hspace{1cm} &  \SI{200}{\micro\second} \hspace{1cm} &  \SI{340}{\micro\second} \hspace{1cm}\\
        \hspace{0.5cm} $T_1$ tailedness\hspace{1cm} & \SI{71}{\micro\second} \hspace{1cm} & \SI{23}{\micro\second}  \hspace{1cm} & \SI{23}{\micro\second}  \hspace{1cm}\\
         \hspace{0.5cm}Single-qubit gate error\hspace{1cm}  & 0.0004\hspace{1cm}   & 0.0002 \hspace{1cm}  & 0.00012\hspace{1cm}\\
         \hspace{0.5cm}Two-qubit gate error\hspace{1cm}  & 0.003\hspace{1cm}  & 0.0005 \hspace{1cm} & 0.00029 \hspace{1cm}\\
         \hspace{0.5cm}State preparation error\hspace{1cm}  & 0.02\hspace{1cm}  & 0.01 \hspace{1cm} & 0.00588 \hspace{1cm}\\
         \hspace{0.5cm}Measurement error\hspace{1cm}  & 0.01\hspace{1cm} & 0.005 \hspace{1cm} & 0.00294 \hspace{1cm}\\
         \hspace{0.5cm}Reset error\hspace{1cm}  & 0.01\hspace{1cm} & 0.005 \hspace{1cm} & 0.00294 \hspace{1cm}\\
         \hspace{0.5cm}Single-qubit gate time\hspace{1cm}  & 25 ns\hspace{1cm}  & 25 ns \hspace{1cm} & 25 ns \hspace{1cm}\\
         \hspace{0.5cm}Two-qubit gate time\hspace{1cm}  & 25 ns\hspace{1cm}  & 25 ns \hspace{1cm} & 25 ns \hspace{1cm}\\
         \hspace{0.5cm}State preparation time\hspace{1cm}  & \SI{1}{\micro\second} \hspace{1cm}  & \SI{1}{\micro\second} \hspace{1cm} & \SI{1}{\micro\second} \hspace{1cm}\\
         \hspace{0.5cm}Measurement time\hspace{1cm}  & 200 ns\hspace{1cm} & 100 ns \hspace{1cm} & 100 ns \hspace{1cm}\\
         \hspace{0.5cm}Reset time\hspace{1cm}  & 200 ns\hspace{1cm} & 100 ns \hspace{1cm} & 100 ns \hspace{1cm}\\
         \hline \rule{0pt}{10pt}
         \hspace{0.4cm}Error suppression rate $\bar\Lambda$\hspace{1cm}  & 2.12\hspace{1cm} & 7.5 \hspace{1cm} & 13.5 \hspace{1cm}\\
         \hspace{0.5cm}Error suppression rate $\Lambda$\hspace{1cm}  & 2.34\hspace{1cm} & 9.3 \hspace{1cm} & 18 \hspace{1cm}\\
         \hline\hline
    \end{tabular}
    \caption{Hardware specifications for three sets of parameters:
    baseline, target, and desired hardware. 
    The baseline set represents the state-of-the-art values; the target set is envisioned to be a promising near-term goal; the desired set of synthetically generated hardware specifications corresponds to a noise model with about twice the error suppression rate, $\Lambda$, of the target hardware extracted from the exponential suppression law $\mu d^2 \Lambda^{-(d+1)/2}$ for quantum memory. Our benchmarking studies resulted in $\Lambda\approx 2.34$ 
    for the baseline set and $\Lambda\approx 9.3$ for the target set, respectively. For comparison, the exponential suppression rate $\bar\Lambda$ following the commonly adopted model $\bar\mu d \bar\Lambda^{-(d+1)/2}$ is also provided (see \Cref{sec:hardware-noise-modeling} for more details).
    The $T_1$ tailedness characterizes the weight of poor-quality qubits with respect to variations in  coherence times across the qubit chip. As discussed in \Cref{sec:impact_of_fat-tails},
    the standard deviation is used as the metric for tailedness in this paper, while the effects of higher moments (such as skewness and kurtosis) can also be crucial given that realistic distributions of $T_1$ values have significantly heavier tails compared with the associated  approximating Gaussian distributions. }
    \label{tab:physical_params}
\end{table*}

\begin{figure}
    \centering
    \includegraphics[width=1\linewidth]{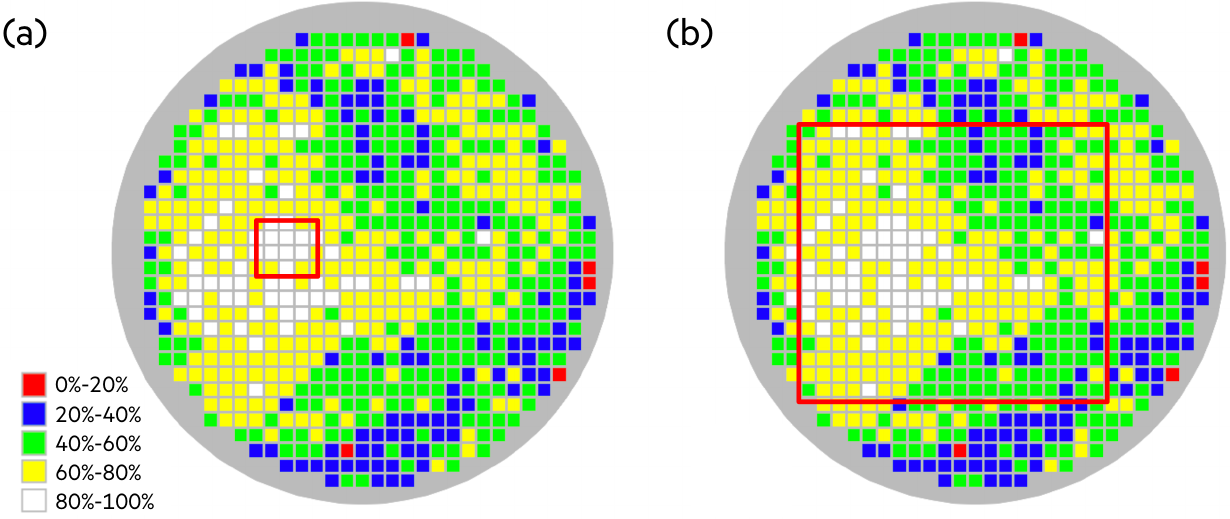}
    \caption{Illustration with simulated data on how the effect of fabrication uniformity and qubit size on median qubit error. (a) When selecting from a small number of qubits, it is possible to cherry pick the best qubits on the wafer and avoid outliers. (b) When selecting from 300+ qubits, it is impossible to avoid fabrication outliers. }
    \label{fig:Qubit Yield}
\end{figure}

Superconducting qubits have achieved coherence times of $T_1 \gtrsim \SI{100}{\micro\second}$ and two-qubit errors of 0.1\%; however, it is not currently possible to achieve such results uniformly across a large wafer, as illustrated in \Cref{fig:Qubit Yield}. A 300-mm wafer can accommodate roughly 20k superconducting flux tunable qubits with adjustable couplers. Furthermore, there is evidence that the two-qubit errors are sensitive not to the average $T_1$ times, but likely the worst $T_1$ times across an entire wafer. In Figure \ref{fig:IBM data distribution}, we see the $T_1$ spread of qubits on an IBM Heron processor and the associated spread in two-qubit error rates. A direct correlation between $T_1$ and two-qubit error spreads  is yet to be properly studied. However, it is known that as the TLS density increases, the harder it becomes to calibrate the device to avoid TLS and the higher the probability a TLS occurs in the adjustable coupler. Therefore, it is imperative to leverage advanced fabrication to improve the uniformity of performance for each qubit in a large wafer. 

Based on this challenge, qubit metrology should be driven by
system performance metrics based on
the worst 1\% of devices. For quantum computers, this is a useful design rule since an $\sim$1\% qubit dropout rate for the surface code will cause the system to degrade or fail.  Our goal is thus similar to fabricating complex CMOS electronics: make every qubit identically good.  
Our discussion is based on the categories of quality, quantity, speed, and connectivity introduced in the previous section, but organized according to hardware subsystems.  Fortunately, we have found that many of the systems engineering constraints can be solved concurrently.  For example, we will explain how qubits can be fabricated in a manner to simultaneously improve both quality and scaling. 

\textbf{Quality.} Both single- and two-qubit error rates are targeted to be in the $10^{-4}$ range for both NISQ and error-corrected quantum computers.  As adjustable couplers have achieved two-qubit error rates in the low $10^{-3}$ range with modest coherence times of $\SI{20}{\micro\second}$, it should be possible to meet this metric with reasonable 10$\times$ improvements in the $T_1$ coherence time. \label{sec:hardware_perf}This coherence requirement has indeed been met with tunable and single-ended qubits made from Al in the academic laboratory of R. McDermott at the University of Wisconsin.  Average $T_1$ times are in the 100--200 $\SI{}{\micro\second}$ range, with a “hero” device showing $T_1$ as long as $\SI{800}{\micro\second}$. This improvement came from identifying a source of TLS defects, then minimizing its contribution in the qubit design and fabrication.  In  particular, recent results indicate that dominant $T_1$ dropouts are due to discrete dielectric TLS that reside within 500~nm of the qubit junction and that are intrinsic to the additive junction liftoff process \cite{Weeden2025}. This finding motivates the development of novel subtractive processes for junction fabrication that are compatible with state-of-the-art processes at 300~mm wafer scale; see below.  

\textbf{TLS modeling.} Developing sophisticated machine-learned models to optimize and characterize the structural properties of the Si-Al interface, with particular attention placed on potential oxygen incorporation, will create the best possible conditions for superconducting-based qubits. Leveraging a moment tensor potential trained on Kohn-Sham DFT data, the model can elucidate how variations in inter-facial structure directly influence the electronic behavior of aluminum, a critical factor in superconducting applications. Additionally, NEGF methods can be employed to provide a thorough analysis of electronic properties, allowing for exploration of the impact of fabrication processes on device performance. By linking structural characteristics, process variables, and electronic outcomes, TLS modeling can drive innovation in the design of materials optimized for quantum computing environments.

\textbf{Fabrication.}  Additive liftoff processing of Josephson junctions \cite{Dolan1977} has supported steady progress in the superconducting qubit field for the last two decades; however, liftoff leaves behind organic residues that harbor a high density of dielectric TLS, leading to long tails in qubit $T_1$ distributions that statistically cannot be avoided in large-scale multiqubit arrays. Approaches that minimize liftoff show a much lower density of TLS defects, and even a significant tightening of the spread of $T_1$ coherence times \cite{Weeden2025}. We believe the improvements in qubit fabrication needed to achieve utility scale are only possible using modern semiconductor tools and processes.  

\begin{figure}
    \centering
    \includegraphics[width=1\linewidth]{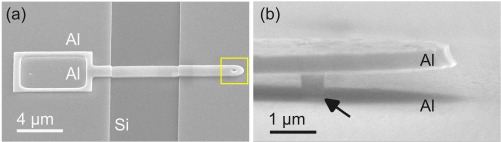}
    \caption{Enter (a) Tilted-angle scanning electron microscope (SEM) image of the window junction (WJ).  (b) Magnified view of the junction structure, corresponding to the yellow square in (a); the arrow indicates the junction location.}
    \label{fig:WJSEM}
\end{figure}

As one example, it is desirable to eliminate liftoff entirely from the qubit fabrication process. The subtractive ``window junction'' process first demonstrated in \cite{Ke2025} is one possible approach to the fabrication of next-generation superconducting qubits \ref{fig:WJSEM}. Another example is to leverage modern cluster tools to execute multiple process steps without breaking vacuum. This approach minimizes qubit loss coming from amorphous interfaces that are only a few nanometers thick.

\begin{figure}
    \centering
    \includegraphics[width=0.5\linewidth]{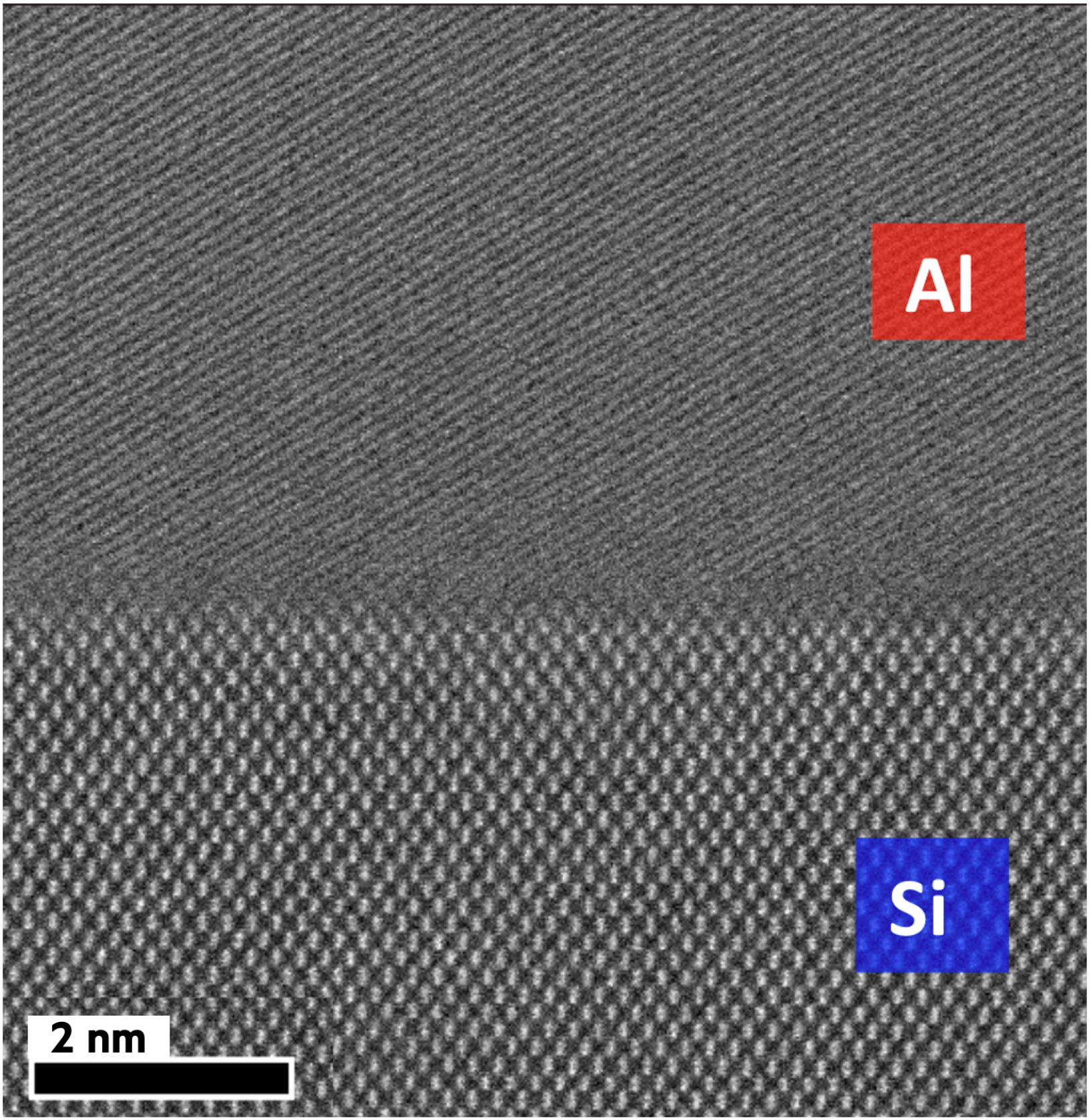}
    \caption{Atomically sharp aluminum-to-silicon substrate interface from Applied Materials' cluster tools used for qubit fabrication \cite{Visser24}.}
    \label{fig:in-situ}
\end{figure}

Working together, authors have developed a novel process flow that improves on the current state of the art at every process step. Figure \ref{fig:in-situ} shows how an \textit{in situ} clean and deposition process for Al on Si yields an atomically sharp metal-substrate interface, the most important interface in the qubit structure based on its dielectric participation ratio \cite{Martinis2022}.  We have also developed a novel process to reduce loss at the substrate-air interface, next in importance. 

In order to maintain a tight feedback loop from fabrication to characterization, it is critical to develop room-temperature diagnostics that can act as proxies for qubit coherence measured at cryogenic temperatures. For this to happen, we must establish correlations between the results of inline and postline metrology performed at the foundry and qubit performance measured in the dilution refrigerator and to build appropriate predictive physics-based models of coherence. For example, we calculate that extra loss from TLS will occur with 0.1-$\SI{}{\micro\meter}$-diameter particle defects, which are detectable with \textit{in situ} optical defect metrology. The modern CMOS foundry provides access to a vast suite of inspection tools including CDSEM, ellipsometry, optical defectivity, XSEM, and TEM that can be used to monitor process stability, reduce defects, and increase device yield. 

A key issue preventing progress in the field is that existing groups do not publish their die yields or qubit yields for larger processors (for example, there is no published data on the $T_1$ distributions of the IBM Condor 1000+ qubit device). We encourage the community to collect and publish detailed metrics on die and qubit yield.

\textbf{Qubit design.}  Our fabrication process is designed to be flexible and thus compatible with a variety of qubit designs.  We are  building adjustable transmon qubits and couplers since it is possible to have both fast gates (30--40 ns) and long coherence times ($>\SI{100}{\micro\second}$).  Ideally, the error per gate is approximately the ratio of the gate to coherence time. Our analysis of the adjustable coupler system indicates that intrinsic control errors (disregarding $T_1$ and $T_2$ decoherence) should allow two-qubit errors in the $10^{-4}$ range. Another aspect of the qubit design is to engineer robustness to gamma and cosmic rays \cite{jmquasiparticle,Iaia2022,quasiparticlepoisoning}.

\textbf{PDK.} A process design kit (PDK) is a central feature in the design of conventional circuit chips.  This is generally supplied by the foundry that will produce the chip, and is based on a particular fabrication process supported by the foundry. In short, the PDK provides all of the information a design engineer would need to architect the chip to the specifications required. However, there may be separate third-party libraries or other information which would supplement the materials provided in the PDK.

One could create an analogous PDK  for superconducting qubits as one develops the technology as described in this document. Initially, the PDK will provide information needed to perform the mask layouts for the initial qubit test chips. The main component will be a technology file for the layout editor. This would define the layers used in fabrication and their purpose. Additionally, device models for the Josephson junctions for use in a circuit simulator, based on parameters measured from the tech chips, will be provided. The PDK would also provide basic physical and electrical information, such as minimum linewidth, minimum spacing, etc. as supplied by the facility performing qubit fabrication. As the designs become more complex, one could add additional descriptions for design rules allowing automated design rule checking (DRC) and circuit connectivity so layout vs. schematic (LVS) testing can be performed. This is entirely analogous to initial stages for PDK development for a standard digital process, a task Synopsys has performed on innumerable occasions.

Note that PDK setup files are dependent on the tools used in the design flow. In some cases, one could support multiple tools that might be in use at different sites within our group. Synopsys has industry-standard tools for layout, DRC, and LVS, and others, which would be brought to bear on the project. As the technology develops further, additional tools more specific to quantum will come into use, and the corresponding technology files will be added to the PDK. For example, unlike in conventional digital circuits, extraction of precise values for capacitance and crosstalk will be needed. This will require the use of a field solver. There will be additional interfaces to specialized software used for modeling and simulating qubits and quantum components at a higher level. The results from the field solver will be back-annotated schematics which can then be simulated to yield results that include parasitic elements, analogous to parasitic extraction of a conventional design.

One can add parameterized cells (PCells) for elements that are used multiple times in our designs. Parameterized cells “draw” themselves (as mask patterns) according to the values of one or more parameters provided. This is for convenience when one needs multiple instances of devices with varying parameters. Most conventional PDKs provide a library of PCells for different types of device supported by the process. It is likely that we will have analogous needs.

Finally, the PDK will provide an area for documentation of all the process steps and other useful information developed along the way as it relates to specific procedures using the supported tools, as well as “golden” results that can be used for comparison purposes.

\subsection{Wafer-scale integration}
\label{sec:wafer-scale}

Present superconducting qubit devices have yield issues even at fifty to a few hundred qubits.  As with conventional electronics, the current solution is to dice the wafer into many dies, test the dies, then assemble the working ones into a larger system.  This solution is not ideal for qubits due to communication bottlenecks and coherence loss between dies.  

A promising solution is fabrication on 300-mm wafers with high-quality processes, which naturally allows for qubit scaling using wafer-scale integration.  This concept only works for low defect densities, which we believe is possible for three reasons.  First, one should use CMOS-type process and tools that are known to give high yields. 
Second, the critical area of the qubit devices are many orders of magnitude lower than typical electronic devices, with only a few 0.2-$\SI{}{\micro\meter}$-sized junctions/mm$^2$, and critical lithography dimensions typically $\sim$$\SI{1}{\micro\meter}$.  Third, a process should be developed in a cleanroom that has extensive metrology tools for automatic detection and optimization of defects. 

Using the center portion of the wafer, 140 mm by 140 mm, and a 1-mm qubit spacing, the number of qubits per wafer is about 20k.  Note these qubit devices can be patterned using conventional deep-UV optical lithography.

\begin{figure}
    \centering
    \includegraphics[width=1.0\linewidth]{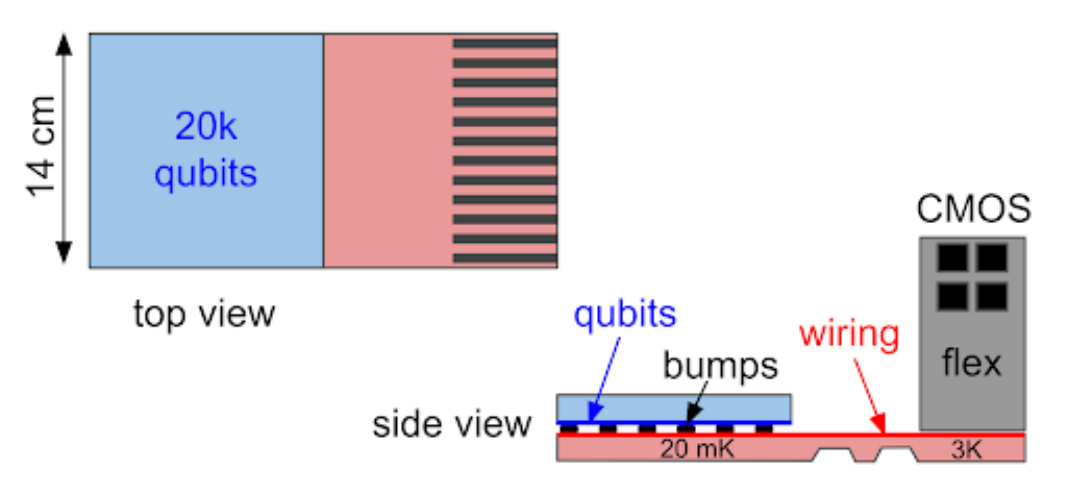}
    \caption{Qubit wafer (blue) bump-bonded to a 300-mm wiring wafer (red).  The wiring wafer is thinned by micromachining for thermal isolation between the qubit temperature (20\,mK) and 3\,K.  The wiring wafer connects via spring contacts to flex circuity (gray) for the control wiring and CMOS electronics.  }
    \label{fig:Wafer}
\end{figure}

\begin{figure}
    \centering
    \includegraphics[width=1.0\linewidth]{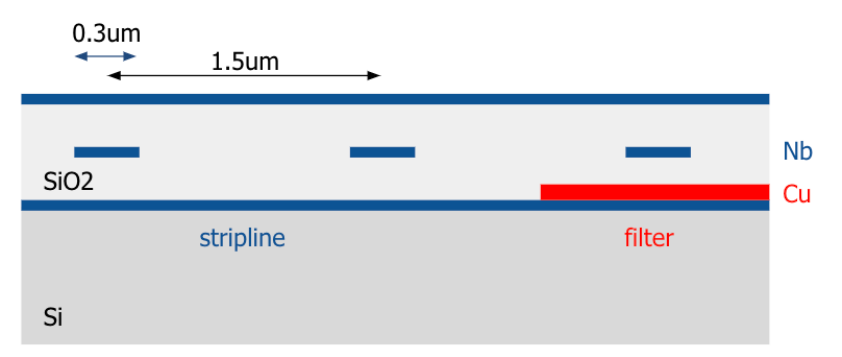}
    \caption{Cross section of wiring wafer, showing stripline width $\SI{0.3}{\micro\meter}$ and pitch $\SI{1.5}{\micro\meter}$.  The right stripline shows design of a low-pass transmission-line filter using a copper damping film. }
    \label{fig:Wiring}
\end{figure}

\textbf{Superconducting wiring.}
One of the most difficult engineering tasks when scaling to a large number of qubits is connecting the qubits to their analog control signals or readout.  This escape wiring is especially difficult at the qubit temperature of 20\,mK because the wiring is typically made using a shielded transmission line such as coax or, for adjustable qubits with DC connections, expensive superconducting NbTi coax.  
 
A solution for scalable wiring is to use wafer-scale integrated-circuit superconducting wiring from 20\,mK to the 3--4\,K stage, as shown in \Cref{fig:Wafer}.  The qubit chip described previously is indium bump-bonded over its entire wafer to the wiring wafer.  The Nb wiring are stripline transmission lines for good isolation and a 20--50$\,\Omega$ impedance.  With a $\SI{0.3}{\micro\meter}$ center width and a $\SI{1.5}{\micro\meter}$ pitch, about 92k wires can be routed across the wafer in a single wiring layer.  As shown in \Cref{fig:Wiring}, transmission-line low-pass filters integrate into these wires to be compatible with present-day designs.  The wiring layer uses micromachining processing to thin the wafer between the 20\,mK and 3\,K stages, with multiple thinned sections for connection to intermediate-temperature heat sinks.

The fabrication of the wiring wafer assumes low defects on a single wafer, which fortunately requires a relatively simple multilayer metal and insulator processes with vias.  Such processes already exist for classical Josephson electronics.  The sensitivity to defects in the wiring wafer is clearly higher than for the qubit wafer, but with \mbox{0.3-$\SI{}{\micro\meter}$-wide} wires, modern processing should provide good yields. Over the last few years, there have been significant advances in wafer-scale packaging. In particular, TSMC has developed a wafer-scale integration solution of Cerebras Systems' AI processors which utilized the entire 300-mm wafer \cite{Cerebras}, and has developed new packaging solutions for Nvidia's Blackwell processors \cite{blackwell}. Applied Materials has also developed processes and tools for wafer-to-wafer bonding and heterogeneous integrations \cite{amat-fanout}.

\textbf{Subsystem modularity.}
The wiring wafer is connected to a flex circuit board and CMOS control electronics via spring connectors at 3\,K.  These connections need not be superconducting because the acceptable heat load at 3\,K is much higher than at the qubit stage (20\,mK).  Spring connections allows this qubit+wafer subsystem (\Cref{fig:Wafer}) to be readily modularized; it can be tested separately then installed in a larger system.  This integrated design is useful since this qubit system can be thought as being controlled at 3\,K, or virtually at 3\,K, with only lower-temperature thermal connections needed to cool the chip.

\begin{figure}
    \centering
    \includegraphics[width=1.0\linewidth]{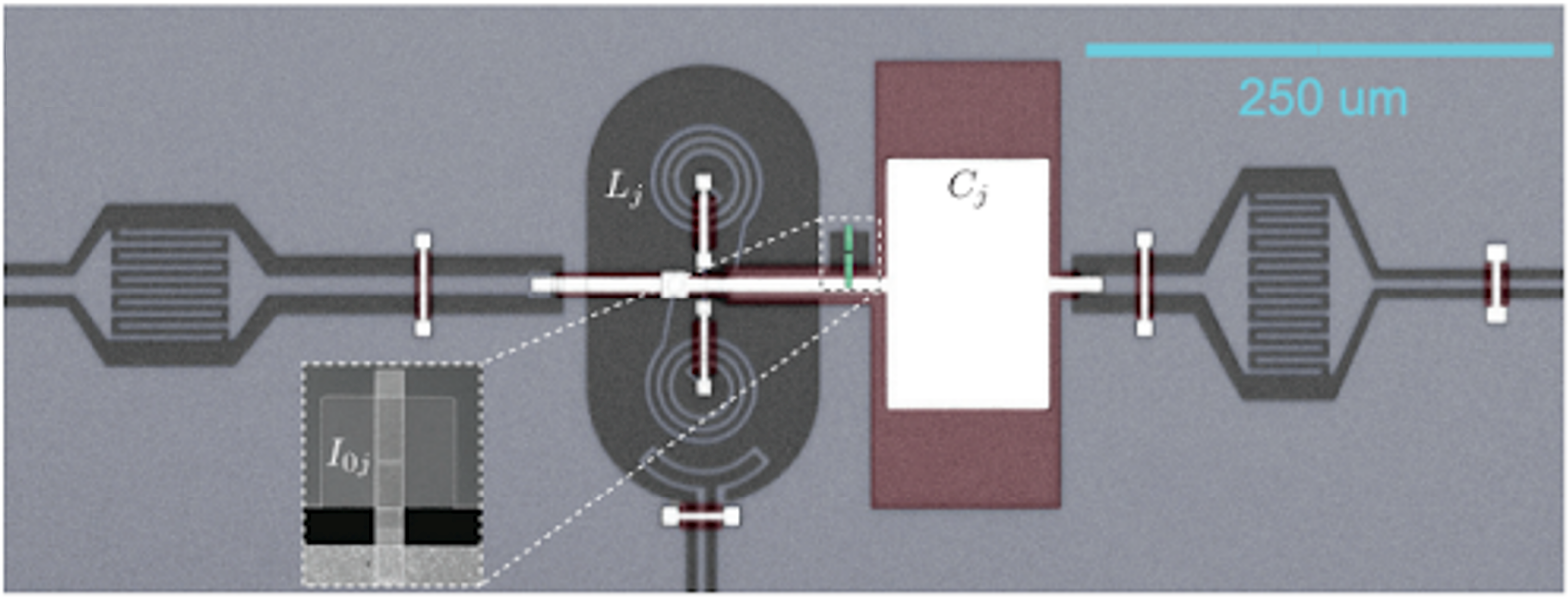}
    \caption{Qubit readout using a Josephson photomultiplier circuit which can be integrated into the qubit wafer.  This design eliminates the need for large circulators and parametric amplifiers.}
    \label{fig:CompactJosephson}
\end{figure}

\begin{figure}
    \centering
    \includegraphics[width=1.0\linewidth]{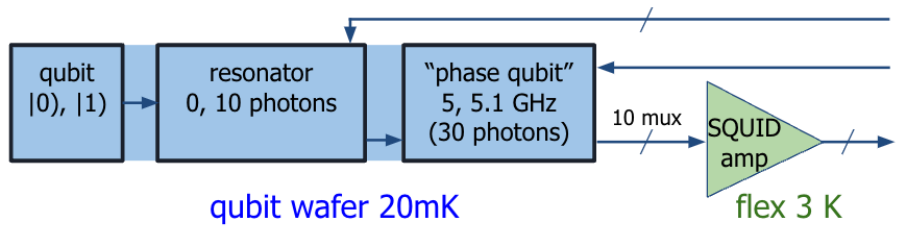}
    \caption{Signal flow for readout.  Conventional dispersive readout maps the $|0\rangle$ or $|1\rangle$ state of the qubit to 0 or 10 photons.  With 10 photons, a flux-baised phase qubit is driven to change its flux state, which changes its small-signal oscillation frequency from 5.0 to 5.1\,GHz. This classical state can be measured with a large number of photons in a multiplexed manner with a low-power SQUID amplifier at 3\,K.}
    \label{fig:Readout}
\end{figure}

\textbf{Measurement and readout.}  Another scaling bottleneck is qubit measurement and readout, as typical systems require circulators and parametric amplifiers that have a volume of  $\sim$1\,cm$^3$ or more.  The proposal uses on-chip readout technology that can be readily integrated into the qubit or wiring wafer, as described in \Cref{fig:CompactJosephson,fig:Readout}. Readout using the Josephson photomultiplier, developed by McDermott \cite{opremcak2021readout}, has achieved acceptable fidelity of 99\% and can further be improved. The Josephson photomultiplier is a fraction of the size of the resonators used for nondestructive qubit measurement and can be integrated into the qubit array with minimal increase in size. The measurement projects the qubit to a classical state of the integrated Josephson detector, which can be read out in a multiplexed manner with a simple superconducting SLUG (SQUID) amplifier at 3\,K, without the need for circulators \cite{Thorbeck2017}.  CMOS drivers at 3\,K have been developed to bias the large number of SLUG amplifiers. 


\textbf{Scaling through tiling.}
For quantum computers larger than 20k qubits, one could create baseline system that uses subsystem tiling where, each qubit wafer is precisely mounted onto an invar frame so the modules can be mechanically assembled with precise capacitive coupling between the wafers. These ``edge couplers'' communicate between each tile and can have higher error rates than qubits within each tile.

In addition to linear tiling, serpentine pattern is possible for a more compact area.  A system with 5--10 tiles can fit into a large dilution refrigerator using present-day designs.  Scaling up to many more tiles would clearly require new designs for a large cryostat. Optical connects \cite{optical-interconnect} could provide an alternative solution allowing easier scaling with modular cryostats.  In this case, it would be particularly useful to separate the $T$-gate distillation factory from the main processor.  These ideas will be incorporated as soon as optical-to-microwave quantum transducers are available. Resource estimates for both capacitive interconnects and optical interconnects is described in \Cref{sec:distributed-ftqc}

\textbf{Scaling cross-talk simulations.}
\label{sec:cross-talk}
Numerical electromagnetic (EM) field solvers are currently used to determine electrical parameters for qubit circuits.  They work well on a small number of qubits, but become prohibitive at scale due to the difficulty of meshing from the submicron film thickness to the centimeter-or-greater size of the chips.  Although qubit parameters can be determined well by isolated simulations, the evaluation of crosstalk is particularly difficult since it requires simulation of the entire circuit, including the chip mount.  

\begin{figure}
    \centering
    \includegraphics[width=0.7\linewidth]{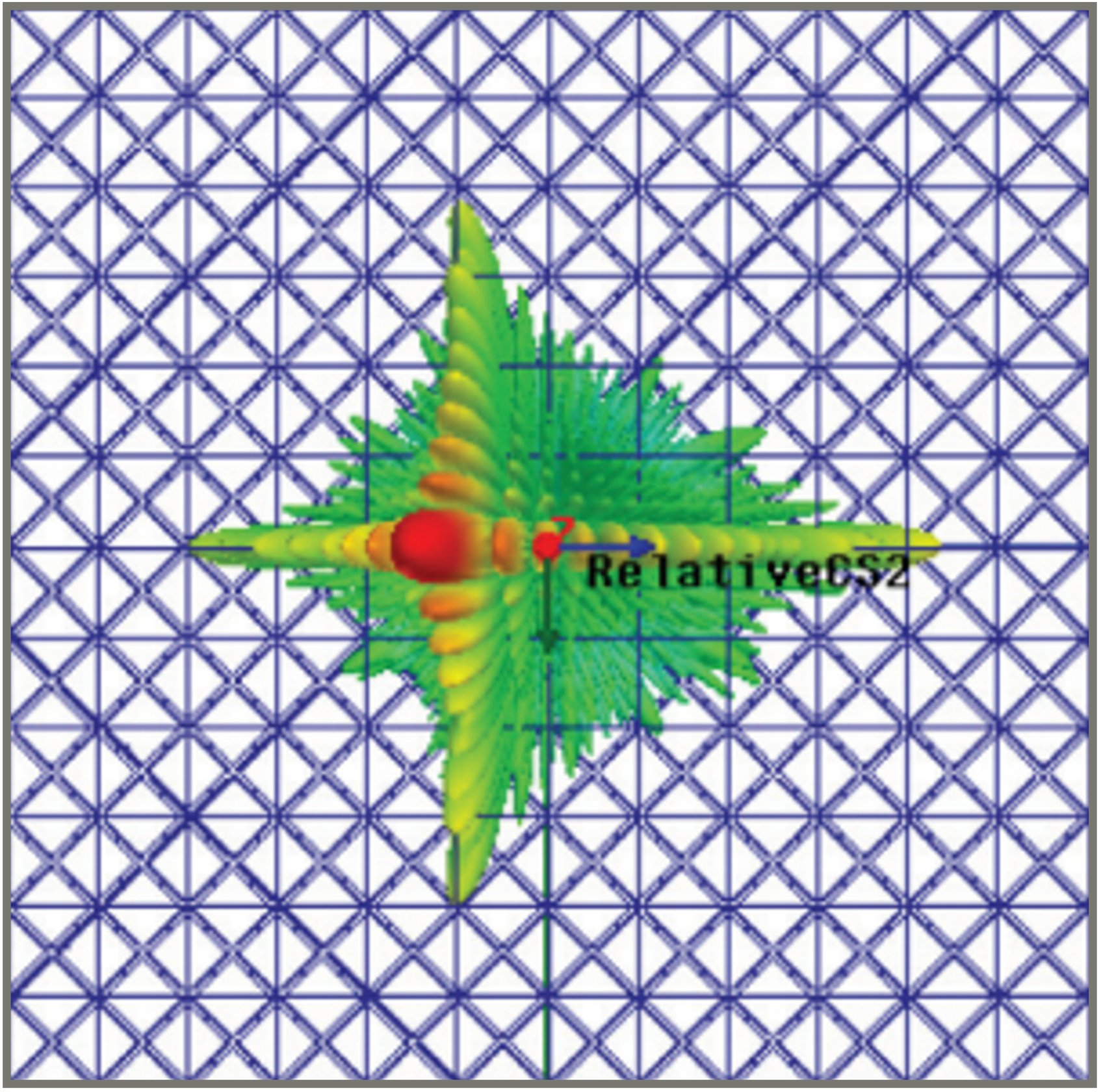}
    \caption{ Domain decomposition method (DDM) for a 1024-element antenna array.}
    \label{fig:EMSimulation}
\end{figure}

Running brute-force EM simulation even on a powerful computer will eventually run into the capacity limitation, especially when the number of superconducting qubits grows to 100--1000 for wafer-scale integration.  To solve for a large number of superconducting qubit layout geometries with a rigorous numerical EM simulation approach, domain decomposition method (DDM) techniques could offer the ability to use a distributed network of compute nodes and leverage larger blocks of distributed memory. A DDM decomposes a mesh representation of a model into a series of non-overlapping mesh domains that, when each matrix is individually solved with a traditional direct matrix solver, could collectively be used as a preconditioner for an iterative matrix solution to the full model.  A generalized scheme, in which a given geometry for simulation is meshed in whole, is developed, resulting in a mesh that is automatically subdivided into equal sized mesh domains for balanced parallel computing.  Figure \ref{fig:EMSimulation} shows an example where a DDM is successfully applied to a 1024-element antenna array.

For an antenna array solved with this general approach, the meshing processes can be quite expensive for the entire array. However, in the approach discussed here, one could leverage the repetitive geometry of an array: only a single unit cell is meshed, then it can be repeated along the array lattice to develop a set of mesh domains for the entire finite-sized array. Each cell of this array will have a unique solution depending on its location, and the resulting full solution takes into account the effects along the edge of the array. The approach is efficient as individual cells can be solved in parallel. Further efficiencies are realized by repeating matrices that result for certain cells residing in identical environments. We believe this technique can be leveraged to solve a superconducting qubit crosstalk simulation at the scale of a logical qubit comprising of order 1000 physical qubits.

\subsection{Control hardware}
\label{sec:control-hw}

Engineering a high-performance quantum control system is necessary to manage the quantum processor unit, execute calibration and application sequences, and interface with additional classical compute resources such as CPU/GPU servers. Such a control system is responsible for:
\begin{enumerate}
\item Generating and orchestrating the precise pulses that drive the quantum system dynamics
\item Reading out qubit states (including digital signal processing and state discrimination)
\item Processing data and making real-time decisions, including conditional operations, control flow, and control operation parameters updates
\item Integrating with additional classical resources
\item Providing suitable software interfaces for productive development and for integration with software components that are higher level in the stack
\end{enumerate}

Generating pulses within the coherence time of the qubits, acquiring qubit measurements, and processing these measurements for real-time feedback and efficient data transfer all require a unique digital processor architecture, which we refer to as the pulse processor unit. Then, the system's analog front-end, which includes the digital-to-analog and analog-to-digital converters as well as the analog signal chains (amplifiers, filters, attenuators, etc.), generates and acquires analog signals. As the system scales, maintaining analog performance, particularly in terms of noise, stability, and crosstalk, becomes crucial to achieving the fidelity targets necessary for NISQ computing or QEC.

\textbf{Scaling quantum control.} With current qubit quality and scale up to 1k qubits, the control system architecture should focus on performance and flexibility. Such performance and flexibility are needed so the control system does not limit overall performance or the speed of testing, research, and development iterations, even at the cost of overall design. As the system scales, the density of control electronics and wiring presents challenges in terms of space and heat dissipation. Moreover, the monetary cost of the control system will become a significant burden unless a substantial reduction in the cost per qubit control is achieved.
Hence, once the desired fidelities are achieved and the requirements for errors and scale-up are well understood, some control system requirements can be relaxed. This would allow optimizing for cost, power consumption, and size while maintaining target fidelities. To reach 20k qubits, one could employ a combination of low-power, high-density, room temperature analog front-end development in conjunction with cryo-CMOS components for moderate 1:4, or 1:8 frequency and/or time division multiplexing. To reach 1M qubits, one could develop dedicated, cryogenic integrated  CMOS connected by a digital interface to the room-temperature control. Effective co-design of room-temperature and cryogenic control electronics will be critical for seamless integration as the system scales.

Another possibility for scalable qubit control might arise from our advanced fabrication, assuming an improvement in qubit quality and reproducibility: if the qubits can be fabricated nearly identically and with a low probability of TLS dropouts, then a single control signal can be split to many qubits, with simple variable amplitude and phase adjusters at 3 K to fine-tune signals. Such techniques have been proposed \cite{zhao2024multiplexed} and demonstrated \cite{asaad2016independent, matsuda2025selective} for controlling small arrays of transmon qubits. Extending this approach to systems with up to 10 qubits is conceptually straightforward, with the primary limitation arising from power delivery constraints.

\begin{figure*}[htbp]
\centerline{\includegraphics[width=2\columnwidth]{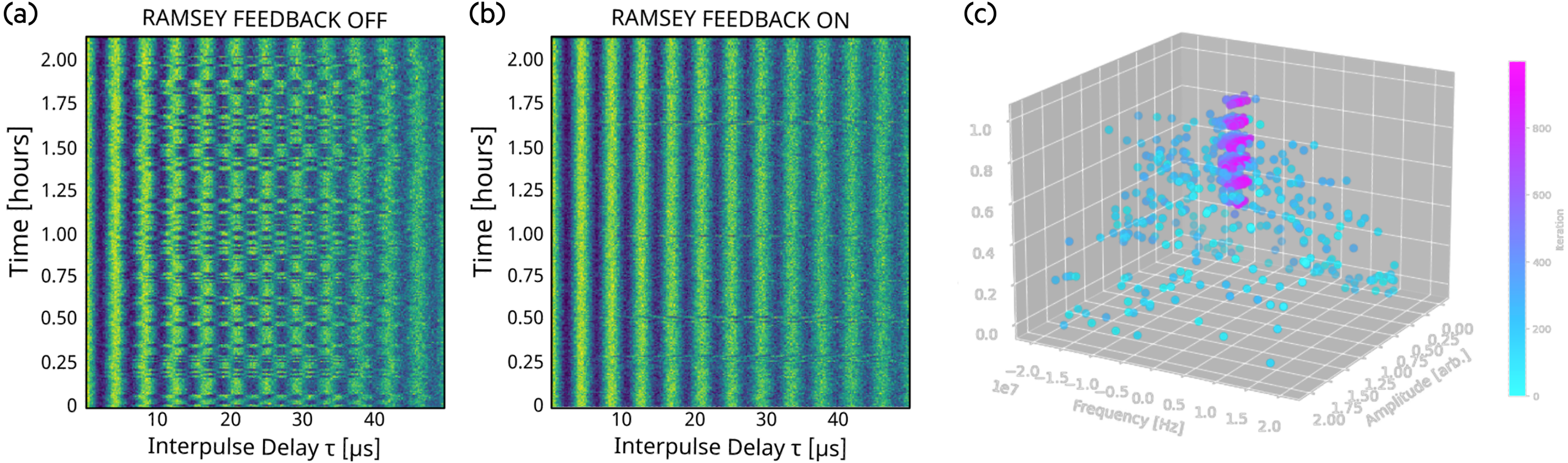}}
\caption{(a--b) Example demonstrating the impact of frequent calibration, showing Ramsey scans over time performed without and with real-time tracking of the qubit frequency~\cite{ella2023quantumclassical}. Each row represents a measurement of the Ramsey oscillations at a fixed detuning. Slow frequency drifts appear in figure (a) as variations of the Ramsey periods. By tracking the frequency drifts and updating them in real time, the Ramsey fringes are stable over an extended period of time.  (c) \(\Pi\)-pulse amplitude and frequency 2D optimization demonstrated on a DGX Quantum system.}
\label{fig:ctrl-calib}
\end{figure*}

\textbf{Room temperature and cryogenic control.} As quantum systems scale, the benefits of cryogenic control increase. Wafer-scale integration enables the connection of flex cables at 3 K connecting control at 77 K or 300 K. We identify several trade-offs between room temperature control at 300 K and cryogenic control at either 77 K or 3 K:
\begin{enumerate}
\item Power consumption: Operating at cryogenic temperatures with lower $V_{dd}$ and reduced signal amplitude, proportional to the reduced thermal noise, leads to significantly lower power consumption, potentially orders of magnitude lower compared to room temperature systems.
\item System complexity and size: Cryogenic control, such as at 77 K, results in a more compact system, housed entirely within the cryogenic refrigerator. This not only eliminates the need for multiple racks but also reduces the number of flex cables required between the cryogenic and room-temperature stages, improving overall efficiency.
\item Control functionality and flexibility: Room-temperature control benefits from the flexibility of incorporating additional computational resources, such as logic circuits, filters, and classical computing, with the option to increase rack space as needed. In contrast, cryogenic systems may face limitations in computational capacity, though emerging cryogenic-compatible technologies could alleviate this.
\item Process and cost: Room-temperature electronics utilize established CMOS processes, benefiting from mature manufacturing ecosystems, stability, and relatively low costs. Cryogenic control can leverage the same advanced CMOS technology nodes, but requires a dedicated cryogenic PDK with updated device models, design rules, and reliability constraints that capture the shifts in threshold voltage, transconductance, leakage, and variability at 4 K. Incorporating these effects into Cryo-DTCO enables low-voltage, low-power operation that is essential for scalable quantum control within the limited cooling budget of dilution refrigerators~\cite{Chiang2023CryoCMOS}. Broader adoption of these processes in industries like aerospace could reduce costs and improve availability, facilitating wider adoption.
\item Signal noise and stability: Operating at lower and more stable temperatures may reduce noise and drift of the signal properties. The reduced thermal noise allows for lower signal amplitudes, potentially reducing the complexity of amplification stages. On the other hand, the limited space and power in the cryogenic control introduces challenges such as crosstalk, noise introduced by analog up-conversion required to avoid high DAC clocks and limited power budget for amplification.
\end{enumerate}

Based on the current analysis, low-power digital-to-analog converters (DAC) along with their analog chains are crucial components for enabling low-temperature control. A power target of 1 mW per channel should allow the control to operate at 77 K. Once the target qubit quality is achieved, a comprehensive end-to-end trade-off analysis should be conducted to determine the optimal architecture for practical implementations, considering both cryogenic and room-temperature control.

For systems of 10k--100k qubits, local optimization of the QPU, packaging, room temperature control, and cryogenic electronics become inadequate. Instead, the control system must be optimized as part of a holistic solution. For instance, the predistortion capabilities of the control should be designed to address the given channels distortion and crosstalk should be addressed in the QPU and chip mount design with the control handling only the residual errors. Similarly, analog characteristics of the control system, such as phase stability, should be optimized based on gate implementation to meet the system’s architectural requirements and avoid unnecessary complexity and cost.

\textbf{Management and monitoring.} In large-scale systems, both control systems and quantum processors are prone to failure. Consequently, it is essential to incorporate mechanisms for testing functionality at both the system and component levels. 

One example of system-level functionality testing arises naturally in the context of quantum error correction (QEC). In the surface code approach, the quantum processor is partitioned into multiple “surfaces”---structured groups of physical qubits arranged to enable the correction of both phase-flip and bit-flip errors. Within each surface, a subset of qubits is continuously measured to generate error syndromes, which are processed by decoders to determine whether a logical error has occurred. The frequency and pattern of these syndromes can reveal local drifts, miscalibrations, and provide insight into the overall operational health of the system.
 Benchmarks evaluating the fidelity of physical qubits can offer information on qubit calibration status as well as potential hardware issues. 
Furthermore, component-specific benchmarks such as those for amplifiers are necessary to isolate and identify exact points of failure. These benchmarks should be optimized for execution in minimal time to enable frequent testing, and may also be used to trigger calibration procedures when necessary. Because large-scale control hardware is susceptible to local errors, robust management and monitoring features such as telemetry and self-testing are required to detect, diagnose, and address such issues effectively. Upon detection of an error, the logic circuit orchestrator is notified, ensuring continued platform function while debugging and calibration processes resolve the faults.

To facilitate debug and control software updates while maintaining an operational system, the control platform is structured with multiple clusters. Each cluster is managed independently with the clusters operating in synchronization. This provides the isolation of individual clusters for maintenance or error handling without disrupting the broader system. 

\begin{figure*}[htbp]
\centerline{\includegraphics[width=1.9\columnwidth]{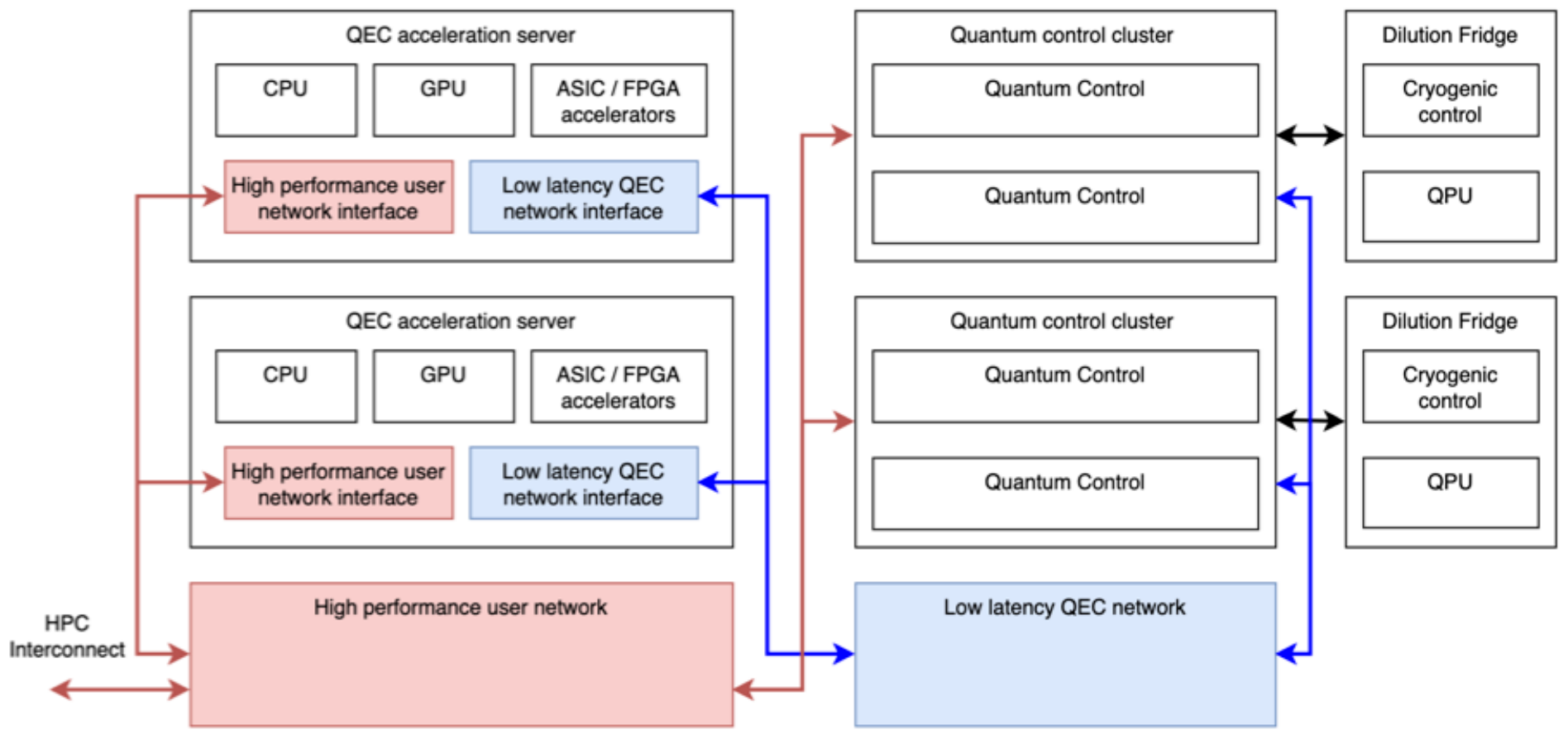}}
\caption{Illustration of an FTQC platform. Multiple quantum control clusters are connected to multiple QPUs. The platform includes optional cryogenic control located adjacent to the QPU. The low-latency QEC network connects the clusters and acceleration servers that run the QEC decoders, logic circuit orchestration and calibration and optimal control routines. A high-performance user network connects the control and servers and is used for tight integration with HPC compute resources.}
\label{fig:ctrl-arch}
\end{figure*}

\textbf{Minimizing quantum--classical feedback latency.} The interaction between the quantum processor and classical control systems is characterized by quantum--classical feedback loops, which can be grouped into three primary timescales \cite{ella2023quantumclassical}. The first is the quantum real-time (QRT) feedback loop, executed within the coherence time of the qubits. An example of QRT feedback is active state preparation, where rapid response is critical to maintain qubit fidelity. Minimizing QRT feedback latency is essential because it directly influences the fidelity and overall performance of the quantum computer. The second feedback loop is the system real-time (SRT) loop, which occurs over one or more iterations beyond the qubit coherence time but is still dictated by the physical system. The SRT loop can be used to track qubit parameter drifts over time, and its speed impacts the system's ability to adapt to noise and environmental fluctuations, thus affecting the quantum computer's long-term fidelity. Lastly, the near-real-time (NRT) feedback loop is used for classical--quantum interactions that are not constrained by immediate physical properties, but rather determined by the convergence time of quantum algorithms and calibration routines. Near real-time feedback often requires intensive classical computations and is typically carried out on the application level, outside the quantum control system. Efficient implementation of all these feedback loops is vital for both NISQ and QEC applications.

\textbf{Control system integration with classical compute resources.} By their nature, quantum computing workloads require quantum--classical iterations. These include calibration workflows, NISQ hybrid applications, and quantum error correction. Quantum--classical iterations require a low-latency, high-throughput interface with the right division of responsibilities that allows the transfer of a compact representation of the data across systems.  For applications that require significant compute resources, tight integration to HPC clusters is required to efficiently utilize the quantum and classical resources. The integration between the control system and classical compute resources may be characterized by the feedback loop timescale.
QRT feedback loops, executed within the coherence time of the qubits, require low-latency feedback and are therefore executed on the pulse processor, a dedicated real-time processor within the control system. For multi-qubit based feedback, the control system must support scalable, low-latency data distribution.
SRT calculations may be performed by the control pulse processor or by classical resources connected to the control system such as CPUs or GPUs. SRT compute resources do not need to be physically located within the same rack as the quantum controller and may be connected through low-latency optical links.
NRT calculations may require significant computational power, particularly for hybrid applications such as circuit knitting, and are typically performed on the client resources, outside the control system. The quantum control may be directly connected to a cluster high-performance network that provides scalable, high-bandwidth, and low-latency connectivity to the compute servers.
In addition to hardware integration, we require a unified software framework that supports co-development, compilation, and execution of the quantum and classical portions of the application.

\textbf{Efficient calibration.} Calibration impacts qubit fidelity and plays a critical role in determining the error correction code distance, which in turn directly affects the scalability and complexity of the decoding process. A significant challenge is variability in qubit fidelity; the system’s overall performance is often constrained by the lowest-performing outlier qubits. To maximize the performance of these outliers, advanced calibration strategies such as optimal control, reinforcement learning (demonstrated in \Cref{fig:ctrl-calib}), and model-based simulations may be used. To meet the fidelity goals for large quantum processors, it will be necessary to have scalable and efficient calibration routines that enable concurrent calibration across qubits. Reducing the calibration time allows to repeat the calibration more frequently and to allocate a larger portion of the time to optimizing outlier qubit performance. \Cref{fig:ctrl-calib} demonstrates the fidelity improvement from frequent calibration, tracking the drift in a qubit frequency.
A key requirement for calibration flows is minimal execution overhead and feedback latency. A typical calibration node requires executing thousands or more shots. A shot is structured with initialization, executing a pulse sequence, and measurement, and typically takes hundreds of nanoseconds. As such, the calibration routine overhead should be minimized accordingly to few milliseconds from loading to gathering the measurements statistics.

\textbf{Architecture of a quantum--classical integrated platform for FTQC.} FTQC workflows require a quantum--classical integrated platform to execute the logic circuit, orchestrate QEC decoding and surface operations, and control physical qubits. The main components of the platform illustrated in Figure \ref{fig:ctrl-arch} include the quantum control, QEC acceleration servers, and dilution refrigerators that hold the quantum processor and cryogenic control. The quantum control is divided into clusters; typically each cluster controls a set of surfaces, and the independent clusters simplify the control and provide resiliency to errors while maintaining synchronized execution. The QEC acceleration servers are HPC systems that are responsible for real-time decoding and logic circuit orchestration. The servers leverage general-purpose GPU/CPU systems with the option for additional ASIC or FPGA accelerators. The low latency QEC network provides an efficient interface for transferring syndromes and surface operations. A high-performance user network is used for data and program loading as well as connectivity to HPC resources required, for example, for logic circuit compilation.

\textbf{Real-time calibration and control with AI using low latency link to an accelerator.} Characterizing, tuning, controlling, and optimizing quantum devices is a time-consuming process requiring various physical resources and particular expertise. AI can help automate these tasks, reducing experts' time and effort while improving accuracy. Methods such as those based on neural networks and Bayesian optimization are particularly useful because they can produce accurate results from a small amount of data without relying on detailed physical models. Various AI methods have already been applied to characterize quantum devices \cite{ArshadPRapp2024,craig2024bridging,berritta2024PRApp,wozniakowski2020boosting}, automate various tuning steps \cite{schuff2024fully,severin2021cross,kalantre2019ml,teske2019ml}, optimize qubit control parameters such as gate fidelity and coherence times \cite{sivak2022model,kurman2024benchmarkingabilitycontrollerexecute,berritta2024realtime,scerri2020extending,ArshadPRapp2024}, and enhance qubit readout~\cite{seif2018machine,phuttitarn2024enhanced}. These methods can identify optimal settings more efficiently than traditional trial-and-error approaches. For example, neural networks can predict the ideal control parameters for qubits based on previous tuning data, while Bayesian optimization can be used to navigate the large parameter spaces involved in device calibration.
The use of GPUs and AI infrastructure as envisioned in open architectures such as NVQlink \cite{NVQlink2025} and those available with DGX Quantum - QM's low-latency link between a quantum control system and GPU and CPU accelerators (see Appendix \ref{sec:dgx} for details) is key to accelerating AI methods, and thus to real-time calibration and control.

Another important related concept is that of digital twins for quantum devices. A digital quantum twin can accurately capture the stochastic and non-Markovian behavior of various types of qubits and operates as a parameterized, data-updated model. Using this approach, a digital quantum twin is built first, then used for characterizing, tuning, controlling, and optimizing real quantum devices \cite{muller2025towards}. Building an AI version of a digital quantum twin would require a massive number of simulations for generating synthetic data for pre-training, performing pre-training, and inference. As an example of the hardware-software integration, DGX Quantum~\cite{dgx_quantum}, which combines GPUs, quantum integration through CUDA-Q~\cite{cuda_q}, and the Nvidia AI framework, could accelerate these simulations significantly. The detailed description of DGX Quantum system and CUDA-Q platform are provided in Appendices \ref{sec:dgx} and \ref{sec:cudaq} respectively.

\textbf{Integration with high-performance computing resources.} In addition to dedicated classical resources for tasks such as calibration, optimal control, and QEC decoding, the quantum workload may require high-performance compute resources. Examples for HPC use-cases include simulations and modeling, optimal control, and circuit knitting. Efficient integration with HPC requires a low-latency, high-throughput connectivity to HPC systems based on HPC standard interconnects such as Infiniband or Cray’s Slingshot or a dedicated protocol. Next, The control system should also be designed to process data locally, transmitting compact data streams to HPC servers and receiving compact instructions to maximize efficiency. A key requirement for ensuring efficient HPC or cloud integration is co-scheduling of the resources to prevent delays caused by one resource waiting for others to become available. See \Cref{sec:hpc-qc} for a more thorough discussion of these topics.

\section{Distributed fault-tolerant quantum computing: sensitivity to hardware improvements, error distributions, and real-time decoding}
\label{sec:ftqc}

\begin{figure}[b]
    \includegraphics[width=0.637\columnwidth]{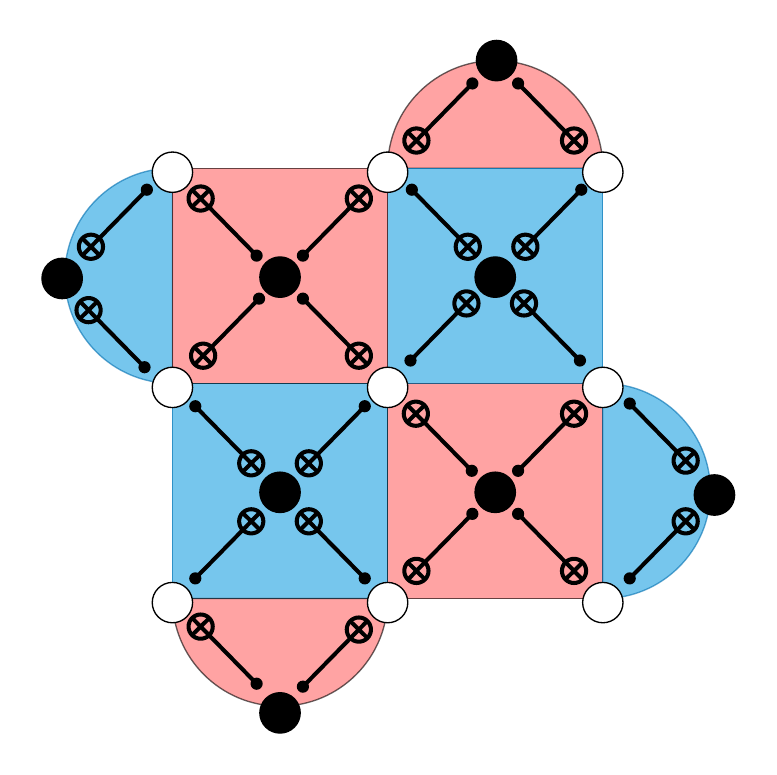}
    \caption{Illustration of the rotated surface code of distance $d=3$. It uses  $d^2=9$ physical data qubits (located
    at the circles of the lattice filled in white) to encode one logical qubit. In addition, it uses 8 ancillary syndrome qubits (located at the circles of the lattice filled in black) to measure the stabilizers. There are two types of syndrome qubits, used for measuring the different types of stabilizers: the ones located in the blue faces measure the $Z$-type  stabilizers, and the ones in the red faces measure the  $X$-type  stabilizers. The controlled-NOT gate symbol between the circles filled in white and those filled in black indicate the local physical interactions between the data qubits and neighbouring ancilla qubits used to implement the measurements of the stabilizer generators. The outcomes of these measurements (also called parity checks) can be used to detect and correct a single error on any data qubit in the patch.}
    \label{fig:surface_code_patch}
\end{figure}

To reliably execute quantum algorithms at utility scale, they must be implemented using nearly noise-free logical qubits, with logical error rates far below the physical error rates of qubits and gates of the QPU. To this end, QEC codes are leveraged to combine many low-fidelity physical qubits into fewer high-fidelity logical qubits \cite{shor1995scheme}. Since the logical quantum state of computation must not be observed during computation, QEC relies on measurements of ancilla qubits to detect the most-probable errors afflicting the code. This introduces additional circuitry to be executed which on its own creates further opportunities for error events. The goal of FTQC is to utilize QECCs in such a way that the rate of production of errors is suppressed by the rate of their correction~\cite{Shor1996fault,preskill1997faulttolerantquantumcomputation}. Fortunately, the {\em threshold theorem} guarantees that the overhead of FTQC scales as $\text{polylog} (1/\epsilon)$ with respect to the desired precision $\epsilon$ for computation when the probability of physical erroneous events is below a certain {\em accuracy threshold}. This was shown by observing that the probability of undetectable errors is exponentially suppressed if the QECC is concatenated iteratively with itself below threshold~\cite{aharonov1997fault-tolerant,knill1998resilient,Kitaev_2003}.

For superconducting qubits, which are constrained to a 2D topology and nearest-neighbor interactions, a promising family of QECCs is the topological QEC codes~\cite{kitaev2003fault, Dennis_2002} in two dimensions, such as surface codes~\cite{Fowler_2012} or color codes~\cite{Bombin2006,landahl2011faulttolerantquantumcomputingcolor}. Interestingly, for these codes the accuracy threshold is identical to the order--disorder phase transition critical point of certain classical Hamiltonians with quenched disorder \cite{terhal2015quantum,Dennis_2002,wang2003confinement}. For the surface code, this Hamiltonian is the 2D random-bond Ising model. Therefore, the threshold can be calculated using Monte Carlo simulations of the code at large sizes (i.e., at large distances). Below threshold, increasing the code distance exponentially suppresses the chance of undetectable errors, which allows us to quantify the performance of the QECC using a single parameter, $\Lambda$, representing the rate of this exponential error suppression~\cite{googlequantumai2023suppressing, Acharya2024}.

In this paper, we discuss FTQC architectures based on the rotated surface code~\cite{Bombin2007topological, Horsman_2012}, a $\llbracket d^2,1,d\rrbracket$ stabilizer code with physical qubits arranged on a 2D lattice, as illustrated in \Cref{fig:surface_code_patch} for distance $d=3$. However, our analyses can be easily adapted to other types of topological 2D codes. The rotated surface code consists of $d^2$ physical data qubits located on the vertices of a 2D square lattice and $d^2-1$ ancillary qubits located inside different types of plaquettes (depicted as red and blue faces) of an alternating checkerboard pattern within the lattice. The total number of physical qubits needed to implement the code of distance $d$ is thus $2d^2-1$.

There are two types of syndrome qubits that are used for measuring the different types of stabilizers associated with the adjacent data qubits. Those located in the blue (red) faces measure the weight-four stabilizers $Z^{\otimes 4}$ ($X^{\otimes 4}$) within the bulk and weight-two stabilizers $Z^{\otimes 2}$ ($X^{\otimes 2}$) on the boundaries. Therefore, all stabilizer generators have a weight of either two or four regardless of the size of the lattice. The error-free logical qubit state is a superposition of the joint eigenstates corresponding to the eigenvalue $+1$ of all the code stabilizer generators. Only local physical interactions between the data qubits and the neighboring ancilla qubits are needed to implement the measurements of the stabilizer generators. The outcomes of these measurements (also called $Z$-type and $X$-type parity checks) can be used to detect errors of weight at most $\frac{d-1}2$ across the code patch. The logical $X$ ($Z$) gate can be realized by chains of Pauli-$X$ (-$Z$) operators with boundaries on the top and bottom (left and right) edges.

Since universal quantum computation cannot be realized solely by executing transversal gates on a single QEC code~\cite{eastin2009restrictions}, a well-established technique for achieving universality is to implement non-transversal gates by consuming resource states, commonly referred to as magic states, that are distilled with sufficiently high fidelity in magic state distillation factories. In particular, for QECCs with transversal Clifford gates, a non-Clifford gate is implemented by preparing and consuming resource states, such as $|T\rangle=\left(|0\rangle+e^{i\pi/4}|1\rangle\right)/\sqrt{2}$ for the case of the $T$ gate. Such magic states can then be used to implement any multi-qubit $\pi/8$ rotation $\exp(-i\pi P/8)$ for $P\in\{I,X,Y,Z\}^{\otimes n}$ acting on an $n$-qubit system~\cite{litinski2019game}.
The creation of high-fidelity logical magic states is an expensive procedure requiring a protocol for their preparation~\cite{singh2022, Gidney2023cleaner, Gidney2024magic} that first produces low-quality, low-distance logical magic states, followed by several stages of magic state distillation units
(along with code growth steps between them), each of which filters many noisy magic states of low quality into fewer magic states of higher quality.

For the surface code, comprehensive techniques have been developed to perform universal quantum computation. Central among these techniques is {\em lattice surgery}~\cite{Horsman_2012}, a method for performing multi-qubit operations on topological QECCs. By performing only physically local operations, the collection of physical qubits comprising different logical qubits (called patches) are merged and split to realize any desired logical operation, where long-range entanglement is facilitated via the use of auxiliary topological patches. Any quantum computation can be compiled down to a scheduled sequence of lattice surgeries~\cite{litinski2019game,silva2024lattice}. However, to implement non-Clifford gates, the lattice surgeries typically require the consumption of high-quality magic states. A continual supply of high-quality magic states is essential for this scheme. As described above, these are produced with a certain rate in a magic state factory (MSF) which generates a few high-quality magic states from many noisy ones. The continual production and consumption of magic states requires optimizing the various trade-offs between the space and time costs needed to execute large-scale quantum circuits~\cite{silva2024optimizing}. During this entire process, any logical data qubits not being acted upon need to be preserved using a quantum memory protocol.

When the code is used as quantum memory, after the projective measurement of all the syndrome qubits in the lattice, the logical quantum state associated with all the data qubits is either stabilized or mapped into a different code word that can be tracked in software by updating the Pauli frame. In contrast, when the lattice surgery is implementing a non-Clifford gate, such a passive error correction strategy cannot be used. In this case, the overhead of decoding and implementing real-time feedback becomes consequential for FTQC compilation.

In what follows, we describe a comprehensive framework for FTQC compilation and execution based on a concept 2D surface code architectures. Our aim is to provide insights as to how the performance of such architectures can be affected by various sources of physical noise, and how improvements in quantum hardware can enhance the performance. In particular, we analyze the performance of various FTQC protocols for several specifications of quantum hardware. These benchmarks are then used for our quantum resource estimation (QRE) studies, presented in \Cref{sec:resource-estimates-pbenzyne}. Our benchmarking studies include additional analyses addressing various open questions. Specifically, we analyze the sensitivity of the performance of quantum memory to different subsets of hardware noise parameters. We also investigate the impact of QPU fabrication process variability (i.e., the tailedness of coherence time and error distributions) on logical infidelity. We then discuss a promising platform for realizing high-performance real-time decoding. Finally, we discuss distributed FTQC involving a quantum network of QPUs, and demonstrate the robustness of lattice surgeries spread among separate capacitively coupled QPU wafers or even separate dilution refrigerators, assuming access to as many weak interconnects as the code distance of the surgery.

\subsection{Fault-tolerant circuit compilation}
\label{sec:ftqc-compile}

Fault-tolerant compilation of quantum algorithms is more complicated than that of classical computer programs because the final physical circuit depends on the specific noise characteristics of the quantum processor. It is commonly understood that the number of non-Clifford logical operations (e.g., the $T$ count of the algorithm) is a good indicator of the approximate cost of running the quantum algorithm. However, assembling the quantum program for exact physical circuits to run in hours, days, or even months on a quantum computer with millions of qubits and sophisticated coprocessors for control and decoding is much more involved. A quantum OS for the fault-tolerant computer must therefore perform offline orchestrating real-time QPU and decoder characterization, modelling, and performance analysis, and incorporate this information into the compilation pipeline for FTQC execution. In what follows, we discuss three main modules for such a quantum OS: the FTQC compiler, the emulator (including a noise profiler), and the assembler, as summarized in \Cref{figure1_Arc}.

At the highest level of abstraction, an FTQC compiler is responsible for circuit transpilation, decomposition, and parallelization of multi-qubit lattice surgeries on logical data qubits ~\cite{1qbit2024topqad,silva2024lattice,silva2024optimizing}. At the lowest level, the emulator receives noise models from qubit arrays provided by various QCVV experiments to emulate fault-tolerant protocols at lower distances (typically $d < 30$) and extrapolates logical error rates to higher distances (sometimes 100 or more, depending on the algorithm; see \Cref{tab:resource_estimates_physical_trotter}). The results from the compiler and emulator are provided to the assembler, which is responsible for allocating various zones within the architecture's layout (e.g., for magic state preparations at lower distances, distillation factories at increasing distances, and code growth and switching) and placement of logical qubits in the algorithm zone and scheduling lattice surgeries.

\begin{figure}[t!]
    \centering
    \includegraphics[width=1.0\linewidth]{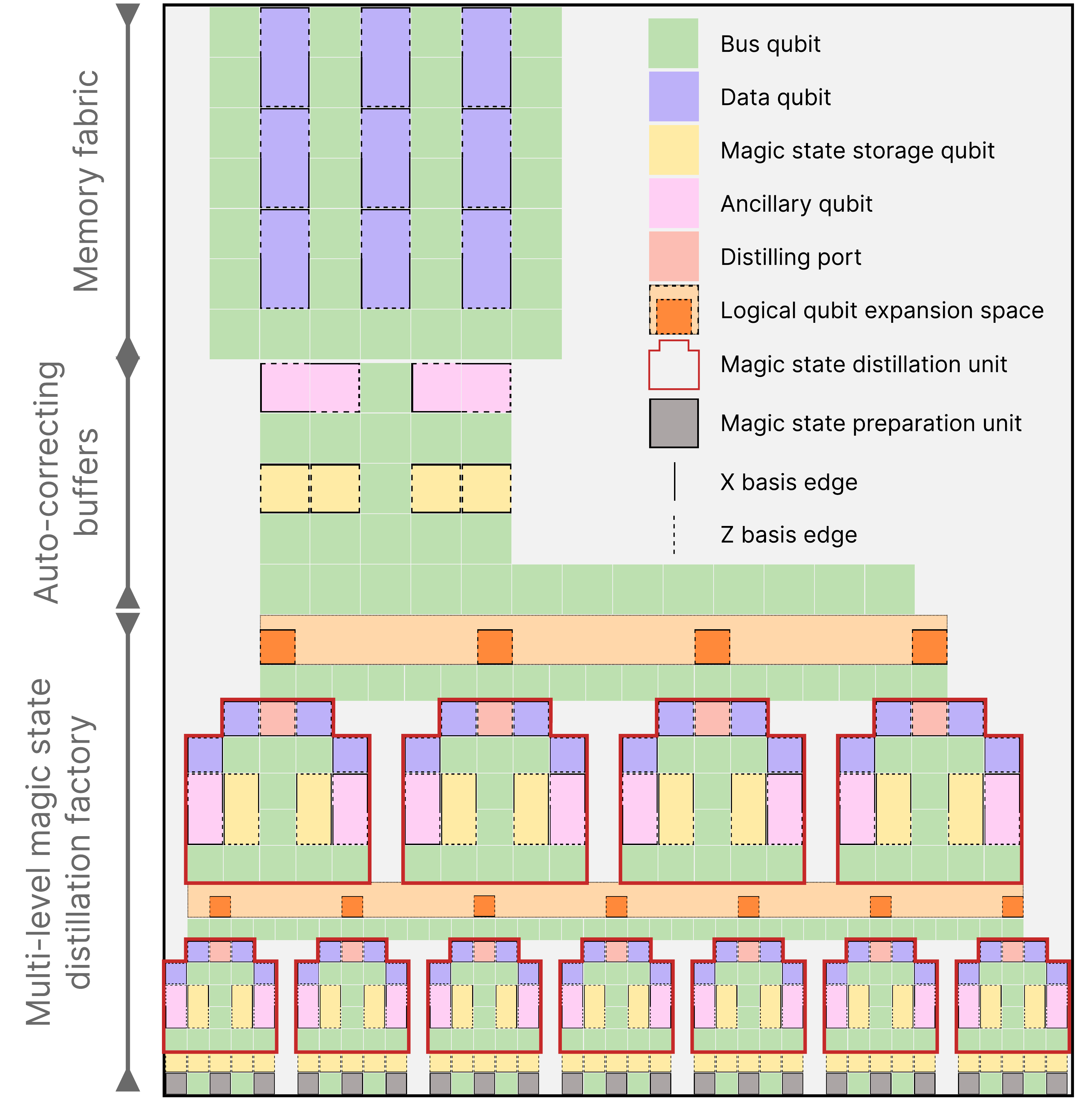}
    \caption{Example of a logical layout of a modular fault-tolerant quantum architecture of the sort designed by TopQAD~\cite{1qbit2024topqad,silva2024lattice,silva2024optimizing}. In this example, two distillation levels are used, composed of four and two distillation units each from lowest to highest.}
    \label{fig:full_layout}
\end{figure}

A basic schematic of such a modular quantum architecture layout is presented in \Cref{fig:full_layout}. In this example, a core processor containing 18 data qubits used to process the algorithmic data is distributed across nine two-tile, two-qubit patches. The core processor also contains a buffer register that allows performing auto-corrected $\pi/8$ rotations by simultaneously connecting the data qubits to a magic state storage qubit and the storage to an ancillary qubit initialized in a $\ket{0}$ state using lattice surgery. The core is connected to a multi-level MSF where the high-fidelity magic states that are consumed in the core are distilled. In the MSF, magic states are first prepared using dedicated preparation units following a magic state preparation protocol. These lower-fidelity magic states are consumed by distillation units to produce higher-fidelity ones in the distilling port. The layout depicted for the distillation units is an example of a feasible layout for the most commonly studied 15-to-1 distillation protocol~\cite{bravyi2005universal}, where 15 lower-fidelity magic states are consumed to produce one higher-fidelity magic state at each distillation cycle. Distillation is conducted in a designated zone, with a sufficient number of distillation units placed side-by-side to facilitate parallel distillation processes, ensuring a continuous supply of magic states between different levels. Once prepared in the distilling ports, the magic states are teleported to a space reserved between levels for expanding the code distance of the magic states since different qubit encodings can be used throughout the architecture. This process repeats until magic states with the required fidelity are produced at the highest level and sent to the core processor, where they are consumed.

Eventually, the procedures performed by the OS, including compilation, emulation, and assembly, deliver the exact sequence of instructions for all the stabilizer measurement rounds, logical operator measurements, and conditional recovery operations to be performed by the 1--10M+ physical qubits system to the controller. This information is also provided to the decoder, since it must keep track of the logical protocols being executed (e.g., memory, teleportation, or code growth). We use this framework to conduct detailed resource estimations as presented in~\Cref{sec:resource-estimates-pbenzyne} for real-world quantum chemistry problems as well. Furthermore, we study the sensitivity of the performance of the fault-tolerant quantum computer to various hardware parameters in \Cref{sec:sensitivity-analysis}, which helps guiding the design and fabrication of QPUs.

\begin{figure}
    \centering
    \includegraphics[width=0.75\linewidth]{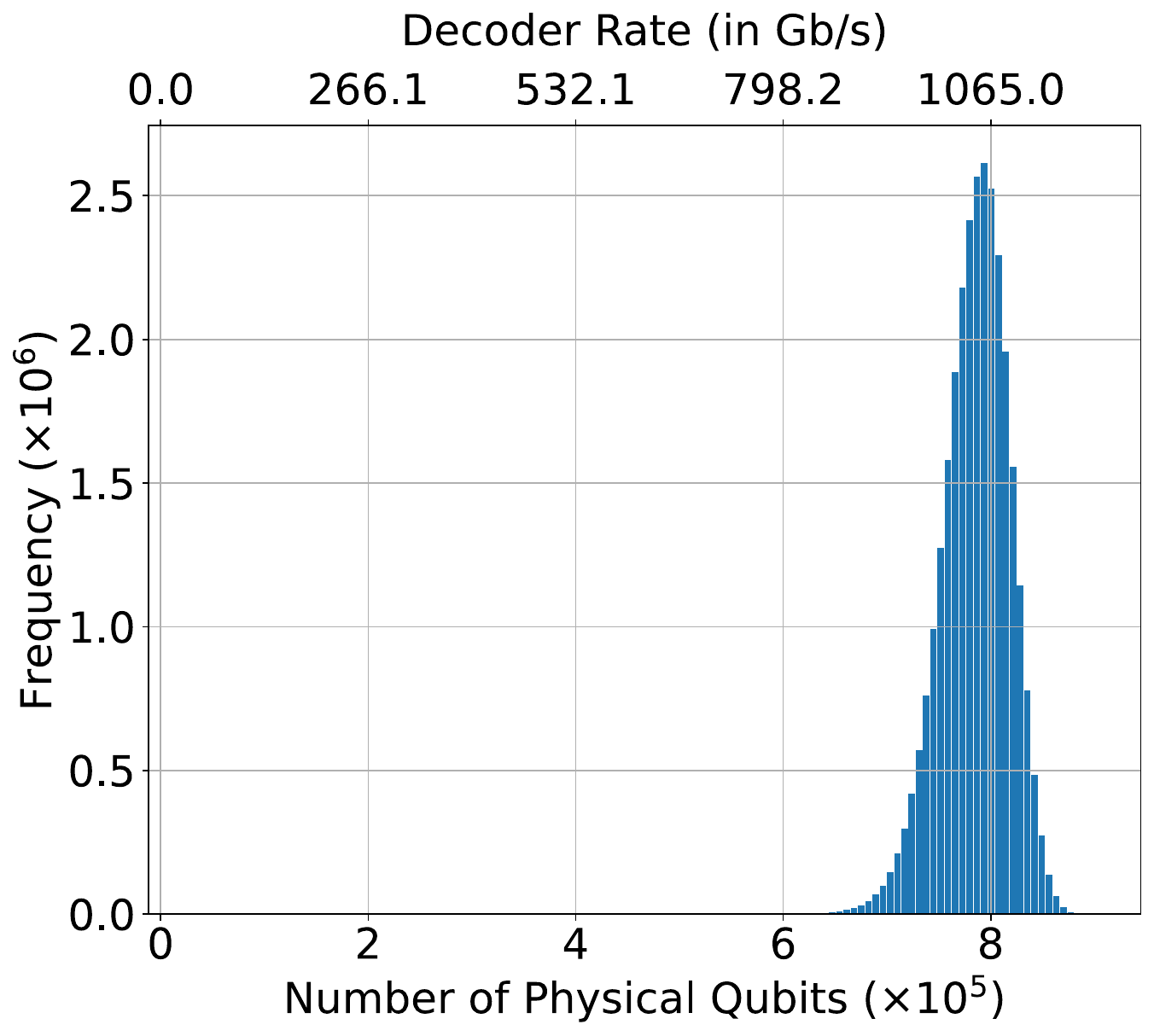}
    \caption{Decoder requirements for electronic-structure quantum simulations of the $p$-benzyne molecule for an active space involving 26 \texttt{6-31G} basis set orbitals, using Trotterization based on the second-order Trotter--Suzuki product formula, with rigorous analytic error bounds (see~\Cref{sec:resource-estimates-pbenzyne}). The histogram illustrates the scale of independent lattice surgery procedures that must be performed within the memory zone of the studied topological architecture to execute the quantum simulation circuits, with the top horizontal axis displaying the size of the independent decoding problems that must be solved by decoders. Independent decoding tasks require processing terabytes-large decoding graphs per second. Moreover, independent surgeries can involve 1M+ qubits spread across tens of DRs. }
    \label{fig:decoder-requirements}
\end{figure}

Resource estimation analyses  discussed in \Cref{sec:resource-estimates-pbenzyne} also provide profiles of all the independent lattice surgeries required to be performed on the concept architecture illustrated in \Cref{fig:full_layout}, and described in more detail in Refs.~\cite{1qbit2024topqad,silva2024lattice,silva2024optimizing} and also in \Cref{sec:resource_estimation}. \Cref{fig:decoder-requirements} shows an example histogram illustrating the sheer scale of independent decoding problems that must be solved by the decoders. The enormous problem sizes and decoding speed required for a successful execution of FTQC demands a tightly integrated high-performance decoding system. We describe such a decoding system in \Cref{sec:decoder}. 

\subsection{Benchmarking quantum hardware for the quantum memory experiment}
\label{sec:hardware-noise-modeling}

We aim to evaluate how improved physical qubits and gates
affect the efficiency of QEC and, consequently, the
overall resource requirements of 
FTQC at a   scale of practical use.
We focus on 2D lattice surgery using rotated surface codes as our
FTQC scheme, although the techniques we developed to this end are applicable
to other types of 2D topological QEC codes as well. 
In what follows, we describe how quantum hardware can be characterized with respect to its performance in realizing FTQC. We do this by emulating fault-tolerant protocols using well-established methods to model quantum hardware noise.

Our benchmarking analyses include three parts. In this section, we show how improved quantum hardware quality affects the efficiency of QEC in suppressing errors. In \Cref{sec:sensitivity-analysis}, we present the results of a sensitivity
analysis investigating which hardware parameters are expected to have the most-significant impact on the performance of FTQC. 
In \Cref{sec:resource-estimates-pbenzyne}, we demonstrate 
to what extent  improvements in the quality of quantum hardware affect the overall resource 
requirements of FTQC at utility scale. These benchmarking studies were conducted using the
TopQAD toolkit~\cite{1qbit2024topqad}.

 For the purpose of these analyses, we compare three sets of hardware parameter specifications, which are  summarized in \Cref{tab:physical_params}: the {\em baseline} set,  which is considered the state of the art for superconducting qubit technologies; the {\em target} set representing an achievable near-term goal; and a set of synthetically generated
specifications that corresponds to a {\em desired} hardware model associated with the value $\Lambda\approx 18$. The parameter $\Lambda$ represents the asymptotic error suppression rate when increasing the code distance by 2 (introduced in Refs.~\cite{google2021exponential,googlequantumai2023suppressing} to characterize the QEC performance of FTQC schemes).  
For these three sets of hardware specifications, in benchmarking the QEC performance for the baseline and  target hardware specifications, we obtain the values  $\Lambda\approx 2.34$ and $\Lambda\approx 9.3$, respectively, as discussed below. 
The motivation for considering a desired hardware model yielding the value $\Lambda\approx 18$ is that it is approximately twice as effective as the value of the target set in suppressing errors.

We conduct Clifford circuit simulations to emulate the fault-tolerant protocols
required for performing FTQC. The simplest such protocol is the \emph
{quantum memory} experiment, which involves only iterative rounds of stabilizer
measurements in a single rotated surface code patch representing the
fault-tolerant idling of a logical qubit. For this purpose, we employ two open
source libraries: Stim~\cite{gidney2021stim} for simulating stabilizer circuits
and PyMatching for decoding using the minimum-weight perfect matching
(MWPM) algorithm~\cite{higgott2022pymatching}. The emulation of other FTQC protocols, such as magic state preparation and teleportation, that are required for a fault-tolerant
implementation of an actual quantum algorithm are discussed in \Cref{sec:other-FTQC-protocols}.

We implement a prototypical quantum memory experiment by setting the number of
parity-check circuit rounds to match the code distance. Gate and qubit errors
are modelled using circuit-level noise with idling errors. \emph{Active noise
channels} are applied to the qubits participating in a gate while
\emph{idling noise channels} are applied to qubits not engaged in a gate. A brief description of the circuit-level noise model  is provided in \Cref{sec:Circuit-Level-Noise-Model}.

Preparation, measurement, and reset gates are executed in the $Z$-basis with
single-qubit Pauli-$X$ channels used to model their errors. Hadamard and CNOT
gate errors are modelled using single- and two-qubit depolarizing noise
channels, respectively.
Single-qubit depolarizing channels are used as idling noise channels. Parameters
of the noise channels are determined based on the reference hardware
parameters, specifically, by matching the fidelity of the noise channel and the
corresponding gate. For active noise channels, the fidelity of the
corresponding gate is obtained. For idling noise channels, the target fidelity is
that of the dephasing noise channels determined by the concurrent gate
duration and the $T_1$ and $T_2$ parameters.

\begin{figure}[h]
    \centering
    \includegraphics[width=0.95\columnwidth]{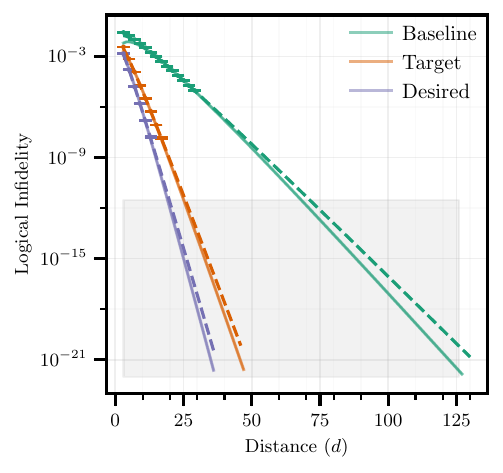}
    \caption{Prediction of high-distance error rates from numerical simulations at lower distances. The numerical dataset is plotted alongside extrapolations derived from the two-parameter fits to the models \Cref{eq:topqad-lambda} (shown using solid lines) and \Cref{eq:non-topqad-lambda} (shown using dashed lines). The dataset is obtained using Clifford simulations based on the noise model described in this section. The fitting parameters are provided in \Cref{tab:lambdas-and-mus}. The shaded region represents the regime of logical infidelities required to implement the electronic-structure simulations' quantum circuits used in our QRE studies, presented in \Cref{sec:resource-estimates-pbenzyne}.}
    \label{fig:benchmarking-hardware}
\end{figure}

The results of our simulations are illustrated in~\Cref{fig:benchmarking-hardware}. We use numerical simulations at lower distances and extrapolate the logical infidelities at the higher distances in the regime of interest for utility-scale FTQC. We regress the lowest-order term of the logical error model of the surface code
\begin{equation}
\mu d^2\Lambda^{-(d+1)/2} \label{eq:topqad-lambda}
\end{equation}
to our numerical data, where $d$ is the code distance and $\mu$ and  $\Lambda$ are fitting parameters. We refer the reader to Ref.~\cite{silva2024optimizing} for further details on the choice of this error suppression model.

To mitigate the bias introduced by small distances, data points with a logical infidelity
below $10^{-2.5}$ are ignored in the fitting. The extracted $\Lambda$ value is an important hardware characteristic, as it determines the rate of logical error suppression with distance~\cite{google2021exponential}. We note that the extracted $\Lambda$ values for baseline and target hardware parameters are,
respectively, $2.34(1)$ and $9.3(3)$, showing an improvement by roughly a factor of $4$. The extracted value of $\Lambda$ for the desired model is $18(1)$, demonstrating an additional improvement factor of $2$ in the error suppression rate as compared to the target parameter set.

We note that previous works~\cite{fowler2012surface, googlequantumai2023suppressing, Acharya2024} use the model $\bar\mu \bar\Lambda^{-(d+1)/2}$ to demonstrate an exponential suppression in the surface code error rates \emph{per cycle}. Converting this per-cycle measure to an error model for the entire fault-tolerant protocol results in the model
\begin{equation}
\bar\mu d \bar\Lambda^{-(d+1)/2} \label{eq:non-topqad-lambda},
\end{equation}
which has a linear coefficient $d$, as opposed to $d^2$ as in our model. In \Cref{fig:benchmarking-hardware}, we show the discrepancies between the two choices in predicting logical error rates, especially at greater distances and with less-performant hardware specifications. The values of all fitting parameters for both models and our three hardware parameters are summarized in \Cref{tab:lambdas-and-mus}.

 \begin{table}[!ht]
 \renewcommand{\arraystretch}{1.4}
 \centering
 \begin{tabularx}{\linewidth}{>{\centering\arraybackslash}X|>{\centering\arraybackslash}X|>{\centering\arraybackslash}X|>{\centering\arraybackslash}X}
 \hline\hline
  & \textbf{Baseline} & \textbf{Target} & \textbf{Desired} \\ \hline
 $\Lambda$ & 2.34(1) & 9.3(3) & 18.0(1) \\ 
 $\mu$ & 0.0038(2) & 0.019(4) & 0.04(1) \\ 
 $\bar\Lambda$ & 2.119(6) & 7.5(1) & 13.5(3) \\ 
 $\bar\mu$ & 0.0259(8) & 0.055(6) & 0.082(9) \\ \hline\hline
 \end{tabularx}
\caption{Fitting parameters obtained for the models displayed in \Cref{eq:topqad-lambda} and \Cref{eq:non-topqad-lambda} when regressed to the numerical data of our three hardware parameter sets, specified as baseline, target, and desired. 
The parentheses and the digits they enclose at the end of the numerical values for the fitting parameters is a notation used to indicate the standard uncertainty of the numerical values in terms of the least significant digits, respectively.}
\label{tab:lambdas-and-mus}
\end{table}

\subsection{Sensitivity of FTQC performance to specific hardware improvements}
\label{sec:sensitivity-analysis}

It is often unclear {\em a priori} which types of noise and errors have the most significant impact on the performance of FTQC, and which hardware parameters are the most critical for achieving improved logical performance. It is unclear whether the coherence time of qubits or the two-qubit error rates matter most, or the state preparation and measurement (SPAM) errors are most crucial. In this section, we report results of a sensitivity analysis addressing this uncertainty.

\begin{figure}[ht]
    \centering
    \includegraphics[width=0.9\columnwidth]{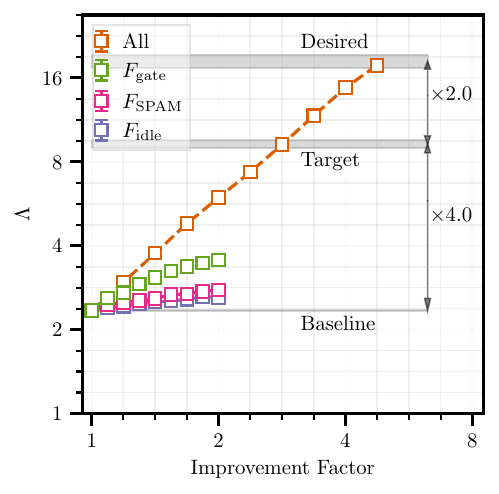}
    \caption{Sensitivity of logical infidelity to hardware
     improvements. The extracted fit parameter
     $\Lambda$, representing the asymptotic error suppression rate for a quantum memory experiment, varies with a multiplicative improvement factor used to scale a
     particular subset of physical parameters, indicated in the legend,
     relative to the baseline hardware parameters. 
    }
    \label{fig:sensitivity_analysis}
\end{figure}

We study the performance sensitivity of FTQC to specific hardware characteristics, as specified in \Cref{tab:physical_params}, by assessing the logical error suppression factor $\Lambda$
as a function of individual hardware parameters. We analyze the following categories of
hardware parameter improvements in the operations that implement the quantum memory experiment:
(i) coherence improvements (for idling physical qubits) involving $T_1$ and $T_2$ times;
(ii) gate-control improvements affecting the Hadamard and CNOT gate infidelities; 
(iii) SPAM improvements concerning preparation, measurement, and reset
errors; and 
(iv) a combined class encompassing all three groups.
We determine the improvement in the error suppression rate $\Lambda$ when each of these parameter sets are improved separately, while keeping the others constant, as well as when all hardware parameters are improved simultaneously.  \Cref{fig:sensitivity_analysis} illustrates
our findings. We observe that improvements in the gate-control errors yield the most significant impact, whereas improvements in SPAM errors and coherence enhancements are significantly less effective for achieving higher $\Lambda$ values. 
The $\Lambda$ value increase resulting from improving all parameters simultaneously is higher than the sum of individual increments of  $\Lambda$. Our findings suggest that quantum gate fidelity improvements are more important than SPAM and idling qubit error rates for achieving greater logical performance. 

\subsection{Impact of qubit and gate quality distributions on logical error rates}
\label{sec:impact_of_fat-tails}

As our QRE studies in \Cref{sec:resource-estimates-pbenzyne} show, a utility-scale quantum computer is expected to require millions of qubits. Any manufacturing process that produces such large QPUs, or clusters of QPUs, will inevitably create qubits and gates of varying quality. In this section, we analyze possible impacts of process variability on the logical error rates of the rotated surface code.

\begin{figure}[tbh]
\centerline{\includegraphics[width=\columnwidth]{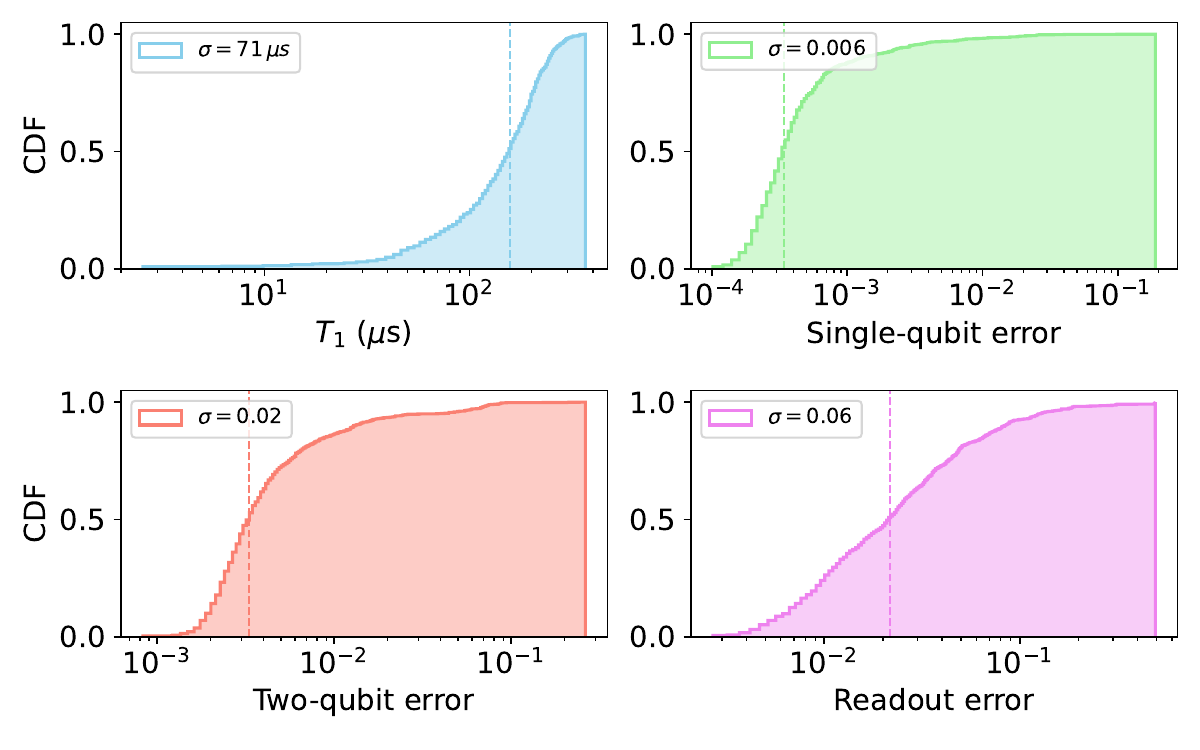}}
\caption{Cumulative distribution functions for $T_1$ times, and the single-qubit, two-qubit, and readout errors accumulated over nine days for the \texttt{ibm\_torino} processor~\cite{ibmtorino}. Dashed vertical lines indicate the mean values. Legends indicate standard deviations.}
\label{fig:cdf}
\end{figure}

In order to obtain a realistic distribution of qubit and gate qualities, we use publicly available calibration data obtained for  the \texttt{ibm\_torino} quantum processor~\cite{ibmtorino}. In particular, we focus on the distributions of the $T_1$ times, as well as single-qubit, two-qubit, and readout errors.
In \Cref{fig:cdf} we plot the cumulative distribution functions (CDF) of this data.
Physically, we expect some correlations between these distributions, for example, a longer coherence time for qubits should allow for higher fidelity or faster quantum control on the gates and therefore higher single-qubit and two-qubit fidelities.

To capture these correlations, we employ a random-forest model \cite{ho1995random}, which is a common choice of machine learning model for small sets of training data. We use the $T_1$ time as an input feature, and train three models for the conditional generation of single-qubit, two-qubit, and readout errors, respectively.

\begin{figure}[tbh]
\centerline{\includegraphics[width=\columnwidth]{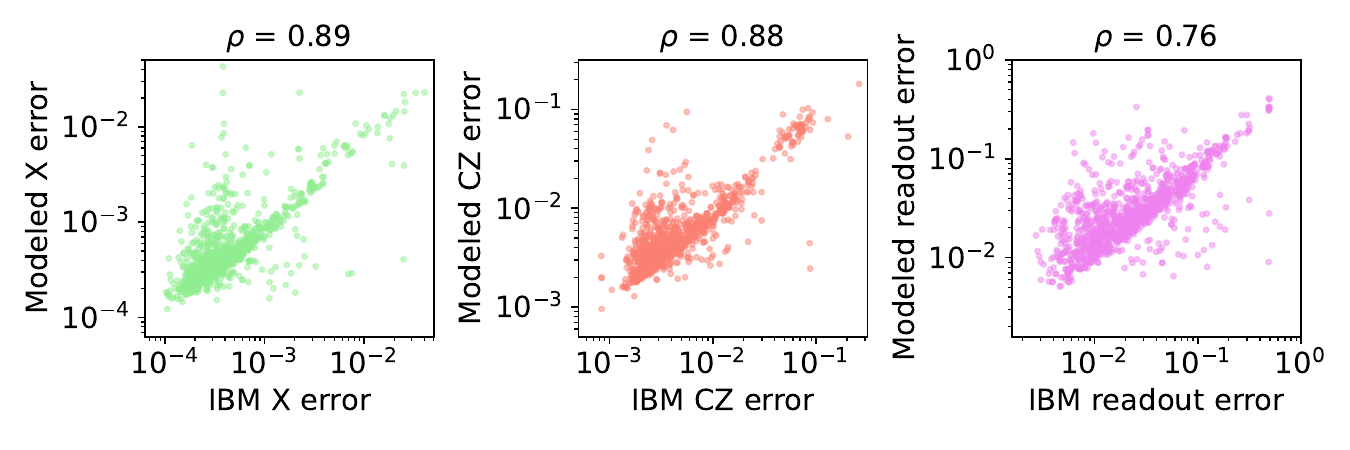}}
\caption{Correlations between the true and predicted single-qubit, two-qubit, and readout errors as a function of the $T_1$ time. The Pearson correlation coefficients $\rho$ are shown at the top of each figure.}
\label{fig:ml_pred}
\end{figure}

The single-qubit and readout errors are predicted from the $T_1$ time of the corresponding qubit, while the two-qubit error model uses the $T_1$ time  of both qubits as input. \Cref{fig:ml_pred} shows that our random-forest models adequately estimate the gate and readout errors.

We use these models to study the impact of process variability on logical error rates of QEC codes. This variability is characterized by the standard deviation $\sigma$ of the distribution. Therefore, we construct several synthetic $T_1$ distributions with varying values of $\sigma$, by rescaling the IBM $T_1$ distribution,
\begin{equation}\label{eq:empircaltransformation}
    T_1 \to \mu + a (T_1 - \mu),
\end{equation}
where $\mu$ is the mean of the original distribution and $a$ is the rescaling factor. This transformation ensures that only $\sigma$ varies while the mean and the higher standardized moments of the distribution remain fixed.

\begin{figure}[htb]
\centerline{\includegraphics[width=.93\columnwidth]{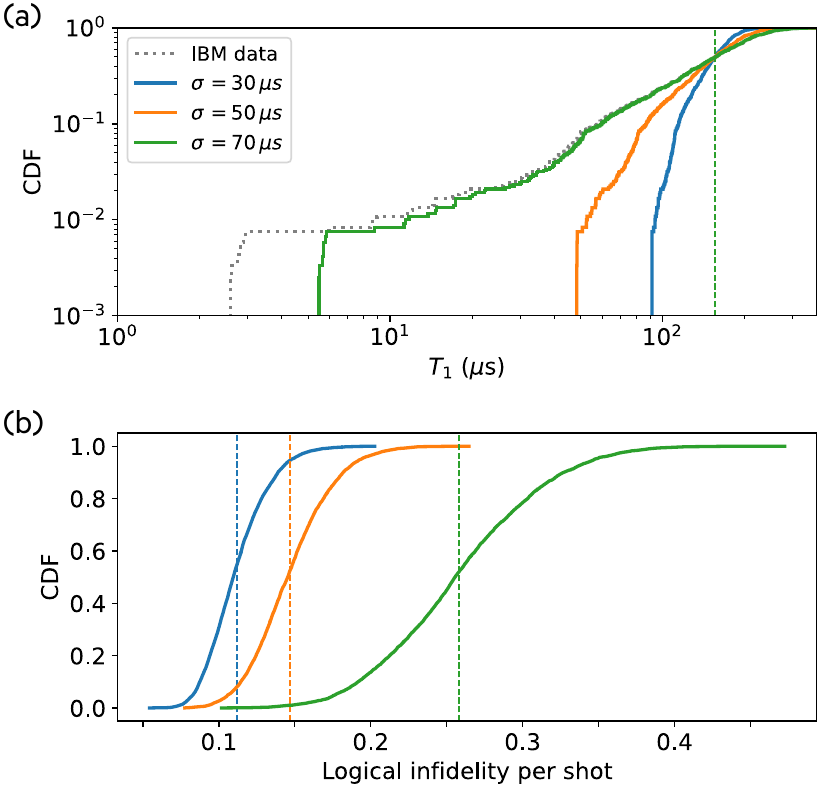}}
\caption{(a) Transformed $T_1$ distributions with different standard deviation, $\sigma$. The dotted grey curve shows the CDF of the original $T_1$ distribution for comparison. (b) Cumulative distribution functions of the logical infidelities of a rotated surface code of distance $d=9$. Dashed vertical lines indicate the respective mean values.}
\label{fig:T1-variance}
\end{figure}

In \Cref{fig:T1-variance}(a), we show the CDFs of three distributions generated by applying the transformation~\eqref{eq:empircaltransformation}. The three values of $\sigma$ are chosen to represent different amounts of reductions in process variability from that of the studied QPU.

We investigate the impact of these distributions on the performance of the logical memory experiment on a distance-9 rotated surface code. To do this, we perform simulations where the gate and measurement errors on each physical qubit on the rotated surface code patch are distinct, better reflecting the experimental reality. For each of the distributions constructed above, we repeat the following process $5000$ times:
\begin{enumerate}
    \item Sample a $T_1$ time for each physical qubit on the rotated surface code patch.
    \item Use the machine learning models to generate synthetic gate and measurement error rates on each physical qubit (and each adjacent pair of qubits in the case of two-qubit gate errors).
    \item Simulate the fault-tolerant memory protocol with assigned gate fidelities, and assuming gate times from the \texttt{ibm\_torino} device data (see~\Cref{tab:ibm_physical_params}), to determine the logical error rate for the code.
\end{enumerate}

\begin{table}
    \centering
    \begin{tabular}{cc}
        \hline\hline
       \hspace{0.7cm} \textbf{Parameter}\hspace{1.5cm}  & \textbf{IBM Values}\hspace{1.6cm}\\
        \hline
         \hspace{0.7cm}Single-qubit gate time\hspace{1.5cm}  & 32 ns\hspace{1.5cm}\\
         \hspace{0.7cm}Two-qubit gate time\hspace{1.5cm}  & 68 ns\hspace{1.5cm}\\
         \hspace{0.7cm}State preparation time\hspace{1.5cm}  & 780 ns \hspace{1.5cm}\\
         \hspace{0.7cm}Measurement time\hspace{1.5cm}  & 780 ns\hspace{1.5cm}\\
         \hspace{0.7cm}Reset time\hspace{1.5cm}  & 780 ns\hspace{1.5cm}\\
         \hline\hline
    \end{tabular}
    \caption{Gate times of the \texttt{ibm\_torino} QPU.}
    \label{tab:ibm_physical_params}
\end{table}

The output distributions of the logical error rates are shown in \Cref{fig:T1-variance}(b). We observe that higher values for $\sigma$ result in a higher logical error rate.  This analysis suggests that QPU manufacturing should not only focus on improving the mean quality of qubits and gates, but also on achieving more-robust fabrication processes so as to avoid heavy tails of poor-quality qubits. Finally, we emphasize that this study is confined to analyzing the impact of the variance of the $T_1$ distribution. However, it is speculated that the higher moments, such as skewness and kurtosis, might carry valuable signatures for such benchmarking and should be investigated in the future.

It is worth emphasizing that even with improved mean performance and a narrower distribution, a large device will contain a nonzero population of defective qubits or couplers. Known bad components should be treated as localized defects rather than as ordinary stochastic errors. At the hardware and compilation layers, defective sites can be disabled and logical patches can be placed to avoid clustered defects. At the QEC layer, surface-code patches can be deformed around isolated defects, neighboring checks can be merged, and syndrome-extraction schedules can be recompiled for the effective connectivity. The overhead depends strongly on defect density and clustering: isolated defects can often be handled with a local reduction in effective distance, whereas extended defect regions may require larger patches, rerouting, or rejection of the affected tile. Quantifying this overhead using the measured defect map is an important extension of the hardware-distribution analysis developed here \cite{815q-xjrb,Debroy2025luciinsurfacecode}. Understanding qubit performance could also help in the co-design of algorithms and robust QEC codes for early fault-tolerant quantum computers~\cite{Kiss:2024sep, Lin:2021rwb, EFTQC}.

\subsection{Emulation of other FTQC protocols}
\label{sec:other-FTQC-protocols}

\begin{figure}[b]
    \centering
    \includegraphics[width=0.9\linewidth]{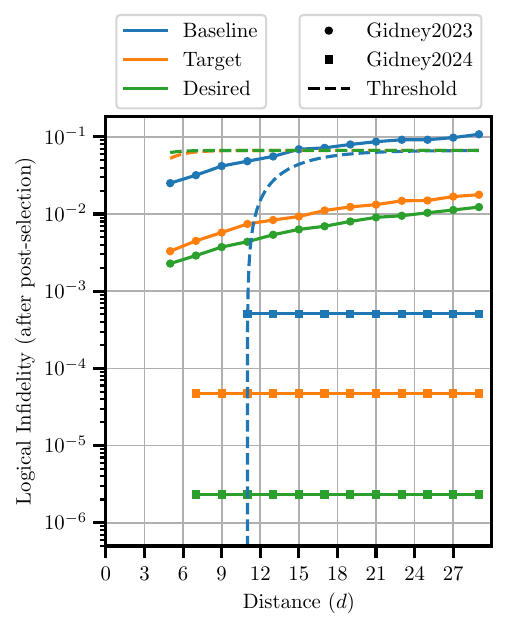}
    \caption{Performance of the magic state preparation protocols Gidney2023~\cite{Gidney2023cleaner} and Gidney2024~\cite{Gidney2024magic} for three parameter sets. Here, we fix $d_1=3$, because it yields the best performance for the baseline parameter set. The dashed curves indicate the distillation unit input threshold \eqref{eq:msduthreshold} for the baseline, target, and desired parameter set, respectively, where the difference is due to the fact that $P_{\text{Cliff}}$, estimated from the memory protocol logical error rate, depends on hardware noise. We observe that the baseline set is below the respective 15-to-1 distillation threshold only for the Gidney2024 protocol; it is above the threshold for the Gidney2023 protocol.  }
    \label{fig:mspuperformance}
\end{figure}

{\bf Magic state preparation.}
A critical process in FTQC is the preparation of high-fidelity logical magic states. These states are produced by first employing a magic state preparation protocol~\cite{singh2022, Gidney2023cleaner, Gidney2024magic}, which produces relatively low-fidelity logical magic states at small distances by employing physical $T$ gates. A large number of such logical states are then grown to greater distances and fed to magic state distillation units to prepare a lower number of higher-fidelity magic states. These magic state distillation units are  able to perform only if they are fed logical magic states of sufficient fidelity. It has been estimated that the 15-to-1 distillation units have an acceptance probability
\begin{equation}
    1 - 15P_{\text{magic}} - 356P_{\text{Cliff}}, \label{eq:msduacceptanceprob}
\end{equation}
where $P_{\text{magic}}$ and $P_{\text{Cliff}}$ are the logical error rates of input logical magic states and Clifford operations, respectively~\cite{beverland2022assessing}. Hence, we need to produce logical magic states with error probability
\begin{equation}
    P_{\text{magic}} < \frac{1 - 356P_{\text{Cliff}}}{15}. \label{eq:msduthreshold}
\end{equation}
Whether magic states of this fidelity can be produced depends both on the specific magic state preparation protocol used and the hardware noise profile. We studied a protocol that cleverly exploits hook-injection errors to create high-fidelity relatively low-distance magic states~\cite{Gidney2023cleaner} using Clifford simulations. This protocol first uses a physical $T$ gate to create a magic state on a small rotated surface code patch of distance $d_1$. It then post-selects the states for which no errors are detected and grows them to a larger code distance $d$. The simulation results reported in \Cref{fig:mspuperformance} show that the error rates increase with $d$, suggesting that the protocol is not fault tolerant, and explains why distillation units are needed instead of directly growing the magic states to a target distance. For simplicity, we substitute the logical Clifford error rate with the logical error rate of the memory protocol in \Cref{eq:msduthreshold} and draw the respective threshold curves for each of the hardware parameters shown in \Cref{fig:mspuperformance}. We observe that both the target and desired parameter sets are significantly below their respective \mbox{15-to-1} distillation thresholds, unlike the baseline set.

A more recent protocol, magic state cultivation \cite{Gidney2024magic}, demonstrates significant improvements in the logical error rates for magic state preparation. This protocol introduces a number of improvements over past protocols, such as cleverly designed gradual growth stages and appropriate post-selections to ensure that error rates drop when increasing distances in practical regimes of error rates. We have used the code \cite{gidney_2024_13777072} developed by the 
inventors of the magic state cultivation protocol to estimate the resulting magic state preparation error rates for our three hardware parameter sets. Their code accepts the circuit-level noise model which we derive from the hardware parameter specifications (baseline, target, and desired hardware) according to the method outlined in \Cref{sec:Circuit-Level-Noise-Model}. Our emulation results based on magic state cultivation are also shown in \Cref{fig:mspuperformance}, in direct comparison to prior work based on hook injection. We observe that all three parameter sets are significantly below the 15-to-1 distillation threshold for the magic state cultivation protocol, and for this reason the QREs in this paper use this protocol. 

\begin{figure*}
    \centering
  \includegraphics[width=0.9\textwidth]{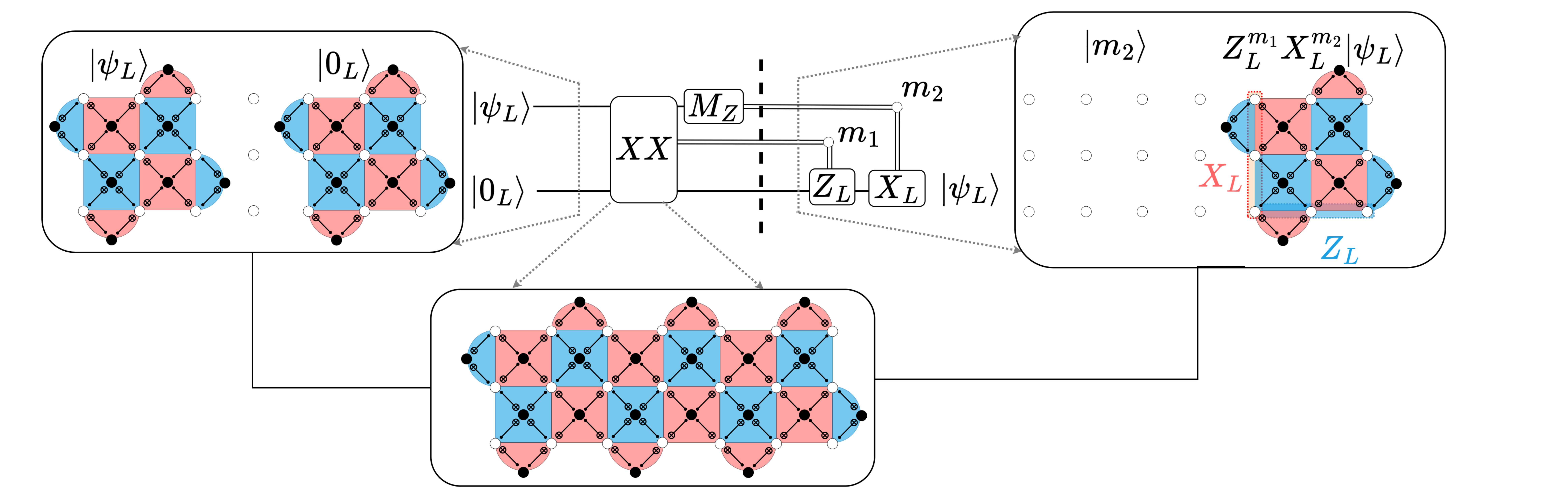}
  \caption{Logical teleportation via lattice surgery for a rotated surface code of distance $d = 3$ and bus width $b= 1$. Illustrated are the states prior to merging (left), during merging (middle), and after splitting (right). The wire diagram of the teleportation quantum circuit, including the relevant classical feedback and the needed recovery operations, is also shown. We simulate the portion of the circuit before the dashed line using a circuit-level noise model. Note that, at the dashed line, invoking a decoder may flip the values of $m_1$ and $m_2$. We do not simulate the circuity of the recovery operations $Z_L$ and $X_L$; instead, we measure all qubits at the dashed line to determine the performance of the teleportation.}
\label{fig:logical_teleportation}
\end{figure*}

{\bf Teleportation of logical qubits.}
To perform lattice surgery fault tolerantly, both space-like and time-like errors must be corrected. As discussed and numerically demonstrated in  Ref.~\cite{Chamberland_2022}, space-like errors are exponentially suppressed by increasing the code distance of the logical qubits. Similarly, time-like errors are exponentially reduced by increasing the number of stabilizer measurement rounds during the merge operation (see \Cref{fig:logical_teleportation}). In this context, the number of parity-check cycles conducted while the two logical patches are merged is referred to as the temporal code distance. To evaluate the overall success rate of lattice surgery, we assess logical qubit teleportation under varying space--time parameters.

\begin{figure}
    \begin{center}
  \includegraphics[width=0.40\textwidth]{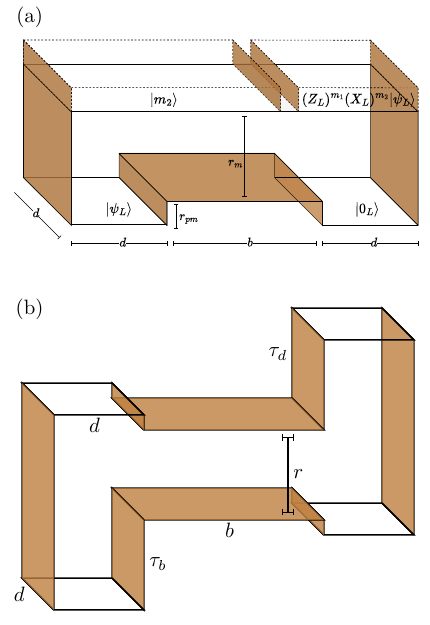}
    \end{center}
    \caption{Space--time diagrams of  teleportation via lattice surgery.
    (a) Space--time diagram of teleportation where $r_\text{s}=0$ QEC cycles are performed after splitting. The spatial dimensions are the same as shown in \Cref{fig:long_bus_with_cuts}, and the temporal dimension is divided into $r_{\text{pm}}$ pre-merge rounds and $r_\text{m}$ merge rounds. 
    (b) Effect of buffer and decoder delays on lattice surgery fidelities. For each logical operation in the core processor, a teleportation is implemented involving a magic state (represented by the left code patch) and data qubits (represented by the right code patch). The magic state may incur a delay $\tau_b$ in the buffer before the surgery is performed. The magic state and the bus are measured after the merge operation, but target patches must await the decoder decisions, available after 
    a decoder delay time $\tau_d$. These idling patches must be protected by a quantum memory protocol using further stabilization rounds during this period; therefore, the accumulation of further errors is inevitable and must be taken into account by the compiler, FTQC emulator, and resource estimators.}
    \label{fig:teleportation_space--time_diagrams}
\end{figure}

To determine the success rate of logical qubit teleportation, we estimate the average state infidelity defined by $ P_a = \frac{1}{6}\sum_{\psi} P_{\psi} $, where $ P_{\psi} $ represents the infidelity of teleporting each logical state $ | \psi \rangle \in$ \mbox{$\{|0_L\rangle, |1_L\rangle, |+_L\rangle, |-_L\rangle, |+i_L\rangle, |-i_L\rangle \} $.} Note that the process fidelity is then given by $ F_p = \frac{(D + 1)(1-P_a) - 1}{D} = 1 - 3/2 P_a$, where $ D = 2 $ is the dimension of the Hilbert space of the single qubit being teleported \cite{Nielsen_2002,ryananderson2024}. Since accessing the $Y$ operator of the surface code is cumbersome \cite{gidney2024inplace}, we teleport the logical states $ | \psi_L \rangle \in \{|0_L\rangle, |+_L\rangle \} $ using an $ XX $ (rough) merge, as illustrated in \Cref{fig:logical_teleportation}. From these simulations, we approximate the average state infidelity as $ P_a  \approx \frac{2}{3}(P_+ + P_0) $, providing an overestimate of the infidelity. The teleportation protocol we study is illustrated in \Cref{fig:logical_teleportation,fig:teleportation_space--time_diagrams}(a) and outlined as follows:
\begin{enumerate}

\item Preparation (\Cref{fig:logical_teleportation}, left side): We begin by preparing the logical source state $ |\psi_L\rangle \in \{|0_L\rangle, |+_L\rangle\} $ and the target patch in the state $ |0_L\rangle $. In this pre-merging step, these states are stabilized for $ r_{\text{pm}} $ rounds.

\item Merging (\Cref{fig:logical_teleportation}, middle): After the $ r_\text{{pm}} $ rounds, the bus data qubits are respectively initialized in the physical $ |0\rangle $ state, and the rough edges of the two surfaces are merged. Then, the entire surface is stabilized for $ r_\text{m} $ rounds to determine (a possibly erroneous) measurement of the $X\otimes X$ observable with an outcome $m_1$.

\item Splitting (\Cref{fig:logical_teleportation}, right side): After the $ r_{\text{pm}} + r_\text{m}$ rounds, the bus data qubits are respectively measured in the $ Z $ basis (splitting) and the remaining patches are stabilized for $ r_\text{s} $ rounds. Note that this also results in a perfect round of syndrome measurements on the bus patches.

\item Projection (\Cref{fig:logical_teleportation}, right side): After the $ r_{\text{pm}} + r_\text{m} + r_\text{s}$ rounds, the source data qubits are respectively measured in the $ Z $-basis, to determine the (possibly erroneous) value of an outcome $m_2$.

\item Recovery (\Cref{fig:logical_teleportation}, right side): At this point, by invoking a decoder, the values of $m_1$ and $m_2$ may be corrected, and the recovery operations $Z_L$ and $X_L$ are conditionally applied.
\end{enumerate}

To estimate $P_a$ at large distances, we simulate the mentioned teleportation protocol for varying spatial code distances $d$, temporal code distances $r_\text{m}$, and bus widths $b$. We then regress the following predictive models from the obtained numerical results of $P_0$ and $P_+$ values to predict the fidelity of the protocol at large distances~\cite{Chamberland_2022, silva2024optimizing}:

\begin{align}
    P_0 &= \mu_X (2d+b)r_\text{m} \Lambda_X^{-(d+1)/2}, \label{eq:p0_model} \\
    P_+ &= \mu_Z d \Lambda_Z^{-(d+1)/2} + \mu_T d b \Lambda_T^{-(r_\text{m} +1)/2} \label{eq:p+_model}.
\end{align}

In our simulations, the number of pre-merge stabilization rounds is fixed at $r_{\text{pm}} = 1$, during which the two logical states are prepared. For the remaining rounds, we incorporate the circuit-level noise model detailed in \Cref{sec:Circuit-Level-Noise-Model}  using the baseline noise parameters. We also set $r_\text{s} = 0$. For our benchmarking purposes, the recovery step is not performed; instead, the target data qubits are also measured in the basis corresponding to the initial source state $|\psi_L\rangle$ to determine the teleported logical state at the target patch.

\begin{figure*}
    \centering    \includegraphics[width=2\columnwidth]{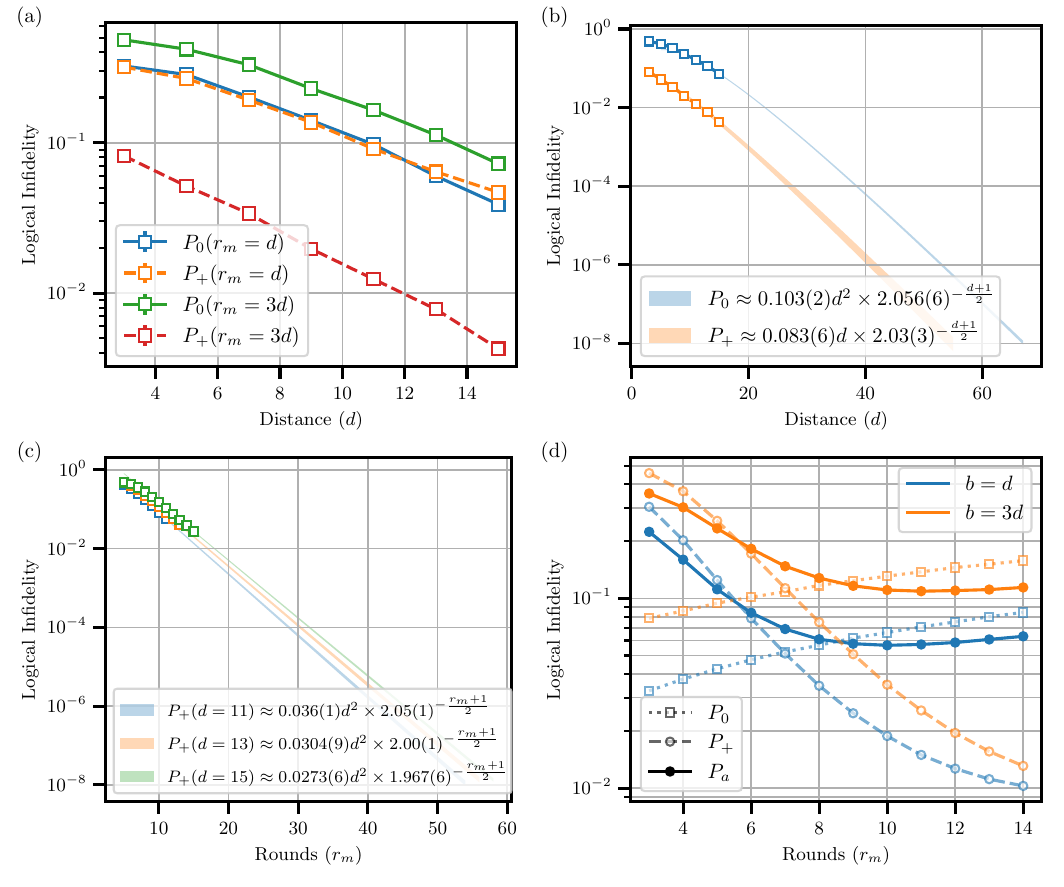}
    \caption{(a) Error suppression as a function of code distance $d$ for a bus width  $b = 3d$ and for $r_\text{m} \in \{d, 3d\}$ merge rounds, for the teleportation of quantum states $|0_L\rangle$ ($P_0$) and $|+_L\rangle$ ($P_+$). (b) Exponential fit for $P_0(r_\text{m} = 3d)$ and $P_+(r_\text{m} = 3d)$ using exponential functions of the form $P_0 \approx \mu_X d^2 \Lambda_X^{-( d + 1)/2} $ and $P_+ \approx \mu_Z d \Lambda_Z^{-( d + 1)/2} $, where the latter equation results from the leading-order term of \Cref{eq:p+_model} for $r_\text{m} = 3d.$ (c) Exponential suppression of $P_{+}$ as a function of the number of merge rounds for $r_\text{m} < d$, for which we expect the logical error suppression of the form $P_+ \approx \mu_T d^2 \Lambda_T^{-(r_\text{m} + 1)/2}$ resulting as the dominant term in \Cref{eq:p+_model}. (d) Logical teleportation error rates for teleporting the states $|0_L\rangle$ and $|+_L\rangle$ as a function of the merge stabilization rounds (temporal distance) for  a code distance $d = 7$ over a bus width $b \in \{d, 3d\}$ and the corresponding average state infidelities calculated using $\frac{2}{3}(P_0+P_+)$.}
\label{fig:dist_and_rounds_scaling}
\end{figure*}

\Cref{fig:dist_and_rounds_scaling}(a) shows the logical error rates in teleporting the states $|+_L\rangle$ and $|0_L\rangle$ (labeled $P_+$ and $P_0$, respectively) as a function of the code distance for $3\leq d\leq 15$, with a corresponding bus width $b = 3d$ and a temporal distance $r_\text{m} \in \{d, 3d\}$. We highlight two observations from \Cref{fig:dist_and_rounds_scaling}(a):

\begin{itemize}
    \item The error rates are suppressed as a function of code distance for both states. This indicates that the noise parameters are below the threshold.
    \item Increasing $r_\text{m}$ decreases the teleportation fidelity of $|0_L\rangle$ while increasing the fidelity of teleporting $|+_L\rangle$ (see also \Cref{fig:dist_and_rounds_scaling}(d)).
\end{itemize}

\Cref{fig:dist_and_rounds_scaling}(b) shows the fitting for the $X$- and $Z$-type terms of~\Cref{eq:p0_model,eq:p+_model}. This model predicts the $X$- and $Z$-type errors better in the high-$r_\text{m}$ regime (hence our choice of the $r_\text{m} = 3d$ data). Similarly, \Cref{fig:dist_and_rounds_scaling}(c) shows an estimate of the time-like error suppression term in \Cref{eq:p+_model} in the high-distance regime, 
yielding the values $\mu_T \approx 0.0273(6)$ and $\Lambda_T \approx 1.967(6)$. This information is sufficient to estimate the average error rate $P_a$ of high-distance teleportations.

As another application, \Cref{fig:dist_and_rounds_scaling}(d) shows the optimal number of merge stabilization rounds $r_\text{m}$ for a given distance $d$. Here, we choose $d=7$ and consider two sizes for the bus $b\in \{d, 3d\}$. We plot the estimated teleportation fidelity $P_a \approx \frac{2}{3}(P_0+P_+)$ as a function of $r_\text{m}$ and note that the minimum of each curve is at values of $r_\text{m} > 7$, highlighting the fact that the optimal number of QEC rounds for a lattice surgery operation with code distance $d$ may actually deviate from the commonly assumed value $d$.

For the teleportation of magic states in the core processor, the magic state has resided in the buffer for some average expected buffer delay time $\tau_b$, which can be as low as 1 clock cycle for balanced production and consumption of the magic states. The targets of teleportation are logical data qubit patches in the core processor for which further QEC rounds are executed until the decoder outcome becomes available. We denote this delay by $\tau_d$. Inclusion of the buffer and decoder delays and assuming an average rate for all types of surgeries results in the model

\begin{equation} \label{eq:cliff-error-model}
    \mu \big[d (2r + \tau_b + \tau_d + 1) + br \big] \Lambda^{-(d+1)/2} + \mu_T d b \Lambda_T^{-(r +1)/2},
\end{equation}

\noindent which still distinguishes time-like and space-like errors but ignores the type of surgery (see \Cref{fig:teleportation_space--time_diagrams}(b)). {A thorough analysis of decoder delays based on parallel space- and time-window decoding methods and taking into account communication latencies drawn from a concrete quantum execution environment comprising a high-speed network of quantum processing units, controllers, decoders, and HPC nodes has been presented in a recent work~\cite{khalid2025impacts} by a subset of the authors of this manuscript.

\subsection{High-performance real-time decoding platform}
\label{sec:decoder}

{\bf Challenges and requirements.}
The high speed of superconducting processors, a great advantage for utility-scale applications, requires well-engineered control and decoder architecture, both on the software and hardware levels. A key technical challenge is that decoding simultaneously requires peta-scale computation and low latency for real-time feed-forward. Furthermore, at stage, it is not known yet what algorithms are most effective at decoding; therefore, there is a systems engineering trade-off between performance and flexibility.
 
Performing universal fault-tolerant quantum computation using QEC mandates feedforward-based implementation of certain quantum gates (e.g., $T$ gates) with low latency \cite{kurman2024benchmarkingabilitycontrollerexecute}. In these implementations, a conditional operation is applied based on the result of a logical measurement as well as the decoding of syndromes of many previous QEC cycles. The classical feed-forward latency is measured from the physical execution of the logical measurement until the controller executes a conditional gate ($L^0$ and $L^1$ in \Cref{fig:qec_realtime_req}(c)).
 
For efficient execution of fault-tolerant feed-forward gates, the decoder needs to be ready in time for the conditional gate execution. We note that on average, $ d $ (distance) cycles are allowed for the decoder result for multiple reasons. First, when the conditional gate is followed by gates that commute, it may be deferred after the gates that do not depend on the decoding result, as shown in \Cref{fig:qec_realtime_req}(b). Second, the gates that follow the conditional gate may require synchronization with other surfaces, allowing to defer the conditional gate without impacting the circuit. In addition, we note that sporadic delays caused by the decoder have a small impact on the overall performance as long as the delay remains within tens of microseconds and on average the decoder result is ready on time \cite{kurman2024shor}. Therefore, we target an end-to-end average decoding latency shorter than $ d $ QEC cycles, which for superconducting qubits implies a target latency of approximately $\SI{10}{\micro\second}$; {see Ref.~\cite{khalid2025impacts} for a more elaborate analysis on decoder requirements in a general FTQC setting.
To meet these latency targets and implement QEC decoding efficiently, it is essential to optimize the performance of both the controller--decoder communication channel and decoding task. For the controller--decoder channel, throughput must exceed the data generation rate. The controller should locally perform state discrimination including optional soft readout indicators, encoding each qubit state with a minimal bit representation. For 20k qubits, a 4-bit state representation per qubit, and a QEC cycle time of 550 ns, a minimum net bandwidth of 150 Gb/s is required. Additionally, data sent from the decoder to the controller must be efficiently compacted to communicate only the necessary logical instructions. In addition, the overall latency should be minimal, including readout state discrimination, data aggregation from multiple controllers and transmission to and from the decoder.

The decoding process, which includes multiple concurrent decoding tasks \cite{kurman2024shor}, should be designed to minimize the resulting latency. Scalable hardware is required, as decoding for circuits with 10k--100k qubits demands extensive computational capacity. Some decoding algorithms, such as Fusion Blossom, may exhibit variability in the decoding time dependent on the error pattern. The QEC implementation should be designed to accommodate this variability. For the decoder to not limit performance, the average decoder throughput should exceed the syndrome generation rate. In addition, the average decoder latency, including the roundtrip communication time, should be shorter than the $d$ QEC cycles.

\begin{figure*}[htbp]
\centerline{\includegraphics[width=1.7\columnwidth]{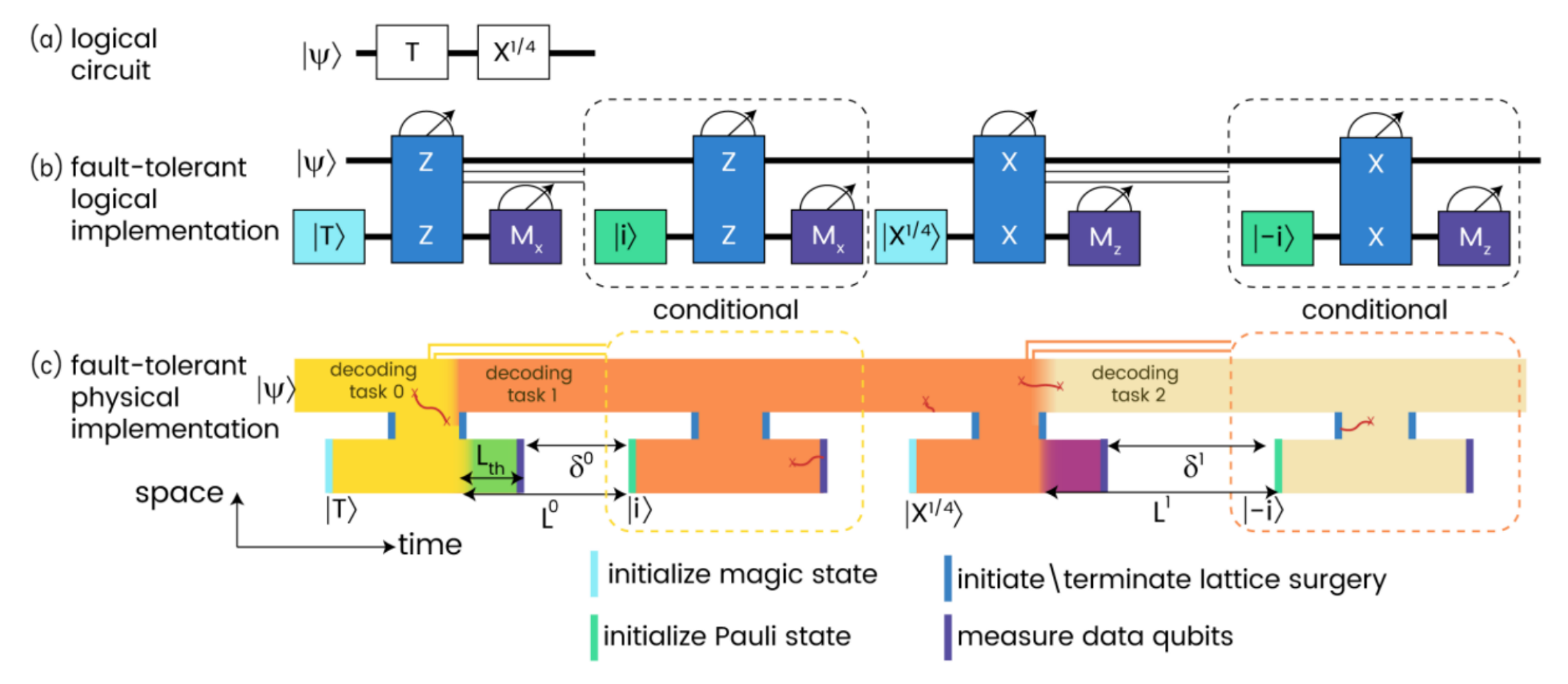}}
    \caption{Example of non-Clifford computation with surface codes adapted from Ref.~\cite{kurman2024benchmarkingabilitycontrollerexecute}. (a) Example of a logic circuit containing two non-Clifford gates. (b) Fault-tolerant logic circuit that implements the circuit in (a) with surface codes with only a single ancillary surface. The dashed square denotes the feed-forward conditional logical gates that verify that the planned circuit is executed. (c) Space--time view of the circuit in (b) with surface codes. Each colour denotes a separate decoding task, chosen to end each task with a logical measurement. The decoding outcome of the lattice surgery between a magic state surface and the computation surface determines a feed-forward circuit, which delays the circuit by  if the feed-forward latency (\(L\)) is larger than a threshold latency (\(L_{\text{th}}\)).}
    \label{fig:qec_realtime_req}
\end{figure*}
{\bf Real-time decoding architecture.} 
To address the FTQC requirements, a proposed architecture, illustrated in \Cref{fig:ctrl-arch} is designed to allow a high-performance decoding platform along with low latency communication between the quantum control clusters and the decoding platform. The control clusters control one or more quantum surfaces, operating in synchronization to execute QEC cycles based on the state of each surface. In addition, the control clusters are responsible to benchmark the qubits performance and maintain qubits calibration. The acceleration servers provide a high-performance platform for the execution of decoder instances and the logic circuit orchestration. They are based on CPU/GPU processors with direct data transfer capabilities to and from the control system. We note that GPUs are beneficial for real-time decoding of QECCs thanks to their massive parallelism and their high-bandwidth / low-latency interfaces. In addition, the servers may incorporate specialized ASIC or FPGA acceleration cards and dedicated hardware offload capabilities. In large-scale systems, multiple acceleration servers may operate in parallel, running multiple decoder instances. The control clusters and accelerator servers are connected by a low-latency network. The network facilitates the aggregation of readout data from the control clusters and distribution of the data to the appropriate decoders. To meet the end-to-end latency requirements, the total time for data aggregation, roundtrip communication and decoding should be in the order of $\SI{10}{\micro\second}$, which require the design of a specialized communication protocol for QEC. 
{The \mbox{end-to-end} decoder requirements for an FTQC architecture have been thoroughly analyzed in a recent work~\cite{khalid2025impacts} by a subset of the authors
of this manuscript.}

\noindent

{\bf Real-time decoding using OP-NIC within the NVQlink architecture.}  QM's OP-NIC provides the low-latency link between the quantum control system and classical accelerators, aligned with NVIDIA’s NVQlink architecture\cite{NVQlink2025}, and offers an effective platform for FTQC, supporting both logic circuit orchestration and decoding processes (see Appendix \ref{sec:dgx} for details). NVIDIA DGX Quantum is the first implementation example, integrating QM's OPX1000 controller with Nvidia's Grace Hopper (GH) superchip. The close integration of CPU and GPU resources enables real-time, parallel execution of various decoding algorithms, including deep-learning-based distributed decoders. The system is connected to the low-latency QEC network and transfers data from the control system to the high-performance CPU--GPU and vice versa over a PCIe interface. A round-trip feed-forward latency (not including the decoding task) from measurement to the decoder and back to conditional gate control, has been benchmarked at less than 3.8 µs.

The OP-NIC-based platform is designed to be connected to a hierarchical, scalable QEC network for data aggregation and distribution, ensuring low latency across systems with 10K qubits and beyond. The future OP-NIC-based platform, based on the Blackwell architecture B200 GPUs, utilizes PCIe 6.0, which provides up to 128 GB/s bidirectional bandwidth in an x16 configuration. The PCIe 6.0 specification incorporates Forward Error Correction (FEC) to maintain data integrity and limit additional latency to under two nanoseconds, ensuring high-speed data transfers with low latency. As scaling extends to systems with 10K--100K qubits, a dedicated, optimized interface connecting the QEC network directly to the decoders may be desired to further reduce the latency. B200 GPU achieves up to 18 petaflops (PFLOPS) for sparse operations. This is made possible through advanced sparsity techniques and architectural enhancements, doubling the effective performance in scenarios where sparsity can be exploited effectively, which is often the case for quantum circuit simulations given the sparsity of data.

The OP-NIC-based platform leverages the extensive parallel processing capabilities, large memory, and high memory bandwidth of the GPU, providing a robust platform for QEC decoding and runtime execution. Moreover, as a software-based solution, NVQlink provides a platform for flexibility and rapid development that is desired in the early stages of FTQC.

Preliminary evaluation of a software-based implementation of the Fusion Blossom algorithm with batch decoding on QM's low latency link to accelerator demonstrated that a serial implementation could sustain the necessary decoding throughput for a distance d=11, with a basic error model with error probability $P_{error}=10^{-3}$. Future iterations will focus on implementing Fusion Blossom in stream mode, in addition to leveraging parallel processing and utilizing the large memory capacity for potential caching and optimization for common error patterns.

The low-latency link between a controller and an HPC server is particularly effective for QEC schemes that benefit from local decoding, as these can be efficiently accelerated by GPUs. Local decoders, such as Steiner tree-based decoders, are prime examples of algorithms that perform well in GPU-accelerated environments due to their parallelizable structure. In addition to standard local decoding, there are more-complex classes of quantum codes, such as entanglement-assisted quantum error-correcting codes (EAQECC) and entanglement-assisted quantum low-density parity-check (EA-QLDPC) codes, which offer enhanced error-correction capabilities by leveraging entanglement~\cite{panteleev2022asymptotically,swaroop2024universal}. Some of these codes, however, have varying degrees of GPU acceleration potential, depending on the complexity and structure of their decoding algorithms. Some EA-QLDPC codes, particularly those with irregular or non-local error syndromes, require more-sophisticated scheduling but can still benefit from GPU acceleration with optimized parallelization strategies. Recent research efforts have focused on discovering new classes of quantum codes that not only offer high error-correction rates but are also capable of full utilization of GPUs.

AI-assisted decoders and quantum algorithms can also significantly improve real-time decoding performance by leveraging rich AI infrastructure and GPU acceleration~\cite{dgx_quantum}. A great example of AI's application to QEC is Google’s AlphaQubit approach, which leverages machine learning to enhance the decoding process for QEC codes~\cite{pan2024effective}. AlphaQubit uses reinforcement learning and neural networks to identify optimal error-correction strategies by learning from a large amount of synthetic and experimental data.

End-to-end testing of the QEC system, including control, decoders, and runtime, is desirable prior for qubits availability at this scale. The OP-NIC-based system can emulate a larger-scale setup by generating synthetic syndromes based on a given error model and loading them to the control system. To minimize an impact on the server under test, a separate server could be dedicated to the emulation, leveraging the system's support for multiple server instances. In this setup, the control system streams syndrome data to the QEC server under test using the low-latency communication interface, which then updates the control state. This
 emulation enables measurement of end-to-end latency, providing a comprehensive engineering perspective on system bottlenecks and opportunities for architectural optimization.

\subsection{Distributed FTQC across multiple dilution refrigerators}
\label{sec:distributed-ftqc}

A fault-tolerant quantum computer with 1M+ physical qubits may require multiple dilution refrigerators (DR) with quantum interconnects between the DRs. 
In this section, we discuss 
the inter-DR and intra-DR architecture for such a distributed quantum computer.  The assembler prioritizes intra-DR lattice surgeries (involving less than 120k qubits) over multi-DR surgeries, as logical teleportation of states between different DRs is much slower and of lower fidelity than intra-DR operations.  Multi-DR surgeries involving code distance $d$ will require at least $d$ optical interconnects between nearest-neighbour DRs, which is a demanding requirement \cite{lee2020high,youssefi2021cryogenic,lecocq2021control,awschalom2021development,delaney2022superconducting}.

{\bf Assembling large FTQC programs among multiple DRs.} 
Embedding FTQC architectures across multiple DRs involves solving a complex embedding problem to determine the connectivities between DRs, teleporation sites adjacent to the interconnects, and appropriate areas across the multi-DR system for the core processor and the MSF zones of required code distances. The embedding prioritizes intra-DR connectivity to ensure the robustness of FTQC protocols against noise introduced by the imperfect interconnects. This remains an important consideration even within individual DRs when lattice surgeries span across the edge couplers of the QPU (i.e., the weaker capacitive coupling between the 20k qubit wafers; see \Cref{sec:wafer-scale}).

Designing the layout of the core processor and the MSFs, including the shapes of the distillation units, is critical for fitting them within the available space. Figure \ref{fig:multi_dr_example} illustrates an example of an embedded multi-DR architecture designed for executing the quantum circuit associated with qubitized ground-state energy estimation of \mbox{$p$-benzyne} with an active space characterized by HL$\pm 2$ (involving six molecular orbitals; see \Cref{sec: QPE mole resource,tab:resource_estimates_logical_trotter}) based on the data generated in \Cref{sec:sensitivity-analysis}.

\begin{figure}
    \centering
    \includegraphics[width=\linewidth]{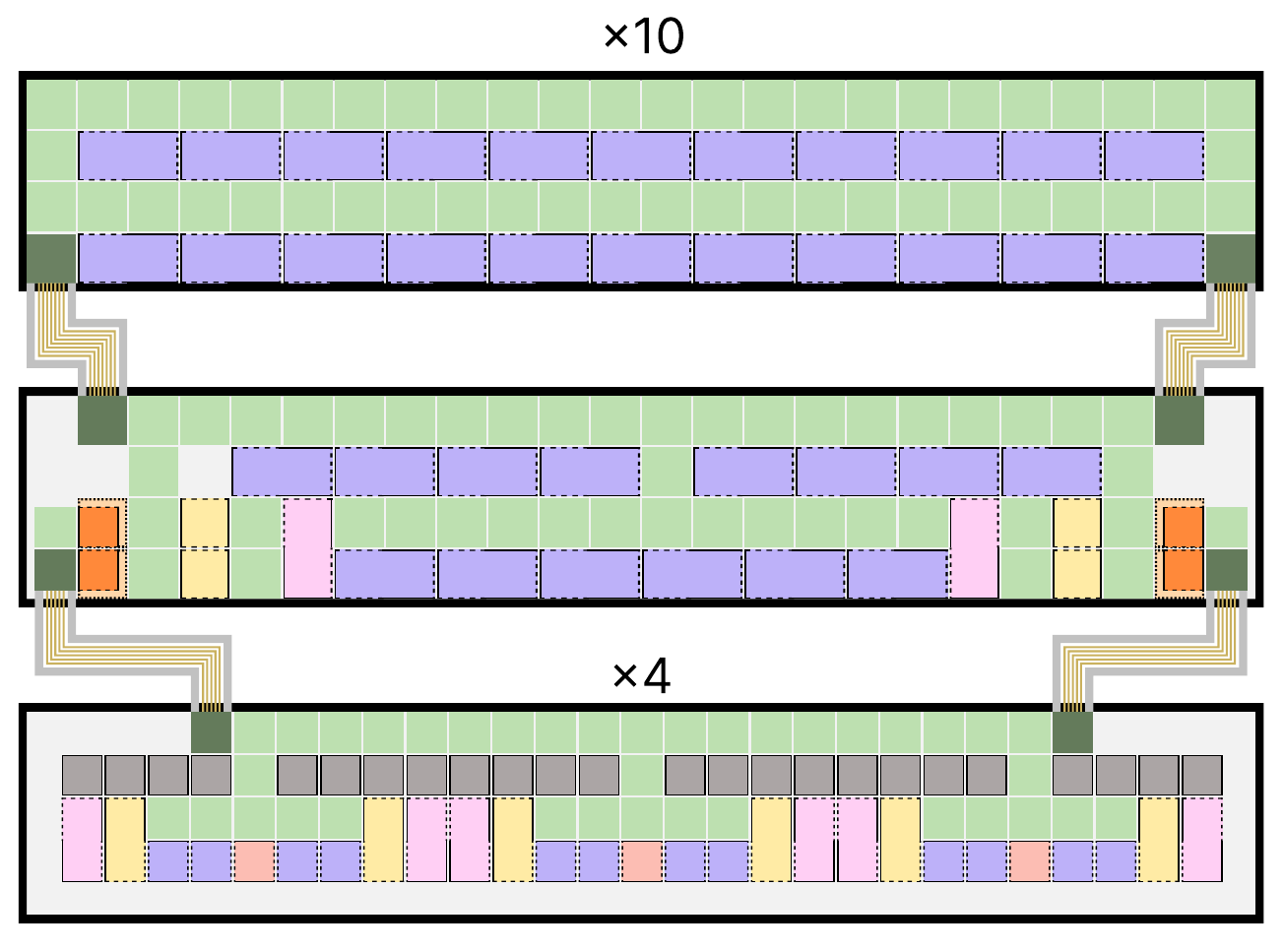}
    \caption{Example of an embedded multi-DR architecture for executing a quantum circuit associated with electronic-structure quantum simulation with a target error of 0.01 mHa based on qubitization of the $p$-benzyne molecule for an active space involving six molecular orbitals (see~\Cref{sec:resource-estimates-pbenzyne}).  
     Our estimates indicate that this circuit requires at least 11 distillation units in the MSF with a code distance of 21 to ensure a continuous supply of magic states to the core processor, which requires 341 data qubits with a code distance of 25. The architecture shown consists of 15 DRs, each containing 120,000 physical qubits. Four DRs are configured with three distillation units each, 10 DRs accommodate 33 data qubits each, and the remaining DR includes additional data qubits along with dedicated zones for magic state growth and storage. Multi-qubit lattice surgery is performed using the quantum bus, while magic states are transferred between DRs using as many optical interconnects as the code distances involved.}
    \label{fig:multi_dr_example}
\end{figure}

{\bf Impact of noisy optical interconnects.} 
To study the impact of both types of weak couplers described above, we have rerun the teleportation experiments of~\Cref{sec:other-FTQC-protocols} by incorporating columns of weaker CNOT gates (called ``cuts'') between the surface code patches as illustrated in \Cref{fig:long_bus_with_cuts}.
\begin{figure}
    \includegraphics[width=0.5\textwidth]{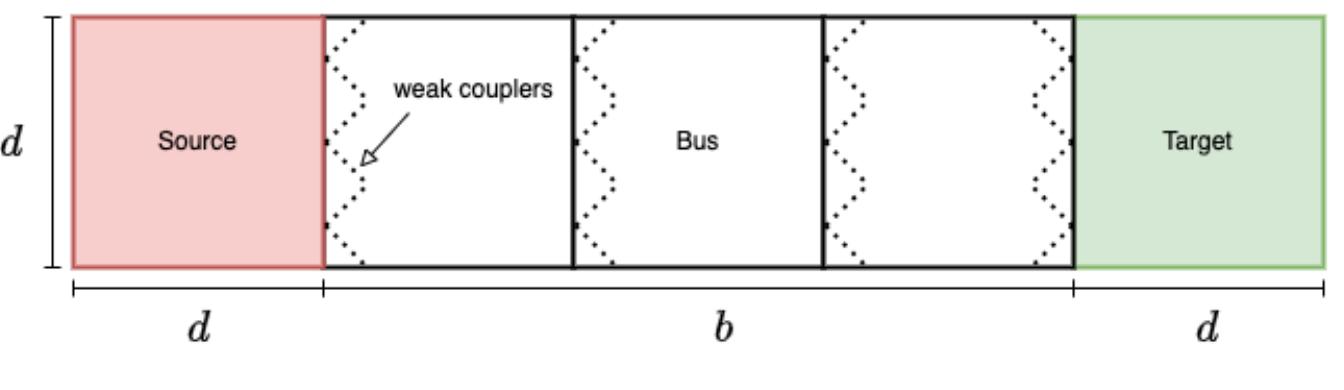}
   \caption{A time slice of the logical teleportation protocol via lattice surgery in a multi-DR distributed architecture where the cuts (weak couplers) are illustrated in broken lines and placed regularly along a bus of dimension $b \times d $.}
    \label{fig:long_bus_with_cuts}
\end{figure}
In Figs.~\ref{fig:dist_extrap_with_cut}(a) and (b), we show that using weaker CNOT gates of infidelity $p_{\text{link}}= 0.01$ and fixing all other coupler infidelities to the baseline value $p_{CX}= 0.003$ does not significantly affect the teleportation fidelity even for four cuts within the bus. However, much weaker interconnects can be problematic as shown in Figs.~\ref{fig:dist_extrap_with_cut}(c) and (d) where we observe a thresholding behavior at about $p_{\text{link}} \approx 0.06$. 

We conclude that distributed surface code architectures across multiple dilution refrigerators can tolerate two-qubit errors on the order of $1\%$ arising from noisy optical interconnects between the DRs. However, the logical performance rapidly deteriorates as these errors become significantly worse, and for error rates beyond $5\%$, achieving fault tolerance becomes problematic. Our numerical results are consistent with an analogous analysis performed for neutral-atom computers~\cite{sinclair2024faulttolerantopticalinterconnectsneutralatom}.

We note that the approach based on multi-DR lattice surgeries described in this section competes with the entanglement-distillation based quantum networking approach proposed in~Ref.~\cite{sunami2024scalablenetworkingneutralatomqubits}. In this approach, time-multiplexed remote physical Bell pairs are used to encode noisy logical Bell pairs by state injection, followed by entanglement distillation using logical operations on the code blocks to reach a desired error rate for the teleported state in the target QPU. However, the entanglement distillation units create a resource trade-off between the number of optical interconnects and the number of qubits used to complete the teleportation scheme in the target QPU. 

\begin{figure*}
    \centering
     \includegraphics[width =0.9\textwidth]{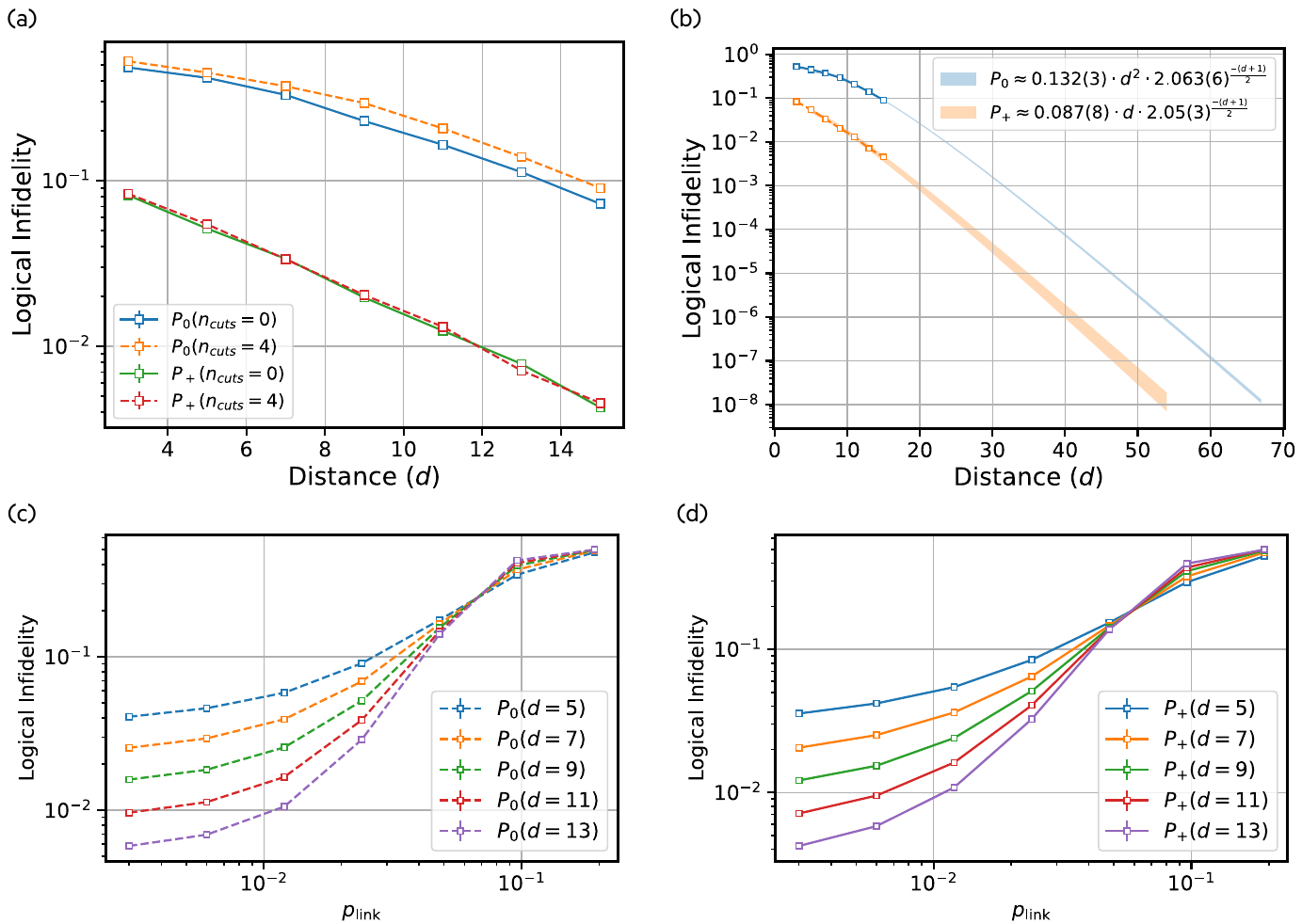}
    \caption{(a) Logical infidelity as a function of code distance for teleportation of $|+_L\rangle$ and $|0_L\rangle$ states with and without cuts for corresponding values of bus width $b = 3d$ and temporal code distance $r_\text{m} = 3d$. (b) Exponential fits for $P_0$ and $P_{+}$ for $n_{\text{cuts}} = 4$. The obtained $\mu$ and $\Lambda$ values are close to that of \Cref{fig:dist_and_rounds_scaling}(b). (c)--(d) Logical infidelity of teleporting $|0_L\rangle$ states and $|+_L\rangle$ states, respectively, as a function of the infidelity of the weak $CX$ gates in the cuts.}
    \label{fig:dist_extrap_with_cut}
\end{figure*}

{\bf Emulation of inter-QPU logical Bell-pair generation via quantum state transfer.} To illustrate the possible impacts of slower and noisier inter-QPU operations in a distributed FTQC setting, we consider a realistic implementation scheme based on a concrete realization of quantum interconnects. 
Quantum interconnects are typically used for the deployment of quantum entanglement, which then serves as a resource in implementing nonlocal two-qubit (or multi-qubit) quantum gates via state teleportation or gate teleportation~\cite{Barral2024cef}. At the physical level, various approaches exist for generating entangled states (such as Bell states) shared among remote stationary qubits of two or more distributed QPUs. For example, in neutral-atom quantum computing, a typical setup for heralded entanglement generation involves an atom--photon interface, an optical link, and a measurement scheme comprising a non-polarizing beam splitter and single-photon detectors. Such an atom--atom entanglement generation scheme is typically based on erasing \lq\lq{}which-path\rq\rq{} information of the photons emitted from two remote quantum network nodes, first proposed in Ref.~\cite{PhysRevA.71.060310} and recently analyzed for neutral-atom quantum computing~\cite{sunami2024scalablenetworkingneutralatomqubits,sinclair2024faulttolerantopticalinterconnectsneutralatom}. 

For superconducting qubit technologies, a viable approach for realizing quantum interconnects is by means of a superconducting transmission line designed to realize a {\em quantum state transfer} using a single-photon microwave pulse. An experimental demonstration of a tunable coupling scheme for coherent transfer of microwave pulses at the single-photon level with more than 99\% efficiency was first reported in Ref.~\cite{wenner2014}. A quantum state transfer between stationary qubits of two distributed QPUs that is mediated by bosonic modes typically relies on the principle of \mbox{time-reversal} symmetry between resonator energy absorption and emission and realizing a dark-state condition which guarantees destructive interference occurring between signals reflected off the couplers and re-emitted (leaked out) off the resonators. The original scheme for a deterministic quantum state transfer goes back to the proposal in Ref.~\cite{cirac1997quantumstatetransfer} in the context of quantum networking of distributed optical cavities. Several other works have analyzed deterministic quantum state transfer between stationary superconducting qubits using microwave photons in recent years (see, e.g., Refs.~\cite{Kurpiers_2018,Axline_2018}).

Mathematically, the ideal deterministic quantum state transfer between two qubits $q_1$ and $q_2$ 
can be effectively  described by the following  transformation (cf.~Ref.~\cite{cirac1997quantumstatetransfer}):
\begin{equation} \label{eq:quantum-state-transfer}
|\phi\rangle_1 \otimes |0\rangle_2\otimes|\text{vac}\rangle_{\text{bos}} \mapsto |0\rangle_1 \otimes |\phi\rangle_2 \otimes|\text{vac}\rangle_{\text{bos}},
\end{equation}
where $|\phi\rangle$ is any single-qubit state, and $|\text{vac}\rangle_{\text{bos}}$ represents the vacuum state of the collection of bosonic modes involved in realizing the state transmission channel. 
Its realization relies on intermediate transduction processes that convert the quantum state between the local qubits ($q_1$ and $q_2$) and the  \lq\lq{}flying qubit\rq\rq{} carried by a single-photon microwave pulse. After tracing out the bosonic degrees of freedom and disregarding their intermediate non-vacuum states used to mediate the state transmission,  the ideal quantum state transfer between $q_1$ and $q_2$ can be effectively achieved by the two-qubit unitary transformation  $U_{\text{ST}}=\left(H\otimes \mathds{1}\right)_{1,2} \text{CZ}_{1,2}\left(H\otimes \mathds{1}\right)_{1,2} \text{CX}_{1,2}$, which effectively results in swapping the states of qubits $q_1$ and $q_2$ if qubit $q_2$ is initialized in $|0\rangle_2$. 

Once a setup for (nearly) deterministic quantum state transfer interconnects has been established, it can be employed in two ways: (i) to generate physical Bell-pair states shared among distributed QPUs, which can subsequently be used to construct logical Bell pairs via state injection following the scheme described in Ref.~\cite{sunami2024scalablenetworkingneutralatomqubits}; or (ii) to directly generate inter-QPU logical Bell states through lattice surgery across two surface-code patches hosted on two distributed QPUs. Here, we focus on the second approach, providing a detailed description of its implementation and presenting the results of our emulation studies.

For the 2D surface code architectures considered in this work, an inter-QPU logical Bell-pair state can be generated by performing an inter-QPU $X\otimes X$ (or $Z\otimes Z$) lattice surgery involving two surface code patches (with the same code distance) 
placed on the boundaries of two distributed QPUs, as illustrated in \Cref{fig:Bell-Pair-generation_quantum-state-transfer}.  Each boundary syndrome qubit of the first patch in $\text{QPU}_1$ is quantum-interconnected via quantum state transfer with the corresponding boundary syndrome qubit of the second patch in $\text{QPU}_2$. Intra-QPU stabilizer measurements within each QPU are performed in the standard way at their regular speed. The stabilizer measurements that extend across the two QPUs are performed at a slower pace using the quantum interconnects. Once an inter-QPU logical Bell-pair state has been generated, it is consumed as a resource in implementing inter-QPU multi-qubit quantum gates, as illustrated in \Cref{fig:Bell-Pair-generation_quantum-state-transfer} for the case of an inter-QPU logical CNOT gate.

An inter-QPU $X$-type stabilizer measurement via quantum state transfer and the impact resulting from latencies associated with the quantum interconnect are illustrated in \Cref{fig:interQPU-stabilizer-measurement}. 
First, the two data qubits in $\text{QPU}_1$ interact with the syndrome qubit via intra-QPU two-qubit gates. Then, the state of that first syndrome qubit in $\text{QPU}_1$ is transferred (using a microwave photon pulse) to the corresponding second syndrome qubit in  $\text{QPU}_2$, which {\em effectively} results in swapping their states. Note, however, that the second syndrome qubit must be initialized in $|0\rangle $ prior to executing the state transfer. 
Finally, the other two data qubits located in $\text{QPU}_2$ interact with the second  syndrome qubit, and the syndrome qubit is measured. 
An inter-QPU $Z$-type stabilizer measurement can be performed in the same fashion. Typically, the two-qubit gates used for $X$-type and $Z$-type stabilizer measurements are oriented perpendicularly to each other, using an N-shaped (or \reflectbox{N}-shaped)  ordering of CNOT gates for $X$-type stabilizers and a Z-shaped (or \reflectbox{Z}-shaped) ordering for $Z$-type stabilizers, in order to avoid hook errors~\cite{litinski2018latticesurgery}. In our setting, however, this would lead to an additional $3\times$ slowdown of the inter-QPU stabilizer measurement, which would result in both a slowdown in the computational speed and an increase in the idling errors. Hence, we use the same two-qubit gate orientation for the inter-QPU $Z$-type stabilizer measurements as for the $X$-type ones.

Such inter-QPU stabilizer measurements will typically be slower and noisier than regular intra-QPU stabilizer measurements, 
due to latencies and the additional noise and errors introduced by implementing the quantum state transfer interconnect. 
To analyze how the computation time  and logical error rates will be affected by these  additional latencies and sources of errors 
associated with the quantum interconnect, we have conducted detailed emulation studies using the {\em target} hardware specifications from \Cref{tab:physical_params}. For our emulation studies, we assumed that 
the effective SWAP operation 
resulting from the quantum state transfer is 10 times slower and 10 times more erroneous than an intra-QPU two-qubit gate.

\begin{figure}[t]
    \centering
    \begin{tabular}{c }
    \text{(a) Example two-DR distributed architecture}\\
    \includegraphics[width=0.79\linewidth]{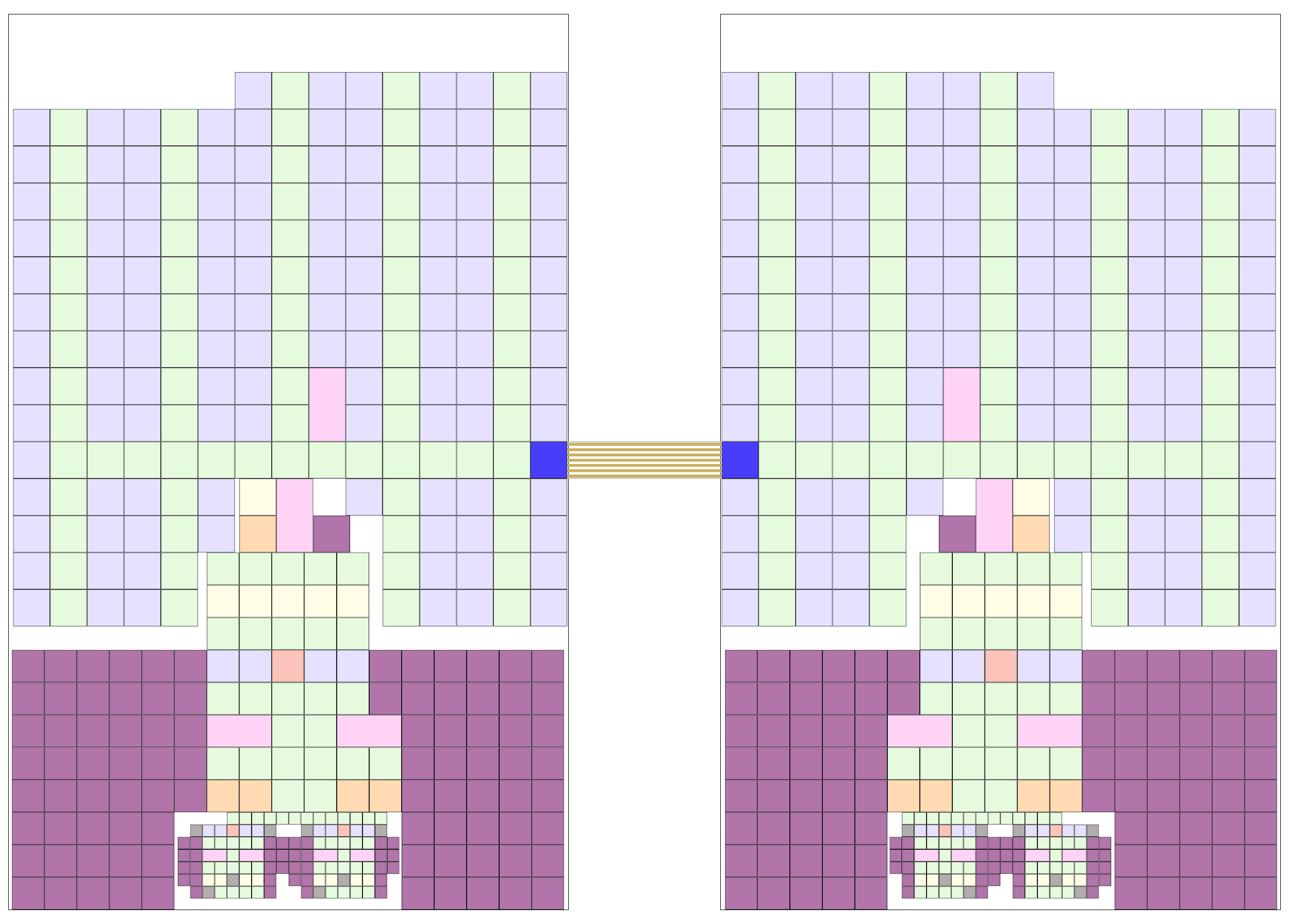}\\
    \\
    \text{(b) Inter-QPU CNOT between qubits $Q_1$ and $Q_2$}\\
    
    \includegraphics[width=0.61\linewidth]{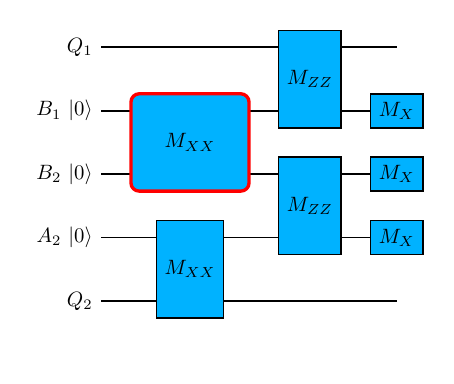}\\
     \text{(c) Inter–QPU logical Bell-pair generation}\\
    
    \includegraphics[width=0.63\linewidth]{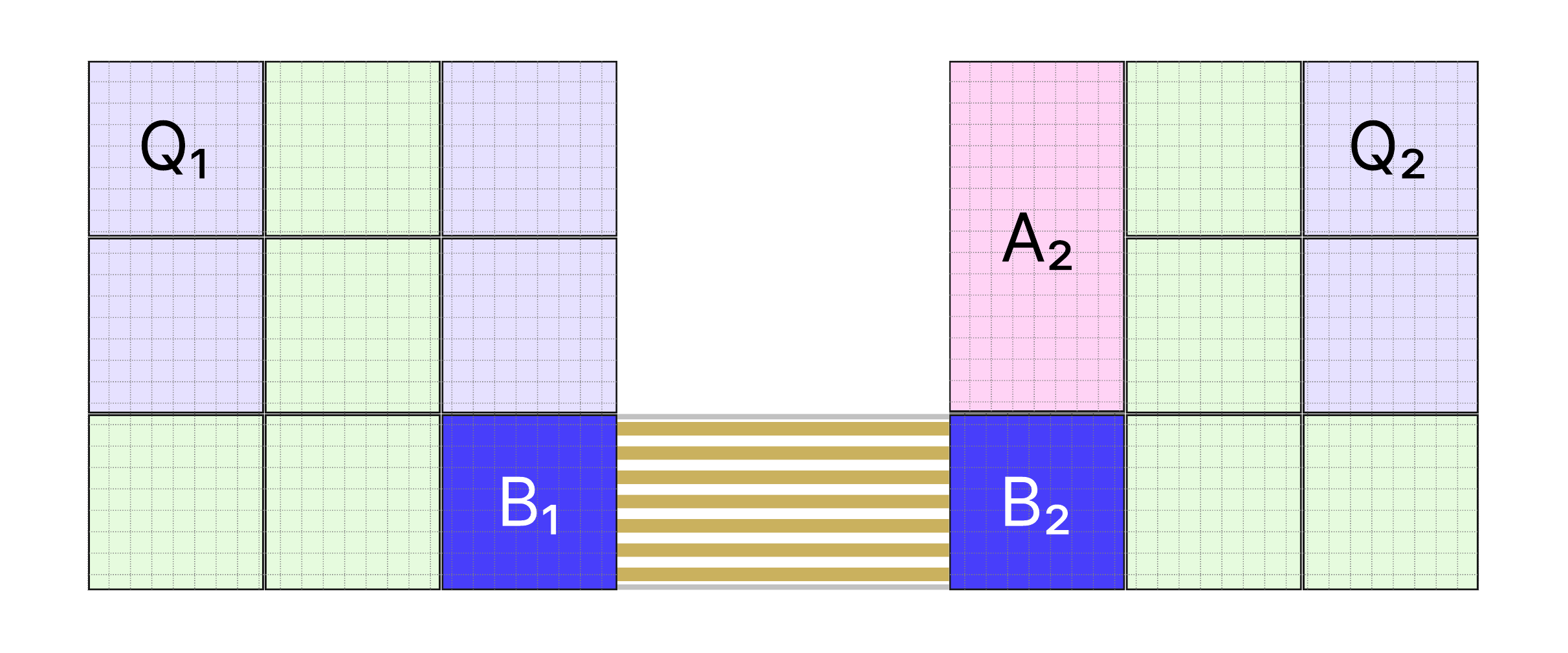}
    \end{tabular}
\caption{Illustration of an inter-QPU CNOT implementation for a distributed architecture consisting of two DRs with quantum interconnects. (a) Example of an embedded two-DR architecture for executing a quantum circuit associated with a qubitized quantum simulation of the 2D Fermi--Hubbard model with $|U/J|=8$ and lattice size $10\times 10$; see \Cref{subsec:FTQC_FH2d}. (b) Quantum circuit for implementing an inter-QPU CNOT gate between two logical qubits $Q_1$ and $Q_2$ 
remotely located in two DRs (hosting QPU$_1$ and QPU$_2)$ and acting as control and target qubits, respectively. The inter-QPU CNOT is implemented using inter-QPU and intra-QPU lattice surgeries; see  Ref.~\cite{fowler2019lowoverheadquantumcomputation} for a similar circuit implementing a CNOT using two lattice
surgeries and one ancilla. Here, the inter-QPU CNOT gate is realized by generating a  logical Bell pair across the two QPUs followed by consuming it using  intra-QPU lattice surgeries involving an additional ancilla $A_2$ in QPU$_2$ that is initialized in $|0\rangle$. The inter-QPU logical Bell pair can be created via an $X\otimes X$ lattice surgery between ancilla qubits $B_1$ and $B_2$ both initialized in $|0\rangle$. (c) Inter-QPU logical Bell-pair generation via quantum state transfer interconnects between the two surface code patches representing $B_1$ and $B_2$. The required number of physical quantum interconnects is equal
to the code distance of the surface code patches used to form a Bell pair.  
    }
    \label{fig:Bell-Pair-generation_quantum-state-transfer}
\end{figure}

\begin{figure}[bt]
    \centering
    \includegraphics[width=1.0\linewidth]{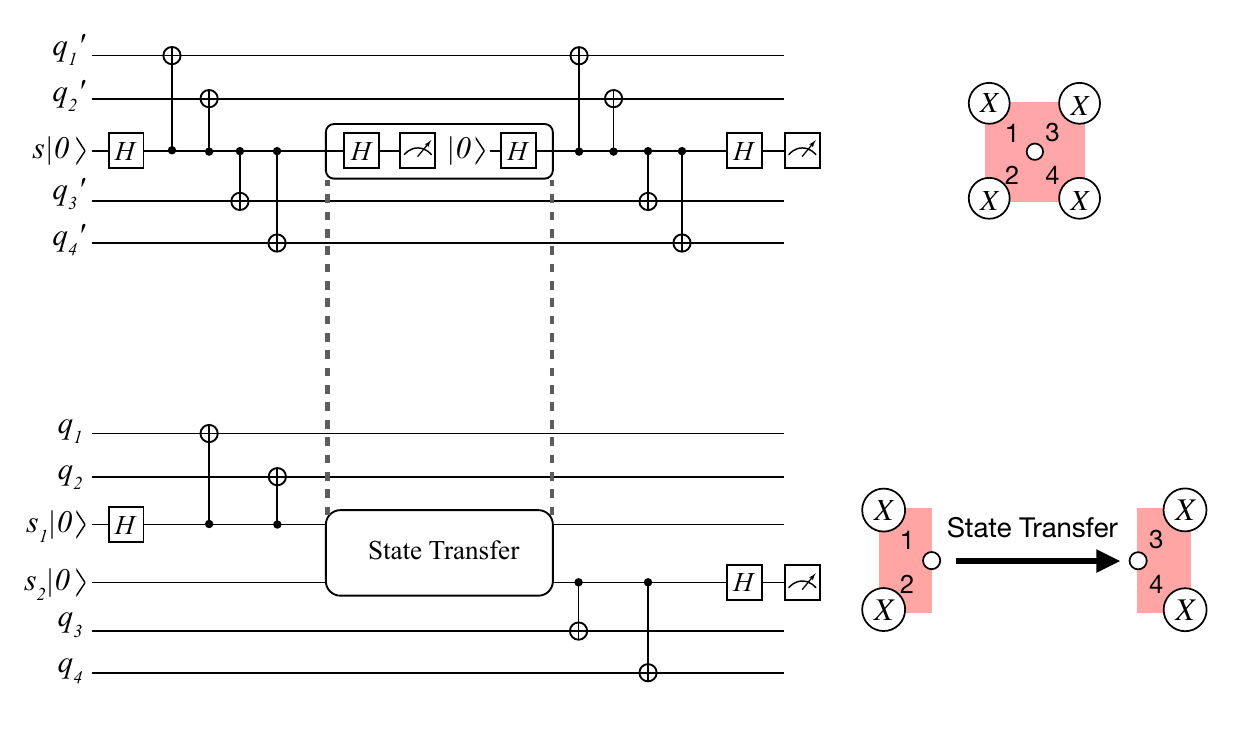}\\
\caption{Illustration of an inter-QPU stabilizer measurement via quantum state transfer, and of the impact resulting from latencies associated with the quantum interconnect. 
To implement an inter-QPU $X$-type stabilizer measurement across two QPUs (bottom circuit), two data qubits $q_1$ and $q_2$ interact with a  syndrome qubit $s_1$ (initialized in $|0\rangle$, and subsequently prepared in $|+\rangle$) at the boundary of $\text{QPU}_1$. The quantum state of syndrome qubit $s_1$ is transferred via the state transfer channel to the corresponding second syndrome qubit $s_2$ in $\text{QPU}_2$, where $s_2$ must be initialized in $|0\rangle$. Two other data qubits $q_3$ and $q_4$ located in $\text{QPU}_2$ interact with that second  syndrome qubit, and $s_2$ is measured in the $X$ basis. Such an inter-QPU stabilizer measurement will typically be slower than a regular intra-QPU stabilizer measurement (top circuit) due to latencies introduced by the quantum state transfer channel. Assuming the effective SWAP operation 
resulting from the quantum state transfer between the two syndrome qubits to be 10 times slower than an intra-QPU two-qubit gate, its execution can be scheduled concurrently with a measurement and a reset of a syndrome qubit (along with two Hadamard gates acting on it) for intra-QPU stabilization rounds, as indicated by the dashed-line window.  Thus, the inter-QPU stabilization (parity check) time is only twice as long as the intra-QPU stabilization time.}
    \label{fig:interQPU-stabilizer-measurement}
\end{figure}

To minimize the impact of the slower inter-QPU state transfer operation, its execution should be scheduled in parallel to intra-QPU operations with a similar time-scale. For example, its execution can be scheduled concurrently with the measurement and reset operations (along with Hadamards) for intra-QPU stabilization rounds (as illustrated in  \Cref{fig:interQPU-stabilizer-measurement}), as these are typically slower than quantum gates for superconducting qubits. Indeed, for the {\em target} hardware parameter specifications (see \Cref{tab:physical_params}) used in our emulation studies, both measurement and reset times are four times longer than two-qubit gate times. Along with two Hadamards, measurement and reset together amount to an execution time equivalent to that of the quantum state transfer operation, namely, 10 times the execution time for single- and two-qubit gates. As a result, the inter-QPU stabilization (parity check) time is only twice as long as the intra-QPU stabilization time. 
Thus, slower quantum state transfer interconnect operations 
do not significantly slow down the computation. In particular, 
the computation speed may remain completely unaffected, especially when taking into consideration the significantly larger impacts associated with decoder latencies that have been recently analyzed and reported~\cite{khalid2025impacts}.

To study the impact of noisier and more erroneous \mbox{inter-QPU} state transfer operations 
on the  logical error rate of logical Bell-pair generation, we emulate the inter-QPU $X\otimes X$ lattice surgery  (based on state transfer interconnects) for varying spatial code distances $d$, and a varying number of bus stabilization rounds  $r$, for a bus width $b=1$. Here, stabilization of the bus plaquettes refers to stabilizer measurements effecting 
the suppression of logical errors associated with the two-qubit logical parity measurement of the $X\otimes X$ observable in implementing the lattice surgery. 
Each bus stabilization round consists of executing both $X$- and $Z$-type stabilizer measurements across the bus, conducting one parity check for each plaquette type, respectively. 
Our emulations are based on the circuit-level noise model detailed in~\Cref{sec:Circuit-Level-Noise-Model}, where the effective SWAP operation 
resulting from the microwave-pulse quantum state transfer is assigned a depolarizing noise rate that is 10 times higher than that of intra-QPU two-qubit gates (i.e., $p_\text{MW} = 10p_{CX}$).

Our validation strategy is as follows. The ideal outcome of the $X\otimes X$ lattice surgery in the absence of errors is the logical Bell state $\left(\mathds{1} + (-1)^m X\otimes X)|00\rangle\right)/\sqrt{2} = \left(|00\rangle + (-1)^m |11\rangle\right)/\sqrt{2}$, where $m$ is the lattice surgery measurement outcome, obtained as the product of all $X$ stabilizers in the bus. In our simulations, to validate whether we have obtained the correct Bell state, we respectively measure each patch in the logical $X$ basis, that is, we perform a measurement of the logical operator $X_{L, 1}$ on the logical patch associated with ancilla qubit $B_1$ (see \Cref{fig:Bell-Pair-generation_quantum-state-transfer}), yielding a measurement outcome $m_1$, and we perform a measurement of the logical operator $X_{L, 2}$ on the logical patch associated with ancilla qubit $B_2$ yielding a measurement outcome $m_2$. It is easy to see that, in the absence of errors, if the outcome of the lattice surgery is $m=0$, then either 
$(m_1,m_2)=(0,0)$ or $(m_1,m_2)=(1,1)$, and if the lattice surgery outcome is $m=1$, then either 
$(m_1,m_2)=(1,0)$ or $(m_1,m_2)=(0,1)$. In all four cases, we obtain $m \oplus m_1 \oplus m_2 = 0 $ \mbox{(mod 2)}. This relation may no longer hold if logical errors occur. It is straightforward to show that logical $X_{L}$ errors on either of the two patches do not have a detectable impact, because the outcome $m \oplus m_1 \oplus m_2 = 0 $ remains unaffected.  However, if the logical error $Z_{L}$ occurs on either of the two patches, we obtain $m \oplus m_1 \oplus m_2 = 1$. Yet, logical $Z$ errors are extremely unlikely during the merging of the two code patches, because the associated error chain is of length $2d$. 
However, logical $Z$ errors may also occur with a higher probability during the preparation of the initial state, or during the measurement of the final state. Because the lattice surgery starts with the $|00\rangle$ state, the occurrence of a logical $Z$ error during preparation has no impact. It is thus only during the final measurement stage that a logical $Z$ error has a detectable significant impact. The area of the boundaries of these shortest-length $Z$-type errors grows as $\mathcal{O}(d)$ (cf. Ref.~\cite{silva2024optimizing}). Time-like errors occur when the values of the stabilizers used to infer the lattice surgery outcome are incorrectly determined. The degeneracy of the shortest-length undetectable time-like errors scales with the area $bd$ of the measured bus, that is, as $\mathcal{O}(d)$ for a bus width $b = 1$.

Assuming unbiased physical $X$ and $Z$ errors, we can infer the following predictive model for the logical error of the logical Bell state from the area laws associated with shortest-length undetectable $X$- and $Z$-type errors and time-like errors (cf. Ref.~\cite{silva2024optimizing}):
\begin{equation} \label{eq:error-model_logical-Bell_XX-Surgery}
    \mu_{S} \left(2d +3\right) d \,\Lambda_{S}^{-(d+1)/2} + \mu_T d \,\Lambda_T^{-(r +1)/2}.
\end{equation} 
Because logical $X$ errors are not detectable with our validation strategy as discussed above, we consider only the $\mu_S (2d)\Lambda_S^{-(d+1)/2}$ (associated with the logical $Z$ errors) and $\mu_T d\Lambda_T^{-(r+1)/2}$ (associated with time-like errors) terms. We independently regress these predictive models from the
obtained numerical results.

Our emulation and regression results are presented in \Cref{fig:distance_rounds_scaling}. These results indicate 
that a noisier quantum interconnect does not significantly affect the space-like logical error rate suppression compared to the case where the quantum interconnect noise is equivalent to that of a standard intra-QPU two-qubit gate operation. However, a noisier interconnect substantially impacts the time-like logical error rate suppression. Consequently, for a fixed code distance, reducing quantum interconnect noise would significantly decrease the number of stabilization rounds required to achieve a given logical error rate. The numerically inferred overall logical error according to \Cref{eq:error-model_logical-Bell_XX-Surgery} also allows us to find the  optimal number of bus stabilization rounds from the expression $r\approx\log \Lambda_S/\log \Lambda_T$, which yields $r\approx 1.05d$.

\begin{figure*}[t]
    \centering
    \centering
    \begin{tabular}{c @{\hskip 0.37in} c}
    \text{(a) Logical error scaling with respect to code distance}
    & \text{(b) Logical error scaling with respect to bus rounds} \\
    \includegraphics[width=0.437\linewidth]{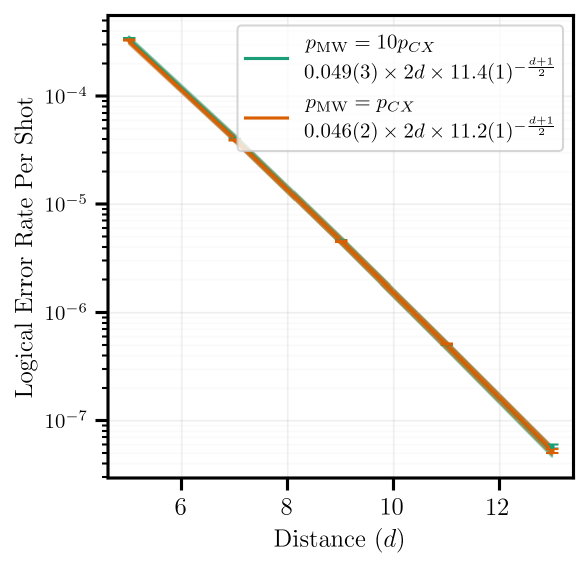}
    & \includegraphics[width=0.437\linewidth]{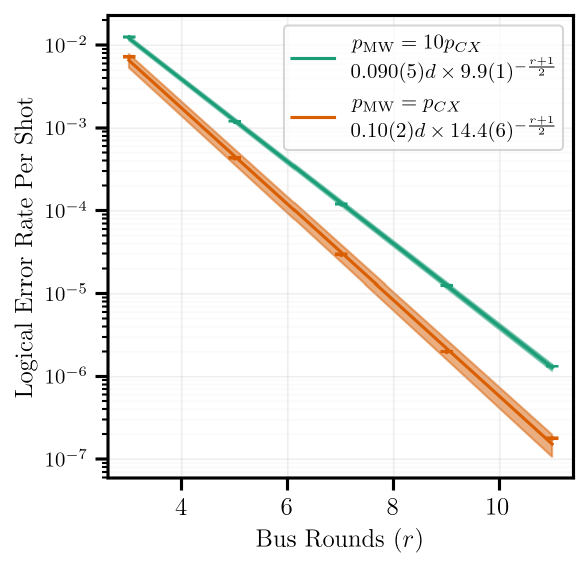}
    \end{tabular}
\caption{Logical error scaling with respect to (a) code distance $d$ and (b) the number of stabilization rounds $r$ of the bus plaquettes for inter-QPU logical Bell-pair generation based on inter-QPU lattice surgery by exploiting quantum state transfer interconnects. Here, stabilization of the bus plaquettes refers to the suppression of  logical errors associated with the two-qubit parity measurement of the logical $X\otimes X$ observable resulting from  the lattice surgery process, for a bus width $b=1$.  The green lines represent the logical error rate scaling when the effective inter-QPU SWAP operations realized via the quantum state transfer interconnects between two QPUs are modeled to be 10 times noisier and 10 times slower than an intra-QPU two-qubit gate operation, while the orange lines represent the logical error rate scaling for 10 times slower inter-QPU SWAP operations but using equivalent noise levels for intra-QPU and inter-QPU two-qubit operations. Here, $p_{CX}$ represents the intra-QPU CNOT operation error rate, and $p_\text{MW}$ represents the inter-QPU operation error rate resulting from microwave-pulse interconnect. For our emulations, we assumed the {\em target} hardware specifications from \Cref{tab:physical_params}. The legends indicate the corresponding fitting parameters and their comparison for each case. The shaded regions indicate the uncertainty in the fit for each line.}
    \label{fig:distance_rounds_scaling}
\end{figure*}

\section{Resource estimation for utility-scale quantum applications}
\label{sec:resource-estimates-pbenzyne}

The ultimate goal of developing a utility-scale quantum computer (USQC) is to deliver valuable computations that justify the device’s cost. Achieving this goal for any hardware–software architecture requires satisfying several criteria: (1) the existence of well-defined, real-world  computational applications with measurable value; (2) verification that the total cost of ownership of a quantum computer is lower than the value it generates for at least a subset of these applications; and (3) assurance that no alternative approaches---classical, analog, quantum-inspired, or otherwise---can execute these applications more cost-effectively. Evaluating proposals for a USQC is inherently challenging, as it requires simultaneous considerations of technical feasibility, domain-specific use cases, device architecture and system integration, and the full spectrum of available computational methods.

To assess the requirements of a USQC, it is essential to conduct detailed physical-level {\em quantum resource estimation} (QRE). Conventional algorithmic analyses based on big-O complexity characterize algorithmic efficiency only in terms of the asymptotic leading-order behavior and  therefore fail to provide a detailed accounting of the concrete resources required for a specific problem instance and target accuracy. As a result, rigorous QRE analyses are critical for evaluating the practicality, feasibility, and cost of implementing quantum algorithms for real-world applications at utility scale. 
}

In this section, we examine how attaining an improved quality of physical qubits and gate operations  affects the overall resource requirements of FTQC for selected real-world applications. 
Our numerical QREs aim to compare the concrete FTQC resource requirements for the three sets of hardware parameter specifications summarized in \Cref{tab:physical_params}.  These QRE studies were conducted using automated tools of TopQAD~\cite{1qbit2024topqad,silva2024lattice,silva2024optimizing} and AzureQRE~\cite{microsoft2022azureQRE}.

\subsection{Quantum computation of electronic spectra as a representative high-utility application}

One of the most promising applications of quantum computing is solving quantum chemistry problems. An important representative computational task in quantum chemistry is estimating ground-state energies of molecules. While this task is classically tractable for small molecules using advanced classical algorithms developed in the traditional quantum chemistry~\cite{szabo1996modern}, electronic-structure simulations of large molecules are widely considered to be intractable for classical computers. Here, for the purpose of demonstrating the practicality of solving such problems on a quantum computer, we present QRE studies of electronic-structure quantum simulations for two molecules of high practical interest. The first molecule analyzed is the biradical \mbox{{\em para}-benzyne} molecule (\mbox{$p$-benzyne}), which has the molecular formula $\mbox{C}_6\mbox{H}_4$. Its energetically lowest configuration is formed by a singlet biradical. Among other applications, its reactivity has the potential to play an important role in the design of enediyne drugs with high antitumour or anticancer
properties~\cite{crawford2001problematic,kraka2000computer}. 
The second molecule analyzed is the iron-molybdenum cofactor (FeMoco) of nitrogenase, which acts as a crucial catalyst in the process of biological nitrogen fixation. This molecule is among several that have been used as representative targets for future quantum simulators in several recent  works~\cite{reiher2017elucidating,motta2021low,li2019the,lee2021even,otten2023qrechem}.

For our QRE studies, we first generate the logical quantum circuits associated with electronic-structure simulations of the \mbox{$p$-benzyne} and FeMoco molecules. More concretely, we analyze the resource requirements associated with estimating the energy of the ground states of these molecules using the well-established quantum phase estimation (QPE) algorithm~\cite{kitaev1995quantummeasurementsabelianstabilizer,nielsen2010quantum}. For the \mbox{$p$-benzyne} molecule, we analyze the singlet ground state using a variety of active spaces that are specified below; for the FeMoco molecule, we analyze the active-space model proposed in Ref.~\cite{li2019the}. 

Our studies rely on electronic-structure simulations in the second quantization framework of quantum theory. Numerous software tools exist to derive the second-quantized Hamiltonian from a molecule's specifications that fully characterize the quantum system. Basic molecular specifications include the types of atoms that constitute the molecule and the molecule's geometry (typically summarized in an \texttt{xyz} file), total charge, and the total spin. In addition, a basis set $\{\phi_\alpha(\boldsymbol{x})\}$ must be selected to represent the fermionic orbitals, which in the language of second quantization are occupied or unoccupied, represented by occupation number states and fermionic creation and annihilation operators ($\hat{a}_\alpha$ and $\hat{a}^\dagger_\alpha$) acting upon them. Furthermore, to reduce the computational cost, a common approach is to restrict computations to a reasonably chosen active space involving only a subset of the chosen orbital basis set.  The model Hamiltonian associated with an active space is translated from the second quantization framework to a framework suitable for the quantum circuit model. This fermion-to-qubit mapping is typically accomplished via either the \mbox{Jordan--Wigner}~\cite{Jordan1993} or the Bravyi--Kitaev~\cite{BRAVYI2002210} transformations. To derive the model Hamiltonians for the various active spaces associated with \mbox{$p$-benzyne}, we used Tangelo~\cite{tangelo}, an open source Python software package for end-to-end chemistry workflows. For the FeMoco molecule, we used the \texttt{FCIDUMP} file provided as part of the data and code repository~\cite{FCIDUMP_FeMoco} of Ref.~\cite{lee2021even}. 

The standard QPE algorithm~\cite{nielsen2010quantum} samples in the eigenbasis of the molecular Hamiltonian $H$ by measuring the phase that is accumulated on an initial input quantum state through multiple controlled executions of the time-evolution operator $\exp(-iHt)$. Its most resource-intensive part consists in implementing the unitary $\exp(-iHt)$ by a quantum circuit, along with repeating this circuit a number of times that scales as $\mathcal{O}(1/\epsilon)$ for an allowable target error $\epsilon$ in phase estimation. An alternative approach to sampling the spectrum of the molecular Hamiltonian $H$ via phase estimation is based on the framework of qubitization~\cite{low2019hamiltonian}. Indeed, most of the recent QRE studies on electronic-structure quantum simulations rely on the qubitization framework (see, e.g., Refs.~\cite{berry2019qubitization,vonBurg2021quantum,lee2021even,goings2022reliably,beverland2022assessing, otten2024quantum,bellonzi2024feasibility,watts2024fullerene,nguyen2024quantum}). This approach uses a new operation called {\em qubiterate} that is akin to the quantum walk operator $e^{i\arccos(H/\lambda)}$ (where $\lambda$ is typically the sum of the absolute values of the weightings in the molecular Hamiltonian). Since the qubiterate's eigenvalue spectrum can be obtained from that of the unitary $\exp(-iHt)$ via an $\arccos$ transformation, the former can be used in QPE in place of the latter, as proposed in Refs.~\cite{berry2018improved,poulin2018quantum}. An advantage of this approach is that steps of a quantum walk can be implemented exactly, assuming access to arbitrary single-qubit rotations, in contrast to all approaches that are based on Hamiltonian simulation of the time-evolution operator $\exp(-iHt)$, which can only be approximated.

Moreover, aiming to reduce algorithmic complexity, the majority of recent literature on electronic-structure quantum simulations and the associated resource requirements has focused on combining the technique of qubitization applied to molecular systems with various tensor factorization techniques for the Coulomb operator. State-of-the-art algorithms of this type include the single low-rank factorization algorithm of Berry et al.~\cite{berry2019qubitization}, the double low-rank factorization algorithm of von Burg et al.~\cite{vonBurg2021quantum}, and the tensor hypercontraction algorithm of Lee et al.~\cite{lee2021even}, resulting in a continual improvement of the $T$-gate or Toffoli-gate complexity from $\mathcal{O}\left(N^5/\epsilon^{3/2}\right)$ for the Trotter-based approach to $\mathcal{O}\left(N\lambda/\epsilon\right)$, where $\lambda$ is the 1-norm of Hamiltonian coefficients which typically has a scaling between $\mathcal{O}\left(N\right)$ and $\mathcal{O}\left(N^3\right)$, and $\epsilon$ specifies the target error in ground-state energy estimation.

\subsection{Quantum resource estimation for \mbox{$p$-benzyne} and FeMoco}
\label{sec: QPE mole resource}

In this section, we present QRE studies for two algorithmic approaches: the Trotter-based approach, and the qubitization-based double low-rank factorization algorithm originally proposed in Ref.~\cite{vonBurg2021quantum} and further analyzed in Ref.~\cite{lee2021even}. Moreover, for the Trotter-based algorithm, we report QRE results for two methods to ensure quantum computations within a target precision (for a discussion in greater detail, see \Cref{sec:error-analysis}): the first approach is based on using rigorous analytic bounds on the errors resulting from the use of Trotter--Suzuki approximation and Trotterization, thus yielding a worst-case number of Trotter slices; the second approach relies on more-realistic Trotter numbers obtained through extrapolation from empirical studies of the Trotter--Suzuki errors for small circuits. In \Cref{sec:logical_circuit,sec:error-analysis,sec:error-analysis-qubitization}, we elaborate on the workflow for generating the associated logic circuits and analyze the various bounds on the errors incurred in this process, as well as how we choose these bounds to ensure that quantum simulations achieve a given target accuracy. 

We report estimates for concrete physical resources required for a fault-tolerant implementation of the QPE algorithm for the \mbox{$p$-benzyne} and FeMoco molecules based on either the second-order Trotter--Suzuki formula used to approximate $\exp(-iHt)$ for the molecular Hamiltonian or the double-factorized (DF) qubitization algorithm of von Burg et al.~\cite{vonBurg2021quantum}. In both cases, we assume access to a quantum state with significant overlap with the ground state as input to QPE, for example, a Hartree--Fock state. We do not include the cost of preparing this initial state in our resource estimations. It is worth emphasizing, however, that QPE is only provably fast for problems when the initial state is an eigenstate. When it is not an eigenstate of the Hamiltonian, there is a sampling overhead because the initial state is a superposition of eigenstates. This challenge is problem-dependent but can be ameliorated by classical preprocessing, that is, by running calculations on a classical computer to generate a better initial state which is then loaded into the quantum computer with a short quantum circuit. For example, one recent work~\cite{berry2024rapidinitialstatepreparation} presents an estimate that by using a matrix product state of bond dimension 4000, an overlap of 0.96 can be obtained for the ground state of the FeMoco molecule by implementing a circuit composed of nearly $10^9$ Toffoli gate. 

The overall cost (in terms of, e.g., $T$-gate or Toffoli-gate count) of implementing the QPE algorithm can be bounded by (see Ref.~\cite{babbush2018encoding})
\begin{equation}
    \mathcal{O}\left(\frac{g(\epsilon_{\text{\tiny QPE}})\Omega}{\epsilon_{\text{\tiny QPE}}}\|W'(H)\|^{-1}\right).
\end{equation}
Here, $\epsilon_{\text{\tiny QPE}}$ is the desired error tolerance in phase estimation; $W(H)$ represents the unitary operator (as a function of the Hamiltonian $H$) used in the QPE algorithm (e.g., $W(H)=\exp(-iH\tau)$ in the case of Hamiltonian simulation for some time $\tau$, or $W(H)=e^{i\arccos(H/\lambda)}$ in the case of qubitization); $\Omega$ denotes the cost of  a primitive circuit used to realize the implementation of $W(H)$ (such as a Trotter step in the Trotterization approach, or the LCU oracles associated with qubitization); and $g(\epsilon_{\text{\tiny QPE}})$ denotes the number of times that the primitive circuit must be repeated to ensure that the error in the spectrum of $H$ resulting from phase estimation of the eigenphases of $W(H)$ is at most $\epsilon_{\text{\tiny QPE}}$. Note that, in the qubitization approach, the operator $W(H)=e^{i\arccos(H/\lambda)}$ can typically be implemented as a quantum circuit exactly  without approximations beyond those required for the synthesis of arbitrary-angle rotation gates; this implies
$g(\epsilon_{\text{\tiny QPE}})=\mathcal{O}\left(1\right)$. 
Hence, due to $\|W'(H)\|^{-1}\le \lambda $, the overall cost of QPE in qubitization-based approaches becomes $\mathcal{O}\left(\Omega\lambda/\epsilon_{\text{\tiny QPE}}\right)$. Various versions of QPE  have been analyzed in the literature, aiming to reduce its cost. For example, the standard QPE algorithm~\cite{nielsen2010quantum} allows the estimation of eigenvalues  within a target error $\epsilon_{\text{QPE}}$ with probability at least $1/2$ using $ \lceil 16\pi/\epsilon_{\text{QPE}}\rceil $ applications of the unitary  $\exp(-iHt)$. However, more optimized QPE strategies  can achieve the multiplying factor (see Refs.~\cite{reiher2017elucidating,babbush2018encoding,lee2021even})
\begin{equation}
 M:=\lceil \pi/(2\epsilon_{\text{QPE}})\rceil.
 \label{eq:multiplying-factor_QPE}
\end{equation} 
The QRE analyses presented in this section are based on using this repetition factor. For example, 
the gate cost of the qubitization-based QPE algorithm is computed as $M\lambda \Omega$, where $\Omega$ comprises the costs of the LCU oracles \textsc{select} and \textsc{prepare}. 

In~\Cref{tab:resource_estimates_logical_trotter,tab:resource_estimates_physical_trotter}, we summarize our logical and physical QRE results for a number of circuits specified by the active space with sizes  characterized by the number of orbitals $N_{\text{orb}}$, the number of logical qubits involved in the computation, and the overall allowable target error for QPE. 
To achieve chemical significance, the overall target error in ground-state energy estimation should be at least within ``chemical accuracy'', that is, 
$\epsilon \le 1.6 
 \text{ mHa}$ (see, e.g., Ref.~\cite{RevModPhys.92.015003}).
Energy estimations within chemical accuracy are often sufficient to predict important chemical properties such as chemical reaction rates, but even higher accuracies may be required for quantitatively precise predictions. 
Here we report resource estimates for two different precisions specified by either the allowable target error $\epsilon = 1.6\text{ mHa}$ (for qualitative accuracy) or the much lower target error $\epsilon = 0.1\text{ mHa}$ (for quantitative accuracy), using a circuit-level error budget of 0.01, respectively. 
The chemical basis set \texttt{6-31G} is used to represent the spin orbitals in the case of \mbox{$p$-benzyne}, while for the FeMoco molecule we use the active-space model proposed by Li et al.~\cite{li2019the}.

In the case of the Trotter-based approach, physical runtimes for a complete implementation of QPE are obtained by multiplying the physical runtime for a single Trotter slice by the number of Trotter slices, and then by the number of controlled applications in QPE given by the value of $M$ in \Cref{eq:multiplying-factor_QPE}. For the overall error budget of $\epsilon=0.1$ mHa, we obtain a value of $\epsilon_{\text{QPE}}=0.065$ mHa as an optimal choice (see \Cref{sec:error-analysis}), yielding $M=$ 24,167; for the overall target error $\epsilon=1.6$ mHa, we use  $\epsilon_{\text{QPE}}=1.04$ mHa, yielding $M=$ 1511. The error budget allocation in the qubitization approach is discussed in \Cref{sec:error-analysis-qubitization}. 

For FTQC, the critical figure of merit characterizing the cost of running a quantum algorithm is the number of non-Clifford $T$ gates. In \Cref{tab:resource_estimates_logical_trotter}, the number of the $T$ gates resulting from circuit synthesis and decomposition  over the Clifford+$T$ gate set is reported for each circuit. Efficient circuit synthesis tools to compute approximations of arbitrary-angle single-qubit $Z$-rotations over the Clifford+$T$ gate set include the well-established Solovay--Kitaev (SK) decomposition that has a $T$-count scaling of $\mathcal{O}\left(\log^c(1/\varepsilon)\right)$, with the exponent $c>3$, and the software package gridsynth~\cite{gridsynth} based on the algorithm by Ross and Selinger~\cite{selinger2015efficient,ross2015optimal} achieving $T$-gate counts that are typically on the order of $4\log_2(1/\varepsilon)+\mathcal{O}\left(\log(\log(1/\varepsilon))\right)$, for a given allowable per-gate synthesis error $\varepsilon$. We use the latter method in our QRE studies due to its superior scaling.

As explained in Refs.~\cite{litinski2019game,silva2024lattice,silva2024optimizing}, Clifford operations can be efficiently commuted to the end of the logic circuit by tracking a Clifford frame along the circuit. Some of the resulting non-Clifford gates can be merged into Clifford gates. Therefore, the process may be repeated until convergence. We call this procedure {\em transpilation} as explained in \Cref{sec:topqad-compiler}. The outcome of transpilation is a sequence of non-Clifford gates in the form of multi-qubit $\pi/8$ Pauli rotations that must be executed using magic state injection. Therefore, the design of a fault-tolerant architecture that efficiently implements a given quantum algorithm reduces to constructing magic state factories (MSF) that can distill magic states of a target distance and fidelity at a rate on par with the fault-tolerant execution of non-Clifford gates. More details about the design of MSFs and the additional components of the layout are provided in \Cref{sec:topqad-assembly}. In \Cref{tab:resource_estimates_physical_trotter}, we report the expected physical runtime and the number of physical qubits required for a fault-tolerant implementation of a quantum circuit when using hardware with baseline, target, or desired parameter values. 

\begin{table*}[t]
    \centering
    \begin{tabular*}{0.67\textwidth}{|c|cc@{\extracolsep{\fill}}c@{\extracolsep{\fill}}cccc@{\extracolsep{\fill}}}
        \hline\hline \multicolumn{2}{c}{}&\\
     \multicolumn{1}{c}{}  &
       \multicolumn{2}{c}{Molecule Specification} & \multicolumn{4}{c}{Logical Resources}\\
        \cmidrule{2-3} \cmidrule{4-7}
        \multicolumn{1}{c}{} & \multirow{2}{*}{\shortstack[c]{Active space}} & \multirow{2}{*}{\shortstack[c]{Number of \\ orbitals, $N_{\text{orb}}$}}  & \multicolumn{2}{c}{$\epsilon = 1.6$ mHa} & \multicolumn{2}{c}{$\epsilon = 0.1$ mHa} \\
        \cmidrule{4-5} \cmidrule{6-7}
       \multicolumn{1}{c}{} &&& \shortstack[c]{\# Qubits }   & \shortstack[c]{\# $T$ gates} & \shortstack[c]{\# Qubits }   & \shortstack[c]{\# $T$ gates} \\
        \hline\hline
         \rule{0pt}{10pt}\parbox[t]{3mm}{\multirow{6}{*}{\rotatebox[origin=c]{90}{Rigor.\ Trotter}}} &&&&&&\\
        & \mbox{$p$-benzyne}, HL$\pm 2$ & 6 & 12 & $4.5\times 10^{9}$ & 12 & $4.1\times 10^{11}$\\
        & \mbox{$p$-benzyne}, HL$\pm 6$ &14 & 28 & $4.6\times 10^{11}$ & 28 & $3.8\times 10^{13}$\\
        & \mbox{$p$-benzyne}, HL$\pm 8$ & 18 & 36 & $2.0\times 10^{12}$ & 36 & $1.6\times 10^{14}$\\
        & \mbox{$p$-benzyne}, HL$\pm 12$ & 26 & 52 & $1.7\times 10^{13}$ & 52 & $1.4\times 10^{15}$\\
         \rule{0pt}{10pt}\\
         \hline
          \rule{0pt}{10pt}\parbox[t]{5mm}{\multirow{6}{*}{\rotatebox[origin=c]{90}{Empir.\ Trotter}}} &&&&&&\\
        & \mbox{$p$-benzyne}, HL$\pm 2$ & 6 & 12 & $2.7 \times 10^{8}$ & 12 & $2.5\times 10^{10}$\\
        & \mbox{$p$-benzyne}, HL$\pm 6$ & 14 & 28 & $3.5 \times 10^{9}$ & 28 & $4.0\times 10^{11}$\\
        & \mbox{$p$-benzyne}, HL$\pm 8$ & 18 & 36 & $1.1 \times 10^{10}$ & 36 & $8.0\times 10^{11}$\\
        & \mbox{$p$-benzyne}, HL$\pm 12$ & 26 & 52 & $2.7 \times 10^{10}$ & 52 & $2.3 \times 10^{12}$\\
         \rule{0pt}{10pt}\\
         \hline
        \rule{0pt}{10pt}\parbox[t]{5mm}{\multirow{6}{*}{\rotatebox[origin=c]{90}{DF\ Qubitization}}} &&&&&&\\
        & \mbox{$p$-benzyne}, HL$\pm 2$ & 6 & 291   & $4.9\times 10^{6}$ & 341 & $9.0 \times 10^{8}$\\
        & \mbox{$p$-benzyne}, HL$\pm 8$ & 18 & 502  & $2.3\times 10^{9}$ & 568 & $4.6 \times 10^{10}$\\
        & \mbox{$p$-benzyne}, HL$\pm 12$ & 26 & 668   & $9.0\times 10^{9}$ & 748 & $1.8 \times 10^{11}$\\
        & FeMoco~\cite{li2019the} & 76 & 1789   & $7.7 \times 10^{11}$ & 1972 & $1.4 \times 10^{13}$\\
        \rule{0pt}{10pt}\\
        \hline\hline
    \end{tabular*}
    \caption{Logical resource estimates for electronic-structure quantum  computations for two molecules, \mbox{$p$-benzyne} and FeMoco, and for two precisions in energy estimation: qualitatively accurate computation within a target error \mbox{1.6 mHa}, and quantitatively accurate computation  within a target error \mbox{0.1 mHa}, respectively, using a circuit-level error budget of 0.01. We report estimates for the number of logical qubits and the number of $T$ gates required for fault-tolerant implementations of the QPE algorithm on electronic spectra associated with various molecular active spaces for \mbox{$p$-benzyne} specified by HL$\pm 2, 6, 8, 12$ (using HL$\pm n$ to denote ``HOMO$-n$ and LUMO$+n$''; see \Cref{sec:logical_circuit} for an explanation of these terms) using the \texttt{6-31G} basis to represent the fermionic orbitals, and the active-space model for FeMoco proposed in Ref.~\cite{li2019the}. The sizes of the active spaces are characterized by the number of orbitals $N_{\text{orb}}$. Logical resources are reported for three quantum algorithmic approaches to implement QPE: a Trotterization approach based on using rigorous analytic bounds on the error resulting from the use of second-order Trotter--Suzuki approximation (Rigor.\ Trotter); a Trotterization approach relying on empirically obtained Trotter numbers (Empir.\ Trotter); and the double-factorized qubitization algorithm (DF Qubitization) of von Burg et al.~\cite{vonBurg2021quantum}. The reported $T$-gate counts are obtained after circuit synthesis over the Clifford+$T$ gate set.}
    \label{tab:resource_estimates_logical_trotter}
\end{table*}

\begin{table*}[t!]
    \centering
    {\scriptsize
    \begin{tabular*}{\textwidth}{|c|c|c|@{\extracolsep{\fill}}ccc@{\extracolsep{\fill}}ccc@{\extracolsep{\fill}}ccc}
        \hline\hline
       \multicolumn{3}{c}{}&&&&&&&&\\
       \multicolumn{3}{c}{} & \multicolumn{3}{c}{\bf Baseline Parameter Set } & \multicolumn{3}{c}{\bf Target Parameter Set } & \multicolumn{3}{c}{\bf  Desired Parameter Set } \\
        \cmidrule{4-6} \cmidrule{7-9} \cmidrule{10-12}
       \multicolumn{1}{c}{} & \multicolumn{1}{c}{\multirow{2}{*}{\shortstack[c]{Target \\ error $\epsilon$}}} & \multicolumn{1}{c}{\multirow{2}{*}{\shortstack[c]{$N_{\text{orb}}$}}} & \multirow{2}{*}{\shortstack[c]{\# Phys.\\ qubits}} & \multirow{2}{*}{\shortstack[c]{Phys. \\ time}}& \multirow{2}{*}{\shortstack[c]{QEC code\\ distances}} &  \multirow{2}{*}{\shortstack[c]{\# Phys.\\ qubits}} & \multirow{2}{*}{\shortstack[c]{Phys. \\ time}} & \multirow{2}{*}{\shortstack[c]{QEC code\\ distances}} & \multirow{2}{*}{\shortstack[c]{\# Phys.\\ qubits}} & \multirow{2}{*}{\shortstack[c]{Phys. \\ time}} & \multirow{2}{*}{\shortstack[c]{QEC code\\ distances}}\\
        \multicolumn{3}{c}{}&&&&&&&&\\
        \hline\hline
         \parbox[t]{4mm}{\multirow{11}{*}{\rotatebox[origin=c]{90}{{Rigor.\ Trotter}}}}&
          &&&&&&&&&\\ 
        & \multirow{4}{*}{\rotatebox[origin=c]{90}{$1.6$ mHa}} & 6 &  3.5$\times$10\textsuperscript{7} & 2.5 days & 15, 33, 75 | 87 & 1.6$\times$10\textsuperscript{6} & 13.7 hours & 11, 29 | 31 & 4.4$\times$10\textsuperscript{5} & 11.1 hours & 21 | 25 \\
        & & 14 & 3.7$\times$10\textsuperscript{7} & 282.8 days & 17, 37, 87 | 97 & 2.2$\times$10\textsuperscript{6} & 64.9 days & 13, 33 | 35 & 7.6$\times$10\textsuperscript{5} & 50.1 days & 25 | 27 \\
        & & 18 & 3.9$\times$10\textsuperscript{7} & 3.5 years & 17, 39, 91 | 101 & 2.4$\times$10\textsuperscript{6} & 293.7 days & 13, 35 | 37 & 9.2$\times$10\textsuperscript{5} & 230.2 days & 27 | 29 \\
        & & 26 & 4.5$\times$10\textsuperscript{7} & 33.5 years & 17, 43, 97 | 111 & 3.2$\times$10\textsuperscript{6} & 7.5 years & 15, 37 | 39 & 1.1$\times$10\textsuperscript{6} & 6.0 years & 29 | 31 \\
         &&&&&&&&&\\
        \cline{2-12}
         &&&&&&&&&\\ 
        & \multirow{4}{*}{\rotatebox[origin=c]{90}{$0.1$ mHa}} & 6 &  3.5$\times$10\textsuperscript{7} & 248.1 days & 17, 37, 87 | 95 & 2.3$\times$10\textsuperscript{6} & 58.2 days & 13, 33 | 35 & 8.4$\times$10\textsuperscript{5} & 44.9 days & 25 | 27 \\
        & & 14 & 4.7$\times$10\textsuperscript{7} & 76.6 years & 21, 41, 97 | 117 & 3.4$\times$10\textsuperscript{6} & 16.3 years & 15, 37 | 39 & 1.8$\times$10\textsuperscript{6} & 12.9 years & 11, 29 | 31 \\
        & & 18 & 4.3$\times$10\textsuperscript{7} & 323.2 years & 19, 41, 103 | 117 & 3.3$\times$10\textsuperscript{6} & 72.1 years & 15, 39 | 41 & 1.9$\times$10\textsuperscript{6} & 58.0 years & 11, 31 | 33 \\
        & & 26 & 4.7$\times$10\textsuperscript{7} & 2839.9 years & 19, 45, 107 | 119 & 3.4$\times$10\textsuperscript{6} & 683.4 years & 15, 41 | 45 & 2.6$\times$10\textsuperscript{6} & 501.2 years & 13, 31 | 33 \\
         &&&&&&&&&\\
        \hline
          \parbox[t]{4mm}{\multirow{11}{*}{\rotatebox[origin=c]{90}{{Empir.\ Trotter}}}}&
          &&&&&&&&&\\ 
        & \multirow{4}{*}{\rotatebox[origin=c]{90}{$1.6$ mHa}} & 6 &  1.1$\times$10\textsuperscript{7} & 3.2 hours & 29, 69 | 77 & 6.7$\times$10\textsuperscript{5} & 46.5 minutes & 25 | 29 & 3.7$\times$10\textsuperscript{5} & 36.9 minutes & 19 | 23 \\
        & & 14 & 4.1$\times$10\textsuperscript{7} & 2.1 days & 19, 31, 75 | 95 & 1.6$\times$10\textsuperscript{6} & 10.7 hours & 11, 29 | 31 & 4.9$\times$10\textsuperscript{5} & 8.6 hours & 21 | 25 \\
        & & 18 & 3.1$\times$10\textsuperscript{7} & 6.4 days & 15, 33, 79 | 95 & 1.5$\times$10\textsuperscript{6} & 1.5 days & 11, 29 | 35 & 6.3$\times$10\textsuperscript{5} & 1.1 days & 23 | 25 \\
        & & 26 & 3.7$\times$10\textsuperscript{7} & 16.0 days & 15, 35, 81 | 93 & 2.0$\times$10\textsuperscript{6} & 3.8 days & 13, 29 | 33 & 6.7$\times$10\textsuperscript{5} & 2.7 days & 23 | 25 \\
         &&&&&&&&&\\
        \cline{2-12}
         &&&&&&&&&\\
        & \multirow{4}{*}{\rotatebox[origin=c]{90}{$0.1$ mHa}} & 6 &  3.4$\times$10\textsuperscript{7} & 14.4 days & 15, 35, 81 | 89 & 2.1$\times$10\textsuperscript{6} & 3.4 days & 13, 29 | 33 & 6.6$\times$10\textsuperscript{5} & 2.6 days & 23 | 25 \\
        & & 14 & 3.5$\times$10\textsuperscript{7} & 239.2 days & 17, 37, 87 | 95 & 2.4$\times$10\textsuperscript{6} & 56.1 days & 13, 33 | 35 & 8.9$\times$10\textsuperscript{5} & 43.3 days & 25 | 27 \\
        & & 18 & 3.6$\times$10\textsuperscript{7} & 1.4 years & 17, 37, 89 | 99 & 2.3$\times$10\textsuperscript{6} & 120.3 days & 13, 33 | 37 & 8.9$\times$10\textsuperscript{5} & 94.3 days & 25 | 29 \\
        & & 26 & 3.8$\times$10\textsuperscript{7} & 4.2 years & 17, 39, 91 | 103 & 2.7$\times$10\textsuperscript{6} & 346.7 days & 13, 35 | 37 & 1.1$\times$10\textsuperscript{6} & 271.7 days & 27 | 29 \\
         &&&&&&&&&\\
        \hline
        \parbox[t]{4mm}{\multirow{11}{*}{\rotatebox[origin=c]{90}{{DF\ Qubitization}}}}&
          &&&&&&&&&\\
        & \multirow{4}{*}{\rotatebox[origin=c]{90}{$1.6$ mHa}} & 6 & 1.4$\times$10\textsuperscript{7} & 3.3 minutes & 25, 59 | 73 & 1.3$\times$10\textsuperscript{6} & 1.2 minutes & 23 | 27 & 8.2$\times$10\textsuperscript{5} & 35.8 seconds & 17 | 21 \\
        & & 18 & 4.8$\times$10\textsuperscript{7} & 1.4 days & 15, 31, 75 | 93 & 3.6$\times$10\textsuperscript{6} & 7.3 hours & 11, 27 | 33 & 1.7$\times$10\textsuperscript{6} & 5.6 hours & 21 | 25 \\
        & & 26 & 5.5$\times$10\textsuperscript{7} & 5.4 days & 15, 33, 79 | 95 & 4.7$\times$10\textsuperscript{6} & 1.3 days & 11, 29 | 35 & 2.6$\times$10\textsuperscript{6} & 23.6 hours & 23 | 27 \\
        & & 76 & 1.2$\times$10\textsuperscript{8} & 1.4 years & 17, 37, 89 | 107 & 1.3$\times$10\textsuperscript{7} & 121.8 days & 13, 33 | 39 & 7.7$\times$10\textsuperscript{6} & 96.8 days & 25 | 31 \\
          &&&&&&&&&\\ 
          \cline{2-12}
         &&&&&&&&&\\
        & \multirow{4}{*}{\rotatebox[origin=c]{90}{$0.1$ mHa}} & 6 & 2.3$\times$10\textsuperscript{7} & 12.6 hours & 29, 77 | 91 & 3.0$\times$10\textsuperscript{6} & 2.7 hours & 11, 27 | 31 & 1.4$\times$10\textsuperscript{6} & 2.2 hours & 21 | 25 \\
        & & 18 & 5.8$\times$10\textsuperscript{7} & 30.8 days & 15, 35, 91 | 105 & 5.1$\times$10\textsuperscript{6} & 6.5 days & 13, 31 | 35 & 2.6$\times$10\textsuperscript{6} & 5.4 days & 23 | 29 \\
        & & 26 & 7.9$\times$10\textsuperscript{7} & 132.1 days & 19, 35, 85 | 115 & 6.8$\times$10\textsuperscript{6} & 28.5 days & 13, 31 | 39 & 3.4$\times$10\textsuperscript{6} & 21.2 days & 25 | 29 \\
        & & 76 & 1.5$\times$10\textsuperscript{8} & 28.5 years & 17, 41, 99 | 119 & 1.8$\times$10\textsuperscript{7} & 6.5 years & 15, 35 | 43 & 9.8$\times$10\textsuperscript{6} & 5.0 years & 29 | 33 \\
         &&&&&&&&&\\
        \hline\hline
    \end{tabular*}
    }
    \caption{Physical resource estimates generated using the  TopQAD toolkit~\cite{1qbit2024topqad} for implementing 
    the QPE algorithm on electronic-structure quantum circuits associated with the \mbox{$p$-benzyne} and FeMoco molecules, for two precisions in energy estimation: qualitatively accurate computation within a target error \mbox{1.6 mHa}, and quantitatively accurate computation within a target error \mbox{0.1 mHa}, respectively, using   
    a circuit-level error budget of 0.01. 
    The corresponding logical resource requirements are reported in \Cref{tab:resource_estimates_logical_trotter}. 
    We report estimates for the physical wall-clock time and the number of physical qubits required for fault-tolerant implementations of the QPE algorithm for electronic spectra associated with various molecular active spaces with sizes specified by the number of orbitals $N_{\text{orb}}$. The data for $N_{\text{orb}}=6, 14, 18, 26$ correspond to active space selections HL$\pm 2, 6, 8, 12$ (using HL$\pm n$ to denote ``HOMO$-n$ and LUMO$+n$''; see  \Cref{sec:logical_circuit} for an explanation of these terms) for \mbox{$p$-benzyne} using the \texttt{6-31G} basis to represent the fermionic orbitals; the data for $N_{\text{orb}} =76$ pertains to the active-space model for FeMoco proposed in Ref.~\cite{li2019the}. In addition, we also report the QEC code distances that are required for running the corresponding circuits fault-tolerantly. For example,  \mbox{[17, 37, 87 | 97]} means that the code distances $d=17$, $37$, and $87$ are required for the first, second, and third magic state distillation levels, respectively, while the QEC code distance $d=97$ is needed to encode the logical qubits of the core processor. These choices are determined by the architecture's assembler~\cite{1qbit2024topqad} based on optimizations of the various trade-offs between the space and time costs proposed in Ref.~\cite{silva2024optimizing}. Physical resources are reported for the same three quantum algorithmic approaches as in \Cref{tab:resource_estimates_logical_trotter}. The associated resource requirements are reported for three hardware specifications, namely, baseline, target, and desired hardware, as summarized in \Cref{tab:physical_params}. The results of this table are plotted in \Cref{fig:QRE_Plots}.}
    \label{tab:resource_estimates_physical_trotter}
\end{table*}

To test the usefulness of parallelization and other optimization techniques, we ran smaller sample circuits through the resource estimation pipeline (see \Cref{sec:resource_estimation}) and estimated the resource requirements at various stages during the optimization.  We found that the number of $\pi/8$ rotations before and after optimization differed only by a small factor, and that the dependency graph of the operations was nearly linear, indicating that there is no significant
parallelization potential when this circuit is routed on the 2D layout.
Consequently, for all circuits for which we provide resource estimates in \Cref{tab:resource_estimates_logical_trotter} and \Cref{tab:resource_estimates_physical_trotter}, a
single auto-correcting buffer is used in the last stage of the MSF (see \Cref{sec:topqad-assembly} for details).
We note that more parallelizable circuits can be synthesized for the same
quantum simulation via re-orderings of the terms in the product formula.
However, this can saturate the error bounds in the circuit  decomposition and
therefore may create nontrivial trade-offs that are interesting avenues for
future research.

We plot the results of our QRE studies, including both the runtime and the number of physical qubits, in \Cref{fig:QRE_Plots}, alongside estimates for the runtime for two classical algorithms, the variational numerically exact full configuration interaction (FCI) computation and the heuristic density matrix renormalization group (DMRG) method~\cite{PhysRevLett.69.2863}, which were calculated by extrapolating the results of recent classical calculations~\cite{bellonzi2024feasibility,gao2024distributed}. See \Cref{sec:classical_re} for details on this extrapolation.

\begin{figure*}[t]
\centering
\begin{tabular}{c @{\hskip 0.34in} c}
    \text{(a) Qualitatively Accurate Simulation  ($1.6$ mHa)}
    & \text{(b) Quantitatively Accurate Simulation  ($0.1$ mHa)} \\
    \includegraphics[width=0.43\linewidth]{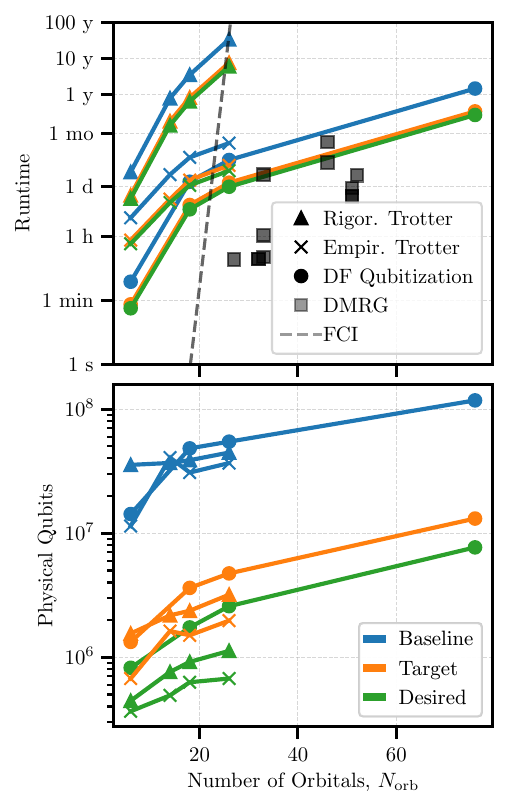}
    & \includegraphics[width=0.43\linewidth]{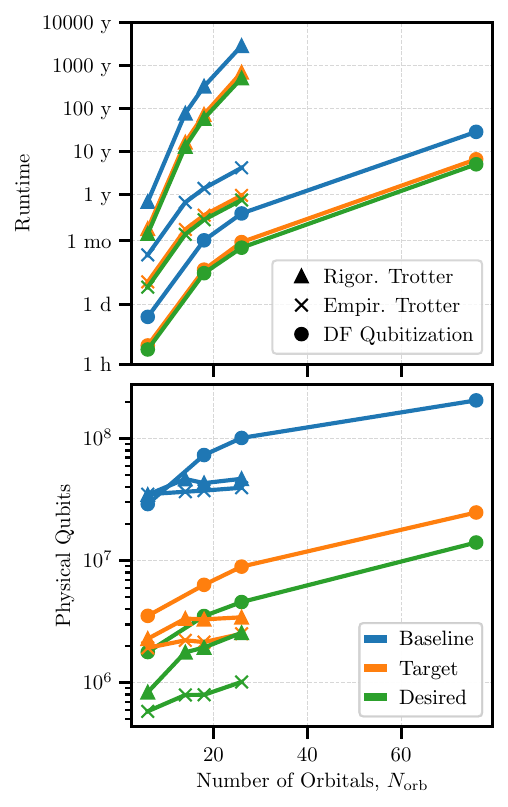}
\end{tabular}
\caption{Physical resource estimates for electronic-structure quantum computations for two molecules, \mbox{$p$-benzyne} and FeMoco, and for two precisions in energy estimation: (a) qualitatively accurate simulation within a target error \mbox{1.6 mHa}, and (b) quantitatively accurate simulation  within a target error \mbox{0.1 mHa}, respectively, using  
a circuit-level error budget of 0.01.
For both target precisions, we report estimates for the physical wall-clock time (runtime) and the number of physical qubits required for fault-tolerant implementations of the QPE algorithm on electronic-structure quantum circuits associated with various molecular active spaces with sizes specified by the number of orbitals $N_{\text{orb}}$. The data for $N_{\text{orb}}=$ 6, 14, 18, 26 correspond to the active-space specifications HL$\pm 2, 6, 8, 12$ (using HL$\pm n$ to denote ``HOMO$-n$ and LUMO$+n$''; see \Cref{sec:logical_circuit} for an explanation of these terms) for \mbox{$p$-benzyne} using
the \texttt{6-31G} basis set to represent the fermionic orbitals; the data for $N_{\text{orb}}=76$ pertains to the active-space model for FeMoco proposed in Ref.~\cite{li2019the}.  
Runtime and physical qubit counts are reported for three  quantum algorithms: Trotterization based on using rigorous analytic bounds on the error resulting from the use of second-order Trotter--Suzuki approximation (Rigor.\ Trotter), thus yielding a worst-case number of Trotter slices in approximating the Hamiltonian evolution; Trotterization relying on more-realistic, empirically obtained Trotter numbers (Empir.\ Trotter); and the double-factorized qubitization algorithm (DF Qubitization) of von Burg et al.~\cite{vonBurg2021quantum}. Furthermore, the associated resource requirements are reported for three hardware specifications: baseline, target, and desired hardware, as summarized in \Cref{tab:physical_params}. 
For comparison, for energy estimations within the target error \mbox{1.6 mHa}, predictions of CPU times are provided for classical algorithms based on either the full configuration interaction (FCI) method or the  density matrix renormalization group (DMRG) method run on a classical computer. These predictions were obtained by extrapolating the results of recent classical calculations~\cite{bellonzi2024feasibility,gao2024distributed}. 
}
\label{fig:QRE_Plots}
\end{figure*}

This study demonstrates that ground-state energy estimation for molecules involving active spaces with orbital numbers in the range of 10 to 76 require a number of physical qubits ranging from approximately $10^6 $ to $10^8$ and physical runtimes ranging from a few hours to several years. Both the quantum algorithm used and the hardware quality can have a significant impact on the resource requirements. On the algorithmic level, substantial space--time trade-offs can be observed. Implementations of the QPE algorithm based on Trotterization typically yield  high $T$ counts resulting in long  runtimes, while the required number of qubits to run the algorithm is low, whereas implementations based on qubitization result in much lower $T$ counts and higher qubit counts. For both algorithms, improving the hardware quality from baseline to target results in a reduced runtime and qubit count by up to  a factor of 5. Better algorithms run on better hardware can result in a reduction in runtime of up to two orders of magnitude. For example, an implementation of QPE with qubitization and target hardware  results in runtime reduction by a factor of 50 compared to running QPE based on Trotterization (using empirical bounds) on baseline hardware. We also provide results using AzureQRE in \Cref{sec:azure_qre}. 

Furthermore, we observe that, for $N_{\text{orb}}\gtrapprox 25$, quantum simulations begin to outperform  classical FCI computations. 
Linear variational post-Hartree--Fock approaches based on the FCI method are designed to provide numerically exact solutions, but their practical use is known to be limited to molecular systems with few electrons and small basis sets. Although some molecular systems are classically tractable  even at scales up to 100 orbitals, in general, 
exact classical computations become nearly impossible beyond 25 orbitals, especially for highly correlated systems. However, powerful classical heuristic algorithms can push the quantum advantage to much greater molecular sizes. For example, the DMRG method~\cite{PhysRevLett.69.2863} run with parallel processing on an HPC system can be significantly faster than quantum simulations for molecular active spaces involving up to $N_{\text{orb}}\approx 50$ spatial orbitals, as can also be observed in \Cref{fig:QRE_Plots}(a). 
The most recent cutting-edge petaFLOPS performance results for a hybrid CPU--multi-GPU parallel implementation of the spin-adapted \textit{ab initio} DMRG method on a state-of-the-art Nvidia  DGX-H100 architecture suggest that heuristic classical simulations of molecular systems are tractable even for sizes well above 75 spatial orbitals~\cite{menczer2024parallelimplementationdensitymatrix}. Nevertheless, DMRG methods eventually become increasingly unreliable for molecular systems involving a number of spatial orbitals far beyond 50, as such systems are typically too strongly correlated, requiring calculations with large bond dimensions, and thus intractable runtimes~\cite{RevModPhys.92.015003,reiher2017elucidating}. While there is no such sharp transition line between what is classically tractable and classically intractable (which is highly dependent on the extent of quantum correlations in the studied systems), a quantum advantage gradually appears for spatial orbital numbers beyond $N_{\text{orb}}\approx 50$. These insights also motivate future research, namely, developing quantum heuristic algorithms that could bring the transition to a quantum advantage down to smaller problem sizes. This naturally follows the development of classical algorithms, where the transition from the guarantees of FCI to the heuristics of DMRG greatly reduced the necessary resources.

The FeMoco molecule has been representatively used as an important utility-scale use case for fault-tolerant quantum simulations in quantum chemistry in several recent QRE  studies~\cite{reiher2017elucidating,motta2021low,lee2021even,otten2023qrechem}. Reiher et al.~\cite{reiher2017elucidating} presented the first thorough QRE analysis for FeMoco for an active space involving $54$ spatial orbitals, reporting a $T$-gate count on the order of $10^{14}$ for qualitatively accurate computations (\mbox{1.0 mHa}), and on the order of $10^{15}$ for  quantitatively accurate computations (\mbox{0.1 mHa}), for quantum simulations based on the Trotter approach. Assuming that a logical $T$ gate can be implemented within a time of $\sim$$\SI{100}{\nano\second}$, which is a highly optimistic assumption, they conclude that runtimes on the order of several months to several years would be required to estimate ground-state energies of FeMoco within a precision of 1.0 mHa or 0.1 mHa. However, the active space analyzed by Reiher et al.\  (involving $54$ spatial orbitals) was later pointed out not to be a suitable model for 
representing the ground state of FeMoco, because it does not correctly capture the open-shell character of the underlying molecular system, and a more appropriate active space (involving $76$ spatial orbitals) was proposed by Li et al.~\cite{li2019the}. Hence, our QRE study pertains to this much larger FeMoco active space. The reported runtimes for the DF qubitization algorithm are on the order of four months (for an accuracy of \mbox{1.6 mHa)} to a few years (for 0.1 mHa), when using the target hardware specifications. Yet, more-optimistic logical resource counts and associated physical runtimes have been recently reported for quantum simulations of the Li et al.~FeMoco. In particular, Lee et al.~\cite{lee2021even} propose  more-efficient electronic-structure quantum simulations based on qubitization through tensor hypercontraction (here referred to as \lq\lq{}THC qubitization\rq\rq) and demonstrate  that the Li et al.~FeMoco can be simulated within a qualitative chemical accuracy of \mbox{1.6 mHa} with a Toffoli count of $3.2\times 10^{10}$, implemented with four million physical qubits and a runtime of only four days, using an FTQC architecture based on CCZ magic state factories and a self-correcting CCZ (AutoCCZ) state consumption~\cite{gidney2019flexiblelayoutsurfacecode}. 
We note that the runtime for FeMoco simulations could possibly be reduced even further through the recently proposed symmetry-compressed double factorization approach~\cite{rocca2024reducing}. Such recent developments in advancing quantum simulation algorithms and in architectural design considerations are a promising avenue for further 
reducing the expected runtime for estimating the ground-state energy of FeMoco. 

It is insightful to understand the discrepancy between the runtimes reported here for the DF qubitization algorithm and the four days reported for THC qubitization in Ref.~\cite{lee2021even}, for quantum simulations of the Li et al.~FeMoco within the chemical accuracy 1.6 mHa. For instance, the 122 days and 97 days that we report as runtimes for the target and desired hardware specifications, respectively (see \Cref{tab:resource_estimates_physical_trotter}), are longer by the factors of approximately 30 and 24 than the four days reported in Ref.~\cite{lee2021even}. We note that, on the logical level, the $T$-gate count reported here for DF qubitization (see \Cref{tab:resource_estimates_logical_trotter}) is higher by a factor of only 6 over the $T$-gate count resulting from converting the Toffoli gate count reported for THC qubitization in Table III of Ref.~\cite{lee2021even}, assuming that at least four $T$ gates are required for decomposing each Toffoli gate into gates from the Clifford+$T$ gate set. Apart from using different representations of the Hamiltonian in the two algorithms (both based on qubitization), the use of different criteria for determining the truncation thresholds in the double factorization and tensor hypercontraction approaches can significantly contribute to the difference in the reported $T$-gate counts.
Thus, the discrepancy between the reported runtimes is not entirely due to running a different quantum simulation algorithm (DF qubitization versus THC qubitization), but also to differing assumptions on the physical and architectural levels. The resource requirements reported in Ref.~\cite{lee2021even} are based on using AutoCCZ states (and the associated CCZ factories) to implement the Toffoli gates, while the resource estimates reported here are based on $T$ magic states (and the associated factories) to implement the $\pi/8$ rotations. We expect that supporting CCZ factories in our architecture will allow for up to a  four times faster consumption of magic states depending on the input logic circuit. Together, the factor of 4 (on the architectural level) and factor of 6 (on the logical level) result in a factor of 24 for the expected difference between the runtime for DF qubitization reported in this work and the runtime for THC qubitization in Ref.~\cite{lee2021even}, which explains the 
difference in 122 days (or 97 days) versus four days.
Further differences arise  due to using  different hardware specifications. The physical noise model assumed in Ref.~\cite{lee2021even} results in the values $\Lambda \approx 6.1$ and $\bar\Lambda \approx 5.0$ for the error suppression rates associated with the error suppression models discussed in \Cref{sec:hardware-noise-modeling}.
In addition, different physical space--time trade-offs can even be obtained for the same logic circuit and the same noise model, as is illustrated in \Cref{fig:space-time_trade-off_diagram_FeMoco} for the circuit associated with quantum simulation of the Li et al.~FeMoco based on THC qubitization, given the architecture discussed in this paper. Such \mbox{space--time} trade-off considerations are  discussed in more detail in Ref.~\cite{silva2024optimizing}.

\begin{figure}[ht]
    \includegraphics[width=1\columnwidth]{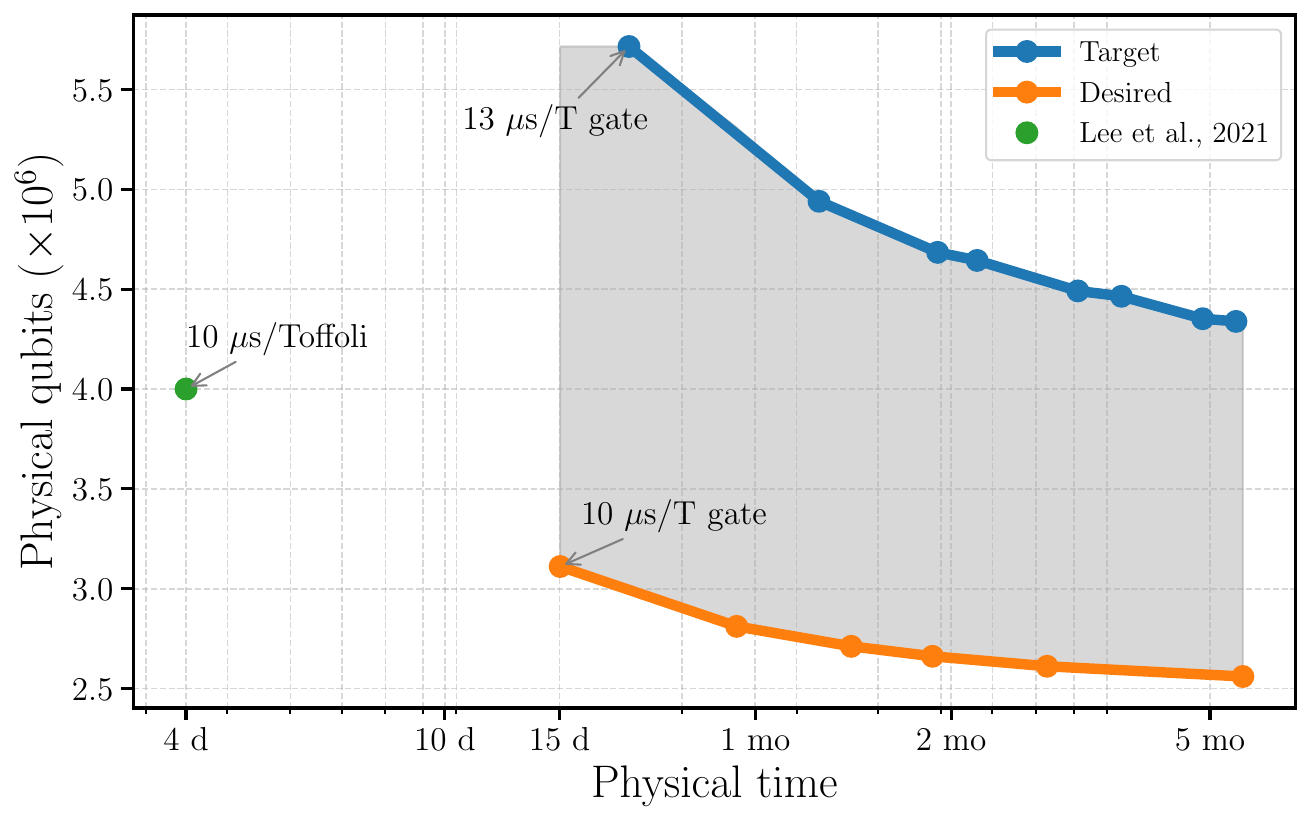}
   \caption{Space--time trade-offs for implementing a quantum circuit associated with quantum simulations of the Li et al.~active space~\cite{li2019the} for FeMoco based on the THC qubitization algorithm~\cite{lee2021even}, using a chemical accuracy of 1.6 mHa. The shown space--time requirements are based on the architecture designs considered in our work. Here, we assume that each Toffoli gate is converted to four $T$ gates, meaning that the $3.2 \times 10^{10}$ Toffoli count reported in Ref.~\cite{lee2021even} is equivalent to executing $1.3 \times 10^{11}$ $T$ gates for a quantum circuit involving 696 logical qubits. Each point in the plots represents a Pareto front solution for the space--time costs considering different MSF sizes, with the leftmost point representing the case where the MSF is large enough to allow serial execution without any idling~\cite{silva2024optimizing}. The shaded area represents the region for the expected estimates for quantum hardware with specifications between the target and desired ones. We note that Lee et al.~\cite{lee2021even} report estimates comprising four  million physical qubits and a runtime of four days assuming an architecture design based on CCZ factories~\cite{gidney2019flexiblelayoutsurfacecode} and hardware parameter values pertaining to $\Lambda= 6.1$ (or $\bar\Lambda= 5.0$) to implement the same logic circuit. The physical runtime of implementing the circuit is determined primarily by how fast magic states can be consumed in executing the non-Clifford gates. 
   A control system reaction time of $\SI{10}{\micro\second}$ was assumed for consuming a CCZ magic state in Ref.~\cite{lee2021even}, and  $\sim$$\SI{10}{\micro\second}$ is also required for consuming a $T$ magic state in the architecture considered in our work.}
   \label{fig:space-time_trade-off_diagram_FeMoco}
\end{figure}

\subsection{Catalyst ground-state energy estimation}
We now elaborate on the problem of accurate ground-state energy estimation for industrially relevant catalyst systems as a high-utility application of FTQC~\cite{von2021quantum, gratsea2025achieving} and present additional QRE studies for such systems. 

The importance and challenges of ground-state energy estimation for large and complex catalyst systems have been extensively discussed in the literature  (see, e.g., Refs.~\cite{von2021quantum, bellonzi2024feasibility}) and could have a significant positive real-world impact. As discussed in Ref.~\cite{von2021quantum}, the chemical transformation of carbon dioxide with green hydrogen to methanol offers an extremely favorable path toward limiting rising carbon dioxide levels in the atmosphere, as methanol is considered a renewable fuel that could  be used directly in combustion engines or for power generation~\cite{von2021quantum}. One recent work~\cite{gratsea2025achieving} by a subset of the authors of this manuscript carefully revisited this problem by performing an extended QRE analysis for ion-trap and neutral-atom quantum hardware systems, and benchmarking them against state-of-the-art classical simulations.

In Ref.~\cite{gratsea2025achieving}, the system under study is the complex XVIII with an increasing number of orbital sizes (namely, 56, 100, and 150) that could be used as a computational proxy for benchmarking quantum and classical algorithms from  various intermediate systems involved in the chemical transformation of carbon dioxide to methanol. Using the semistochastic heat-bath configuration interaction (SHCI) classical algorithm~\cite{Li2018}, the classical results reported in Ref.~\cite{gratsea2025achieving} are considered to be state of the art in terms of accuracy compared to other works in the literature. A more detailed analysis on the performance of the SHCI classical method is presented in Ref.~\cite{koh2025systematic}. The reported CPU hours are 7, 27.8, and 294.4 days for orbital sizes of 56, 100, and 150, respectively~\cite{gratsea2025achieving}. 

In Ref.~\cite{gratsea2025achieving}, a quantum benchmarking graph is presented that breaks down the problem instance into  classical and quantum subroutines, and a detailed QRE analysis is presented for implementations based on ion-trap and neutral-atom systems, where a quantum advantage is observed in terms of both runtime and accuracy of the quantum algorithm over the classical SHCI methodology. In the present work, we present the results of our QRE analysis conducted using TopQAD for a superconducting device, using three sets of specifications characterizing the quality of the hardware, namely, baseline, target, and desired hardware parameters, which are summarized in \Cref{tab:physical_params}. In \Cref{tab:catalysis_QRE}, we report the resource requirements comprising the runtime (QPU hours) and the number of physical qubits, for time- and space-optimal implementations. Importantly, in both time- and space-optimal cases, we find a nearly linear scaling of the quantum runtime as the number of spin orbitals increases, compared to the exponential scaling of classical methods, which suggests a quantum advantage in runtime, in accordance with the findings of Ref.~\cite{gratsea2025achieving}. 

\begin{table}[h]
\centering
\begin{tabular}{l|c|c|c}
\toprule
\multicolumn{4}{c}{Complex XVIII System} \\  
\cmidrule(lr){1-4}
\textbf{\# Orbitals} & \textbf{56} & \textbf{100} & \textbf{150} \\
\# Log.\ Qubits       & 924 & 1960 & 2870 \\
\# $T$ gates     & $8.2\times 10^8$ & $4.2\times 10^9$ & $1.1\times 10^{10}$  \\
\midrule
\midrule
\addlinespace
\multicolumn{4}{l}{\textbf{Baseline Parameters Set:}}\\
\multicolumn{4}{l}{\textbf{Time-optimal}}\\
Phys. Runtime       & 12.0 hours & 6.6 days & 2.5 days \\
\# Phys. Qubits      & $5.7\times 10^7$ & $9.7\times 10^7$ & $1.4\times 10^9$  \\

\addlinespace
\multicolumn{4}{l}{\textbf{Space-optimal}}\\
Phys. Runtime      & 7.2 days  & 17.2 days & 54.9 days \\
\# Phys. Qubits      & $3.3\times 10^7$ & $7.5\times 10^7$ & $1.2\times 10^8$ \\
\midrule
\midrule
\multicolumn{4}{l}{\textbf{Target Parameters Set:}}\\
\multicolumn{4}{l}{\textbf{Time-optimal}}\\
Phys. Runtime       & 3.0 hours & 14.4 hours & 1.6 days \\
\# Phys. Qubits      & $5.4\times 10^6$ & $1.1\times 10^7$ & $1.6\times 10^7$ \\

\addlinespace
\multicolumn{4}{l}{\textbf{Space-optimal}}\\
Phys. Runtime      &  1.3 days  & 4.1 days & 3.1 days \\
\# Phys. Qubits      & $4.4\times 10^6$ & $1.0\times 10^7$ & $1.5\times 10^7$ \\

\midrule
\midrule
\multicolumn{4}{l}{\textbf{Desired Parameters Set:}}\\
\multicolumn{4}{l}{\textbf{Time-optimal}}\\
Phys. Runtime       & 2.2 hours & 11.1 hours & 1.2 days \\
\# Phys. Qubits      & $2.9\times 10^6$ & $6.4\times 10^6$ & $9.1\times 10^6$  \\

\addlinespace
\multicolumn{4}{l}{\textbf{Space-optimal}}\\
Phys. Runtime      & 8.1 hours  & 5.1 days & 2.3 days \\
\# Phys. Qubits      & $2.6\times 10^6$ & $6.0\times 10^6$ & $8.9\times 10^6$ \\

\bottomrule
\end{tabular}
\caption{Quantum resource estimates generated using TopQAD for ground-state energy estimation to an accuracy of $1.0$ mHa, for XVIII systems~\cite{von2021quantum}, using a quantum algorithm based on the DFTHC tensor factorization method along with spectrum amplification, which is introduced in Ref.~\cite{low2025fast}. 
The resource requirements are reported for increasing orbital sizes, and for three hardware parameter specifications, namely, ``baseline'', ``target'', and ``desired'', which are summarized in \Cref{tab:physical_params}.
}
\label{tab:catalysis_QRE}
\end{table}

Moreover, for the time-optimal case, we find that the QPU runtimes reported for the superconducting qubits in this work are 2--12$\times$ faster than those reported for trapped-ion and neutral-atom systems in Ref.~\cite{gratsea2025achieving}. The exact speedup depends on the assumed set of hardware parameter specifications (baseline, target, or desired) as well as on the molecular size, quantified by the number of orbitals. Furthermore,  for the time-optimal case the runtime estimates are $\sim$300$\times$ faster, and for the space-optimal case they are $\sim$20$\times$ faster, than the runtime  estimates reported in Ref.~\cite{beverland2022assessing}, while the reported number of physical qubits remains comparable. 

Note that the physical resource estimates we report in this section rely on the use of logical resource counts as input to TopQAD. Future work could involve a more elaborate QRE analysis, which would take as input an intermediate representation of the actual quantum circuit associated with the quantum algorithm presented in Ref.~\cite{low2025fast} and compile it to the level of lattice surgery instructions. This would allow for further optimizations at the logical circuit level, potentially resulting in reduced runtime and physical qubit estimates. 

\subsection{NMR spectral prediction}

In this section, we discuss another potential application of  high-utility FTQC in the field of quantum chemistry, namely, nuclear magnetic resonance (NMR) spectral prediction (see Ref.~\cite{elenewski2024prospectsnmrspectralprediction}). We present a detailed QRE study using TopQAD, reporting the resource requirements pertaining to the baseline, target, and desired hardware parameter specifications summarized in~\Cref{tab:physical_params}.

One recent work~\cite{elenewski2024prospectsnmrspectralprediction} proposes  replacing zero-to-ultralow field (ZULF) NMR by quantum computation, with the aim of achieving practical utility rather than merely a quantum advantage. Therefore, the authors of that work focus primarily on analyzing numerous small molecules that could potentially be tractable with classical methods, and found that such systems correspond to quantum circuits involving approximately 100 logical qubits and $10^7$--$10^8$ $T$ gates per spectrum. We use these logical counts as input to TopQAD to generate the corresponding estimates of the physical qubit count and runtime. Our estimates are reported in~\Cref{tab:NMR_QRE} for the time-optimal case, for the three hardware parameter specifications summarized in~\Cref{tab:physical_params}.
Given the runtime estimates per spectrum, careful benchmarking against classical methods is needed to determine the practical utility of this application.

Importantly, in the same work~\cite{elenewski2024prospectsnmrspectralprediction}, molecular systems are also discussed that are taxing for classical computations, such as mastoparan-X and $\alpha$-conotoxin, which are complex peptides of high relevance for drug-discovery pipelines. NMR spectral prediction for Mastoparan-X requires $6.7 \times 10^{12}$ $T$ gates and 195 logical qubits, while for $\alpha$-conotoxin it requires  $5.1\times 10^{11}$ $T$ gates and 168 logical qubits.

QRE analyses play an important role in helping us to gain a better understanding of the resource requirements 
for this application, along with what is needed for it be of high practical utility. Future work on quantum algorithm design is a critical step towards achieving lower runtimes  for both small molecules and classically taxing large systems.

\begin{table}
\centering
\begin{tabular}{c|c|c|c}
\toprule
 & \textbf{Baseline} & \textbf{Target} & \textbf{Desired} \\
\midrule
\midrule

\multicolumn{4}{l}{\textbf{Small molecules:} 100 Log. Qubits and $10^7$ $T$ Gates}\\
\addlinespace
Phys. Runtime & 6.3 min & 1.5 min & 1.1 min \\
\# Phys. Qubits      & $1.3\times 10^7$ & $8.7\times 10^5$ & $7.1\times 10^5$ \\

\midrule
\midrule
\multicolumn{4}{l}{\textbf{Small molecules:} 100 Log. Qubits and $10^8$ $T$ Gates}\\
\addlinespace
Phys. Runtime & 1.2 hours & 15.8 min & 12.3 min \\
\# Phys. Qubits      & $1.3\times 10^7$ & $1.8\times 10^6$ & $6.0\times 10^5$  \\

\midrule
\midrule
\multicolumn{4}{l}{\textbf{Mastoparan-X:} 195 Log. Qubits and $6.7\times 10^{12}$ $T$ Gates}\\
\addlinespace
Phys. Runtime       & 12.0 years & 2.9 years & 2.3 years  \\
\# Phys. Qubits      & $4.8\times 10^7$ & $3.7\times 10^6$ & $1.6\times 10^6$ \\

\midrule
\midrule

\multicolumn{4}{l}{\textbf{$\alpha$-conotoxin:} 168 Log. Qubits and $5.1\times 10^{11}$ $T$ Gates}\\
\addlinespace
Phys. Runtime       & 315.0 days & 76.4 days & 59.9 days \\
\# Phys. Qubits      & $4.0\times 10^7$ & $2.9\times 10^6$  & $1.3\times 10^6$ \\

\midrule
\bottomrule
\end{tabular}
\caption{Physical quantum resource estimates generated using TopQAD for NMR spectral prediction, for small molecules and classically taxing peptides relevant for drug-discovery pipelines~\cite{elenewski2024prospectsnmrspectralprediction}. For each molecule, resource requirements are given per spectrum, based on logical resource counts reported in Ref.~\cite{elenewski2024prospectsnmrspectralprediction}, for three hardware parameter specifications, namely, ``baseline'', ``target'', and ``desired'' hardware, which are summarized in \Cref{tab:physical_params}. 
}
\label{tab:NMR_QRE}
\end{table}

\subsection{Tight HPC--QC integration for utility-scale applications: simulating Fermi--Hubbard models}
\label{subsec:FTQC_FH2d}
The 2D Fermi--Hubbard (2DFH) model is a prototypical model of strongly interacting fermions capable of capturing a wide range of quantum many-body phenomena, from high-temperature superconductivity to topological phases of matter~\cite{Arovas2022Hubbard}. The ubiquity of this model is not without its challenges, as the presence of several competing phases makes learning the phase diagram computationally nontrivial. Over the years, extensive analytical and numerical effort has been devoted to the development of techniques capable of simulating this model. State-of-the-art techniques include density-matrix renormalization group (DMRG), quantum Monte Carlo (QMC), and others~\cite{Zheng2017,PhysRevX.5.041041,PhysRevX.10.031016}. Generally speaking, most techniques struggle to solve the model at moderate lattice sizes of \(10 \times 10\), equivalent to \(200\) qubits (for the spinful case). DMRG struggles with bond dimension scaling, while QMC struggles with the sign problem. This makes the 2DFH a natural benchmark for progress in quantum hardware since it is both classically nontrivial and its quantum simulation useful for solving practical problems \cite{jiang_quantum_2018,PhysRevA.94.062304,Stanisic2022}. 

In fact, quantum simulation of the 2DFH is widely believed to be one of the first utility-scale applications of quantum computing, along with quantum chemistry. Recent years have witnessed experiments on several NISQ hardware simulating aspects of the Fermi-Hubbard model, such as quantum dynamics and superconducting pairing correlations \cite{alam2025programmabledigitalquantumsimulation,alam2025fermionicdynamicstrappedionquantum,granet2025superconductingpairingcorrelationstrappedion}. However, a careful analysis of the resources required to perform useful quantum simulation of the Fermi-Hubbard model suggests a nontrivial fault-tolerant cost and hence are clearly beyond the reach of current day NISQ hardware \cite{reiher2017elucidating,kivlichan2020improved,lee2021even,yoshioka2024hunting}. 

We will discuss the fault-tolerant resources needed for quantum simulation of the 2DFH model at a scale that is comfortably beyond classical simulation, both in terms of quantum dynamics and ground state properties. For concreteness, the 2D Fermi--Hubbard Hamiltonian is given by:
\begin{align}
\label{eq:2dFermi--Hubbard-hamiltonian}
H_{\mathrm{FH}}=-J \sum_{\langle p, q\rangle, \sigma = \{\uparrow,\downarrow\}}\left[c_{p, \sigma}^{\dagger} c_{q, \sigma}+c_{q, \sigma}^{\dagger} c_{p, \sigma}\right]+ \\ \nonumber
U \sum_{\langle p\rangle} n_{p, \uparrow} n_{p, \downarrow} -\mu \sum_{p, \sigma} c_{p, \sigma}^{\dagger} c_{p, \sigma},
\end{align}
where \(J\) denotes the hopping strength, \(U\) denotes the density-density interaction, and \(\mu\) is the chemical potential. The parameter regime of \(4 \leq |U/J| \leq 12\) is believed to be hard to simulate classically for initial states near half-filling \cite{Zheng2017}. At half-filling, the model has a particle-hole symmetry that avoids the \textit{sign problem} and allows us to efficiently find the ground state by standard quantum Monte Carlo techniques~\cite{Zheng2017,PhysRevX.5.041041,PhysRevX.10.031016}. Such exactly solvable or classically easy points in the parameter point are important from both benchmarking and verification point of view. However, to give a fair representation of the complexity of simulating the 2DFH, we choose Hamiltonian instances that are protected from {\em classical attacks} such as mapping onto free-fermions~\cite{Mann2025freefermion}, Clifford perturbation theory~\cite{begui2023simulatingquantumcircuitexpectation,Begui2024IBMkicked}, and low-rank tensor decompositions~\cite{yoshioka2024hunting,menczer2024two}. Notably, recent work using novel tensor network approaches is slowly expanding the regime of classical simulability for the 2DFH model~\cite{SchebNoack2023,ZhouZhouLiang2024,Liu2025FermionicPEPS}. 

As mentioned previously, the 2DFH is an attractive quantum benchmark since its smaller instances can be solved on early fault-tolerant quantum hardware, while larger instances promise economic utility. In fact, a systematic analysis of the open-science usage of DOE labs HPCs suggests that roughly a third of the computational tasks being solved in these HPC simulations are Fermi--Hubbard type instances ~\cite{Camps2023,agrawal2024quantifyingfaulttolerantsimulation,camps2025quantumcomputingtechnologyroadmaps}. In the long-run, simulations of the Fermi--Hubbard model are expected to deeply impact a number of areas, such as (i) {\em high-temperature superconductivity}, with implications for power grids and magnetic levitation trains~\cite{dong2022mechanism}; (ii) {\em magnetic materials}, with implications for spintronics and magnetic storage~\cite{pasqualetti2024equation}, (iii) {\em quantum materials}, with applications for novel topological insulators; (iv) {\em phase diagram discovery}, with applications in materials search for catalysts and batteries; (v) {\em non-equilibrium phenomena}, allowing  searching for switchable materials (e.g., smart windows); (vi) {\em photovoltaics}, providing insights into charge dynamics, potentially improving transport efficiency~\cite{hofmann2012doublon,tarruell2018quantum}; among many others.

Our resource estimates are focused on a {\em general-purpose} Fermi--Hubbard solver with the ability to solve {\em hard} 2D Fermi--Hubbard instances. Searching for novel materials, e.g., high-temperature superconductivity, will also require solving \textit{variants} of the 2DFH model, e.g., adding next-nearest neighbor hopping, multi-orbitals, triangular/bilayer lattices, etc. These resource estimates can provide useful insights into the ability of quantum computers to study physics of bulk 2D materials. We will separately discuss the quantum resources required for quantum dynamics as well as ground-state energy estimation. Together, these provide a good estimate of the resources required to probe the low-energy subspace of bulk 2D materials, including via non-equilibrium phenomena.

The two prominent methods for quantum simulation are Trotterization \cite{campbell_early_2022} and qubitization/quantum signal processing \cite{babbush2018encoding}. The former originates from physics and directly encodes the locality of the Hamiltonian while the latter treats the Hamiltonian as an oracle and aims to minimize the number of queries to the oracle, similar to the setting of Grover's search. Qubitization enjoys {\em asymptotically optimal} simulation costs but for relative accuracy estimates, Trotterization can be more efficient in the number of $T$ gates~\cite{campbell_early_2022}. To compare qubitization with Trotterization, we discuss resource estimates for two different methods: (i) qubitization-based simulation with the linear-T encoding of Babbush et al.~\cite{babbush2018encoding} and (ii) Trotter-based estimates using the "Plaquette Trotterization" of Campbell ~\cite{campbell_early_2022}. For the qubitization based resource estimation, we use the open-source library pyLIQTR~\cite{pyLIQTR} to obtain logical resource estimates and then translate them into physical resource estimated via TopQAD. For Plaquette Trotterization, we directly follow the analysis of Campbell~\cite{campbell_early_2022} but translate the Toffoli counts to T-counts instead.

\textbf{Resource estimates for quantum dynamics.}
We first focus on simulating quantum dynamics under the 2DFH Hamiltonian. Starting from an arbitrary initial product state which is not an eigenstate of the Hamiltonian (say the \(|0\rangle^{\otimes n}\), without loss of generality), apply the time evolution operator \(e^{-iHt}\) to this state, so as to prepares the quantum state \(|\psi(t)\rangle =  e^{-iHt}|0\rangle^{\otimes n}\). The hardness of simulating this process classically originates from a \textit{universal} feature of quantum many-body systems with locality (in the absence of disorder): when you time-evolve an initial low-entanglement state, the entanglement entropy grows linearly in time for a timescale \(t_{\star} \sim L\), where \(L\) is the system size \cite{Calabrese2005QuenchUniversal,Alba2017QuenchIntegrable}. This allows the system to generate quantum states with volume-law entanglement, that cannot be simulated via classical methods. For example, in terms of tensor networks, the {\em bond dimension} of such a quantum state grows exponentially, ruling out compact representations on a classical computer.

To estimate the resources required to implement \(e^{-iHt}\), we start by block-encoding the Hamiltonian and then using standard quantum signal processing identities to compute polynomial functions of it. Namely, since \(e^{-iHt} = \cos(Ht) - i \sin(Ht)\) for a block-encoded Hamiltonian, we simply need to identify the phases in a QSP algorithm to approximate cosine and sine functions of \(H\). Once we have the quantum circuits that implement the block-encoded polynomials, we remap the circuits using standard Clifford+T gate libraries (and perform quick circuit optimization) to estimate the fault-tolerant resources needed to implement quantum dynamics using pyLIQTR. \cref{tab:resources-qubitized-dynamics-main} lists the \(T\) counts as well as the physical qubit counts and runtimes for the 2DFH model's qubitized dynamics for a total time \(T_{\max} = N\) (scaling with the system size). This ensures that under quantum quenches, the system generates sufficient entanglement to defy classical simulation via tensor network methods.

\begin{table}[!thb]
\centering
\begin{tabular}{||c|c|c|c|c||}
  \hline
  \textbf{\makecell[l]{Lattice \\ size}} & \textbf{\makecell[l]{Logical \\ qubits}} & \textbf{\makecell[l]{$T$ count \\ (\(\left| U/J \right| = 4\))}} & \textbf{\makecell[l]{Physical \\ qubits}} & \textbf{\makecell[l]{Physical \\ runtime}}\\
  \hline
  8  & 148   & $3.01 \times 10^{7} $ & $1.0\times 10^{6}$ & $4.70 \text{ min}$ \\ \hline
  10 & 224   & $9.74 \times 10^{7} $ & $1.4\times 10^{6}$ & $16.50 \text{ min}$  \\ \hline
  12 & 312   & $2.02 \times 10^{8} $ & $1.9\times 10^{6}$ & $36.50 \text{ min}$  \\ \hline
  14 & 416   & $3.90 \times 10^{8} $ & $2.7\times 10^{6}$ & $1.30 \text{ hours}$  \\ \hline
  16 & 536   & $6.17 \times 10^{8} $ & $3.6\times 10^{6}$ & $1.90 \text{ hours}$ \\ \hline
  18 & 676   & $1.20 \times 10^{9} $ & $4.4\times 10^{6}$ & $3.90 \text{ hours}$  \\ \hline
  20 & 828   & $1.91 \times 10^{9} $ & $5.1\times 10^{6}$ & $6.10 \text{ hours}$ \\ \hline
  22 & 996   & $2.96 \times 10^{9} $ & $6.9\times 10^{6}$ & $10.60 \text{ hours}$  \\ \hline
  24 & 1180  & $4.39 \times 10^{9} $ & $7.3\times 10^{6}$ & $14.90 \text{ hours}$   \\ \hline
  26 & 1380  & $6.42 \times 10^{9} $ & $8.3\times 10^{6}$ & $21.80 \text{ hours}$  \\ \hline
  28 & 1596  & $9.06 \times 10^{9} $ & $1.0\times 10^{7}$ & $1.40 \text{ days}$ \\ \hline
  30 & 1828  & $1.26 \times 10^{10}$ & $1.2\times 10^{7}$ & $1.90 \text{ days}$   \\ \hline
  32 & 2076  & $1.64 \times 10^{10}$ & $1.4\times 10^{7}$ & $2.60 \text{ days}$  \\ \hline
\end{tabular}
\caption{Resource estimates for quantum simulation of time dynamics of a 2D Fermi--Hubbard model using qubitization/quantum signal processing techniques and the linear-T encoding of Ref.~\cite{babbush2018encoding}. This is the cost of applying \(e^{-iHt}\) to an arbitrary initial state. We focus on \(|U/J|=4\) and a total runtime, \(T_{\max} = N\), that is, it scales with the system size. The $\epsilon = 10^{-2}$ for the error is the quantum simulation (operator norm) and we use TopQAD to translate logical estimates to physical estimates.}
\label{tab:resources-qubitized-dynamics-main}
\end{table}

\textbf{Resource estimates for ground-state energy estimation.} We now discuss how to perform quantum-phase estimation and learn the low-energy physics of the 2D Fermi--Hubbard model. The largest (experimental) quantum simulation to date is limited to the \(6 \times 6\) 2DFH model \cite{granet2025superconductingpairingcorrelationstrappedion}. Our discussion below outlines how to simulate \textit{spinful} fermions with system sizes up to \(32 \times 32\) with a maximum runtime of \(\sim\) days and a few million physical qubits. Notice that the FTQC resources for Plaquette Trotterization with a relative error scaling are significantly lower, requiring a few million physical qubits and a few seconds of runtime. This is sufficient to extract the low-temperature phase-diagram of the model without running into \textit{finite-size} effects. Following Campbell's "Plaquette Trotterization"~\cite{campbell_early_2022} allows us to significantly save on the \(T\) costs. For resource estimates pertaining to  phase-diagram discovery and studies of general low-energy properties, it suffices to assume a per-site energy uncertainty of \(\epsilon_{\mathrm{site}} = 0.005\). Equivalently, since the total energy scales as \(L^2\), we choose \(\epsilon_{\mathrm{total}} = 0.005L^2\). This constant fraction energy uncertainty is sufficient to resolve the phase diagram for energy-based criteria, for example, kinks in the energy density. For other order-parameters,  such as staggered magnetization, compressibility, \(d\)-wave pairing, these do \textit{not} rely on the energy precision but only require a state in the low-energy subspace which is achieved by this scaling. This nontrivial insight leads to significant savings in the total $T$-cost of the QPE algorithm. In particular, the \(T\) cost saturates at \(\sim\)\(10^6\), for instances of lattice size up to \(32 \times 32\),  and so we pay a \textit{fixed} price in the runtime of the quantum algorithm; see \cref{tab:gsee-campbell-logical} for exact numbers. However, the number of logical qubits still scales with \(L\) and the overlap of the classical ansatz still decays with increasing \(L\). We do not discuss here the overhead due to preparing an exact ground state.

\begin{table}[!th]
\centering
\begin{tabular}{||c|c|c|c|c||}
  \hline
  \textbf{\makecell[l]{Lattice \\ size}} & \textbf{\makecell[l]{Logical \\ qubits}} & \textbf{\makecell[l]{$T$ count \\ (\(\left| U/J \right| = 4\))}} & \textbf{\makecell[l]{Physical \\ qubits}} & \textbf{\makecell[l]{Physical \\ runtime }}\\
  \hline
  8  &  162  & $2.42\times10^6$ & $8.5\times 10^{5}$ & $14.90 \text{ seconds}$   \\ \hline
  10 &  252  & $2.02\times10^6$ & $1.1\times 10^{6}$ & $11.40 \text{ seconds}$  \\ \hline
  12 &  362  & $1.96\times10^6$ & $1.4\times 10^{6}$ & $10.50 \text{ seconds}$  \\ \hline
  14 &  492  & $1.76\times10^6$ & $1.7\times 10^{6}$ & $8.80 \text{ seconds}$  \\ \hline
  16 &  642  & $1.71\times10^6$ & $2.6\times 10^{6}$ & $9.60 \text{ seconds}$    \\ \hline
  18 &  812  & $1.66\times10^6$ & $2.8\times 10^{6}$ & $8.50  \text{ seconds}$   \\ \hline
  20 & 1002  & $1.64\times10^6$ & $3.4\times 10^{6}$ & $7.90 \text{ seconds}$   \\ \hline
  22 & 1212  & $1.65\times10^6$ & $4.0\times 10^{6}$ & $8.40 \text{ seconds}$  \\ \hline
  24 & 1442  & $1.61\times10^6$ & $4.7\times 10^{6}$ & $8.00 \text{ seconds}$  \\ \hline
  26 & 1692  & $1.64\times10^6$ & $6.2\times 10^{6}$ & $8.90 \text{ seconds}$   \\ \hline
  28 & 1962  & $1.65\times10^6$ & $6.2\times 10^{6}$ & $8.10 \text{ seconds}$   \\ \hline
  30 & 2252  & $1.67\times10^6$ & $8.1\times 10^{6}$ & $8.90 \text{ seconds}$  \\ \hline
  32 & 2562  & $1.67\times10^6$ & $9.1\times 10^{6}$ & $8.90 \text{ seconds}$  \\ \hline
\end{tabular}
\caption{Resource estimates using plaquette Trotterization with Hamming weight phasing from Campbell.~\cite{campbell2021early}, and estimated runtimes. This quantum phase estimation scheme assumes access to the true ground state of the Hamiltonian. The scaling of the energy precision is \(\epsilon_{\mathrm{total}} = 0.005L^2\). We use TopQAD for these resource estimates. State preparation, repetitions needed to obtain the ground-state outcome, and subsequent observable measurements are not reported.}
\label{tab:gsee-campbell-logical}
\end{table}

\begin{table}[ht!]
\centering
\begin{tabular}{||c|c|c|c|c||}
  \hline
  \textbf{\makecell[l]{Lattice \\ size}} & \textbf{\makecell[l]{Logical \\ qubits}} & \textbf{\makecell[l]{$T$ count \\ (\(\left| U/J \right| = 4\))}} & \textbf{\makecell[l]{Physical \\ qubits}} & \textbf{\makecell[l]{Physical \\ runtime }}\\
  \hline
  8  & 148   & $3.01 \times 10^{7}$ & $1.0\times 10^{6}$ & $4.70 \text{ min}$  \\ \hline
 10  & 224   & $9.74 \times 10^{7}$ & $1.4\times 10^{6}$ & $16.50 \text{ min}$    \\ \hline
 12  & 312   & $2.02 \times 10^{8}$ & $1.9\times 10^{6}$ & $36.50 \text{ min}$  \\ \hline
 14  & 416   & $3.90 \times 10^{8}$ & $2.7\times 10^{6}$ & $1.30 \text{ hours}$     \\ \hline
 16  & 536   & $6.17 \times 10^{8}$ & $3.6\times 10^{6}$ & $1.90 \text{ hours}$   \\ \hline
 18  & 676   & $1.20 \times 10^{9}$ & $4.4\times 10^{6}$ & $3.90 \text{ hours}$   \\ \hline
 20  & 828   & $1.91 \times 10^{9}$ & $5.1\times 10^{6}$ & $6.10 \text{ hours}$   \\ \hline
\end{tabular}
\caption{Resource estimates for \(|U/J|=4\) for a 2D Fermi--Hubbard model using qubitization/quantum signal processing techniques and the linear-T encoding of \cite{babbush2018encoding}. The total runtime here \(T_{\max} = N\), that is, it scales with the system size.}
\label{tab:resources-qubitized-dynamics}
\end{table}

\begin{figure*}[t]
\centering
\includegraphics[width=0.8\textwidth]{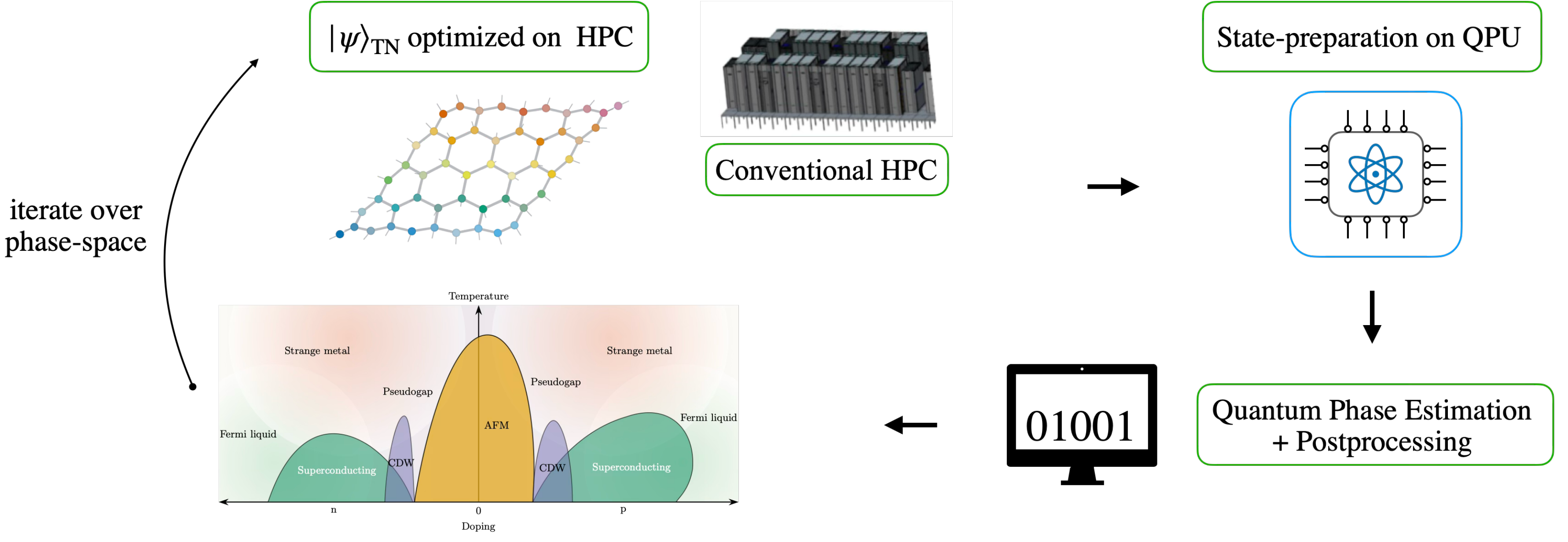}
\caption{Proposed end-to-end workflow to discover the low-temperature phase diagram of the 2D Fermi--Hubbard model using a tightly integrated HPC--QC architecture. We start by optimizing a tensor network ansatz \(|\psi\rangle_{\mathrm{TN}}\) on high-performance computers to target \(1-10\%\) overlap with the true ground state. We then perform quantum phase estimation to learn the ground state energy as a function of the Hamiltonian parameters \(E(J,U,\mu)\). We then measure superconducting order parameters on the ground state to test the presence/absence of superconductivity. The schematic Fermi--Hubbard phase diagram at the end of the flowchart is inspired from the corresponding figure in Ref.~\cite{Zhou2021HTSC}. Here AFM and CDW stands for antiferromagnetic and charge density wave phases, respectively.}
\label{fig:QPE-workflow-FH}
\end{figure*}

\textbf{Workflow for phase-diagram discovery.}
For our application of interest, understanding the phase diagram of minimal models of high-T\(_{c}\) superconductivity such as the 2DFH, we take a hybrid quantum--classical approach. We adapt the quantum phase-estimation algorithm by Campbell \cite{campbell_early_2022}, but do so without assuming access to the ideal ground state of the quantum system. Preparing quantum ground states is \textit{QMA-hard} in general \cite{Kempe2006QMAhard} and forms a crucial subroutine in actual ground-state energy estimation. The intuition behind QPE algorithms is to prepare an initial state \(|\psi_{\mathrm{guess}}\rangle\) with overlap \(\Omega(1 / \operatorname{poly}(n))\) with the true ground state and use amplitude amplification to project onto the true ground state. To prepare this initial guess, we train a tensor network ansatz on high-performance computers. Here and in general, a hybrid quantum--classical workflow partitions computational tasks between classical computers and quantum computers to leverage their natural abilities. We discuss briefly the technical elements involved in the workflow above in \Cref{fig:QPE-workflow-FH}:

\begin{enumerate}
\item {\em Selecting initial states.} To map out the low-temperature phase diagram, the QPE workflow needs a number of initial states~\cite{Zhou2021HTSC}. 
To distinguish between initial states, recall that the filling fraction \(n\) is the average electron/hole density on a single site, \(n=\frac{\langle N\rangle}{L^2}, \quad N=N_{\uparrow}+N_{\downarrow}=\sum_{i}\left(n_{i \uparrow}+n_{i \downarrow}\right)\). The hole-doping fraction is defined as \(\delta=1-n\) and is equivalent to preparing an initial state with a particle number \(N = (1-\delta)L^2\). 
We consider seven nominal doping values $\delta=0,0.05,\ldots,0.30$, with the actual fillings restricted by the allowed integer particle numbers at each lattice size. The \(n=1, \delta=0\) case is special and is referred to as {\em half-filling}.
At half-filling, the ground state has a particle-hole symmetry which avoids the {\em sign problem} detrimental to quantum Monte Carlo (QMC) techniques otherwise. Thus, QMC techniques efficiently find the ground state and such a classical benchmark helps evaluate the performance of the quantum computer. The initial states (with different doping values) can be scaled to \(\sim\) 100 points with little overhead. Moreover, partial information about the phase diagram 
obtained after a few runs and classical post-processing allows one to improve sampling of initial states: greater density around quantum-critical points and lower density in regions within the same phase.

\item {\em Optimizing the tensor-network ansatz} (independently for each \(\delta\)) via imaginary-time evolution and classical gradient descent. This step relies on standard techniques such as iPEPS~ \cite{cirac_matrix_2021}. Importantly, the \textit{hard instances} of the 2DFH model~\cite{Zheng2017} generically have volume-law entanglement, so \textit{cannot} be prepared in a classically efficient manner since the bond dimension of the tensor network grows exponentially. However, for FH instances up to size (\(16 \times 16\)), the true ground state can still be approximated well, with the overlap vanishing as we move to larger system sizes. Novel tensor network methods suggest this optimization can be achieved in a few hours of runtime on classical HPCs \cite{Liu2025FermionicPEPS}.

\item {\em Preparing initial states on a quantum device}. For 1D, translationally-invariant MPSs can be prepared on a quantum computer in $\log$-depth~\cite{Malz2024}. This complexity can be further reduced if one has access to \textit{measurements-and-feedback}. For 2D systems, we use an ansatz based on fermionic isometric tensor networks since they are sufficiently expressive while being classically simulable. Preparing such states on a quantum computer requires additional circuit depth, which needs to be factored in while estimating the total FTQC cost of FH simulation. 

\item {\em Quantum phase estimation} follows the structure outlined by Campbell~ \cite{campbell_early_2022} based on three insights: Hamming weight phasing, plaquette Trotterization, and optimized distribution of the total error rate. For the relative per-site error regime, this achieves the best known bounds for quantum phase estimation in the Fermi--Hubbard model. This is a rare scenario where Trotterized dynamics outperform qubitization and quantum signal processing techniques. For \textit{absolute} energy precision, Trotterization still achieves the best known scaling and so we allow for both algorithmic subroutines based on the precision requirements.

\item {\em The single-ancilla iterative QPE} from Ref.~\cite{campbell_early_2022} allows us to target an absolute energy error of \(\epsilon = 0.005L^2\) with \(\log(1/\epsilon)\) rounds. Here the logarithmic scaling ensures that our resource estimates are not dominated by the choice of \(\epsilon\). Using ($i$) amplitude amplifications based on standard Grover search and ($ii$) the \(1\%\) overlap in the initial state allows us to reach the ground state (approximately) in \(\left\lceil \frac{1}{\gamma} \right\rceil\) iterations, where \(\gamma\) is the overlap. We measure the energy as a function of the Hamiltonian parameters \(E(J,U,\mu)\).

\item {\em Measuring physically interesting observables} such as staggered magnetization, double occupancy, compressibility, and \(d\)-wave pairing, becomes possible once we have read out the energy and projected the quantum state into the corresponding low-energy state. Measuring a physical observable to precision \(\epsilon_{\mathrm{obs}}\) requires \(1/\epsilon^2_{\mathrm{obs}}\) many iterations. One can also use observable estimation techniques that allow for a \(1/\epsilon_{\mathrm{obs}}\) scaling at the cost of increased depth. For simplicity, we consider the \(1/\epsilon^2_{\mathrm{obs}}\) technique.

\item {\em Filling the phase diagram}, we use 30 full QPE circuits (\(3\) energy shots \(\times\) \(10\) Grover amplitude amplification rounds). With \(12\) doping points, as above, this totals \(30 \times 12 = 360\) QPE iterations for each lattice size.
\end{enumerate}

Notably, the T-count of the Fermi--Hubbard model and its variants (in Plaquette Trotterization) remain in the \(10^6 - 10^8\) regime \cite{Kan2025,baysmidt2025faulttolerantquantumsimulationgeneralized}, which allows for this scheme to be compatible with variants of the 2DFH model. This is key to modeling phenomena such as superconductivity since variants of the 2DFH can capture the physics of realistic materials.

\textbf{High-temperature Gibbs state preparation and phase-diagram discovery.}
The workflow above supports constructing phase diagrams at zero and near-zero temperatures efficiently via the QPE algorithm. However, generalizations to high temperatures is challenging. Some approaches rely on preparing  the Gibbs state at inverse temperature \(\beta\), \(\rho_{\beta} = e^{-\beta H}/\operatorname{Tr}\left[ Z(\beta) \right]\). The quantum cost of preparing Gibbs states for both \(\beta \gg 1\) (close to a maximally mixed state) and \(T \rightarrow 0\) (close to the ground state manifold) is reasonable. However, studying high-Tc superconductivity requires a nontrivial, intermediate value of \(\beta\), for example, one corresponding to critical temperature \(T \approx 100\) K~\cite{Keimer2015CuprateHTSC}. 

Two different algorithmic advances suggest that simulating thermal states of Fermi--Hubbard models could be enabled by fault-tolerant quantum computers. {\em First}, the discovery of optimized ``split'' Trotter simulation techniques that can simulate more complex variants of the Fermi--Hubbard model, including those with nearest-neighbor, multi-orbital, and non-square lattices with T-counts scaling as \(10^{6}\) to \(10^{8}\) \cite{kan2024resourceoptimizedfaulttolerantsimulationfermihubbard,baysmidt2025faulttolerantquantumsimulationgeneralized}. This keeps the T-counts in the same order of magnitude as Campbell's original observation but also outperforms qubitization-based methods. With the addition of multi-orbital scalings, our methods can be extended from (pure) cuprates to iron-pnictide superconductors \cite{PhysRevB.77.220503,PhysRevB.79.134502}.

{\em The second advancement} is the introduction of efficient Lindbladian (open-system) simulation schemes for quantum computers \cite{chen_quantum_2023,chen2023efficientexactnoncommutativequantum}. These works have pioneered a master equation for the thermalization of quantum many-body systems, which is not easy to develop since, naively, these systems do not satisfy a weak-coupling assumption. The efficient simulation on a quantum computer was established by imitating its classical analogue: the classical Monte Carlo methods \cite{Ding_2025,gilyen2024quantumgeneralizationsglaubermetropolis}. This entailed designing a Lindbladian that satisfies detailed balance and quasi-locality of the evolution, while still scaling logarithmically in the precision and mixing time. It is hard to overstate the success of \textit{classical} Monte-Carlo techniques and it is believed that these breakthroughs will have the same effect in quantum many-body physics, including for Gibbs state preparation. Building on these results, Refs.~\cite{rouzé2024efficientthermalizationuniversalquantum,rouzé2024optimalquantumalgorithmgibbs} established optimal quantum algorithms capable of preparing Gibbs state (at high-temperatures) in a time that scales only logarithmically with the system size. And finally, Ref. \cite{Smid2025FHGibbs} builds on these results for the specific problem of preparing Gibbs states for the weakly-interacting Fermi--Hubbard Hamiltonian at constant temperatures (that do not scale with the system size). This enables them to show that the purified Gibbs state can be prepared for any constant temperature \(\beta = O(1)\) with \(\tilde{O}\left(n^{3} \text { polylog }(1 / \varepsilon)\right)\) gates on \(O(n)\) qubits, where \(\widetilde{O}\) hides polylog factors.

While fascinating progress on quantum algorithms has enabled the simulation of classically challenging Fermi--Hubbard models on quantum computers, the FTQC cost of these simulation is still beyond anything available in the near-term. Significant progress in both quantum hardware and quantum algorithms is needed to make these simulations practical in the next decade but the progress looks promising.

In this section, we have carefully analyzed the resources required to execute quantum algorithms for computationally hard real-world applications  
at utility scale on fault-tolerant superconducting quantum devices. Our work highlights the challenges that will need to be tackled over the next decade.  
In addition to the necessary advancements  pertaining to the quality and scale of quantum hardware, developing a USQC will benefit from various algorithmic and software-stack improvements, including, but not limited to, (i) improvements in the $T$-counts for quantum algorithms that focus on ground-state energy estimation; (ii) improvements in quantum state preparation and observable estimation overheads to enable more-efficient preparation of good ansatz states on quantum computers and more-efficient extraction of physically relevant observables; (iii) search for more-efficient QEC codes and FTQC schemes to reduce the overheads associated with running quantum algorithms fault-tolerantly;
(iv) advancements in automated QRE and FTQC architecture design tools; and (v) tight integration with HPC, which we address in the next section. The Fermi--Hubbard workflow in Section~IV.E and Figure~36 illustrates the need to coordinate tensor-network preprocessing, fault-tolerant execution, and classical analysis.

\section{High-performance hybrid quantum--classical computing} 
\label{sec:hpc-qc}

Achieving utility-scale quantum computing will require seamless integration with existing heterogeneous HPC infrastructure and the development of a comprehensive hybrid quantum–classical software and hardware stack. In particular, executing utility-scale applications requires real-time decoding, QEC orchestration, recalibration, and circuit compilation, which will place heavy demands on conventional HPC resources; as discussed in \Cref{sec:100k-1M_qubits} and \Cref{sec:ftqc}. Consequently, classical HPC components and subsystems will form the backbone for compiling, scheduling,  orchestrating, and managing all types of hybrid quantum–classical workloads. Classical preprocessing and postprocessing require coordinated resource management, while decoding and feedforward additionally impose real-time latency constraints.

At the same time, quantum computing has generated considerable interest in the high-performance computing (HPC) community as a strategy to improve computational capabilities beyond state-of-the-art Exascale systems \cite{alexeev_quantum-centric_2024,ibmPositionPaper,humbleQuantumClassical}. By performing specialized tasks, quantum computers that are integrated with HPC infrastructure have the potential to significantly improve the performance of key bottlenecks of existing HPC applications or to enable those applications to solve problems beyond what is currently possible. Successful acceleration of an HPC application will require improvements to quantum computing hardware as well as tight integration with classical hardware and software so as to minimize overheads. In particular, such acceleration will not be practical if quantum circuit execution experiences latencies of hours to days as reported in recent experiments on cloud-based quantum hardware access \cite{BECK202411}.

Hybrid quantum--classical computing paradigms are essential not only for algorithms in the NISQ era (as discussed in \Cref{sec:100-1000_qubits}), but also for pre-FTQC and full FTQC quantum algorithms embedded within larger quantum--classical solvers and frameworks. In the previous section, we focused on the quantum resource estimation of utility-scale applications without considering details of the classical compute required for the execution. In general, there could be significant demand for high-performance classical pre-processing, co-processing, and postprocessing in standard quantum subroutines such as qubitization, THC, observable estimation, etc. For example, compute-intensive techniques to prepare a good initial ground state for quantum phase estimation will benefit from HPC acceleration. Furthermore, there are a variety of variational algorithms that iterate between classical and quantum subroutines. For example for Fermi-Hubbard simulation one could require hours of HPC time per iteration (e.g., see \Cref{subsec:FTQC_FH2d}).

Circuit knitting and quantum circuit simulation are additional capabilities that rely on HPC resources to be feasible at scale. Circuit knitting enables the execution of quantum circuits that require more qubits than available on any single quantum processor but requires significant classical computing to do so (see \Cref{sec:100-1000_qubits}). Quantum circuit simulation is necessary during application development to reduce costs and to provide a way to inspect quantum states in a way that is not possible with real qubits. As circuit simulation time grows rapidly in the number of qubits, even for approximation techniques, HPC systems are required for simulations of all but the smallest quantum circuits.

The ability to perform mid-circuit measurements is an important component of a programming framework for hybrid quantum--classical computing, and has various use cases. Dynamic circuit programming is one such use case, where real-time classical feedback is needed to execute a quantum circuit based on mid-circuit ancillary qubit measurements. Another set of use cases are advanced adaptive circuit knitting techniques which could rely on mid-circuit measurements and feedforward operations and are discussed in \Cref{sec:hpc-qc-workload-ack}. 
Furthermore, mid-circuit logical-qubit measurements also play a central role in feedforward-based, fault-tolerant implementations of logical non-Clifford gates, as well as in the associated real-time decoding processes (see~\Cref{sec:decoder}). Finally, at large system scales, it will be critical to ensure that circuit compilation does not become a bottleneck, as discussed in \Cref{sec:100k-1M_qubits}, and thus concurrent compiling and execution might become necessary.

There are several challenges that need to be faced in order to successfully integrate HPC and quantum computing into a single system. We have discussed these challenges in \Cref{sec:hpc-qc-challenges}; they include: hardware integration challenges, algorithmic challenges, programmability challenges, the need for efficient scheduling to optimize quantum and HPC resource utilization and minimize iterative latencies. Solutions to these challenges must be factored into the practical design and implementation of a hybrid quantum--HPC system.

A natural starting point to address the quantum--HPC integration challenges is to integrate tools that program, compile, and execute quantum circuits into current classical HPC programming environments. This approach addresses some of the programmability challenges for HPC users by enabling quantum programming in a familiar environment. Existing infrastructure for HPC (e.g., data and user management, process scheduling, control and networking) can then be leveraged for future quantum--HPC systems. For many end users, access at the HPC programming environment level will be familiar and functional and might be already sufficient for most applications. More-advanced users, however, may want lower level control of the underlying qubits and control pulses. Different levels of abstraction in the quantum--HPC software portfolio should be provided for the different needs of the end users.

The diagram in \Cref{figure1_Arc} illustrates an architecture for a comprehensive HPC software portfolio with extensions toward a full quantum--HPC stack.  The HPE Cray Programming Environment (CPE) is a mature HPC programming system that provides software development toolchains supporting a full range of heterogeneous HPC platforms, hardware architectures, and processors. CPE provides support for multiple compilers including HPE Cray, AMD, Intel, Nvidia, and GNU along with libraries, debugging, and performance analysis tools. We propose to extend CPE to include CUDA-Q; to include other existing quantum SDKs (e.g. Qiskit~\cite{qiskit}, Cirq~\cite{cirq}, Pennylane~\cite{pennylanesw}, and Classiq~\cite{classiq}); to integrate with TopQAD as the quantum OS for FTQC, including compilation (see \Cref{sec:ftqc}).}; and to include features that focus on \mbox{HPC--QC} integration. Building on top of CPE reduces development efforts and enables rapid experimentation with different quantum SDKs on available quantum and quantum-inspired accelerators as well as simulators. We can identify and target modular software capable of adapting to emerging and increasingly powerful QPU technologies, while at the same time leveraging existing NISQ QPUs as well as CPU/GPU cores for high-performance simulation. 

The following sections further describe aspects of the integrated high-performance hybrid quantum--classical computing system. \Cref{sec:hpc-qc-programming} discusses software programming frameworks for hybrid applications; \Cref{sec:hpc-qc-workload-ack} discusses adaptive circuit knitting strategies for combining multiple QPUs; \Cref{sec:hpc-qc-scheduling} discusses resource management and scheduling;  \Cref{sec:alg-design} discusses hybrid algorithm design; and \Cref{sec:perf-benchmarking} discusses performance benchmarking in such systems.

\subsection{Programming frameworks for high-performance hybrid quantum--classical computing} 
\label{sec:hpc-qc-programming}

In this section we present a software integration strategy and outline development efforts for extending quantum computing capability within the HPE Cray Programming Environment (CPE). We adopt a modular hardware/device-agnostic approach to develop components for quantum
programming, dispatching, and compilation within CPE. The purpose is to provide users with a unified programming environment and a full quantum--classical stack built upon existing HPC tools (compilers, libraries, parallel runtime, and process scheduling). The quantum computing capability extension includes four key components:
\begin{itemize}
    \item Quantum interface library with application programming interface (API) extensions to enable seamless invocation of quantum kernels from vendor-specific quantum SDKs within HPC applications
    \item Quantum compiler and runtime extensions to enable performant language-level support for quantum constructs and to address the bottlenecks in compile-time with increasing circuit size
    \item Hybrid quantum-classical programming environment providing both GPU kernels and QPU kernels, e.g. via CUDA-Q
    \item FTQC compiler, emulator, and assembler to form a quantum OS, e.g., via integrating with TopQAD
\end{itemize}

\begin{figure*}[htbp]
\centerline{\includegraphics[width=2\columnwidth]{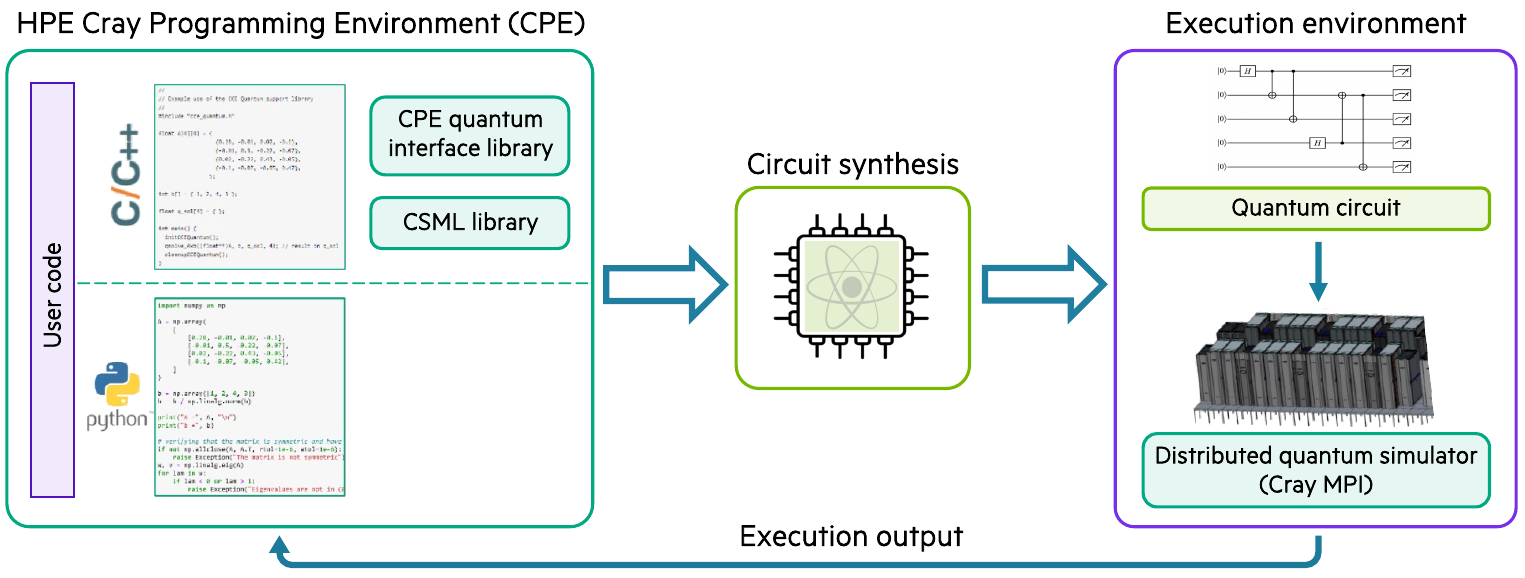}}
\caption{Schematic of a hybrid quantum--HPC application development workflow utilizing the CPE quantum interface library. HPC applications in Python or compiled languages (e.g., C/C++) can call different quantum SDKs (via the Quantum Interface Library) and still link with other libraries, e.g., Cray Science and Math Library (CSML). Circuit synthesis is then available from a variety of quantum SDKs---CUDA-Q, Qiskit, Cirq, Pennylane, Classiq, etc. Synthesized quantum circuits can be subsequently executed on various supported quantum hardware and simulators hosted remotely on cloud or on-premise. Simulation results are sent back as return values to the HPC applications. }
\label{figure3}
\end{figure*}

Circuit synthesis and execution time in quantum computing can vary significantly based on the complexity of the algorithm and the number of qubits involved. Different SDKs offer various levels of optimization and distinct approaches to handling quantum gates, scheduling, and resource allocation, all of which impact the time and efficiency of circuit synthesis and execution. Each quantum SDK (e.g., Qiskit and CUDA-Q) has its native gate set optimized for the underlying hardware. The choice of gate set impacts synthesis complexity since some SDKs might require multiple gates to synthesize specific operations efficiently, while others may directly support them. Decomposition of high-level operations into hardware-native gate sets is usually required. For example, synthesizing a rotation or controlled operation might require multiple CNOT gates and rotations, affecting both gate count and execution time. Many SDKs provide optimization levels to minimize circuit depth, gate count, or overall execution time. These optimizations directly influence both the synthesis time (by making trade-offs for synthesis overhead) and execution time on supported simulators and hardware. An example of time ranges is given in \Cref{sec:TFIM-DQPT}, where circuit synthesis for Trotterization is found to take on the order of milliseconds. Note that circuit synthesis in this case is only for an exact simulation and as such, it does not include FTQC compilation overheads such as decomposing arbitrary-angle rotation gates into Clifford+T gates.

Different quantum SDKs also vary in their approaches to gate scheduling, particularly for parallel gate execution across qubits (when the hardware supports it). Efficient scheduling reduces overall circuit execution time. Some SDKs optimize for parallelizable gates to reduce circuit depth, reducing execution time especially on hardware with limited coherence times. Different SDKs are tailored to different backends, with constraints on qubit connectivity, gate fidelities, and coherence times. Efficient SDKs take these constraints into account during synthesis to minimize gate count and depth based on hardware capabilities. Some SDKs are tightly coupled with specific hardware, while others support multiple hardware backends, including simulators. SDKs that are hardware-agnostic may have longer synthesis times due to added compatibility layers.

\textbf{Quantum interface library: flexible acceleration for HPC applications.}
To meet the needs of HPC users seeking to accelerate their applications with quantum computing, a high-level C library can address programmability challenges by making it easy to call well-known quantum computing algorithms. One such high-level C library is the quantum interface library extension within CPE which has the following advantages: 1) the ability to support and interact with a range of quantum SDKs and backends from different vendors through a standardized interface; 2) data and functionality of other software systems are available while implementation details are abstracted, facilitating efficient and secure application development; 3) support for compiled languages commonly used in massively parallel HPC applications which could allow large-scale system modeling and computation with large datasets; and 4) direct utilization of the existing HPE Cray MPI, which offers GPU-aware MPI support integrates with heterogeneous workload and resource managers such as Slurm and PBS (Portable Batch System) \cite{SLURM, PBS}.

The ability to interact with a range of quantum SDKs makes the quantum interface library flexible. With diverse quantum hardware including superconducting qubits,  neutral atoms, trapped ions, spin-based qubits, and photonic qubits, various quantum software packages focus on different aspects of quantum computing. This may include circuit synthesis or optimizing complex design processes including qubit allocation, auxiliary qubit reuse, error mitigation, or quantum error correction. These quantum software packages are developed in different programming models and have parallelism models supporting different backends. For example, CUDA-Q provides multi-GPU, multi-node state vector and tensor network simulation backends on Nvidia GPUs~\cite{cudaq} (see Appendix \ref{sec:cudaq} for details). Pennylane Lightning, as another example, provides state vector simulators that can be executed on both AMD and Nvidia GPUs~\cite{pennylane}.

To enable seamless invocation of quantum kernels from different quantum SDKs within HPC applications developed in C/C++/Fortran, the Quantum Interface Library is implemented in C, which can be seamlessly interfaced with Fortran applications using the \verb|iso_c_binding| module. Application developers can use high-level, portable invocation of quantum algorithm libraries from a variety of third-party SDKs within a classical HPC application in their programming language of choice. \Cref{figure3} provides a schematic for a hybrid quantum--HPC development workflow within CPE with the quantum interface library.

The CPE quantum interface library has been utilized to deliver two hybrid quantum--HPC applications in both Python and C/C++ with quantum kernels for circuit synthesis and execution provided by Classiq’s Python-based quantum SDK \cite{classiq}. One example consists of solving linear systems of equations with the Harrow--Hassidim--Lloyd (HHL) algorithm~\cite{hhl}. The solution was compared with the CPE BLAS library and showed less than 2\% deviation.  Another example considers the quantum approximate optimization algorithm (QAOA) for solving the MaxCut problem of partitioning a large graph into smaller sub-graphs, which can each be represented on current quantum devices. The workflow is well suited for a hybrid classical--quantum execution on supercomputers, where various sub-problems can also be solved classically if there is an advantage. Initial investigations have been conducted by HPE and Classiq and published at IPDPS24~\cite{esposito2024hybrid}. For a simplified problem, this hybrid workload was demonstrated at ISC24 using a 20-qubit IQM quantum device accessed from the LUMI supercomputer in Finland. Understanding the latency implications in accessing a remote quantum device was one of the main goals of this investigation; the communication overhead was typically found to be on the order of seconds. This latency could be reduced with co-location and tighter integration.

\begin{figure*}[htbp]
\centerline{\includegraphics[width=2\columnwidth]{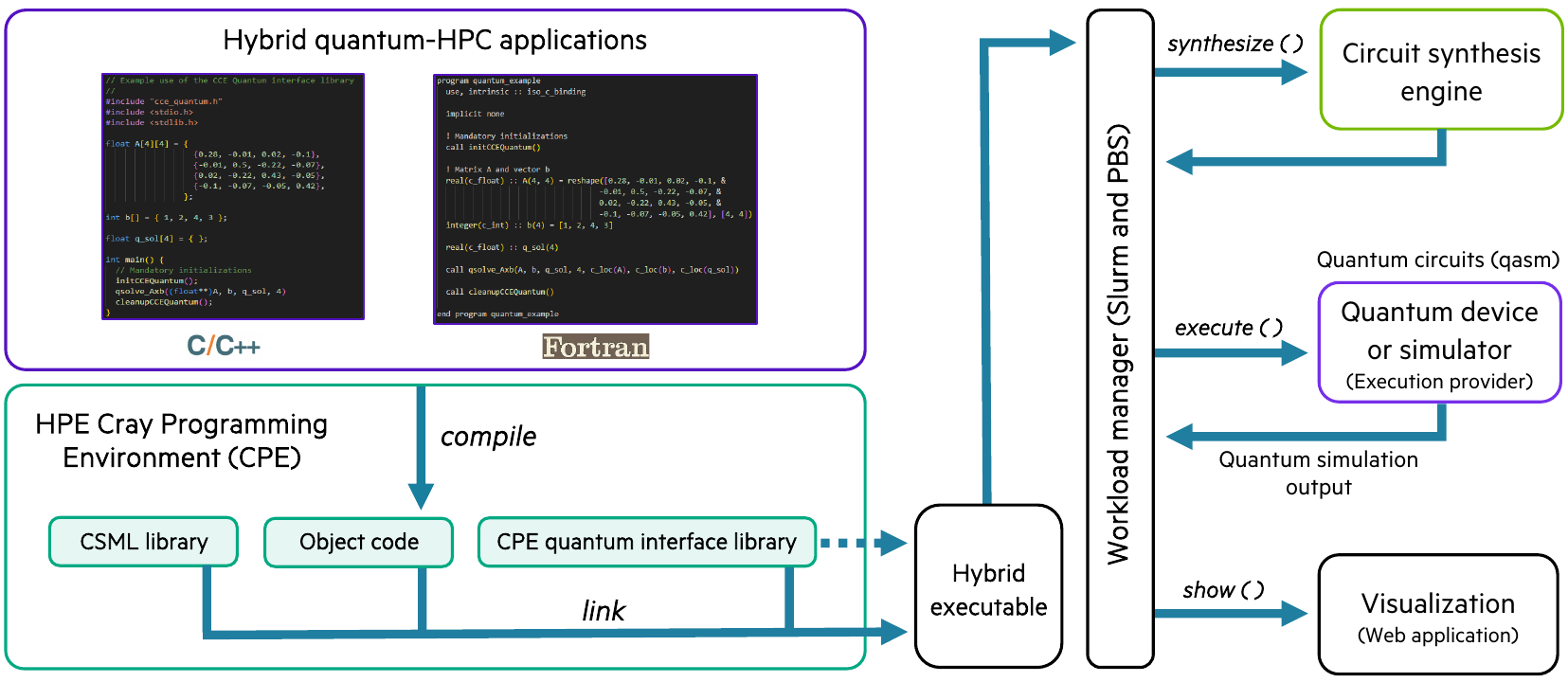}}
\caption{Development and execution of hybrid quantum--HPC applications in C/C++/Fortran through the quantum interface library within the HPE Cray Programming Environment (CPE). A hybrid executable is dispatched to a workload manager (e.g., Slurm or PBS) for parallel execution. Remote visualization through web applications can be enabled with a different protocol and configurations for secure connection to clusters.}
\label{Figure_QIL_last}
\end{figure*}

Figure \ref{Figure_QIL_last} presents a detailed illustration of hybrid quantum--HPC application development and execution within CPE with the quantum interface library and HPC workload management. Within CPE, users can automatically utilize debugging and profiling tools as well as math and communication libraries with chosen compilers. These compilers (for Fortran, C, and C++) are designed to extract maximum performance from a variety of architectures like ARM and x86-64 and devices like AMD and Nvidia GPUs. In addition, users have access to the HPE Cray Message Passing Interface (MPI), a highly scalable implementation for collective communications. Applications developed in C/C++/Fortran are compiled and linked within CPE, potentially including other libraries such as the Cray Science and Math Libraries (CSML) if the HPC application requires it. The quantum interface library is linked as a shared library during the application build process. Hybrid executables can be submitted and scheduled for parallel execution by a workload manager such as Slurm or PBS. At runtime, the interface library routes the quantum API calls from the application to the respective vendor-specific quantum SDKs with appropriate data handling (e.g., parameters and return values). When a quantum API call provided by the interface library is invoked by the application at runtime: 
\begin{itemize}
\item   Arguments to the API are converted for passing on to vendor-specific SDK;
\item   The interface library calls the relevant vendor-specific SDK routines to compile the quantum code into a quantum assembly language (such as OpenQASM~\cite{qasm}, Quil~\cite{quil}, etc.) for a given gate-based quantum device or simulator;
\item   The interface library calls the vendor-specific SDK routines to execute the quantum assembly on compatible QPUs or simulators on-premise or on remote cloud-hosted resources; and
\item   The quantum circuit is executed (either on a quantum device or simulator) and results are converted and passed back to the application as return values of the quantum API.
\end{itemize}

\textbf{Quantum compiler and runtime extensions.} With increasing qubit counts, higher gate fidelity, and improved coherence times and scalability, compilation bottlenecks and latency between classical and quantum components become more apparent. As discussed in \Cref{sec:100k-1M_qubits}, When scaling to large circuit sizes with $\sim$100 or more qubits, declarative frameworks struggle to handle compilation bottlenecks and latency between classical and quantum components in the program, even for NISQ systems. Leveraging classical compilation tool chains such as \mbox{Clang/LLVM} is crucial for developing large-scale hybrid quantum--HPC workloads, and research efforts are moving in this direction \cite{llvmq, plq}.

To enable performant language-level support for quantum constructs and to address bottlenecks in compile-time with increasing circuit size, we have on-going efforts to build extensions on top of classical compilation tool chains. Given the diversity of pulse-level quantum instructions by different quantum hardware vendors, the quantum assembly language OpenQASM and LLVM IR with its extension to Quantum Intermediate Representation (QIR) \cite{qir} are adopted as a middle ground for comparability with different quantum hardware and software. We leverage codegen modules to emit machine-specific native code for target architectures based on the LLVM IR/QIR produced by different quantum software front ends such as CUDA-Q.

This low-level integration at the IR level will ensure hardware support, software compatibility, and optimal circuit compilation and execution performance for large-scale quantum--HPC workload development.

\textbf{A programming framework for heterogeneous quantum--classical supercomputing via CUDA-Q.} A major challenge in heterogeneous quantum--classical computing is the development of software platforms that enable efficient programming of quantum--HPC systems. Many near- and mid-term algorithms and workflows require the simultaneous and efficient utilization of CPUs, GPUs, and QPUs. Achieving this level of performance requires accurate resource estimation and the development of schedulers capable of balancing algorithms and workloads across accelerators to maximize overall system utilization.
One major obstacle is the access to hardware specifications and software libraries developed by quantum hardware vendors, which complicates integration. Additionally, the lack of standardized interfaces and protocols further hinders the seamless integration of QPUs into existing HPC systems.

CUDA-Q provides the NVQ++ compiler for quantum kernels lowering to QIR eventually, as well as a standard library of quantum algorithmic primitives with upcoming support for quantum error correction primitives. Generated code, when linked with appropriate quantum vendor-provided runtime libraries, can be executed on target quantum devices or simulator backends in a quantum-backend-retargetable manner.  
 
Although various quantum--HPC platforms could be developed, those designed for near-term critical applications (e.g., quantum error correction) must be built to inherently support low-latency execution and tight coordination between GPUs and QPUs. A key concept for such software platforms is the use of kernels as units of work that can be executed on either GPUs or QPUs and efficiently transferred between them. CUDA-Q is a particularly promising software solution due to its architecture, which natively supports GPU-based acceleration and kernel execution. This makes it well-suited for a tightly coupled heterogeneous quantum--GPU system, like DGX Quantum, that requires efficient resource allocation and load balancing across quantum and classical resources.

\begin{figure*}[htbp]
\centerline{\includegraphics[width=2\columnwidth]{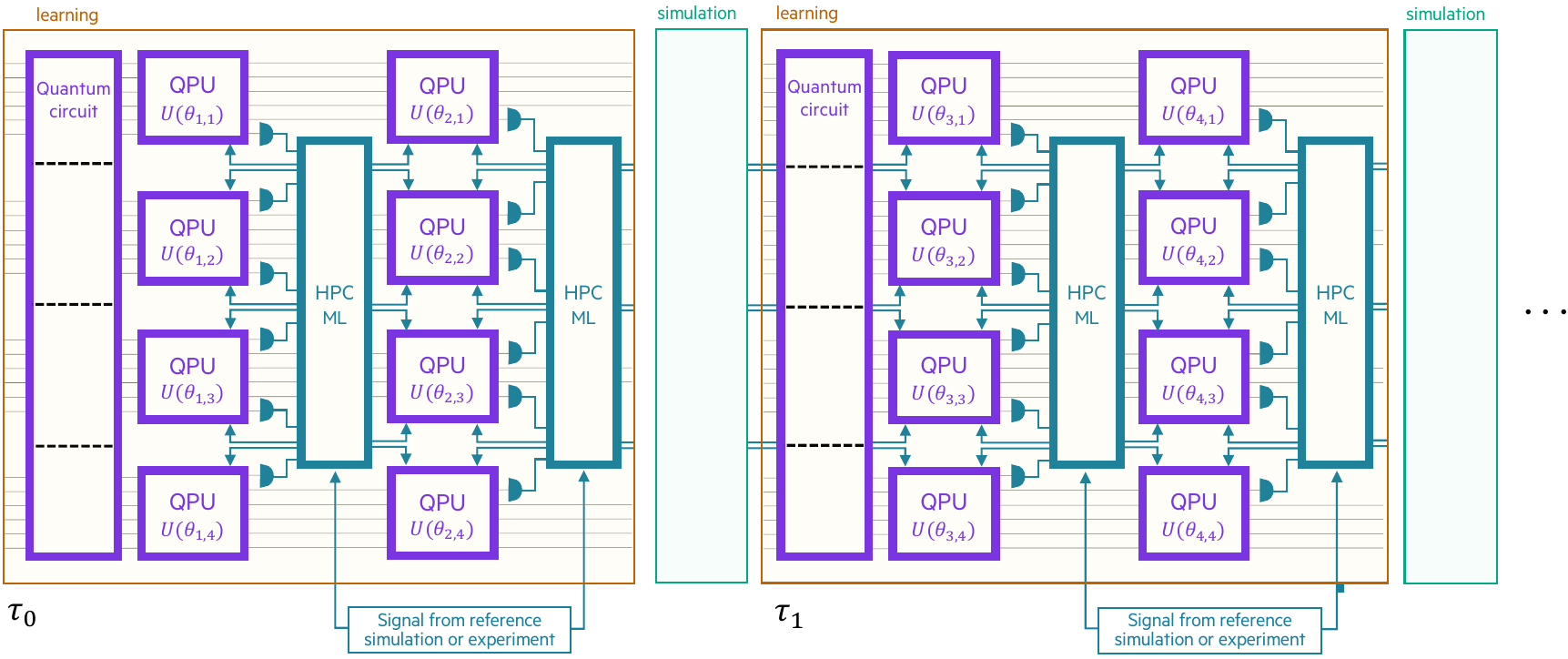}}
\caption{Adaptive circuit
knitting hypervisor for distributed quantum simulation/learning: layer-wise learning of quantum correlations with feedforward mechanism could dynamically reveal an entanglement heat map. This information can guide which correlations to keep and which to ignore (circuit cuts), thus minimizing the exponential classical post-processing corresponding to circuit knitting. The overall pipeline could act as a quantum generative model which outputs quantum data that can be fed to other quantum processors.
}
\label{Figure_DCK}
\end{figure*}

CUDA-Q and NVQlink enable dynamic circuit programming and mid-circuit measurements, and moreover can efficiently program the OP-NIC based solution, which excels in the low-latency communication between QPUs and HPC that are required for these use cases. For further information on CUDA-Q, see \Cref{sec:cudaq}.

\textbf{FTQC compiler, emulator, and assembler via TopQAD.}
TopQAD translates synthesized quantum circuits provided as input from a variety of quantum SDKs into executable machine-level instructions for  controllers and decoders within the heterogeneous \mbox{HPC--QC} platform. Dynamic code generation techniques are adopted to generate code at runtime based on specific characteristics of the target quantum devices. As outlined in \Cref{sec:ftqc}, FTQC compilers, unlike conventional compilers for classical computers, must incorporate detailed knowledge of the hardware noise profile. Given the long execution times of utility-scale FTQC algorithms, frequent mid-execution recalibration of QPUs is required. This recalibration is achieved by dispatching a variety of quantum characterization, verification and validation (QCVV) protocols \cite{blumekohout2025quantumcharacterizationverificationvalidation} to the hardware through the control electronics at an appropriate cadence. The results of these QCVV experiments can then be used to infer noise models for different types of lattice surgeries and also to adjust the control sequences delivered to the hardware for recalibration. A quantum OS for FTQC must therefore perform offline orchestration of real-time QPU and decoder characterization, modeling, and performance analysis, and incorporate this information into the compilation pipeline for FTQC execution. In addition, a QEC orchestrator module, which receives information from the decoder API, is responsible for the efficient refactoring of recovery operations in conjunction with surgery schedules. Dynamic code generation is thus essential for consolidating information generated in real time to enable the compiler and assembler to optimize the execution of quantum programs across different hardware generations or architectures. 

Finally, mid-circuit logical-qubit measurements also play a central role in feedforward-based, fault-tolerant implementations of logical non-Clifford gates, as well as in the associated real-time decoding processes (see~\Cref{sec:decoder}). As discussed in~\Cref{sec:ftqc-compile}, our FTQC compilation framework relies on compiling a given quantum algorithm into a sequence consisting solely of multi-qubit $\pi/8$ Pauli rotations. Mid-circuit logical-qubit measurements are essential for constructing fault-tolerant quantum gadgets that efficiently implement such multi-qubit $\pi/8$ Pauli rotations \cite{litinski2019game}. In these implementation schemes, conditional operations---logical gates or logical measurements---are applied via feedforward of the outcomes of earlier logical measurements, as well as based on real-time decoding of syndrome data accumulated over many preceding QEC cycles; see also Ref.~\cite{khalid2025impacts}. Utility-scale performance is an operational capability and intimately relies on tightly-integrated and performant FTQC components, not just their benchmarking. Dynamic code components such as the FTQC compiler, emulator and assembler are an integral part of these functional components.

\subsection{High-performance quantum workload distribution with Adaptive Circuit Knitting}
\label{sec:hpc-qc-workload-ack}

For quantum computing to operate at scale, it will be necessary to efficiently parallelize over many QPUs, possibly of different hardware types, integrating them as coprocessors within an HPC framework. As an example of quantum workload distribution technique, we develop a novel adaptive circuit knitting strategy serving as a hypervisor for classical and quantum communication between distributed classical and quantum compute nodes. Unlike traditional circuit knitting, this strategy uses machine learning  at multiple levels to learn a quantum circuit decomposition capturing maximal quantum entanglement while enabling high-performance communication at scale. By integrating scalable message passing techniques within an adaptive circuit knitting approach, we could efficiently learn how to perform distributed quantum simulation or distributed quantum machine learning over quantum data generated by quantum processors themselves. Follow up work by a subset of the coauthors expand on this novel approach with application to disordered quantum systems \cite{johnson2026distributed} and electronic structure Hamiltonian \cite{necaise2026distributioncomplexityelectronicstructure}.

Existing quantum processors have relatively few qubits with low gate fidelity. With the current state of technology, it is highly unlikely for NISQ devices to be scaled to tackle utility-scale problems \cite{PhysRevResearch.4.033154}. Moreover decreasing noise (and therefore increasing quantum circuit depth) is crucial to avoid simulability by classical techniques such as tensor networks \cite{Berezutskii2025} and Pauli Propagation \cite{rudolph2025paulipropagationcomputationalframework}. While proof-of-principle demonstrations of  logical qubits are just starting to appear, even upcoming error-corrected quantum computers will be small with respect to the number of logical qubits required for utility, according to all industrial roadmaps. Recent resource estimates for FTQC indicates that the required number of physical qubits are  about 0.5M--2M for quantum dynamics, 1M-6M for quantum chemistry, and 1M--30M for integer factoring \cite{beverland2022assessing,gidney2025factor2048bitrsa,Zhou2025}. Since our focus is on superconducting qubits, we do not discuss the recent improvements in quantum resource estimates for FTQC applications using quantum low-density parity-check (qLDPC) codes~\cite{yoder2025tourgrossmodularquantum,webster2026pinnaclearchitecturereducingcost}. At the same time, the largest quantum processors to date are on the order of hundreds of qubits, and even most optimistic roadmaps do not envision more than 100k qubits at a single QPU level. Consequently, to reach utility-scale quantum computing, efficiently distributing computation across multiple QPUs will be required.

Efficiently partitioning quantum systems has a rich history in quantum science \cite{white1993density, meyer1990multi, manthe2008multilayer}. In recent years, circuit knitting \cite{peng2020simulating, piveteau2023circuit} has emerged as a  method to distribute quantum circuits using only classical communication, thereby avoiding the reliance on quantum interconnect technologies that are still in early stages of development. The primary goal of circuit knitting is to enable simulating large circuits on NISQ devices available today---a measured observable is reconstructed by sampling sub-circuits of the original circuit multiple times. However, circuit knitting can be also applied to early-FTQC and full FTQC architectures at the logical layer. The circuit knitting partitioning was historically introduced as an error mitigation mechanism by using a quasi-probability decomposition to mimic the output of a large noiseless quantum circuit by a number of smaller noisy quantum circuits \cite{Temme_quasi_2017,piveteau_quasiprobability_2022}. However, this reconstruction comes at an exponential cost in the number of samples that depends on the identity and number of gates that have been cut out \cite{piveteau2023circuit}. While recent efforts have focused on reducing this exponential overhead \cite{tang2021cutqc, tang2022scaleqc, basu2024fragqc} and even demonstrated a real-time classical interconnect between two superconducting QPUs \cite{carrera2024combining}, further work is necessary to demonstrate the practical advantage of circuit knitting.

\textbf{Adaptive circuit knitting:} To mitigate the exponential sampling complexity of the current circuit-knitting techniques, we introduce adaptive circuit knitting (ACK), a method based on finding low-entanglement cuts in the circuit. ACK is built upon the theoretical foundation of quasiprobability decomposition (QPD) \cite{Temme_quasi_2017, piveteau_quasiprobability_2022}. QPD allows one to express a unitary channel $\mathcal{U}$ acting across two separate quantum processors $A$ and $B$ as a weighted sum of operations $\mathcal{F}_j$'s, i.e.,
\begin{equation}
    \mathcal{U} = \sum_j c_j \,\mathcal{F}_j,
\end{equation}
where, $\mathcal{F}_j$'s are operations that can be physically realized on the (independent) quantum hardware, for example, local operations (LO), or local operations with one-way/two-way classical communication (LOCC) \cite{piveteau2023circuit}. The coefficients $c_j$ are real numbers which can be negative, but we can absorb the sign and normalization factor into the operation $\mathcal{F}_{A/B,j}$ and turn them into a true probability distribution, hence the name quasiprobability. The cost of this sampling is quantified by the overhead 
\begin{equation}
\gamma = \sum_j |c_j|, 
\label{eq:gamma_factor}
\end{equation}
the gamma factor. The number of samples required to estimate a linear observable with error $\epsilon$ is $O(\gamma^2/\epsilon^2)$. For two-qubit gates, this overhead ranges from $\gamma = 1$ to $\gamma = 7$; for instance, $\gamma(\mathrm{CNOT}) = 3$ and $\gamma(\mathrm{SWAP}) = 7$, while for product unitaries, i.e., $U_{AB} = U_A \otimes U_B$, $\gamma(U) = 1$ \cite{schmitt_2025_cutting_circuits}. The sampling overhead of cutting multiple two-qubit gates (the total gamma factor) is the product of individual gamma factors. In this way, the sample complexity of a typical circuit scales \textit{exponentially} i.e., $O(\gamma^{2n}/\epsilon^2)$ in the number of two-qubit gates cut.

Instead of cutting gates by performing QPD on two-qubit gates one by one, we can also perform a \emph{combined cutting}, where a single quasi-probability decomposition is applied jointly across all gates that cross the partition, rather than decomposing each gate independently. Formally, if $\{ U_1, U_2, \dots, U_n \}$ are gates acting across a cut, individual cutting applies QPD to each $U_i$ separately, leading to an overhead scaling like $\prod_j \gamma(U_j)$. In contrast, combined cutting constructs a joint decomposition over the entire set $\{ U_1, \dots, U_n \}$, which can exploit correlations among these gates to reduce the multiplicative blow-up in the gamma factor \cite{ufrecht2024jointcutting}. In Ref. \cite{schmitt_2025_cutting_circuits}, the authors show that for combined cutting of general two-qubit unitaries, the optimal sampling overhead can be achieved even without classical communication. 
Similarly, Ref. \cite{harrow_2025_optimal_quantum_circuit_cuts} gives a double-Hadamard test-based protocol for achieving the optimal overhead for cutting a generic unitary without classical communication. However, whether classical communication improves the circuit-knitting overhead for combined cutting of such general unitaries remains an open question.

In addition to cutting gates, one can also cut wires (time-like cuts) to simulate circuits beyond a device's coherence time \cite{peng2020simulating, lowe2023fast}. Unlike space-like cuts, where the classical communication doesn't strictly improve the overhead, wire cutting overhead scales strictly multiplicatively without it \cite{brenner2025optimalwirecutting}, whereas communication can significantly improve this via QPD. While utilizing both cut types facilitates distributed computation of generic circuits, the cost is invariably exponential in the number of cuts. Focusing on the former approach, ACK targets gate cutting (space-like cuts) to mitigate the exponential sample overhead for certain quantum systems.

A natural application of ACK is the simulation of disordered quantum many‑body systems. In the low‑disorder regime, the entanglement grows linearly \cite{kim_huse_2013_ballistic_spreading} in time, whereas in the high‑disorder, many‑body‑localized regime, the entanglement entropy grows only logarithmically \cite{serbyn_2013_universal_slow_growth, abanin2019colloquium}. In this strongly disordered regime, the system is expected to be classically simulable. However, a broad intermediate regime exists in which the entanglement growth crosses over between linear and logarithmic behavior. In this crossover regime, we expect heterogeneous structures, with islands of large entanglement connected by regions of low entanglement. The high‑entanglement regions require distributed quantum computing rather than classical tensor‑network methods, while the low‑entanglement regions would be amenable to ACK with manageable sampling overhead. In \cref{sec:adaptive-circuit-knitting-experiments}, we present a concrete example of a distributed simulation of a disordered quantum many-body system enabled by ACK using a tensor network ansatz. 

Although disordered quantum systems may appear to be the primary application of our proposed algorithm, the scope of ACK could extend to more general quantum algorithms. One notable example is the quantum Fourier transform (QFT), a core subroutine in many important quantum algorithms such as Shor’s factoring and quantum phase estimation. The QFT exhibits low (constant in system size) operator entanglement \cite{aharonov_2007_quantum_fft_classically_simulated, chen_2023_quantum_fourier_transform}, making it especially amenable to efficient ACK-based simulation. In addition, our adpative circuit cutting technique is valuable in optimizing distributed quantum computing architectures connected via quantum interconnects, where minimizing the consumption of entangled resources, such as Bell pairs, is essential for achieving scalability.

\begin{figure*}[htbp]
\centerline{\includegraphics[width=2\columnwidth]{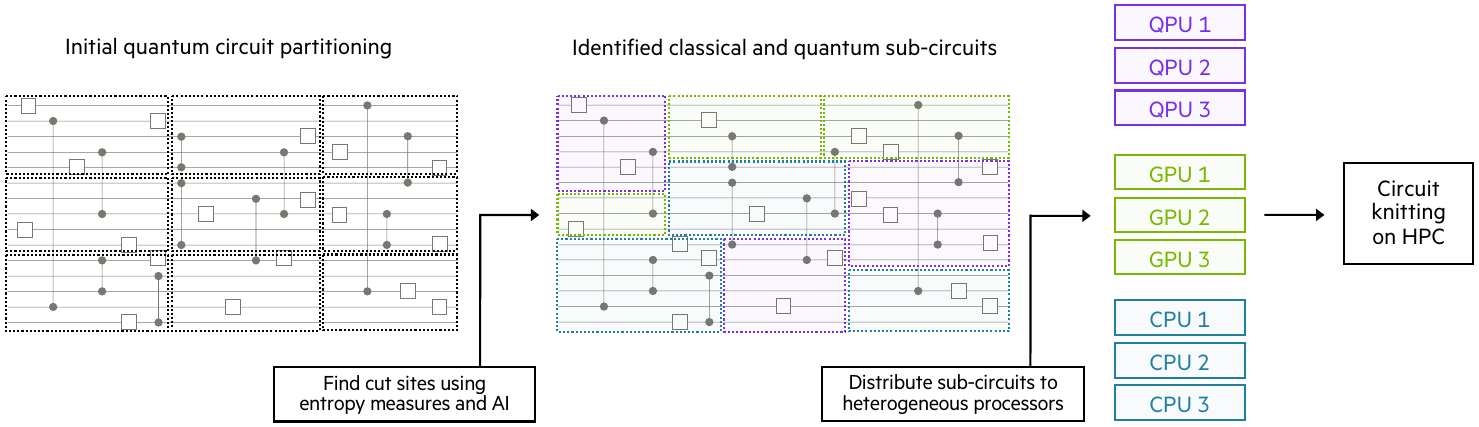}}
\caption{Space-time Heterogeneous Circuit Knitting (SHCK) framework. This diagram shows the main steps in the workflow: initial sub-circuit partitioning, finding optimal cut sites in space and time for heterogeneous execution, performing wire cutting, and assigning sub-circuits to the execution of QPUs, GPUs, and CPUs. The last step is observable reconstruction using circuit knitting, which is accelerated by execution on HPC.
}
\label{Figure_HCK}
\end{figure*}

Inspired by ACK, we introduce a family of hybrid quantum–classical algorithms designed for distributed quantum machine learning. As shown in \Cref{Figure_DCK}, our framework consists of alternating learning layers and simulation layers.
This can be cosidered as a generalization of layer‑wise quantum circuit learning \cite{skolik2021layerwise, carolan2020variational} to layer-wise non-unitary learning to capture relevant quantum correlations. This is achieved through a sequence of distributed positive operator‑valued measurements (POVM) enabled by measurement-and-feedforward operations. Within this layer, the algorithm adaptively determines which quantum correlations are essential to retain and which can be discarded, thereby reducing the sampling overhead associated with ACK.
Furthermore, a feedback mechanism is applied across the distributed variational circuits, allowing the system to iteratively refine the circuit structure. 
This includes optimizing the initial partitioning strategy in an adaptive manner to improve learning efficiency.

The ACK framework can act as a hypervisor that learns an efficient communication decomposition to support execution in a hybrid quantum--HPC ecosystem. As shown in Figure \ref{figure1_Arc}, the hypervisor can be developed within CPE while adopting a hardware-agnostic approach. HPE Cray MPI, which offers "GPU aware" MPI support, can be used to handle message passing and process management for distributed variational quantum circuit execution across multiple nodes.  Integration with PennyLane and CUDA-Q can allow direct utilization of state vector simulators and tensor network simulators with multi-node/multi-GPU support as well as allowing quantum circuits to be dispatched to multiple quantum devices from different vendors.  

We explore the efficiency of the ACK approach for simulating disordered quantum spin-glass systems through integration with CUDA-Q within CPE, see \Cref{sec:adaptive-circuit-knitting-experiments}. Scalability and performance benchmarks can be assessed across a variety of supercomputing systems with different hardware architectures and processors. In principle, ACK could allow approximate simulations of disordered quantum many-body systems for up to thousands of qubits, but the accuracy of such approaches needs to be investigated.

As a generalization of ACK, we also introduce Space-time Heterogeneous Circuit Knitting (SHCK), shown in Figure \ref{Figure_HCK}, as an approach to dynamically cut and merge circuits in both space and time and allocate sub-circuits executions across various quantum and classical accelerators. Key differences in this approach include the use of AI to identify optimal cut sites and the use of quantum wire cutting rather than gate cutting. Wire cutting is a more powerful technique than gate cutting, as it allows for the cutting of a quantum circuit across both time and space. Gate cutting, in contrast, is restricted to decomposing only two-qubit gate operations. The execution of sub-circuits is performed in parallel on QPUs, GPUs, and CPUs. The target for execution of a sub-circuit (i.e. QPU or the type of circuit simulator) depends on the underlying property of a sub-circuit. In this approach, highly entangled sub-circuits are assigned to execution on a QPU and less entangled sub-circuits on a GPU or CPU, e.g. using CUDA-Q simulators. One way to estimate whether a sub-circuit can be simulated classically is to perform a symbolic simulation of the tensor network without actual execution in order to estimate a bond dimension. Finding optimal cut sites could be achieved using von Neumann quantum entropy estimation combined with AI. Effective load balancing between QPU, GPU, and CPU tasks is a critical component of this workflow to ensure efficient utilization of both quantum and classical resources and to minimize overall execution time.

\subsection{High-performance quantum--classical workload scheduling}
\label{sec:hpc-qc-scheduling}

Efficient utilization of quantum computing resources is imperative due to their scarcity. This section explores various factors contributing to the utilization of quantum computers, especially when integrated into a multi-user environment such as an HPC cluster.

Firstly, a significant portion of quantum computer utilization is attributed to the time spent on calibration and readiness for task execution. To ensure continuous calibration, it is essential to efficiently execute the calibration graph, especially on larger scale quantum processors. In addition, it will be necessary to submit quantum benchmarking tasks to provide the quantum computer management software with information regarding its calibration status and necessary re-calibrations.

Secondly, a considerable overhead in running hybrid quantum--classical algorithms is associated with executing and loading quantum tasks, particularly when submitted by users. Quantum computing tasks exhibit significantly different timescales compared to typical HPC jobs. For instance, tasks involving the measurement of parameterized circuits may take only milliseconds for certain modalities like superconducting qubits, while other modalities with longer shot times may take several seconds. Note that a task duration may increase substantially when submitting a task that includes an entire iterative process. Maintaining high utilization for such tasks necessitates the implementation of an ultra-low latency interface between classical and quantum computation systems, exemplified by the DGX Quantum system. Examples of HPC jobs executing numerous short quantum tasks are hybrid algorithms with iterative quantum and classical coprocessing. Such hybrid algorithms perform an iterative process of measuring parametric circuits and subsequently calculating the next set of parameters based on the obtained measured results. During the execution of a single algorithm, the quantum computer often remains idle while HPC nodes retrieve measurement results, perform calculations, and submit new quantum tasks. This idle time between quantum tasks within the same HPC job can be particularly prolonged in an interactive workflow, where user actions trigger the submission of subsequent quantum tasks. 

\begin{figure}
    \centering
    \includegraphics[width=\columnwidth]{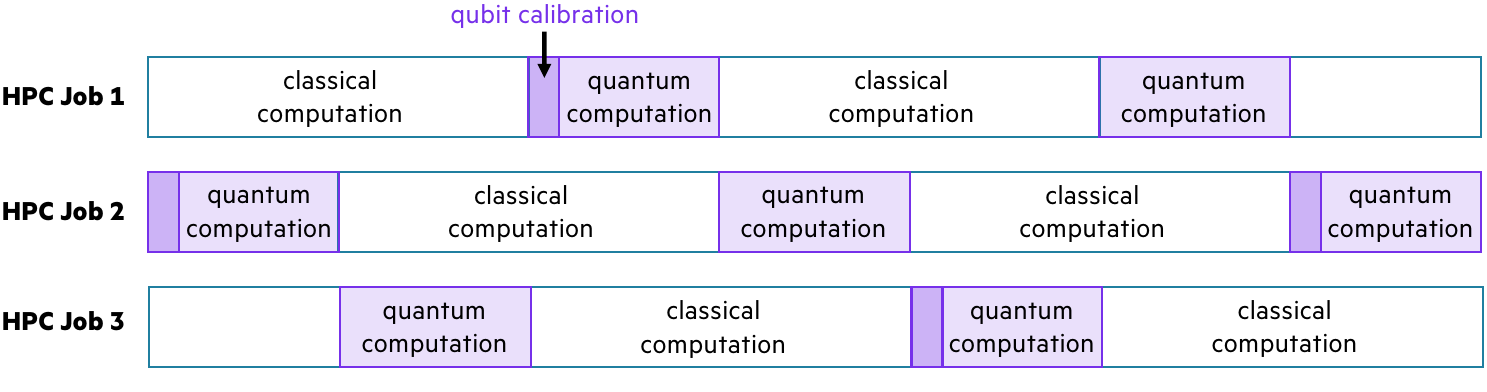}
    \caption{An example of three iterative classical--quantum algorithms concurrently executed by three different HPC jobs with an efficient scheduling of quantum computation tasks. Qubit calibration may be required during execution.}
    \label{fig:scheduling}
\end{figure}

To optimize quantum computer utilization, it is essential to schedule tasks from different algorithms and different HPC jobs concurrently, thereby minimizing idle periods while the algorithm performs the classical computations, as shown in Figure \ref{fig:scheduling}.

In the case where an optimal scheduling of classical and quantum resources is not possible or not necessary, block allocation of a quantum device by a single user is legitimate, especially in the presence of several quantum resources. This can be achieved by workload managers (WLM/Scheduler) such as Slurm that are already present in HPC environments. With a workload manager, quantum resources could be exposed as regular nodes encapsulated in a partition.

\begin{figure}[!t]
\centering
\includegraphics[width=\columnwidth]{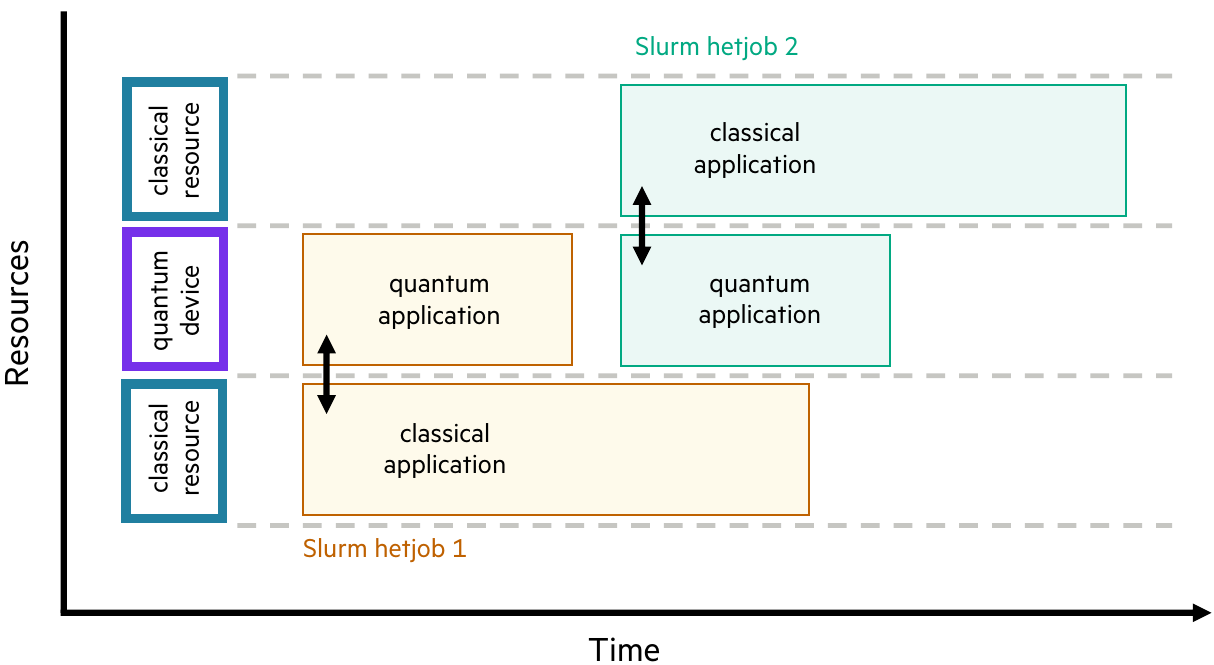}
\caption{Schematic representation of two Slurm heterogeneous jobs requiring a quantum device that is exposed as a compute node. As soon as the quantum device is no longer needed by the first heterogeneous job it can be released while the classical part continues to run. The second heterogeneous job can then start using the quantum device.}
\label{fig:sim3}
\end{figure}

Figure \ref{fig:sim3} represents an example scenario in which one or more classical applications running on multiple nodes share the qubits in all/none fashion. This idea originates from a demonstration at ISC23 where the scheduling and inter-process communication was accomplished via the multiple data multiple program (MPMD) paradigm in Slurm. This model is simple, but has the drawback of blocking a quantum device for the duration of the classical application, potentially wasting resources. A reduction of this idle time can be achieved through the Slurm support for heterogeneous jobs (\texttt{hetjobs}) to split a job across differing hardware~\cite{esposito2023hybrid}. In a simple scenario, two heterogeneous jobs submitted to an execution queue each require classical and quantum computing resources.  Once resources are available, both the classical and quantum parts of a job begin. At a crucial point in the execution, a synchronization requires the classical part to wait on results from its quantum counterpart. Once the quantum computation is finished, the resource is freed and then immediately consumed by the quantum part of the next hybrid job, which has been waiting in the queue for the resource to become available. The contemporary start of the quantum and classical computations is not always guaranteed. 

In order to enable the efficient fine-grained scheduling mentioned above, more intelligent adaptive and heterogeneous task scheduling algorithms must account for quantum resource availability and dynamically adjust task assignments based on runtime feedback. 
A customized {\tt Slurm} plugin to discretize quantum workloads into pulse-level tasks needs to be developed. It will also be necessary to adopt a partition-based approach which allows the application to split available qubits based on the underlying quantum resources. A WLM for hybrid systems should enable the allocation of available qubits among users across different nodes containing QPUs. In the following section (\Cref{sec:NISQ-applications}) we provide a few examples of high-performance distributed quantum simulations for studying dynamics of quantum spin-glasses near quantum phase transitions. These examples could provide useful testbeds for developing high-performance hybrid workload scheduling.

\subsection{Algorithm design for hybrid quantum--classical computing}
\label{sec:alg-design}

The design of quantum computers and the impact of quantum computing on utility-scale applications remain poorly understood, with much room for improvement. 
Despite great strides, quantum processors remain small and fragile. There are now dozens of examples of quantum or hybrid quantum--classical algorithms with a proven or expected quantum advantage, but definitive analysis of their performance in practice is elusive, in part because we lack the quantum hardware necessary and simulation suffers from exponential overhead. As hardware matures, we expect quantum and hybrid algorithms to have greater adoption in applications.

Utility-scale quantum computers will inevitably contain numerous classical components for control, calibration, decoding and Pauli frame updates, and support for quantum information transfer through quantum teleportation. At utility scale, many, if not most, algorithms utilizing quantum computing will also use classical computing in a prominent way, not only supporting quantum computation, but also taking a primary role in the computing itself.  Hybrid algorithms will dominate well into the era of FTQC, and in many cases will remain the best-performing algorithms in the long term. 
For some computational tasks such as sorting \cite{Hoyer2001quantum}, it has been proven that there can be no quantum advantage even for error-free computation. For many quantum algorithms, it remains unclear how to develop quantum compilation strategies that utilize the underlying structure of the problem. Therefore, we do not expect quantum advantage in many conceivable applications, e.g., for quantum search~\cite{ViamontesMH05search}, because existing classical techniques tend to show better-than-worst case performance in important cases. Such tasks will continue to be addressed by classical computing.

Even in the case of a proven asymptotic quantum advantage, this advantage might not survive at the utility scale due to large overhead. For example, it has been shown that proven quadratic speedups of many quantum algorithms based on Grover's amplitude amplification (for computational tasks such as, e.g.,  search, optimization, Monte Carlo methods, and various machine learning tasks), will not result in a quantum advantage on early generations of FTQC devices due to the massive overheads associated with QEC (see Ref.~\cite{babbush2021focus}). 
Furthermore, classical logic gates will likely remain orders of magnitude faster and smaller than logical quantum gates even after significant advances in hardware and QEC. Quantum resources, including logical qubits and gates, are also expected to remain costly and limited well into the FTQC era. All of these factors have implications for algorithm design.

While programs that run quantum algorithms usually comprise multiple gates being executed simultaneously, research on distributed quantum computing hardware is still in its infancy and will be a critical area of research going forward. Some design criteria include advancing classical and quantum algorithms jointly; taking advantage of decades' worth of knowledge in classical algorithms and HPC; considering constants and logarithmic factors that are often ignored due to the use of asymptotic big O complexities; identifying common quantum subroutines for efficiency improvements; carefully assessing the costs of accessing and incorporating classical data and information into quantum subroutines and vice versa; including FTQC considerations when analyzing quantum algorithmic impact; exploring the simplest approaches that might work, with an eye to ruling them in or out; and engaging in the co-design of algorithms, error correction, and hardware, both quantum and classical. We use some of these co-design criteria for electronic-structure quantum simulations on FTQC in Section \ref{sec:resource-estimates-pbenzyne} and for distributed quantum simulations of strongly disordered systems in condensed phase in Section \ref{sec:NISQ-applications}. Here, we briefly discuss additional examples in areas beyond simulation of quantum systems.

{\bf Beyond quadratic speedups in combinatorial optimization.} In principle, subroutines of many classical search algorithms can be replaced by quantum subroutines with a provable quadratic quantum speedup. An example of such a case is when the overall classical algorithm takes advantage of structure but there is an absence of any known structure, or an ignorance as to how to take further advantage of structure, at the subroutine level. However, one must be careful to ensure access to the classical data is well accounted for. As one example, one of the authors of the present paper has analyzed providing various levels of quantum assistance to state-of-the-art approaches to constraint programming \cite{Booth2021}. There is concern that because of large constant factor overheads, a quadratic speedup will not be useful except possibly in the far fault-tolerant regime \cite{babbush2021focus}. Quartic speedups, however, do not suffer the same problem. There is recent analytical work that indicates quartic or even exponential speedups may be possible for some combinatorial optimization problems \cite{chakrabarti2024generalized,jordan2024optimization}, among other numerical observations of possible favorable scaling for hybrid quantum--classical approaches to combinatorial optimization \cite{sankar2024benchmarking}. There is promising further work in this direction: extending the range of techniques providing such speedups,  understanding how to incorporate such techniques in hybrid quantum--classical approaches, and determining how best to do so in a distributed fashion. Many groups continue exploring QAOA and other variational algorithms, which are examples of the simplest algorithms that might work. Some of these techniques have now been shown to be simulable classically and thus ruled out, while hope remains for certain variants. Recent work on iterative quantum algorithms \cite{Brady2023, dupont2023quantum} takes advantage of more-sophisticated classical approaches including noise-directed adaptive remapping, mid-circuit measurements, and inspiration from quantum control theory \cite{magann2022lyapunov, brady2024focqs,stollenwerk2024measurement,maciejewski2024multilevel,maciejewski2024improving}, with ties to analog quantum algorithms \cite{brady2021optimal}. 

{\bf Common subroutines.} Many algorithms that take advantage of quantum computation require quantum versions of classical routines. Such building blocks include the implementation of elementary numerical functions that have well-controlled numerical error, are uniformly scalable and reversible, and can be implemented efficiently. Prior work has provided explicit quantum circuits for some such subroutines for a variety of numerical functions important in scientific computing \cite{hadfield2018quantum,hadfield2021representation}. Developing explicit circuits for other subroutines, creating distributed versions of these subroutines for hybrid architectures, and understanding when to employ classical vs. quantum computing are key directions for realizing the impact of quantum computing. As another example, quantum Fourier transforms over a variety of groups arise in many algorithms. Further advances in distributing quantum Fourier transforms over realistic architectures and generally improving their efficiency \cite{murairi2024highly} are promising directions of study.

{\bf Quantum-assisted and quantum-ready machine learning.} Machine learning and AI are rapidly advancing fields. A hybrid approach can contribute to advancing these fields for use today while providing insight into the potential impact of quantum computing on these approaches in the longer term. This is an area for “quantum-ready” algorithms, with purely classical algorithms that can be run today for which there are identified subroutines that can be substituted by quantum subroutines. Examples of this approach include quantum-assisted variational auto-encoders (QVAE) \cite{Gao2020,Asanjan2023}, quantum-assisted associative adversarial networks (QAAAN) \cite{Wilson2021},  combining energy-based models (EBM) \cite{huembeli_physics_2022} with invertible flows (IF) \cite{oconnor2021rbmflow}, and quantum-compatible discrete deep generative models \cite{templin2023anomaly}. A quantum-ready approach is widely applicable, from optimization \cite{mohseni2021nonequilibrium,denchev2022quantum} (see \Cref{sec:pbits}) to simulating quantum systems (see \Cref{sec:NISQ-applications}). These quantum-ready approaches are natural targets for present-day and near-term quantum-inspired hardware acceleration.

\subsection{Toward performance benchmarking of hybrid quantum--classical computers}
\label{sec:perf-benchmarking}
Here, we discuss how one can approach a holistic benchmarking of hybrid quantum--classical computers, as opposed to the functional benchmarking of critical components of such computers discussed earlier in this paper. Ideally, all performance benchmarking would be hardware-agnostic. This may be satisfied for utility-scale fault-tolerant computers, but, as we will discuss, for mid-scale performance benchmarking some departure from hardware agnosticism may be necessary. We can learn from the field of HPC, where various performance benchmarks aimed at assessing performance relevant to different sorts of computation have been developed. Examples include: dense linear algebra (HPL (High-Performance Linpack) \cite{davies2011high,top500}) that serve as the basis for the TOP500 list~\cite{top500,strohmaier2015top500} as well as the Green500 list~\cite{feng2007green500,top500} which ranks supercomputers based on energy efficiency; sparse linear algebra (HPCG (High-Performance Conjugate Gradient) \cite{hpcg,dongarra2016new});  graph and large scale data analysis (Graph500 \cite{Graph500}); artificial intelligence and machine learning (MLPerf \cite{MLPerf}); physical modeling (SPEC HPC Benchmarks \cite{SPEC}); and scientific workloads (NPB (NAS Parallel Benchmarks \cite{npb,bailey1991parallel})). 

We can consider how to leverage these benchmarks in expanding performance benchmarking of computations that likely require quantum computation, while keeping the benchmarks themselves hardware agnostic, in line with standard HPC benchmarks. Some proposals exist for algorithmic-level benchmarking \cite{proctor2025benchmarking}, but much work remains to be done in that area, as well as for mid-scale benchmarking. Even though it will be some time before we can use utility-scale benchmarks for supercomputer-scale hybrid quantum--classical computers, the design of such benchmarks can inform the design of mid-scale benchmarks and provide targets for resource estimation. Such work can be viewed as a natural extension of identifying utility-scale problems with high impact, such as those in Refs.~\cite{bellonzi2024feasibilityacceleratinghomogeneouscatalyst, penuel2024feasibilityacceleratingincompressiblecomputational, elenewski2024prospectsnmrspectralprediction, agrawal2024quantifyingfaulttolerantsimulation, mozgunov2024applicationsresourceestimatesopen, bärtschi2024potentialapplicationsquantumcomputing, saadatmand2024faulttolerantresourceestimationusing, nguyen2024quantumcomputingcorrosionresistantmaterials, otten2024quantumresourcesrequiredbinding,watts2024fullereneencapsulatedcyclicozonegeneration} from the DARPA QB program \cite{darpaQB}.

In the meantime, as quantum hardware matures, it is useful to have families of benchmarks that can both guide and gauge hardware and algorithm development. Ideally, these benchmarks would have tunable size and difficulty, enabling them to serve as stepping stones to performance benchmarking of fully fault-tolerant quantum hardware. Some examples for such tunable sets of benchmark instances exist~\cite{mandra2025generating,perera2020chook,rieffel2014parametrized}, including those developed in the context of benchmarking quantum-inspired hardware as part of the DARPA Quantum-Inspired Classical Computing (QuICC) program. Such examples can serve as models for the design of families that cover other aspects of computation that will be needed for future applications of quantum computing. As for benchmark families derived from subroutines that we expect to use in hybrid quantum--classical computing (e.g., quantum Fourier transforms, scientific computing subroutines, sampling, search, and error correction), they cannot be completely hardware agnostic in their design, but nevertheless can serve as benchmarks for a diversity of hardware. 

It is critical that these benchmarks can assess the performance of the entire computation, including both quantum and classical components used in solving a problem instance. 
Moving the field from an exclusive focus on quantum processing time to also considering classical processing time has been recently addressed in Ref.~\cite{neira2024benchmarking}, but more work needs to be done in this direction. We can learn from HPC's need to design benchmarks and metrics appropriate for heterogeneous supercomputers of the future; such efforts can inspire benchmarks and metrics for hybrid quantum--classical computers. For both traditional supercomputers and hybrid quantum--classical supercomputers, we expect to see increasing evaluation against multiple metrics such as energy, ratios of computational power to SWaP (size, weight, and power), and amount of computation per time interval. These benchmark sets can not only provide targets for hardware and algorithm development along with assessment of these developments, but can also help identify bottlenecks and set priorities for hardware and algorithm development.

\section{Distributed simulation of quantum many-body systems}
\label{sec:NISQ-applications}

In this section, we examine distributed quantum simulation in the context of two major scalability challenges identified in \cref{sec:challenges-scaling}: the need for more powerful classical simulation techniques as system sizes grow, and the necessity of distributing quantum workloads across multiple QPUs, each with limited number of logical qubits. To demonstrate the importance of distributed computing in addressing these challenges, we present two representative studies. In \cref{sec:TFIM-DQPT}, we conduct an exact statevector simulation of the 2D transverse-field Ising model, revealing a dynamical quantum phase transition enabled by multi-node GPU‑accelerated distributed simulation. In \cref{sec:hpc-qc-workload-ack}, we introduced our adaptive circuit knitting algorithm that we apply in \cref{sec:adaptive-circuit-knitting-experiments} to quantum spin-glass models, showcasing how distributed quantum workload algorithms can extend feasible problem sizes beyond the limits of individual QPUs. Together, these results underscore the critical role of distributed computation—across both classical HPC platforms and quantum workload‑distribution methods such as adaptive circuit knitting—in enabling scalable quantum simulation and motivating further research in this area of active research.

\subsection{Multi-GPU: Dynamical quantum phase transitions of 2D transverse-field Ising models}
\label{sec:TFIM-DQPT}

Here, we demonstrate the use of the multi-node, GPU-accelerated CUDA-Q~\cite{cudaq} state vector simulator to study exotic phenomena in quantum materials. Studying such phenomena can help us better understand materials properties or even control physical/chemical systems in their condensed phase, aiding in the design of new materials. This work shows how CUDA-Q can compile and execute distributed quantum circuit simulations on HPC systems.

The transverse-field Ising model (TFIM), the quantum analog of the classical Ising model, is a well-studied system in the condensed-matter physics community. It describes a lattice of $N$ spins with nearest-neighbor interactions in the presence of
an external magnetic field, with a Hamiltonian given by
\begin{equation}
H=-J\sum_{\langle i,j \rangle}{\sigma_i^z \sigma_j^z} - g\sum_{i=1}^N{\sigma_i^x},
\label{eq:tfim_hamiltonian}
\end{equation}
where $\langle i,j \rangle$ denotes all nearest-neighbor pairs in the lattice, $\sigma^z$ and $\sigma^x$ are the Pauli $Z$ and $X$ matrices, respectively, and $J$ and $g$ are parameters that control the nearest-neighbor coupling and transverse-field strengths, respectively. Despite its simplicity, the Ising model can exhibit complex quantum phenomena that could be difficult to simulate classically. This is especially true for 2D spin lattices, which are thought to be beyond the capabilities of approximate simulation methods like matrix product states (MPS) and quantum Monte Carlo (QMC). Simulating many-body quantum systems (like the Ising model) beyond 1D is an open challenge, especially for non-equilibrium or excited-state properties. One such property is dynamical quantum phase transitions (DQPT), which are non-equilibrium phase transitions of quantum systems in time \cite{heyl2018dynamical}.

Studying these systems is a promising use case for digital quantum computers because their continuous time evolution, given by $|{\psi(t)}\rangle = e^{-iHt} |{\psi(0)}\rangle$,
 can be simulated via Trotterization \cite{trotter1959, suzuki1976}. In principle, this discretization enables scaling up such simulations on fault-tolerant quantum computers (see \Cref{sec:resource-estimates-pbenzyne} for a more thorough discussion on this topic). For Hamiltonians like the TFIM that can be written as the sum of local terms, the time-evolved state can be approximated by 
\begin{equation}
|{\psi(t)}\rangle \approx \left( \prod_j{e^{-i a_j H_j t/r}} \right) ^r |{\psi(0)}\rangle,
\end{equation}
where $t/r$ is the time step in the evolution. For large enough $r$, the product of matrix exponentials is a reasonable approximation for the sum of matrix exponentials. We find that a first-order Trotterization, for example,
$e^{-i(A + B)t} \approx (e^{-iAt/r}e^{-iBt/r})^r$
is sufficient for accurate simulations of DQPTs in the TFIM.

\begin{figure}[ht]
    \centering
    \includegraphics[width=.45\textwidth]{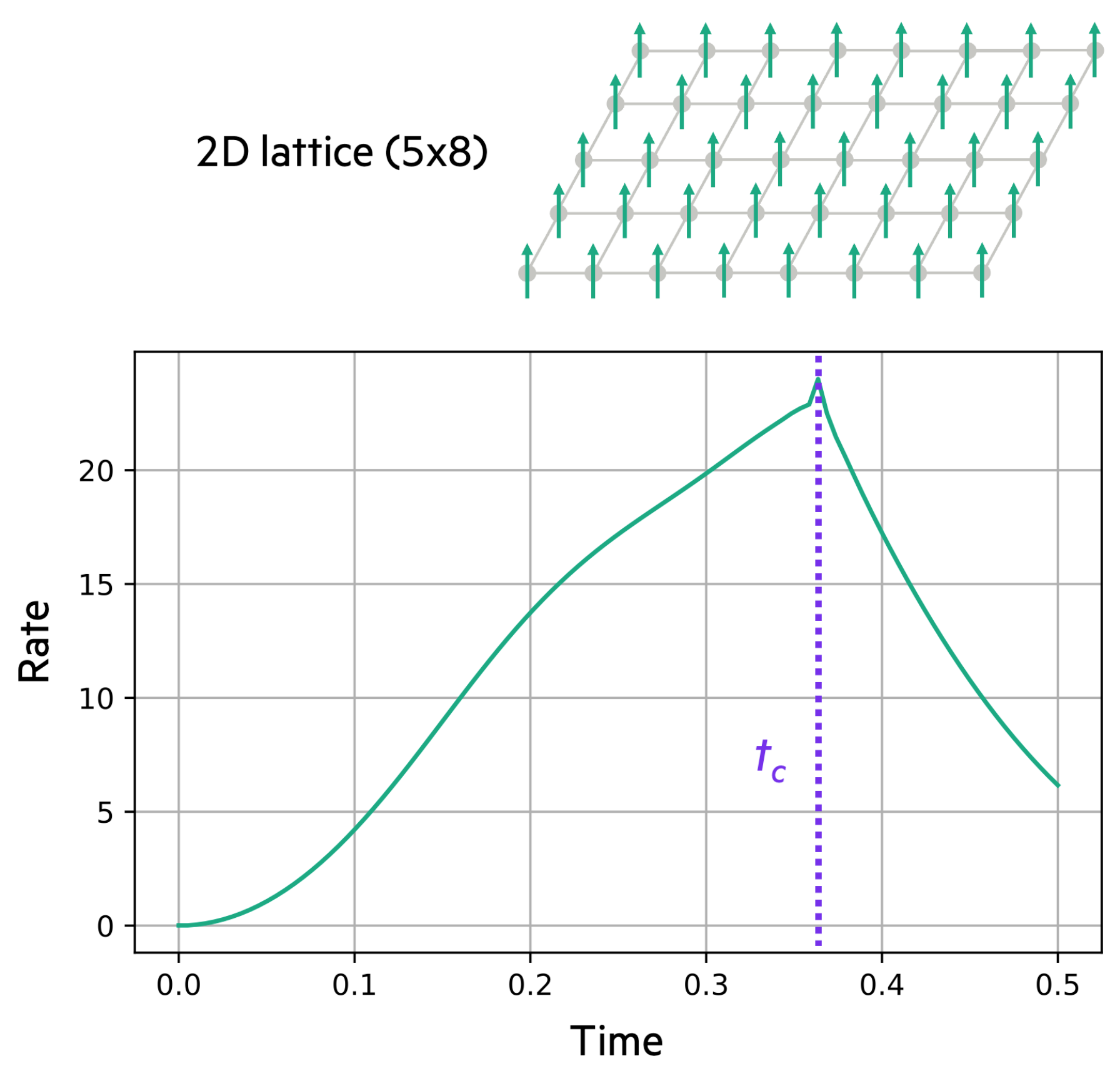}
    \caption{Dynamical quantum phase transition observed at $t_c$ during time-evolution simulation of a 40-qubit 2D Ising model with $J = 1.0$, quenching from $g_0 = 0.0$ to $g = 5.0$. Simulation comprised 100 time steps executed on 512 A100 GPUs across 128 nodes on Perlmutter.}
    \label{TFIM-rate}
\end{figure}

A dynamical quantum phase transition occurs as a result of "quenching" a quantum system. The initial quantum state $|{\psi(0)}\rangle$ represents the ground state of Hamiltonian $H_0$. (For example, a state with all spins up like the one shown in Figure \ref{TFIM-rate} is the ground state of the Hamiltonian with $J_0 = 1.0$, $g_0 = 0.0$.) The time-evolution is then carried out under a different Hamiltonian $H$. This forces the system to undergo a rapid phase transition in time. The quantity of interest when studying DQPTs is the Loschmidt amplitude $\mathcal{G}(t)$, which is the overlap of the time-evolved quantum state with an initial state:
$\mathcal{G}(t) = \langle{\psi(0)} |{\psi(t)}\rangle = \langle{\psi(0)} |e^{-iHt}|{\psi(0)}\rangle$.
The Loschmidt echo $\mathcal{L}(t)$ is the probability associated with the amplitude:
$\mathcal{L}(t) = |\mathcal{G}(t)|^2$.
We can identify DQPTs by tracking the rate function, given by
\begin{equation}
\lambda(t) = - \lim_{N\rightarrow\infty} \frac{1}{N} \log \mathcal{L}(t),
\end{equation}
where $N$ is the number of qubits. A DQPT occurs at critical time $t_c$ where $\lambda(t)$ has a non-analytical peak.

While there have been recent demonstrations simulating DQPTs on both quantum devices (a subset of qubits in a 22-qubit superconducting chip \cite{superconducting23} and 53 qubits in a trapped ion experiment \cite{trappedion53}) as well as using numerical classical simulators \cite{heyl2018dynamical}, these studies have been limited to 1D. Understanding phase transitions in 2D is likely key for designing real devices and materials. \mbox{CUDA-Q’s} \texttt{cuStateVec} backend, which represents the entire $2^N$ state vector and can capture maximum entanglement, enables accurately simulating 2D systems and computing the rate function $\lambda(t)$ throughout the time-evolution.

\begin{figure}[ht]
    \centering
    \includegraphics[width=.45\textwidth]{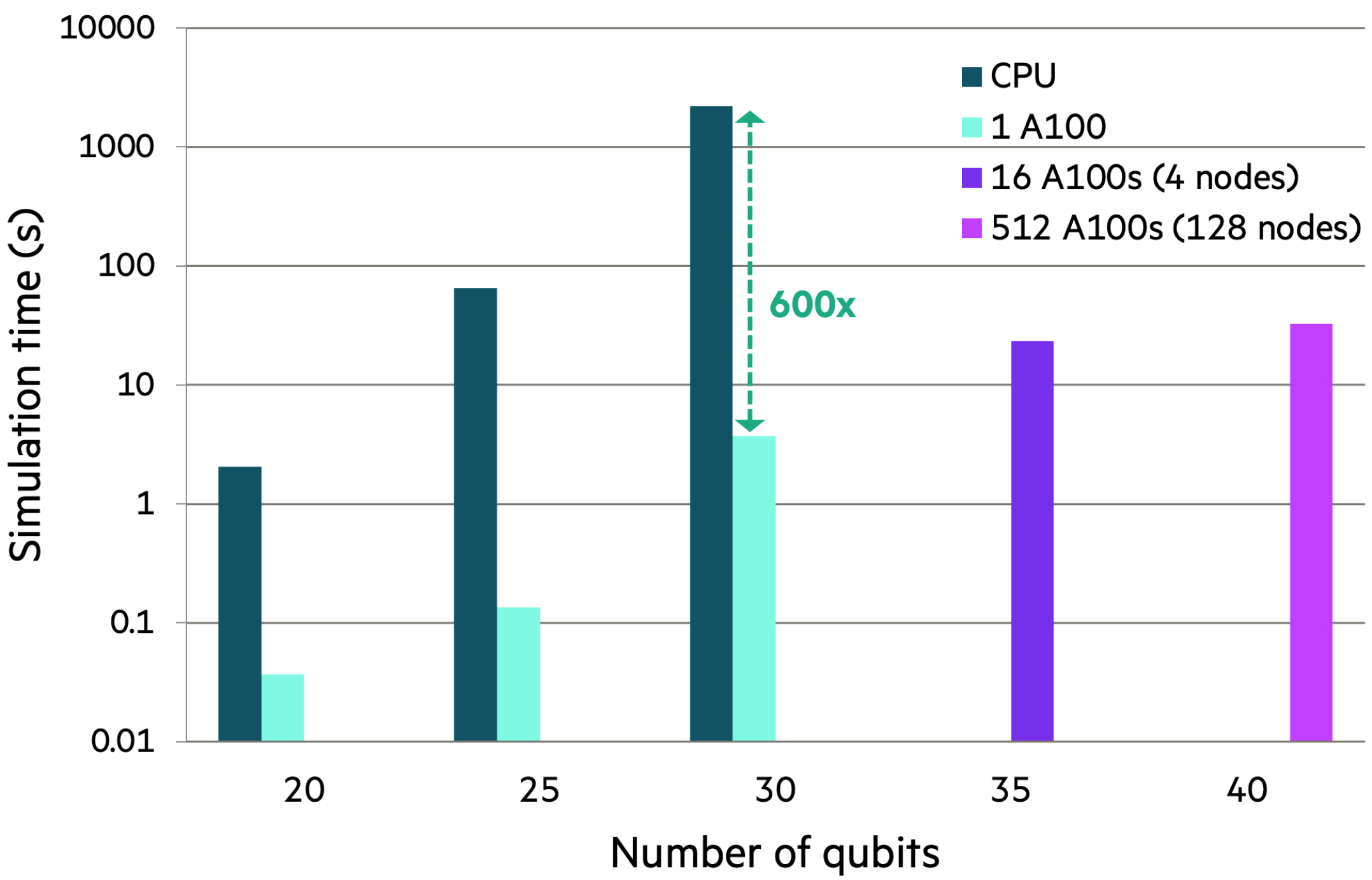}
    \caption{Performance of CUDA-Q simulation using multi-threaded CPU, single GPU, and multi-GPU backends. Systems simulated are 2D lattices: 5$\times$4, 5$\times$5, 5$\times$6, 5$\times$7, and 5$\times$8 qubits. Simulation time reported is for one time step in the time-evolution circuit (100 total time steps).}
    \label{TFIM-perf}
\end{figure}

\begin{figure*}[t]
\centerline{\includegraphics[width=2\columnwidth]{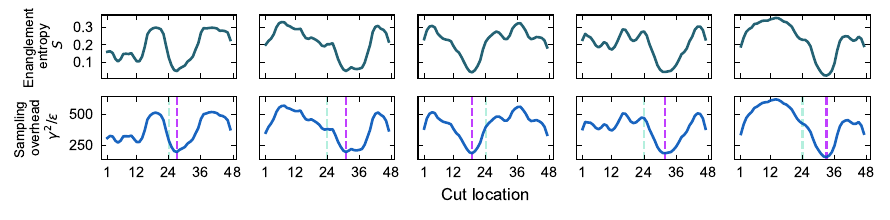}}
\caption{(Top) Bipartite entanglement entropies vs. the cut location for the ground states of the 48-qubit mixed-field Ising chain across five disorder realizations. (Bottom) Sampling overhead for reconstructing expectation values via circuit knitting corresponding to each entanglement entropy plot. The number of samples for circuit knitting an observable is given by $O(\gamma^2/\epsilon^2)$, with $\epsilon = 0.01$. For all five instances, a lower sampling overhead can be achieved by making an adaptive cut (purple) vs. a naive load-balanced cut (light green).}
\label{Figure_mfim_ack}
\end{figure*}

We performed several simulations of 2D spin lattices on various compute configurations ranging from a single CPU to 512 GPUs across 128 nodes. Figure \ref{TFIM-rate} shows a DQPT discovered during the simulation of the largest system studied, an 8$\times$5 spin lattice (40 qubits). During many of phase transitions simulated, the entanglement entropy in the quantum system grows to near its maximum value, making them difficult to simulate classically with tensor network techniques. Although we did not perform one, an MPS time-evolution simulation could provide a useful comparison for classical computing resources.

Circuit synthesis time for one Trotter time step ranges from 0.2--0.4 ms for 5--10 qubits to 2--5 ms for 25--33 qubits on a single A100 GPU. Saving the previous state in GPU memory instead of re-simulating all previous operations at given time step will drastically reduce the circuit synthesis time, keeping it flat with increasing circuit size as time step increases.  

Figure \ref{TFIM-perf} shows the performance comparisons for multi-threaded CPU, single A100 GPU, and multiple A100 nodes. Circuit execution time for one Trotter time step ranges from 0.04--4 s for 20--25 qubits on a single A100 GPU to 23--33 s for 35 qubits (distributed across 64 A100s) and 40 qubits (distributed across 512 A100 GPUs). For 20- to 30-qubit simulations, one A100 provided a 600$\times$ speedup over a multi-threaded CPU. Beyond 30 qubits, the exponential scaling of quantum simulation quickly outstrips the capabilities of a single processor, but scaling to 40 qubits was possible by distributing the simulation across 128 nodes (512 A100 GPUs) on the Perlmutter supercomputer. The 40-qubit simulation took one hour, nearly two orders of magnitude faster than a CPU simulation on 30 qubits. The performance results on these larger qubit systems highlight the multi-node parallel efficiency of both the software and hardware.

Simulations of many-body quantum systems like the ones shown here are important for the near-term benchmarking of quantum computers—a high-performance state vector simulator can enable the accurate study of DQPTs in 2D spin lattices up to 40 qubits, a task currently beyond the capabilities approximate methods. This framework could be used to study other quantum systems as well as study the effects of noise, a crucial element for the development of near-term quantum devices. However, state vector simulations much beyond 40 qubits are out of reach even for the most powerful supercomputers. Approximate methods such as tensor network techniques become crucial. In the following section (\Cref{sec:adaptive-circuit-knitting-experiments}), we provide an example of applying tensor network techniques to an important problem for scaling quantum computing: distributing quantum workloads.

\subsection{Multi-QPU: Strongly disordered quantum spin glasses using adaptive circuit knitting}
\label{sec:adaptive-circuit-knitting-experiments}
\Cref{sec:hpc-qc-workload-ack} introduced adaptive circuit knitting (ACK) approaches for quantum workload distribution that aim to overcome the exponential overhead of circuit knitting. In this section, we present a concrete example of ACK which decreases sampling overhead of circuit knitting by cutting in locations that minimize entanglement between partitions. We describe the method and demonstrate its application simulating the dynamics of quantum spin chains.

This particular ACK method draws from tensor network (TN) approaches developed in the quantum physics and quantum chemistry communities \cite{white1993density, meyer1990multi}. Tensor networks represent quantum states in a compressed form, and can provide structure to characterize entanglement patterns. In the context of quantum circuits, a TN can be efficiently expressed as circuit of linear depth as was shown by Lin et al.~ \cite{lin2021real} for matrix product states (MPS), a type of TN commonly used to study 1D quantum systems. Combining the structure of TNs with linear depth quantum circuits is the basis for this ACK method.

Before presenting our ACK method, we first provide intuition on the applicability of distributed computing for simulating disordered quantum systems. We consider the transverse-field Ising model in \Cref{eq:tfim_hamiltonian}, with an additional disorder term $-\sum_j h_j \sigma^z_j$ resulting in the mixed-field Ising model. Here, $h_j$ is chosen uniformly from $[-0.3, 0.3]$. The top panel of \Cref{Figure_mfim_ack} shows the bipartite entanglement entropy for the ground states corresponding to five disorder realizations for a $48$-qubit chain. We can clearly see islands of high-entanglement separated by low-entanglement boundaries, a prerequisite for distributed quantum simulation as discussed in \cref{sec:hpc-qc-workload-ack}. This heterogeneity allows us to identify low‑entanglement boundaries and adaptively place cuts that minimize circuit‑knitting overhead. Furthermore, if we lower the disorder strength, the entanglement profile increases smoothly from the boundaries toward the center of the chain. In contrast, at stronger disorder, the entanglement becomes more heterogeneous but overall lower in magnitude; hence, suitable for classical simulation. The bottom panel shows the theoretical lower bound on sampling overhead, quantified by $\gamma^2/\epsilon^2$, demonstrating that for all the disorder instances shown, there is always an advantage to adaptively choosing a cut. Leveraging this intuition, we now introduce our ACK algorithm for simulating disordered quantum many-body systems with multiple QPUs.

\begin{figure}[htbp]
\centerline{\includegraphics[width=0.99\columnwidth]{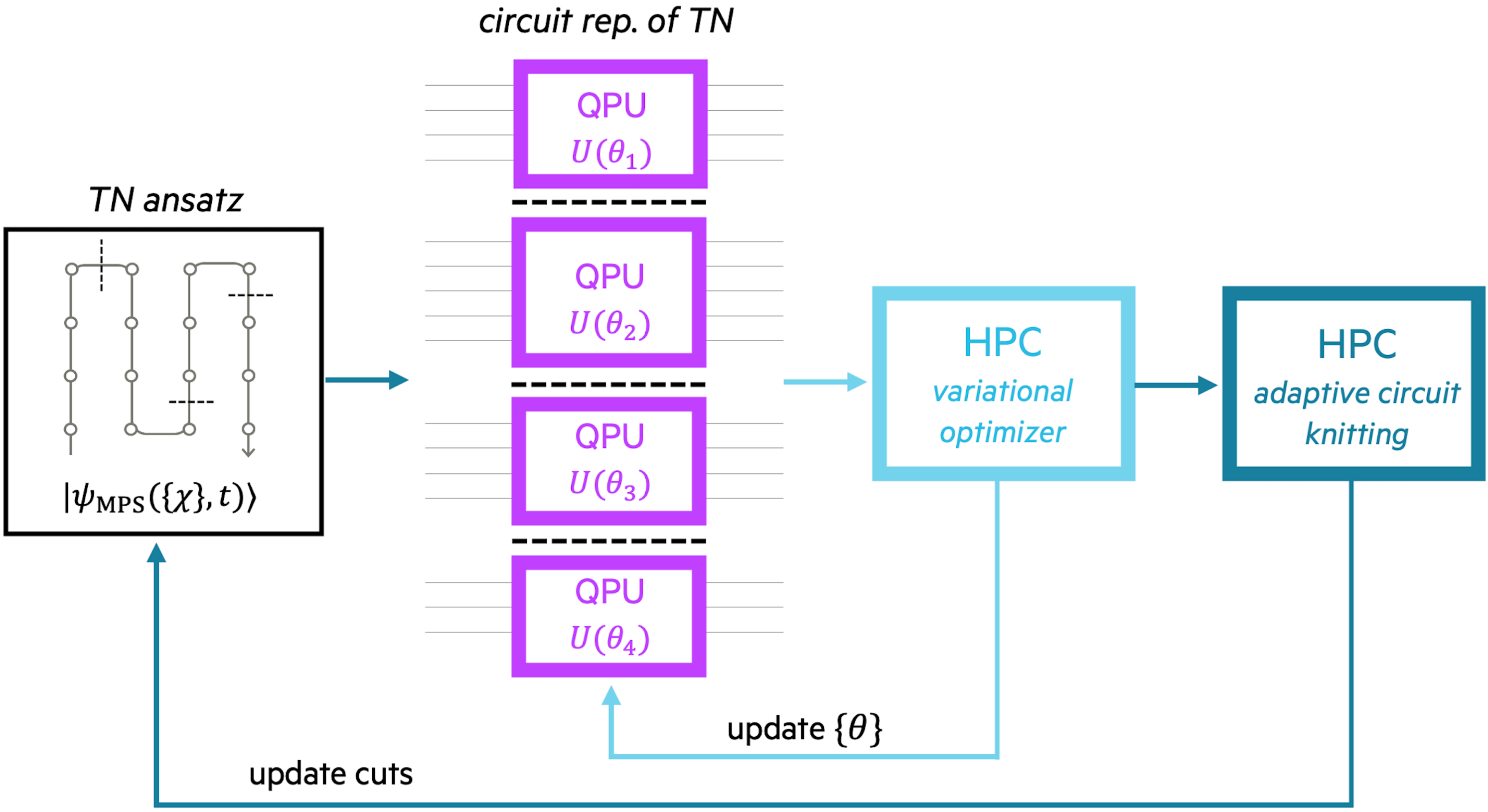}}
\caption{Schematic of an adaptive circuit knitting method using tensor networks. In the inner loop, a variational optimizer finds in parallel circuit parameters for partitions of a quantum system that are induced by an initial TN ansatz. In the outer loop, an adaptive procedure finds cuts which minimize entanglement between partitions. After the best cuts are found, quantum observables are reconstructed via circuit knitting. The procedure is repeated until a desired accuracy of some (global) observables is achieved or the performance of circuit knitting is plateaued.}
\label{ACK-fig2}
\end{figure}

Figure~\ref{ACK-fig2} shows a schematic for this adaptive circuit knitting method. It begins with a TN representation of a quantum state (in the figure, an MPS for a 2D spin lattice). The TN is partitioned into $N$ sub-networks, drawn here with $N=4$.  In an inner loop, for each sub-network the variational procedure outlined by Lin et al. \cite{lin2021real} is followed and gates $U(\theta_N)$ are optimized for an efficient circuit representation of each sub-network. Following this optimization, in the outer loop entanglement measures (e.g., von Neumann entropy) are computed and an entropy heatmap among the qubits is constructed. Although this heatmap is partial (we do not have entropy measures at the cuts between partitions), the information available is used to update cuts to locations with low entanglement. With new cuts on the following iterations, the blind spots are revealed. The adaptive outer loop exits when entanglement between partitions is minimized. The inner loop can be parallelized since each sub-network is independent, and both the inner and outer loop can be executed using classical HPC. Once optimal partitions have been found, a measured observable can be reconstructed from the sub-circuits via circuit knitting \cite{piveteau2023circuit}. As we show below, cutting gates at locations of minimal entanglement can substantially lower the classical overhead of circuit knitting.

Simulating quantum systems for materials science or quantum chemistry is one of the most promising applications for quantum computers. Spin-lattice systems are well-studied in materials science, and despite their simplicity can exhibit complex quantum phenomena that are difficult to simulate classically, for example, the dynamical quantum phase transitions discussed in Section~\ref{sec:TFIM-DQPT}.  As a prototype, we apply this ACK method to simulating the non-equilibrium dynamics of a strongly disordered spin chain evolving under an Ising model with transverse and longitudinal fields given by the Hamiltonian
\begin{equation}
H=-\sum_{i=1}^{N-1}{J_{i,i+1} \sigma_i^z \sigma_{i+1}^z} - \sum_{i=1}^N{g_i \sigma_i^x} - \sum_{i=1}^N{h_i \sigma_i^z}, 
\end{equation}
where $\sigma^z$ and $\sigma^x$ are the Pauli-$Z$ and Pauli-$X$ matrices, respectively, $i$ indexes the lattice site, and $J$s, $g$s, and $h$s are real-valued parameters. We study strongly disordered systems (where parameters are varied at each lattice site) because they are important for understanding exotic states of matter, they can be difficult to study, and because they lead to many-body localization effects which we hope to exploit for more efficient simulation. 

\begin{figure}[htbp]
\centerline{\includegraphics[width=0.95\columnwidth]{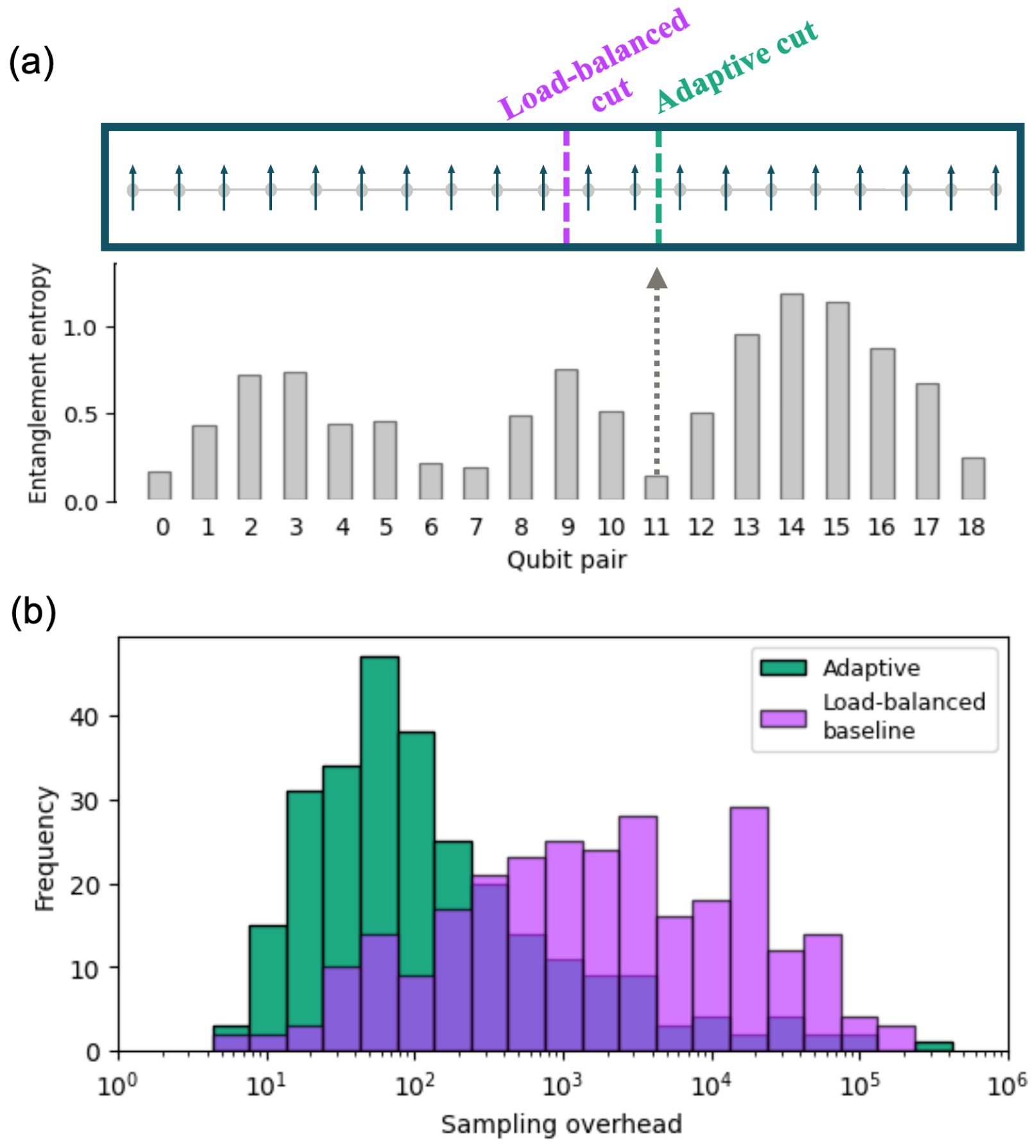}}
\caption{(a) Example of a spin chain where the adaptive cut is chosen at the minimum entanglement entropy. (b) Histogram of sampling overheads resulting from adaptive and load-balanced baseline cuts for an ensemble of 32-qubit strongly disordered spin chains.}
\label{fig3}
\end{figure}

Figure~\ref{fig3} provides a summary of the results for an ensemble of 32-qubit spin chains, each time-evolving under a Hamiltonian with different sets of parameters $J_{i,i+1}$, $g_i$, $h_i$ chosen randomly from a uniform distribution on [$-1, 1$] (excluding 0). For each spin chain, circuit optimization was carried out at eight time steps throughout the dynamics. We partition each system into two sub-circuits and compare the overhead cost of circuit knitting for a na\"ive cut in the middle of the chain (a ``load-balanced'' choice) versus a cut recommended by the entropy heatmap from the adaptive algorithm. Figure \ref{fig3}(a) gives a schematic for a single instance on a smaller 20-qubit system. Figure \ref{fig3}(b) shows the distribution of overheads for reconstructing a magnetization observable via circuit knitting for the adaptive and load-balanced cuts. Both cuts achieve similar accuracy in the observable, but in most cases the adaptive cut results in a much lower overhead---the green distribution is distinctly shifted to the left. On a case-by-case basis, the median reduction in cost was 15$\times$, while the 75th and 95th percentiles were 59$\times$ and 450$\times$, respectively. While not shown in this figure, the gap between adaptive and load-balanced widens during the later time steps, indicating that for longer simulations the benefits of adaptive circuit knitting will increase. As in \Cref{sec:TFIM-DQPT}, the \texttt{cuStateVec} simulator was employed to enable performant execution of the 32-qubit simulations.

Such initial results showing improvements in overhead of one to two orders of magnitude suggest that this ACK method could be used for efficiently partitioning quantum circuits for near-term applications simulating condensed-matter systems. Building on this promising result, we perform a more extensive study of ACK for initial state preparation \cite{johnson2026distributed} as well as, time-evolution \cite{gyawali2026ack}. However, more study is necessary to demonstrate the practical utility of such a method. Future work will include investigating efficient entanglement measures, exploring more sophisticated convergence criteria, and studying 2D systems with higher order TN techniques, as 2D systems are where many classical methods underperform and quantum computers could likely provide the largest advantage (see \Cref{sec:hpc-qc-workload-ack} for a more general discussion of ACK techniques). A high-performance implementation in CUDA-Q could enable fast prototyping of these methods for circuit sizes large enough to capture interesting physics.

Taken together, these results reinforce our case: scalable quantum simulation requires coordinated advances in both classical and quantum distributed methods. On the classical side, more efficient and scalable HPC‑oriented simulation codes are needed to keep pace with growing system sizes and algorithmic complexity. On the quantum side, distributed quantum algorithms will be essential for leveraging networks of limited‑size QPUs. Progress on both fronts is therefore critical for enabling the next generation of large‑scale quantum simulations.

\section{Towards heterogeneous quantum-probabilistic computing for quantum simulation, optimization, and AI}
\label{sec:pbits}

A central thesis of our approach to full-stack quantum-classical computation is that fault-tolerant quantum processors should be integrated as specialized accelerators within heterogeneous high-performance computing infrastructures (\Cref{sec:fullstack}, \Cref{figure1_Arc}). Probabilistic Processing Units (PPUs) are a natural member of this heterogeneous accelerator family: they share the same HPC integration requirements---low-latency interconnects, workload scheduling, and distributed memory management---as QPUs and GPUs, and they occupy a distinct computational regime between deterministic classical processing and fully quantum computation. By handling the large class of industry-scale workloads for which stochastic sampling is the computational bottleneck (for example, combinatorial optimization, Monte Carlo simulation, energy-based generative models), PPUs address a concrete complexity gap. For many problems where quantum advantage is either unproven or where the overhead of quantum error correction is prohibitive (\Cref{sec:100k-1M_qubits}), purpose-built probabilistic accelerators can deliver orders-of-magnitude speedup today while remaining architecturally ready for future quantum integration.

Beyond this architectural complementarity, probabilistic and quantum computing share a deep \textit{physics-level} connection that distinguishes PPUs from other classical accelerators. Both paradigms exploit the computational richness of many-body systems---p-bits sample from classical Boltzmann distributions via Gibbs dynamics, while qubits sample from quantum distributions that additionally encode interference and entanglement. The boundary between these two regimes can be captured by the ``sign problem'' \cite{chowdhury2020emulating}: problems amenable to efficient classical stochastic sampling (sign-problem-free) can be accelerated by p-computers, while problems with intrinsic quantum sign structure may require QPUs. This boundary is not fixed; it shifts as algorithmic techniques such as basis transformations that minimize negativity \cite{murota_2025_local_basis_transformation} expand the reach of classical sampling. A serious scaling roadmap must therefore include the strongest possible probabilistic baselines---both to sharpen the definition of genuine quantum advantage and to ensure that expensive QPU resources are directed at problems where they are truly needed.

This physics-level connection is made concrete by neural quantum states (NQS) \cite{carleo2017solving}, which provide a direct operational bridge between p-computers and quantum computers. In the NQS framework, the amplitudes of a quantum wavefunction are parameterized by a neural network and optimized variationally using Monte Carlo sampling in the computational basis. When the neural-network ansatz takes the form of a Boltzmann machine, the sampling inner loop maps directly onto the native operation of a p-computer. As we demonstrate in \Cref{sec:pbits}, distributed PPU hardware can execute this sampling loop at throughputs exceeding those of CPUs and GPUs by orders of magnitude, enabling large-scale variational quantum simulation on probabilistic hardware. Crucially, the same NQS framework applies to QPU-based quantum simulation: one can deploy PPU-based NQS for sign-problem-free regions of a phase diagram while routing sign-problematic regions to QPUs, enabling a hybrid QPU--PPU exploration of complex quantum many-body systems that neither platform could tackle alone.

The inclusion of p-computing in our architecture directly serves the cost-effectiveness criterion for utility-scale quantum computing. As discussed in \Cref{sec:challenges-intro}, a utility-scale quantum computer must deliver computations whose value exceeds its cost, and no alternative method should accomplish the same task more cheaply. Probabilistic accelerators provide two essential functions in this regard: first, they supply a rigorous cost--performance baseline against which quantum advantage claims must be validated; second, for the many workloads where QPUs do not yet offer a clear advantage, PPUs can deliver practical value within the same heterogeneous infrastructure, improving the overall return on investment of the quantum supercomputer and accelerating the development of quantum-ready software stacks and algorithms.

The vision of dedicated hardware for probabilistic computation has deep historical roots. Feynman, who pioneered the notion of a controllable quantum computer for simulating other quantum systems (in his 1981 talk ``Simulating Physics with Computers'' \cite{feynman1982simulating}), also imagined a probabilistic computer with the same idea: ``The other way to simulate a probabilistic Nature $\ldots$ is by a computer $C$, which itself is probabilistic''. Monte Carlo algorithms, first popularized by Ulam and von Neumann and later developed by Metropolis and Teller, have been widely recognized as one of the top ten algorithms of the 20th century. Yet emulating the requisite stochastic behavior on deterministic hardware is expensive: constructing a single high-quality tunable random number generator can require up to 15,000 transistors \emph{per node} \cite{singh2024cmos}, and emulating many-body emergent probabilistic phenomena can be exponentially hard with sequential Markov chain Monte Carlo techniques due to their long mixing times near non-trivial steady states.

\begin{figure}
 \vspace{-15pt}
\centering
\includegraphics[width=1\linewidth]{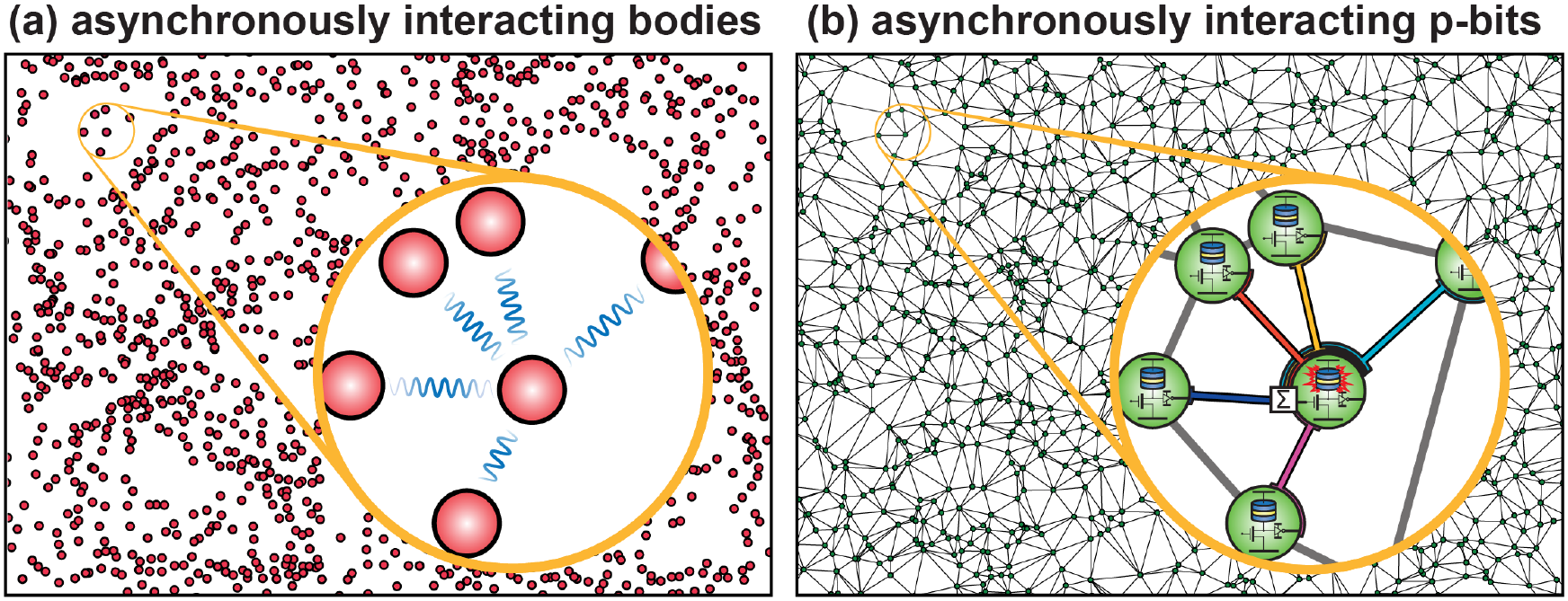}\\
\caption{\textbf{Physical analogy} between asynchronously interacting (a) bodies  and (b) probabilistic bits: both systems are asynchronous,  locally interacting with sparse connectivity and massively parallel.}
 \vspace{-15pt}
\label{fig:physics_pbit}
\end{figure}

Motivated by these observations, a growing community has been developing a full-stack probabilistic computing approach \cite{chowdhury2023fullstack} that draws direct inspiration from physics. Nature employs microscopic stochastic local dynamics, quasi-sparse graph connectivity, strong heterogeneity, non-equilibrium dynamics, and asynchronous updates (without global clocks) in probabilistic networks (see \Cref{fig:physics_pbit}). These networks evolve in a massively parallel manner, leading to emergent many-body phenomena at mesoscopic scales that are computationally rich and hard to simulate with conventional computing paradigms. Interacting probabilistic bits (p-bits) can be engineered to act as native processors for a wide range of such applications, including Bayesian learning and inference \cite{aifer2024bayesian}, training and inference in energy-based models \cite{coles2023thermodynamic,huembeli_physics_2022}, combinatorial optimization, efficient linear algebra \cite{aifer2024thermodynamic}, accelerating certain quantum simulations, and complex reasoning in generative AI \cite{Mert_reasoning_2024}.

Importantly, networks of \textit{hardware} p-bits have been shown to natively represent a wide class of probabilistic algorithms typically implemented in \textit{software}, with significant energy and performance benefits \cite{sutton2020autonomous,aadit2022massively,niazi2024training, singh2024cmos, nikhar2024all, Chowdhury_2025_pushing_the_boundary}. We discuss the integration of these accelerators within the heterogeneous architectures introduced in \Cref{sec:fullstack} in detail in \Cref{sec:hetero_pq}.

\subsection{Probabilistic computing with intrinsic higher-order interactions}

p-computers implement a Markov chain Monte Carlo algorithm called Gibbs sampling, achieved by a stochastic activation and a local field calculation, given by:\vspace{-5pt}
\begin{equation}
m_i = \text{sgn}(\tanh(\beta I_i) - r_U),
\label{eq:pbits}
\end{equation}
\begin{equation}
I_i = \sum_{j} J_{ij} m_j + h_i\,,
\label{eq:synapse}
\end{equation}
where $m_i$ represents the bipolar p-bit state ($\pm 1$), $r_U$ is a uniform random number between $(-1,+1)$ and $[J],\{h\}$ are the weights and biases for a given problem and $\beta$ is the inverse temperature.

There are two immediate generalizations that can be made. One is that p-bits can be extended to have multiple states. These Potts spins have also been implemented in hardware \cite{whitehead2023cmos} and shown to have better embedding than simple p-bits for certain optimization problems such as graph coloring. The other is that the graph connectivity defined by $J_{ij}$ can be generalized to \textit{hypergraphs} where the local field equation becomes (e.g., for \mbox{$k=4$-local} interactions) 
\begin{equation}
I_i = \sum_{j} J_{ij}m_j + \sum_{j<k} J_{ijk}m_jm_k + \sum_{j<k<l} J_{ijkl}m_jm_km_l\,, 
\end{equation}
where \( J_{ijk} \) and \( J_{ijkl} \) denote the interaction coefficients for three and four-local interactions respectively. These can be generally extended to any $k$-local interactions (Ref.~\cite{nikhar2024all} demonstrates an implementation with $k=3$ to solve the XORSAT problem). The binary nature of p-bits significantly eases the implementation of $k$-local interactions. Such $k$-local interactions greatly reduce model embedding and complexity. We are not aware of any programmable multi-qubits entanglement for $k>2$ in quantum computers but for probabilistic computation such hypergraphs can be constructed rather easily. 

\begin{figure}
 \vspace{-15pt}
\centering
\includegraphics[width=1\linewidth]{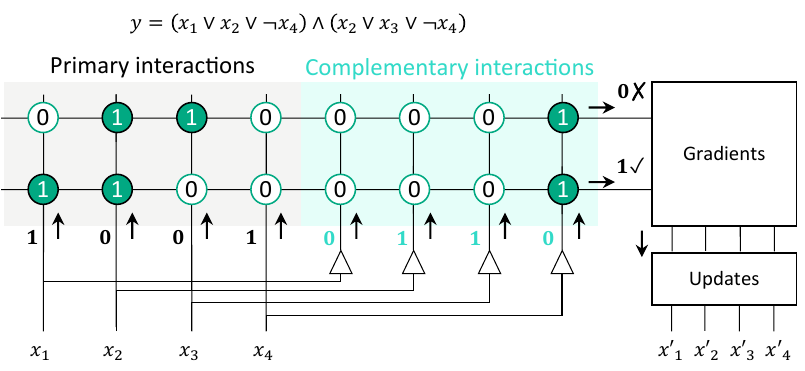}\\
\caption{Accelerating $k$-local interactions with in-memory computing. The interactions in a SAT formula can be evaluated by computing the Hamming distance between the input $x$ and each one of the clauses, mapped such as a Hamming distance of 0 corresponds to a clause violation. Two coupling arrays are used to represent primary and complementary interactions. A positive literal in a clause, such as $x_1$ in the first clause, is mapped as 1|0 in the primary coupling array and complementary coupling array, respectively; negative literal, such as $\neg x_4$ in the first clause is mapped as 0|1; and a non-member literal, such as $x_3$ in the first clause, as a 0|0. After the interactions are computed, the Hamming distance output can be used directly to compute high order gradients. }
 \vspace{-15pt}
\label{fig:klima}
\end{figure}

Recently, higher order $k$-local interactions architecture without limits or scaling dependence on the order $k$ based on in-memory computing have been presented~\cite{pedretti2023zeroth, bhattacharya2024computing}.
In the proposed approach~\cite{pedretti2023zeroth, bhattacharya2024computing}, the interaction between variables is computed before computing the gradients by encoding the clause member variables in an interaction matrix.
In the case of $k$-SAT problem with $N$ variables and $M$ clauses, a $2\times N\times M$ matrix is used for embedding interaction, where the factor $2$ accounts for a primary interaction matrix and its complementary as shown in \Cref{fig:klima}.
Member variables in clauses are encoded with $[1|0]$ and $[0|1]$ for $x$ and $\neg x$, respectively. 
A non-member variable is encoded as $[0|0]$.
Considering the first clause of the 3-SAT formula $y$ in \Cref{fig:klima} with $N=4$, $y_i=(x_1 \vee x_2 \vee\neg x_4)$, its encoding in the interaction matrix $I$ is $I_i = [1,1,0,0|0,0,0,1]$.
Given for example an input $x=[1,0,0,1]$, if the Hamming distance between $x'=[x|\neg x]$ and $I_j$, $\delta(x',I_j)>0$ the clause is satisfied.
In the example $\delta(x',I_j)=1$, meaning the clause is satisfied but only one literal ($x_1$) is positive, thus by flipping it the clause becomes unsatisfied.

Note that compared to traditional quadratic unconstrained binary optimization (QUBO) mapping, such as the one used in conventional Ising machines, the higher-order interactions allow for a native embedding of the problem, without the need for auxiliary variables.
Native mapping leads to improved convergence, no introduction of artificial local minima and settle points due to auxiliary variables \cite{dobrynin2024energy}, and reduced hardware resource utilization by $O(k^2)$.

\subsection{Hardware implementation of p-computers} 

 Physical implementation of p-computers includes a wide range of choices from noisy materials to analog and digital CMOS. State-of-the-art p-computers demonstrate nanodevice (magnetic tunnel junction, MTJ) based prototypes \cite{borders2019integer,singh2024cmos}, where the natural noise of the stochastic MTJ provides computational resources to solve a small-scale problem. This prototype has shown the promise of magnetic RAM technology, as a scalable pathway to build energy-efficient p-computers. The magnetic memory industry has achieved Gigabit densities of magnetic tunnel junctions embedded with CMOS transistors in monolithic integrated circuits \cite{lee20191gbit}. Repurposing these MRAM chips so that their \textit{stable} MTJs become \textit{unstable} (low-barrier) could lead to dedicated probabilistic computers with tens of millions of integrated p-bits. Before an integrated p-computer using millions of stochastic magnetic MTJs, however, digital emulators of p-bits using powerful CMOS-based Field Programmable Gate Arrays (FPGA) have been used to investigate the architectural and algorithmic performance of p-computers at large-scale \cite{sutton2020autonomous,kaiser2021probabilistic,aadit2022massively, niazi2024training,nikhar2024all, singh2024cmos}. Even single FPGA-based implementations of p-computers have shown competitive performance against the state of the art \cite{nikhar2024all}.

\subsection{Scaling up p-computers: A distributed approach}
\label{sec:distp}

 \begin{figure}
\centering
\includegraphics[width=1\linewidth]{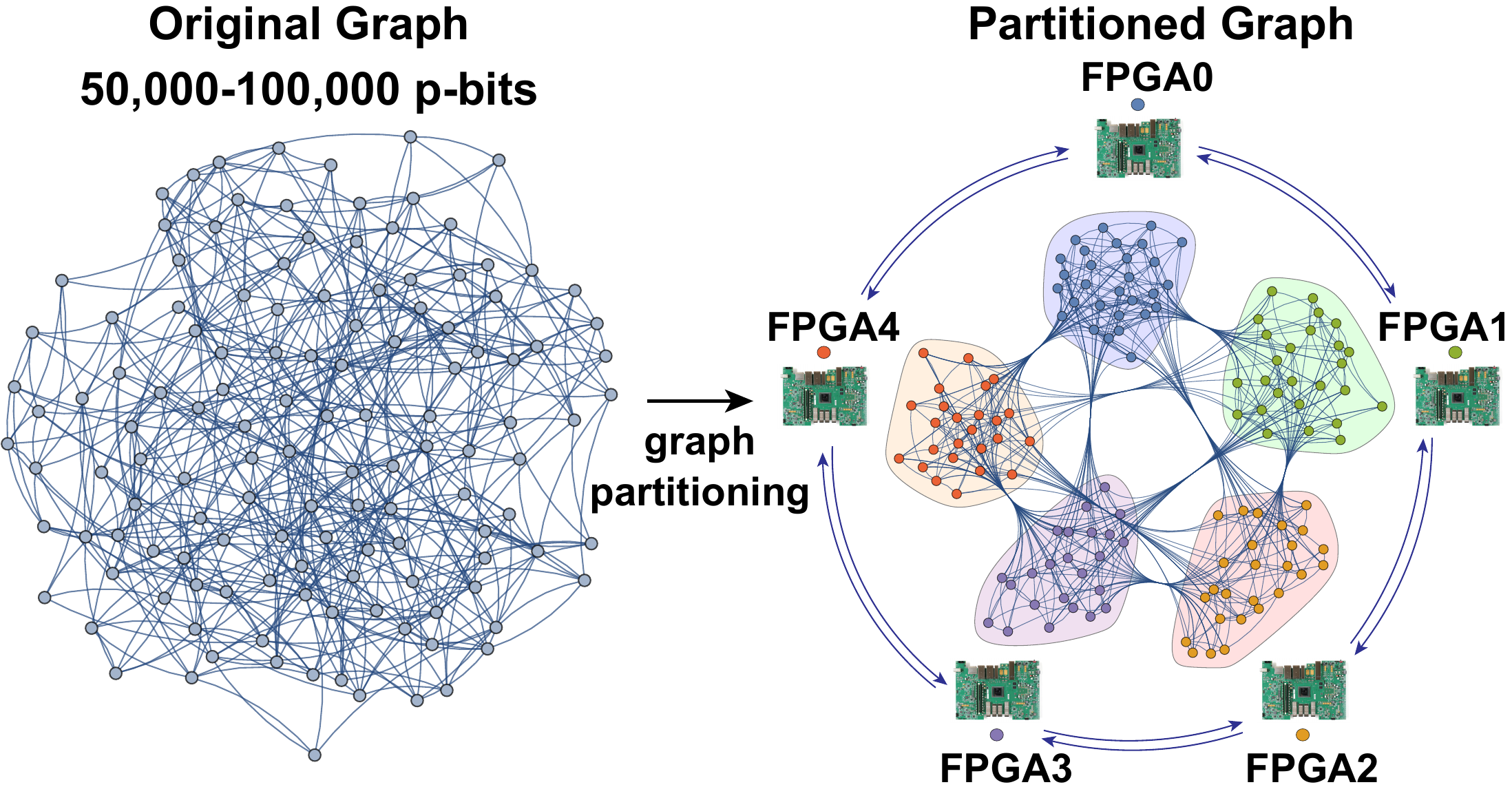}\\
\caption{Distributed p-computers: a single large graph is partitioned into multiple smaller subgraphs to distribute it across multiple processing elements (PE) in the form of FPGAs/GPUs/TPUs. A graph partitioning tool is used to ensure minimum cut across the subgraphs to minimize communication overheads. Preliminary results (in preparation) show over 1000 probabilistic flips per \textit{nanosecond} can be taken in these distributed systems, with about 2 orders of magnitude improvement over single GPU/TPU implementations.}
\label{fig:distributed_p-bits-system}
\end{figure}

Single processing elements (PE), such as FPGAs/GPUs/TPUs, are often not large enough to encode practical problem instances with thousands to millions of variables. To get around this problem, one solution is to design \textit{distributed} architectures where a large problem is partitioned into smaller subgraphs housed in distinct PEs (\Cref{fig:distributed_p-bits-system}). Preliminary results show that as long as the communication links are faster than individual p-bit clocks, distributed probabilistic computers can create the ``illusion'' \cite{radway2021illusion} of a single PE that can house a much larger graph. Sampling rates over   1000 flips per \textit{nanosecond} are feasible (manuscript in preparation), which  bodes well for hard combinatorial optimization and sampling problems. 

\begin{figure}
 \vspace{-5pt}
\centering
\includegraphics[width=0.8\linewidth]{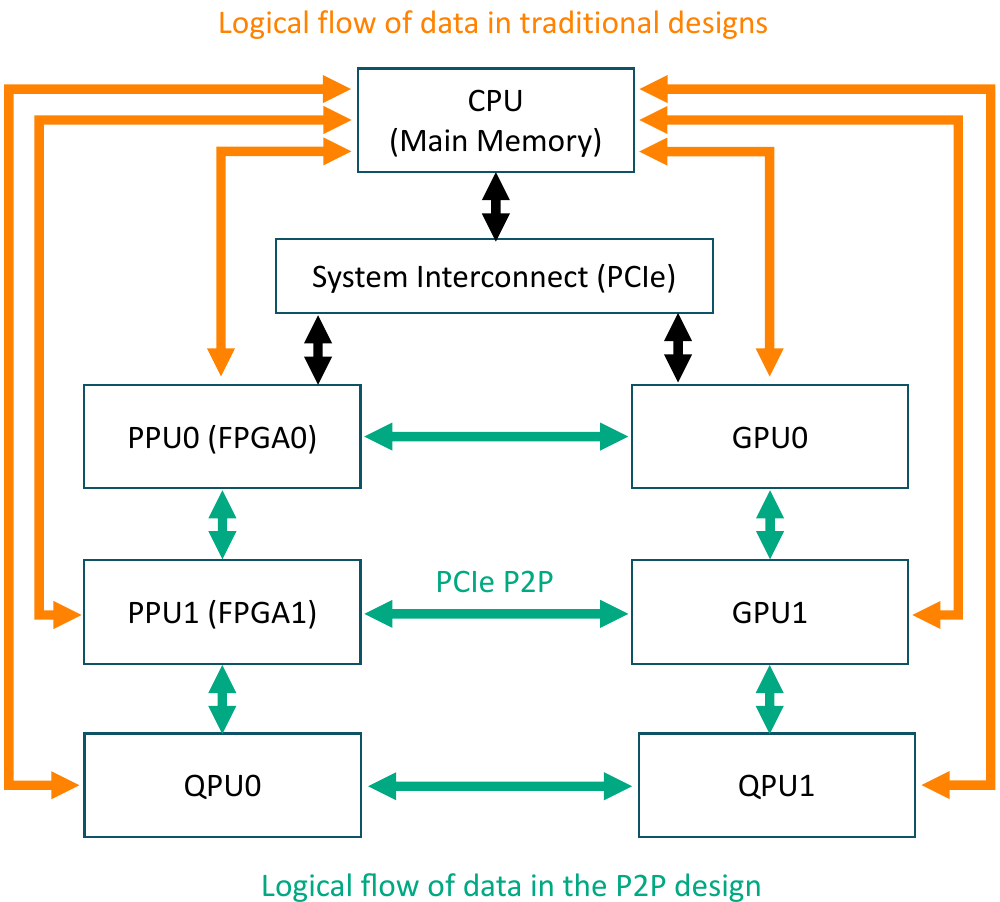}\\
\caption{Heterogeneous system architecture block diagram with conventional approach (orange) and the decentralized peer-to-peer (P2P) communications (green). Removing the required communication to CPU and main memory significantly improves the performance allowing direct communication between multiple accelerators. We have empirically validated this architecture in a heterogeneous probabilistic architecture (HPA) for training deep Boltzmann machines, where the P2P configuration achieves sampling throughputs exceeding $167$~Gflips/s and reduces per-epoch latency by $1.9\times$ compared to host-staged communication (see \Cref{tab:hpa_benchmarks}). Heterogeneous systems including conventional digital accelerators (GPU), quantum processing units (QPU) and probabilistic processing units (PPU) can enable novel classes of heterogeneous classical/probabilistic/quantum algorithms.}
 \vspace{-15pt}
\label{fig:p-bits-system}
\end{figure}

 More generally, many workloads in probabilistic AI models, such as modern energy-based models (EBM)~\cite{liao2022gaussian,niazi2024training}, involve several linear algebra operations, mainly as matrix multiplies.
Thus, to scale up probabilistic architectures a heterogeneous approach involving traditional digital accelerators (e.g., GPU) and \mbox{p-computers} is desirable.
The GPU can be used for a large part of the forward operation, such as computing embeddings, the gradients, and loss optimization while the p-computers (implemented for example on an FPGA) can essentially implement the stochastic neuron operation.
Figure \ref{fig:p-bits-system} illustrates a heterogeneous system architecture with multiple GPUs and FPGAs that can be used for scaling up to a large number of p-bits (e.g., $>$1B) enabling it to train and infer next-generation EBMs.

Two natural concerns arise.
First, given the EBM consists of a lot of matrix operations, careful system-level profiling should be performed to assess if the time for sampling is the one to optimize by Amdahl's law.
Note that in the generalized case of a fully connected model, by scaling linearly the number of neurons, the number of synapses scales quadratically. Thus, at each step of an exponentially long sampling process, linear operations (i.e., activations) are performed on neurons and quadratic operations (i.e., matrix multiplies) on synapses. Second, their PCIe communication might bottleneck the heterogeneous approach with several, e.g.,  GPU2GPU, FPGA2FPGA, and GPU2FPGA calls.
In traditional architecture, every time a GPU communicates with the FPGA it should access the main memory through PCIe, as shown in the orange flow of Figure \ref{fig:p-bits-system} leading to significant overhead due to back-and-forth communications.
Architectures with Peer2Peer (P2P) communication~\cite{bittner2014direct} and disaggregated memory~\cite{lim2009disaggregated,jiang2023chameleon} can mitigate this bottleneck, as shown in the green logical flow of Figure \ref{fig:p-bits-system}, where FPGA and GPU communicate directly through PCIe without routing through the CPU and main memory~\cite{kaiwen2024acceleration}.

We have implemented a concrete instance of a Heterogeneous Probabilistic Architecture (HPA) to accelerate the training of deep Boltzmann machines (BMs) on sparse Pegasus-topology graphs. In our system, an FPGA-based PPU implements massively parallel p-bit sampling using a graph-colored pipelined scheme that enables conflict-free simultaneous updates of all p-bits, while a GPU performs gradient computation and parameter updates. The design uses integer-only arithmetic, avoids DSP blocks, and supports flexible embedding of sparse graphs. Evaluated on MNIST, our HPA achieves sampling throughputs exceeding $167$~Gflips/s with the P2P-enabled FPGA configuration, up to $3585\times$ speedup over GPU-only sampling and $517\times$ over an optimized CPU (Rust) baseline (\Cref{tab:hpa_benchmarks}). Importantly, the same PPU hardware can also be repurposed to solve combinatorial optimization problems, including SAT and MaxCut instances, demonstrating the generality of our approach across both learning and optimization workloads. Our HPA results provide direct empirical evidence that the P2P-enabled FPGA configuration reduces per-epoch training latency from $1397$~s (host-staged PCIe) to $741$~s, a $1.9\times$ improvement attributable entirely to eliminating the CPU-mediated data path. This measured gain validates the P2P architectural principle at the system level and is expected to grow as model sizes increase and communication becomes a larger fraction of total execution time.

\begin{table}[h]
\centering
\caption{Performance comparison of probabilistic training across evaluated platforms in the heterogeneous probabilistic architecture (HPA) for deep Boltzmann machines on MNIST.}
\label{tab:hpa_benchmarks}
\begin{tabular}{@{}lccc@{}}
\toprule
Sampling Hardware & \makecell{Sampling\\Latency (ms)} & \makecell{Latency per\\Epoch (s)} & \makecell{Throughput\\(Gflips/s)} \\
\midrule
CPU (Rust solver) & 1515.8 & 2859.3 & 5.14 \\
GPU (A100) & 10504.07 & 14054.39 & 1.02 \\
FPGA (U250) (PCIe) & 2.93 & 1397 & 31.75 \\
FPGA (U250) (P2P) & 1.18 & 741 & 167.29 \\
\bottomrule
\end{tabular}
\end{table}

\subsection{Neural quantum states bridging quantum and probabilistic computing}
\label{subsec:nqs_bridge}

\begin{figure*}
  \centering
  \includegraphics[width=0.9\textwidth]{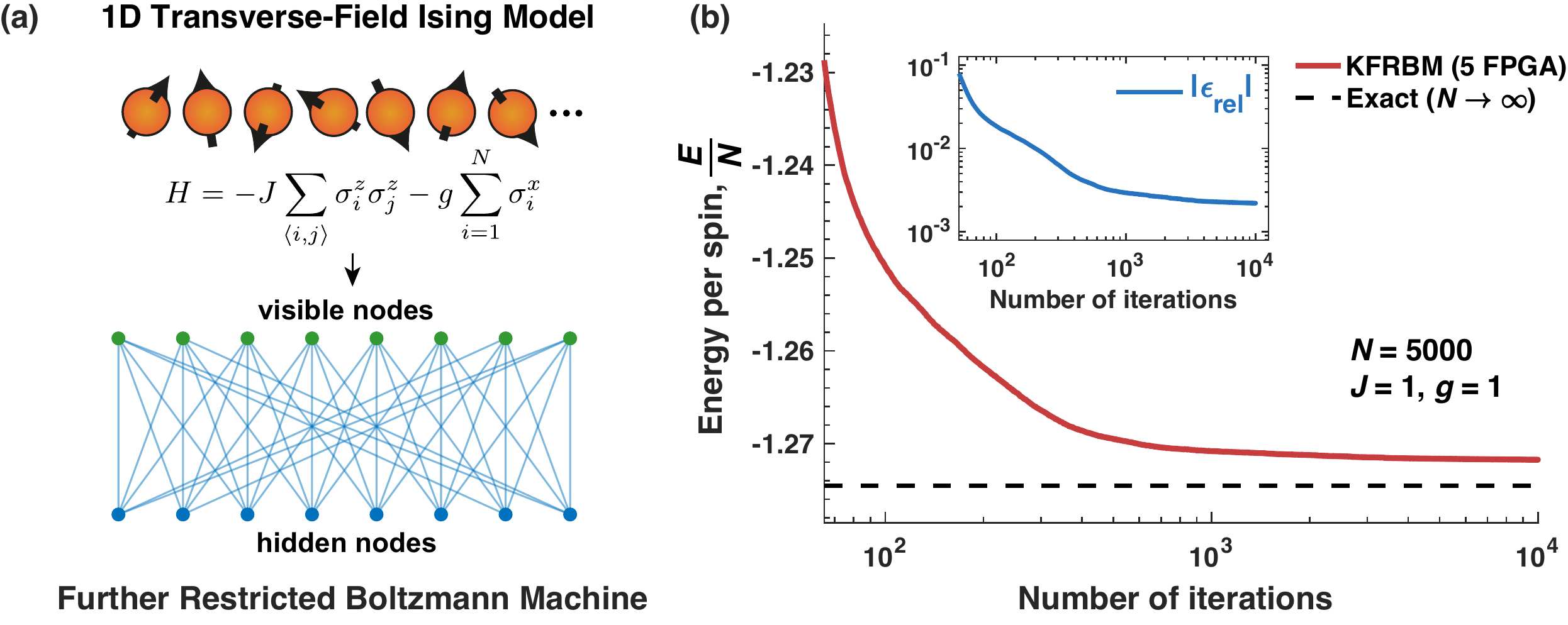}
  \caption{Neural quantum state benchmark on distributed p-computers.  (a) The specific NQS example in this section is a k-FRBM (Further Restricted Boltzmann Machine) ansatz for the 1D TFIM at $h=0$ and $g/J=1$, with $N_v=5000$ visible spins and $N_h=\alpha N_v=5000$ hidden units ($\alpha=1$), and fixed visible degree $K=5$. (b) End-to-end variational training on the five-FPGA p-computer: per iteration we run $10^{6}$ Monte Carlo sweeps at a 15\,MHz p-bit update clock, refresh boundary p-bits at 100\,MHz over FMC (FPGA Mezzanine Card) links, and read out 100 uniformly spaced samples; weights and biases are streamed in signed fixed-point (1 sign, 6 integer, 6 fractional bits). Over $10^{4}$ optimization iterations (SGD learning rate $\eta=0.1$), the energy per spin approaches the exact thermodynamic-limit TFIM ground-state benchmark (dashed line)~\cite{lieb1961two,pfeuty1970one}. The inset shows the relative error $|\epsilon_{\mathrm{rel}}|=\left|\left(E_{\mathrm{NQS}}-E_{\mathrm{exact}}\right)/E_{\mathrm{exact}}\right|$ versus iteration on a log--log scale.}
  \label{fig:tfim_nqs_distributed_kfrbm}
\end{figure*}

A practical way to connect quantum workloads to probabilistic hardware is to focus on quantum simulation tasks that are already bottlenecked by sampling and optimization. Neural quantum states (NQS) are a prominent example: instead of storing an exponentially large wavefunction, one represents its amplitudes with a neural-network model and optimizes the parameters variationally using Monte Carlo sampling in the computational basis~\cite{carleo2017solving,carrasquilla2021use,vivas2022neural,lange2024architectures}. This setting naturally aligns with probabilistic computers because the inner loop of NQS training repeatedly requires drawing samples from a parameterized probability distribution and estimating expectation values, which is precisely what p-bit networks implement natively. In conventional CPU/GPU implementations, this inner loop typically dominates wall time because the variational objective and its gradients require many Markov-chain sweeps and repeated energy and gradient evaluations, often with long mixing times near criticality or in large systems.

To make the connection explicit, we again consider the one-dimensional transverse-field Ising model (TFIM) from \Cref{eq:tfim_hamiltonian}
with periodic boundary conditions. This spin Hamiltonian maps to free-fermions via Jordan-Wigner mapping; hence, it is an exactly solvable model, suitable for benchmarking our novel approach for sampling NQS on distributed p-computers.

In the NQS framework, we write the ground-state wavefunction in the $\sigma^{z}$ basis as $\psi_{\bm{\theta}}(\bm{s})$, where $\bm{s}\in\{\pm 1\}^{N}$. For stoquastic Hamiltonians, those that can be written in a basis where all off-diagonal matrix elements are non-positive, the Perron–Frobenius theorem ensures that the ground-state wavefunction can be chosen real and non-negative \cite{marshall1955antiferromagnetism}. Since the TFIM is stoquastic in the computational basis $\{\boldsymbol{s}\}$, its ground-state amplitudes may be taken to be real and non-negative in this basis.  Consequently, one can focus on modeling the probability $p_{\bm{\theta}}(\bm{s}) = |\psi_{\bm{\theta}}(\bm{s})|^2$ and optimizing $\bm{\theta}$ by minimizing the variational energy $\langle \hat{H}\rangle_{\bm{\theta}}$ using Monte Carlo samples drawn from $p_{\bm{\theta}}$~\cite{carleo2017solving,carrasquilla2021use}. Restricted Boltzmann machine (RBM) ans\"atze are especially illustrative because they induce a Boltzmann-form distribution over visible configurations after marginalizing hidden units, which directly maps to an energy-based model that a p-computer can sample~\cite{deng2017machine,torlai2018neural}. Details of the local-energy estimator and gradient construction used in our implementation follow Ref.~\cite{chowdhury2025probabilistic}. A schematic of the multi-FPGA mapping is shown in \Cref{fig:distributed_p-bits-system}, and the k-FRBM NQS benchmark used in this work is shown in \Cref{fig:tfim_nqs_distributed_kfrbm}a.

\textbf{k-FRBM ansatz and hardware mapping.}
We use a further-restricted Boltzmann machine (FRBM), a structured sparse RBM designed to be hardware-friendly by constraining connectivity. The model has $N_v$ visible units (representing the TFIM spins) and $N_h=\alpha N_v$ hidden units, where $\alpha$ is the hidden-to-visible ratio. A k-FRBM enforces that each visible unit connects to exactly $K$ hidden units, yielding $O(KN_v)$ nonzero couplings rather than a dense $O(N_vN_h)$ bipartite graph. This fixed-degree structure is important for scaling because (i) it matches sparse, local-field p-bit updates, (ii) it makes weight storage and routing predictable, and (iii) it enables balanced graph partitioning with small cut size when distributing the model across multiple processing elements. In our experiment, we set $\alpha=1$ and $K=5$, so the NQS distribution is represented by a sparse bipartite graph over $N_v+N_h$ binary variables (implemented as p-bits). All trainable parameters (visible biases, hidden biases, and sparse weights) are initialized with small independent Gaussian random values (standard deviation $10^{-2}$).

\textbf{Distributed p-computer execution.}
The full k-FRBM graph is mapped onto a distributed p-computer composed of multiple FPGAs (five partitions in our current prototype), where each FPGA updates its local p-bits via Gibbs sampling while exchanging boundary states over inter-FPGA links. The partitioning is performed using KaHIP (Karlsruhe High Quality Partitioning), a multilevel graph partitioning framework that targets high-quality cuts and good load balance~\cite{sanders2013kahip}. KaHIP is used here to keep the p-bit counts balanced across FPGAs while reducing inter-partition couplings. In our setup, the p-bit update clock is 15\,MHz, and the boundary p-bits are refreshed at a 100\,MHz clock. Inter-FPGA boundary exchange is implemented with low-latency FMC (FPGA Mezzanine Card) links, providing short, direct, board-level connections that keep communication delay small compared to the p-bit update time.

\textbf{End-to-end TFIM training benchmark.}
We benchmark the full pipeline on TFIM at $h=0$ and $g/J=1$, a standard critical-point setting where quantum fluctuations are strong and the exact ground-state energy in the thermodynamic limit is known~\cite{lieb1961two,pfeuty1970one}. The visible layer contains $N_v=5000$ spins (qubits in the target quantum model), the hidden layer has $N_h=\alpha N_v=5000$ units, and the sparse connectivity is fixed to $K=5$ hidden neighbors per visible unit. At each optimization step, the distributed p-computer is run to generate samples from the current k-FRBM distribution; these samples are used to estimate the variational energy and gradients following Ref.~\cite{chowdhury2025probabilistic}. Parameters (visible biases, hidden biases, and sparse weights) are then updated by stochastic gradient descent with a constant learning rate $\eta=0.1$. Concretely, each optimization step executes $10^{6}$ Monte Carlo sweeps of the distributed p-bit network and reads out 100 uniformly spaced uncorrelated samples. The weights and biases are quantized and streamed to the FPGAs in a signed fixed-point format with 1 sign bit, 6 integer bits and 6 fractional bits. The full procedure is repeated for $10^{4}$ optimization steps (iterations). We show the optimized energy per spin as a function of the number of iterations in \Cref{fig:tfim_nqs_distributed_kfrbm}b. We observe an asymptotic reduction in the relative energy (see inset in \Cref{fig:tfim_nqs_distributed_kfrbm}b) showing that the NQS wavefunction can achieve the desired accuracy at the critical point by running more iterations. This experiment verifies that distributed probabilistic hardware can execute large-scale probabilistic workloads directly relevant to quantum simulation~\cite{chowdhury2025probabilistic} under realistic hardware constraints on communication, precision, and sampling.

\textbf{Relevance to FTQC.} PPU-based NQS provide a fundamentally new benchmarking approach that overcomes the limitations of classical quantum-simulation methods. Traditional techniques such as tensor networks, Clifford expansions, free-fermion models, and perturbation theory break down in different regimes, including high entanglement, large magic, or strong interactions. Neural Quantum States (NQS) offer greater expressiveness and can represent some volume-law states \cite{lange_2024_from_architectures_to_applications,sharir_2022_neural_tensor_contractions}, but their usefulness is constrained by the number of samples that grows exponentially in the presence of "sign problem."  PPUs help mitigate this bottleneck by delivering orders-of-magnitude faster sampling, enabling rapid benchmarking of QPUs on utility-scale, sign-problem-free workloads. Such benchmarks are critical because utility-scale problems typically involve high entanglement, significant magic, and strong interactions, features that severely limit the utility of most other classical simulation methods.

Furthermore, PPU-based NQS allows us to strategically distribute workloads across QPUs and PPUs to map out complex phase diagrams of practically relevant materials. For regions of the phase diagram where the sign problem can be mitigated by a basis change by minimizing the ``negativity'' \cite{murota_2025_local_basis_transformation},  we can employ PPU-based NQS, whereas we can employ QPU for other regions of the phase diagram.

One of the foundational applications of neural-network wavefunctions is quantum state tomography \cite{torlai2018neural}. Although full state/process tomography is exponentially expensive \cite{Mohseni_2008_QPT}, an NQS can be trained on a limited set of measurement outcomes from a quantum computer to reconstruct a high-fidelity approximation of the underlying state. Once trained, the NQS enables extraction of quantities like entanglement entropy, that would otherwise require an infeasible number of hardware samples. By feeding noisy measurement bitstrings from a QPU into a PPU-based NQS, the probabilistic hardware can rapidly learn the density matrix $\rho$ on the QPU. The PPU's fast sampling capability then allows millions of resamples of the learned state in near real-time, effectively creating a digital twin of the quantum processor. This digital twin can provide a practical environment for testing error-mitigation strategies and performing algorithm co-design without consuming costly QPU runtime. Next, we discuss how hybrid QPUs and PPUs architecture could provide an optimal approach for tackling classically hard sampling and optimization tasks of interest to operations research and AI.

\subsection{Quantum-assisted probabilistic sampling}
\label{sec:hetero_pq}

There are three complementary perspectives that we can envision for the interplay of quantum fluctuations and thermal fluctuations in a probabilistic computing framework:  within the problem space, the algorithm space, and the solution space. 

Within the problem space, as we described in \Cref{sec:distp} (see Figure \ref{fig:distributed_p-bits-system}), we can partition a general dense graph with higher order interactions by sparsification techniques. During the procedure we are iteratively creating conditional probability distributions (i.e., by freezing or clamping a subset of variables) and sampling over the rest of variables residing on a smaller and lower dimensional subgraph. This technique can be used to reduce both the size and complexity of the problem such that it can be easily embedded on finite-size low-dimensional quantum accelerators/solvers. These quantum solvers could be either analog, based on quantum annealing \cite{kadowaki1998quantum,albash2018adiabatic}, or digital, based on the quantum approximation optimization algorithm (QAOA)~\cite{farhi2014quantum,blekos2024review}. Larger, denser, and/or highly structured subgraphs can be sampled via p-computers and the outputs can be used as boundary conditions (e.g., local fields for Ising machines) for potentially quantum-prone subgraphs iteratively. 

Within the algorithm space, we can understand the role of quantum solvers is to provide hot starts/seeds for classical probabilistic framework and vice versa. This was originally introduced in the context of creating non-trivial initial seeds for reverse quantum annealing or MCMC sampling enabling quantum-assisted parallel tempering~\cite{denchev2022quantum} or quantum-assisted genetic algorithms~\cite{King2019-qv}. 

Within the configuration space, we can see the role of quantum fluctuations as a new mechanism to navigate in the saddle regions or regions with shattered configurations space with many unstructured shallow barriers. Such effective quantum walks could in principle lead to a quadratic speedup for diffusing in the configuration space over a classical walk \cite{szegedy2004quantum}. 

It should be noted that these three perspectives are not mutually exclusive, for example in the context of non-equilibrium non-local Monte Carlo algorithm developed~\cite{mohseni2021nonequilibrium}, one can discover backbone or cores of frozen or rigid variables in a configuration space near a phase transition (e.g., for $k$-SAT problems near a computational SAT/UNSAT phase transition). The backbones with a higher degree of connectivity could be sampled with classical probabilistic accelerators to induce large Hamming distance exploration ($O(N)$) at scale. Then the new coordinates can be passed to quantum accelerators to create smaller-scale (in Hamming distance) nonlocal explorations over regions with significant entropic barriers or shallower energy barriers that would be prone to quantum tunneling on finite-size quantum processors. On the other hand, the new local minima found by a quantum processor can be improved by classical fluctuations, especially those induced by quantum many-body localization effects \cite{kechedzhi_efficient_2018}. 

This hybrid algorithm can be incorporated within the heterogeneous HPC computing platforms, with P2P communications among various nodes including CPUs, GPUs, FPGAs, QPUs, or other custom-designed accelerators, see Figure \ref{fig:p-bits-system}. In some implementations, both quantum and classical processors can be placed on the same chip to get additional performance benefits~\cite{mohseni2020chips} (e.g., in hybrid quantum and classical superconducting processors). This quantum-probabilistic framework can enhance the diffusion in configuration and improve the quality and diversity of solutions~\cite{mohseni_sampling_2023,zucca_diversity_2021} given a time or energy budget, as new basins of attractions could be found orders of magnitude faster and more energy efficiently than using either probabilistic or quantum accelerators alone. 

In summary, tight integration of PPUs with QPUs has the potential for bridging the gap between hardware capabilities and utility-scale performance. At the lowest layer, PPUs could accelerate real-time, neural-network-based decoding, ensuring the stability of fault-tolerant operations through rapid inference of error syndromes. At the application layer, the PPU acts as a high-bandwidth co-processor that enhances sample diversity for combinatorial optimization problems (discussed in \cref{sec:hetero_pq}) and accelerates machine learning inference within quantum algorithms. For example,  identifying low-entanglement boundaries in the ACK algorithm (discussed in \cref{sec:hpc-qc-workload-ack}), the PPU enables an intelligent distribution of workloads to minimize the overhead of distributing a quantum workload across QPUs. Furthermore, PPU-based NQS can be used to benchmark against or to share certain workloads with the QPU, where there is no sign problem. Therefore, this hardware-level synergy between PPU and QPU can play a key role in transforming QPU as a standalone processor to a quantum supercomputer.

\section{Concluding remarks}

The ultimate goal of developing a utility-scale quantum computer is to deliver valuable computations 
that justify the device's cost. Achieving this objective for any architecture involves demonstrating 
several non-trivial criteria: (1) the existence of a set of computations that produce measurable 
value when executed; (2) the cost of a quantum computer capable of performing these computations 
is lower than the value they generate; and (3) no alternative methods -- whether classical, analog, quantum, 
experimental, etc. -- can execute these computations more cost-effectively. Evaluating any proposal for a 
utility-scale quantum computer, including the vision we present here, is thus a challenging task involving
answering questions spanning domain-specific use cases, device architecture and integration, and the full 
range of computational methods. This work focuses primarily on the second criterion, providing an initial 
holistic understanding of the size, scale, and cost of a quantum supercomputer and identifying key challenges. 

We have outlined the design of a quantum supercomputer that combines the latest advances in HPC and superconducting qubits. A comprehensive list of known technical challenges has been used to illuminate obstacles for scaling quantum processors that have been overlooked or addressed piecemeal in prior research \cite{alexeev_quantum-centric_2024,ibmPositionPaper,humbleQuantumClassical,SvoreACCM06layered}.  In particular, contemporary practical quantum computing investigations have primarily concentrated on technical hurdles at the hundred-qubit level, constrained by the quality of individual qubits. By anticipating technical scaling challenges from hundreds to millions of qubits, our study outlines a holistic system that could provide practical quantum advantage. 

For qubit hardware, our thesis is that 300-mm semiconductor processing can radically improve qubit quality, reliability, and scaling.  Reproducibility and yield can be greatly improved by replacing junction fabrication from lift-off to a CMOS standard deposition and etch process.  Our architecture uses wafer-scale integration, where 20k-qubit modules are connected to 3 Kelvin control circuitry through a  300-mm wiring die.  Additionally, an on-chip measurement system eliminates the need for expensive and large volume circulators and pre-amplifiers. Tiling multiple 20k-qubit modules allows fitting 160k qubits in a single DR, and reduces the need for optical interconnects between DRs. This integrated and modular approach thus enables a lower-cost path to scaling.

We introduced a design for a classical supercomputing system that integrates fault-tolerant quantum computing to realize hybrid quantum--classical workloads. HPC is changing, becoming more heterogeneous in response to the need to integrate specialized classical processors. Incorporating quantum processors is just a further step in this trend. We described a hybrid quantum--classical full stack that can be extended as these technologies evolve. The stack is modularized within six layers to allow different software vendors to innovate in parallel. We discussed extensions to existing HPC programming environments and explored various circuit knitting techniques for distributing quantum workloads across multiple QPUs as a complementary approach to direct quantum networking. We described approaches to designing algorithms that leverage both quantum and classical computing capabilities in close coordination, and discussed extensions to current HPC performance benchmarking standards. At the lower layers of the stack, we detailed FTQC protocols and tight integration of the HPC system with the quantum control digital logic via low-latency and high-bandwidth PCIe-based interfaces. This allows seamless, standardized, and high-performance integration between QPUs, CPUs, GPUs, PPUs, and any other compute resource in a heterogeneous HPC system. By minimizing communication overheads, our design for a high-performance quantum--classical stack enables calibration, distributed quantum error correction and characterization, adaptive circuit knitting, and any other hybrid process to be performed at the cutting-edge limit of fidelity and runtime.

Our sensitivity analysis revealed that improving the gate fidelities would improve the exponential error suppression of surface codes significantly more than reducing SPAM errors or protecting idling qubits. Another enlightening result of our study is that the logical fidelity of surface code depends considerably on the distribution of qubit errors within the qubit array. Moreover, we demonstrate a threshold behavior at $\sim$6\% for rows of weaker stabilizer measurements representing capacitive tile-to-tile interconnects, or any anticipated future QPU-to-QPU quantum networking technology (such as optical interconnects). Nevertheless, achieving this fidelity especially at the high Bell-pair generation frequencies required for lattice surgeries between high-distance codes is ambitious for the existing optical interconnect technologies for superconducting qubits. We therefore recommend alternative techniques such as partial fault-tolerant compilation and circuit knitting to be explored in parallel \cite{daniel2026partialFTMegaquop}. 

We assess our design with detailed resource estimates of several utility-scale applications including quantum chemistry calculations, catalyst design, NMR spectroscopy, and Fermi--Hubbard simulation. Our analysis leverages detailed experimentally-motivated simulation and benchmarking of fault-tolerant quantum computing protocols using superconducting qubits. We show that for ground-state energy estimation of \(p\)-benzyne and FeMoCo with chemical accuracy $\epsilon =$ 1.6 mHa, one requires hundreds of  millions of physical qubits and a minimum of 1.4 years to run with current hardware quality (the `baseline' hardware in our study). The runtime can be reduced to approximately four months and the qubit count to tens of millions if the hardware quality is significantly improved to our proposed targets.  For the task of catalyst design and baseline hardware specifications, we require tens of millions of physical qubits and a week of runtime. These can be improved to 2.6 million physical qubits and a few hours of runtime for desired hardware specifications. Similarly, NMR spectral prediction of small molecules and peptides relevant for drug discovery pipelines can be achieved with 1.3 million physical qubits and a few months of runtime, for the desired hardware specifications. For the 2D Fermi--Hubbard model, Table~IX gives QPE runtimes of order ten seconds at the chosen energy-density resolution, conditional on access to the ground state. The proposed phase-diagram workflow in Figure~36 additionally requires state preparation, repeated executions, observable estimation, and classical analysis; the lowest of all the quantum applications discussed here. In particular, this task can be easily embedded in a phase-diagram discovery framework that relies on tight quantum-HPC integration for achieving utility as discussed in \Cref{fig:QPE-workflow-FH}.

Our quantum supercomputing vision tightly integrates quantum compute, classical compute, and control components at scale using state-of-the-art, yet existing, underlying technologies. Not all of the challenges we laid out have been addressed by our approach. Our hope is that, by publicizing the obstacles to scaling quantum computers along with a baseline design, we can stimulate the discovery of innovative solutions in both academia and industry. Future revisions of this approach are expected with new discoveries and collaborators.

\textbf{Acknowledgments.} HPE, Qolab, QM, 1QBit, AMAT, Synopsis, UWM acknowledge funding from Defense Advanced Research Projects Agency (DARPA) under Quantum Benchmarking Initiative (QBI) contract no HR00112590116.  The authors from HPE are also supported by DARPA under Air Force Research Laboratory (AFRL) contract no. FA8650-23-3-7313. P.~A.~L. and D.~V. were supported in part through the  NASA Academic Mission Services (contract NNA16BD14C). P.~A.~L. , D.~V., and N. ~A. were supported in part through the Intelligent Systems Research and Development-3  (ISRDS-3) Contract 80ARC020D0010. The authors acknowledge the financial support of Pacific Economic Development Canada (PacifiCan) under project number PC0008525. G.~A.~D. is grateful for the support of Mitacs. P.~R. acknowledges the financial support of Mike and Ophelia Lazaridis, Innovation, Science and Economic Development Canada (ISED), and the Perimeter Institute for Theoretical Physics. Research at the Perimeter Institute is supported in part by the Government of Canada through ISED and by the Province of Ontario through the Ministry of Colleges and Universities. Qolab acknowledges financial support from the Development Bank of Japan. A part of this work was performed for Council for Science, Technology and Innovation (CSTI), Cross-ministerial Strategic Innovation Promotion Program (SIP), "Promoting the application of advanced quantum technology platforms to social issues" (Funding agency: QST), MEXT Quantum Leap Flagship
Program (MEXT Q-LEAP) Grant No. JPMXS0120319794, JST COI-NEXT Grant No. JPMJPF2014, and JST Moonshot R\&D Grant No. JPMJMS2061.

The authors thank Marko Bucyk (1QBit) for editorial review of the manuscript. We are grateful to Alexandre Fleury, Einar Gabassov, Mia Kramer, Huy Anh Nguyen, Kevin Nguyen, Katie Olfert, Valentin Senicourt, Yumeng Wang, Chan Woo Yang, Xiangyi Zhang, Ryan Babbush, Craig Gidney, Matthew Harrigan, N. Cody Jones, Tanuj Khattar, Guang Hao Low, Nicholas Rubin, Akihisa Goban, Masashi Hirose, Shinichi Sunami, Shiro Tamiya, Ming-Zhi Chung, Keisuke Fujii, Yohei Ibe, Toru Kawakubo, Tennin Yan, Andreas Thomasen, Robert Parrish, Bob Sorensen, Jeremy Stevens, Oliver Ezratty, Margaret Martonosi, Taylor Patti, Brucek Khailany, Michael Brown, Filip Wudarski, Andreas Thomasen, Keita Kanno and Sohaib Alam for useful discussions and/or valuable feedback on this manuscript.

\bibliography{all_ref,namit, gaurav}

\clearpage
\begin{appendices}
\appendix

\section{Analysis of logic circuits\\ for quantum resource estimation}
\label{sec:app_logical_circuits}

In this section, we outline our workflow for generating the logical quantum circuits that serve as input to the resource estimation pipeline which computes the associated physical resource requirements. The circuits we generate pertain to quantum simulations for estimating the ground-state energy of molecules. We first specify the quantum simulation algorithm used. We then discuss the workflow for how we obtain the quantum circuits that implement this algorithm from the basic specifications of a molecule. Finally, we analyze the various bounds on the errors incurred in the process of generating the logic circuits, and explain how these bounds must be chosen to ensure that the quantum simulations achieve a given target accuracy. While in our study we use the \mbox{$p$-benzyne} molecule as a concrete example, the described methodology applies to other molecules.

\subsection{Workflows for generating logic circuits} \label{sec:logical_circuit}

The quantum phase estimation (QPE) algorithm~\cite{kitaev1995quantum,abrams1999quantum} is arguably one of the most rigorous quantum computational approaches for estimating ground-state energies in quantum chemistry. Quantum phase estimation is designed to sample in the eigenbasis of the molecular Hamiltonian $H$ by measuring the phase accumulated on an initial input quantum state acted upon by a unitary operator whose eigenvalue spectrum is a function of the spectrum of $H$. The standard approach is to implement QPE with the time-evolution operator $\exp(-iHt)$.  
More-advanced approaches to electronic-structure quantum simulations are typically based on the framework of qubitization~\cite{low2019hamiltonian}. This framework allows taking a Hamiltonian given by a sum of unitaries (which is the typical case for quantum chemistry Hamiltonians) and constructing a new operation called ``qubiterate'' that has a functional dependence on the eigenvalues of the Hamiltonian and thus can be used in QPE in place of $\exp(-iHt)$ (see Ref.~\cite{berry2018improved}). 
Our quantum resource estimation (QRE) studies report estimates for two algorithmic approaches: the Trotter-based approach, and the qubitization-based double low-rank factorization algorithm.  
In what follows, we outline the workflow of QRE for the standard QPE algorithm, with the time-evolution operator $\exp(-iHt)$ implemented 
using Hamiltonian simulation based on the use of product formulae (PF) and Trotterization. The associated analysis of Trotter errors and their propagation into QPE is discussed in \Cref{sec:error-analysis}. We refer the reader to the literature for various qubitization-based approaches~\cite{berry2019qubitization,vonBurg2021quantum,lee2021even,goings2022reliably,beverland2022assessing, otten2024quantum,bellonzi2024feasibility,watts2024fullerene,nguyen2024quantum,rocca2024reducing}). The propagation of errors in qubitization is briefly outlined in \Cref{sec:error-analysis-qubitization}. A brief description of the 
double-factorized quantum chemistry simulations used in our QRE studies is given in  \Cref{sec:Double-factorized_quantum-chemistry}.

In the Trotter-based approaches to QPE,  
the time-evolution operator $\exp(-iHt)$ is 
approximated by a PF along with Trotterization. In general, an operator $\mathcal{S}_p(t)$ is called an order-$p$ product formula associated with the time-evolution operator $\exp(-iHt)$ for a given Hamiltonian $H$ if (see Refs.~\cite{childs2021theory, zkuk2023trotter})
\begin{equation}
\mathcal{S}_p(t)=\exp(-iHt)+\mathcal{O}(t^{p+1})\,.
\end{equation}
Our resource estimation analyses are based on using either the first-order Lie--Trotter formula or the second-order Trotter--Suzuki formula.

In the framework of second quantization,
the electronic model Hamiltonian is typically given as
\begin{equation}
\hat{H}=\sum_{p,q}h_{pq}\hat{a}^\dagger_p\hat{a}_q+\frac{1}{2}\sum_{p,q,r,s}h_{pqrs}\hat{a}^\dagger_p\hat{a}^\dagger_r\hat{a}_q\hat{a}_s\,,
\label{eq:second-quantized-H}
\end{equation}
where $\hat{a}^\dagger_p$ and $\hat{a}_p$ are the fermionic creation and annihilation operators, respectively, associated with a given basis set of spin-orbital basis functions $\{\phi_p(\boldsymbol{x})\}$ (where $\boldsymbol{x}\equiv\{\boldsymbol{r},\sigma\}$ summarizes the orbital and spin degrees of freedom), and the scalar coefficients $h_{pq}$ and $h_{pqrs}$ are the one- and two-electron integrals, respectively, over the basis functions, computed using the kinetic term and the nuclear and electron--electron coulomb potentials.
Numerous software tools exist to derive the second-quantized Hamiltonian from the molecular specifications, which include the basic information for fully characterizing the system, such as the type of participating atoms and the molecule's geometry (typically summarized in an $xyz$ file), total charge, and total spin. For this study, we used Tangelo, which is an open source Python software package for end-to-end chemistry workflows for quantum computation~\cite{tangelo}. The \mbox{$p$-benzyne} molecule $\mbox{C}_6\mbox{H}_4$ (which has zero total charge) exhibits a biradical open-shell singlet ground state (it has zero total spin), with two unpaired electrons. Its geometry is specified by the \texttt{xyz} file configuration shown in \Cref{table:xyz-pbenzyne} (see Section 19 in the supplementary material of Ref.~\cite{keller2015selection}).

\begin{table}[h!]
\centering
 \begin{tabular}{c c c c}
 \hline\hline
$\quad$C & $\quad$$-0.7396$ &   $\quad$$-1.1953$ &  $\quad$0.0000$\quad$\\
$\quad$C & $\quad$$0.7396$ &  $\quad$$-1.1953$ &  $\quad$0.0000$\quad$\\
$\quad$C &  $\quad$$1.3620$ &  $\quad$$0.0000$ &  $\quad$0.0000$\quad$\\
$\quad$C &  $\quad$$0.7396$ &  $\quad$$1.1953$ &  $\quad$0.0000$\quad$\\
$\quad$C &  $\quad$$-0.7396$ &  $\quad$$1.1953$ &  $\quad$0.0000$\quad$\\
$\quad$C &  $\quad$$-1.3620$ &  $\quad$$0.0000$ &  $\quad$0.0000$\quad$\\
$\quad$H &  $\quad$$1.1999$ &  $\quad$$-2.1824$ &  $\quad$0.0000$\quad$\\
$\quad$H &  $\quad$$-1.1999$ &  $\quad$$2.1824$ &  $\quad$0.0000$\quad$\\
$\quad$H &  $\quad$$1.1999$ &  $\quad$$2.1824$ &  $\quad$0.0000$\quad$\\
$\quad$H &  $\quad$$-1.1999$ &  $\quad$$-2.1824$ &  $\quad$0.0000$\quad$\\
\hline
\end{tabular}
\caption{Molecular geometry of $p$-benzyne in \r{a}ngstr\"{o}ms, in terms of the \texttt{xyz} file format; see~Ref.~\cite{keller2015selection}.}
\label{table:xyz-pbenzyne}
\end{table}

In addition to the molecule specifications, we need to select a basis set $\{\phi_p(\boldsymbol{x})\}$. Basis set selection can be a challenging task. While theoretically an infinite basis is required to represent the true molecular multi-body wavefunction, in practice we cannot perform calculations using an infinite number of basis functions and must therefore rely on using a finite basis set. Numerous basis sets have been introduced and extensively studied in quantum computational chemistry. The most-common minimal basis sets are the \texttt{STO-$n$G} basis sets, which are derived from a Slater-type orbital basis set, with $n$ denoting the number of Gaussian primitive functions used to represent each Slater-type orbital. While minimal basis sets are computationally inexpensive, they typically result in insufficiently precise computations. Pople basis sets are a type of split-valence basis sets that use more than one basis function to represent valence orbitals, because it is the valence electrons that typically contribute to the molecular bonding. An entire hierarchy of Pople basis sets have been studied. Importantly, as the basis set grows larger, the resulting approximation gets closer to the true wavefunction; however, increasing the basis set size also results in increasing the required computational resources in both space and time. As a rule of thumb, to achieve semi-quantitative energies, the minimum requirement is to use double-zeta basis sets (such as, e.g., \texttt{6-31G} or \texttt{cc-pvdz})~\cite{tangelo}. For our QRE analysis, we used the \texttt{6-31G} basis set; this basis set yields a good trade-off between accuracy and computation time.

Once a basis set has been selected, we can reduce the size of the system (and thus the computational cost) via {\em active space selection}. This concept relies on the notion that,  when considering only a subset of the full active space, the resulting loss in correlations affecting the energy computation can be small. For example, in the so-called ``frozen-core approximation'',  low-lying occupied core orbitals (which typically do not mix with valence orbitals) are ``frozen'', that is, they are not included in the computation. Choosing which molecular orbitals to freeze is not a trivial task. Again, we used Tangelo, which provides a means to identify the active space specified by
the numbers of molecular orbitals to be included that are energetically next to (i.e., below or above) the highest occupied molecular orbital (HOMO) and the lowest unoccupied molecular orbital (LUMO). For example, the specification ``HOMO$-2$ and LUMO$+1$'' means that we include two additional molecular orbitals below the HOMO and one additional orbital above the LUMO. A common choice is to employ an equal number of additional orbitals to be included next to the HOMO and LUMO; this choice leads to lower energies as opposed to active spaces with unequal numbers of orbitals next to the HOMO and LUMO (see Ref.~\cite{sennane2023calculating}).
For example, for the $p$-benzyne molecule, the full active space involves 68 active spin-orbitals for the \texttt{STO-3G} basis and 124 active spin-orbitals for the \texttt{6-31G} basis, and the frozen-core approximation involves 56 active spin-orbitals for the \texttt{STO-3G} basis and 112 active spin-orbitals for the \texttt{6-31G} basis, while, for instance, an active space selection ranging from HOMO$-5$ to LUMO$+5$ involves only 24 active spin-orbitals for both basis sets. Note that the number of qubits required to encode the system equals the number of active spin-orbitals.

Once the basis set and the active space have been selected, we can generate the associated second quantized Hamiltonian, as given in \Cref{eq:second-quantized-H}. The last step is to translate the model Hamiltonian from the second quantization framework to a framework suitable for the quantum circuit model. This step uses a {\em fermion–to–qubit mapping}, which is typically either the \mbox{Jordan--Wigner}~\cite{Jordan1993} or the Bravyi--Kitaev~\cite{BRAVYI2002210} transformation,  to obtain the Hamiltonian in the Pauli-product form. For a Hamiltonian acting on $n$ qubits, it can be expressed as
\begin{align}
    H&=\sum_{\ell=1}^LH_\ell=\sum_{\ell=1}^L\gamma_\ell P^{(\ell)},\quad\mbox{where} \nonumber\\&\quad P^{(\ell)}:= P^{(\ell)}_1\otimes P^{(\ell)}_2\dots \otimes P^{(\ell)}_n, \quad\; \nonumber\\&\quad P^{(\ell)}_k\in\left\{I,X,Y,Z\right\},
\label{eq:Pauli-product-H}
\end{align}
and $\gamma_\ell\in\mathbb{R}$ are real coefficients. This Hamiltonian can be directly translated into a quantum circuit implementing a single Trotter slice for a PF associated with the time evolution $\exp(-iHt)$ as part of QPE using well-established quantum circuit decomposition methods. This  circuit typically consists of a sequence of single- and two-qubit Clifford gates and $L$ arbitrary-angle single-qubit rotations acting on the qubits involved. This circuit is output as a \texttt{qasm} text file and used as input to the QRE pipeline.
\vspace{0.2cm}

\subsection{Analysis of Trotter errors and\\ their propagation into phase estimation}
\label{sec:error-analysis}
We now discuss the various errors incurred in the process of generating the logic circuits in the Trotter-based approaches to QPE and how we can guarantee  electronic-structure quantum computations to estimate the Hamiltonian eigenvalues within a given target accuracy by satisfying certain error bounds, which in turn must be within given {\em error budgets}. Our analysis of Trotter errors and  their propagation into phase estimation closely follows the approach of Ref.~\cite{reiher2017elucidating}. The propagation of errors in qubitization is discussed in the next subsection.

On the logical level, there are three sources of error in implementing the QPE algorithm
with Hamiltonian simulation based on using PFs. The first error source is the actual use of an \mbox{order-$p$} PF, which results in an additive error $\mathcal{O}(t^{p+1})$ in representing the time evolution $\exp(-iHt)$; the additional Trotterization is a technique for dividing the evolution time into many smaller time steps that reduce this error by a constant referred to as ``the number of Trotter steps'' (or slices). The second error source is associated with circuit synthesis: each term in a PF is implemented by a circuit consisting of single- and two-qubit Clifford gates (such as $H$, $S$, Pauli, and CNOT gates) and some arbitrary-angle, single-qubit rotation $R_Z(\theta_{\ell})$ (with angle $\theta_\ell$ related to the coefficient $\gamma_\ell$ in the Hamiltonian in \Cref{eq:Pauli-product-H}). The latter is approximated by some sequence of the form $HTHST^{\dagger}\dots HS$. This approximation is found using circuit synthesis tools based on either the Solovay--Kitaev (SK) algorithm or the Ross--Selinger (RS) algorithm~\cite{selinger2015efficient,ross2015optimal} that achieves a more favorable scaling in terms of the $T$-gate count. Either of the circuit synthesis methods 
incurs an error associated with the approximation. The third error source is associated with the precision of estimating the phase in the actual QPE algorithm. In what follows, we elaborate on these errors and provide useful analytic error bounds, which serve to guarantee that some error budgets in our QRE analysis are satisfied.

Our circuits pertain to the first-order Lie--Trotter formula or the second-order Trotter--Suzuki formula, defined as~\cite{childs2021theory}
\begin{align}
\mathcal{S}_1(t)&:=\prod_{\ell=1}^{L} \exp\left(-itH_\ell\right)\,,\\
\mathcal{S}_2(t)&:=\left(\,\prod_{\ell=L}^{1} \exp\left(-itH_\ell/2\right)\right)\left(\,\prod_{\ell=1}^{L} \exp\left(-itH_\ell/2\right)\right).
\end{align}
Using Propositions 9 and 10 from Ref.~\cite{childs2021theory}, we  obtain the following analytic error bounds in terms of the spectral operator norm:
\begin{widetext}
\begin{align}
\label{eq:bound-PF-order1}
\left\|\mathcal{S}_1\left(t\right)- \exp\left(-itH\right)\right\|&\le
\frac{t^2}{2}\sum_{l_1=1}^L\left\| \left[\, \sum_{l_2=l+1}^L H_{l_2}, H_{l_1}\right]\right\|
\le t^2\sum_{l=1}^L\sum_{j=l+1}^L C_{lj}|\gamma_l\gamma_j|\\
\left\|\mathcal{S}_2\left(t\right)- \exp\left(-itH\right)\right\|&\le
\frac{t^3}{12}\sum_{l_1=1}^L\left\| \left[\, \sum_{l_3=l_1+1}^L H_{l_3}, \left[\sum_{l_2=l_1+1}^L H_{l_2}, H_{l_1}\right]\right]\right\|+
\frac{t^3}{24}\sum_{l_1=1}^L\left\| \left[ H_{l_1}, \left[H_{l_1}, \sum_{l_2=l_1+1}^L H_{l_2}\right]\right]\right\|\nonumber\\
&\le \frac{t^3}{3}\sum_{l_1=1}^L\sum_{l_3=l_1+1}^L \sum_{l_2=l_1+1}^L C_{l_1l_2l_3}|\gamma_{l_1}\gamma_{l_2}\gamma_{l_3}|+\frac{t^3}{8}\sum_{l_1=1}^L\sum_{l_2=l_1+1}^LC_{l_1l_2}|\gamma_{l_1}^2\gamma_{l_2}|\,,
\label{eq:bound-PF-order2}
\end{align}
\end{widetext}
where $C_{lj}=1$ if $\left[ P^{(j)} , P^{(l)}\right]\not=0$, and $C_{lj}=0$ if $\left[ P^{(j)} , P^{(l)}\right]=0$; similarly, $C_{l_1l_2l_3}=1$ if $\left[ P^{(l_3)}, \left[P^{(l_2)},P^{(l_1)}\right]\right]\not=0$, and $C_{l_1l_2l_3}=0$ otherwise. The first expressions, in terms of commutators, have been proven to be tight bounds for the order-1 and order-2 PFs, respectively~\cite{childs2021theory}.

The standard Hamiltonian simulation based on PFs approximates the unitary time evolution by splitting it into $r$ Trotter slices:
\begin{equation}
\exp\left(-itH\right)= \left[\mathcal{S}_p(t/r)\right]^r+\mathcal{O}\left(r(t/r)^{p+1}\right).
\end{equation}
The smaller the time $\tau:=t/r$ for a single Trotter slice is, the better the approximation associated with Trotterization becomes.
Note that $\lim_{r\rightarrow\infty}\left[\mathcal{S}_p(t/r)\right]^r=\exp\left(-itH\right)$.
Each term
$U_\ell(\tau):=\exp\left(-i\tau H_\ell\right)$ for the order-1 PF (or \mbox{$U_\ell(\tau):=\exp\left(-i\tau H_\ell/2\right)$} for the order-2 PF) is implemented by a circuit consisting of single- and two-qubit Clifford gates along with an additional arbitrary-angle single-qubit rotation; the latter needs to be decomposed and approximated by some sequence of the form $HTHST^{\dagger}\dots HS$ using the SK algorithm (or the RS algorithm). The incurred error of approximation is required  to be bounded by some error budget per gate. More precisely, we let the effective unitary $\tilde{U}_\ell(\tau)$ denote the  approximation of $U_\ell(\tau)$ by a circuit consisting of gates from the standard gate set $\{H,\, S,\, T,\, \mbox{Pauli gates}, \mbox{CNOT}\}$ after running the SK algorithm and using other circuit synthesis tools, and let $\Delta_{\mbox{\tiny synth}}$ denote the maximum circuit synthesis error in this approximation in terms of the spectral norm. We then define the {\em error budget per gate} for the SK algorithm, denoted by $\delta$, to be a given upper bound on the allowable circuit synthesis error:
\begin{equation}
\label{eq:delta-synth}
\Delta_{\mbox{\tiny synth}}:=\max_\ell\|U_\ell(\tau)-\tilde{U}_\ell(\tau)\|\le \delta\,.
\end{equation}
Similar to the approach in Ref.~\cite{reiher2017elucidating},
we define, for any $t\ge 0$, an {\em effective} Hamiltonian associated with the resulting quantum circuit:
\begin{align}
    \tilde{H}_{\mbox{\tiny eff}}(t)&:=i\ln \left(\left[\tilde{\mathcal{S}}_p(\tau)\right]^r\right)/t\,,\quad\mbox{where} \nonumber\\&\quad\quad
    \tilde{\mathcal{S}}_1(t):=\prod_{\ell=1}^{L} \tilde{U}_\ell(t) ,\nonumber\\&\quad\quad\tilde{\mathcal{S}}_2(t):=\left(\prod_{\ell=L}^{1} \tilde{U}_\ell(t)\right)\left(\prod_{\ell=1}^{L} \tilde{U}_\ell(t)\right).
\end{align}
The operator logarithm is well-defined, because $\left[\tilde{\mathcal{S}}_p(\tau)\right]^r$, which is a product of unitary operations, is invertible. The operator $\tilde{H}_{\mbox{\tiny eff}}(t)$ is the effective Hamiltonian associated with the {\em effective} unitary
 $\widetilde{\mathcal{U}}(t):=\exp\left(-it\tilde{H}_{\mbox{\tiny eff}}(t)\right)$ that symbolically represents
the circuit resulting from two procedures: (i) the use of a PF along with Trotterization and (ii) circuit synthesis involving especially the SK algorithm.
When we run the QPE algorithm, we use circuits effectively represented by controlled applications of the unitary $\widetilde{\mathcal{U}}(t)$, that is, the quantum circuit implementation of the QPE algorithm is designed to estimate the energy eigenvalues of the effective Hamiltonian $\tilde{H}_{\mbox{\tiny eff}}(t)$ (rather than those of $H$). Since $\widetilde{\mathcal{U}}(t)$ represents a perfect circuit consisting of gates from the standard gate set, the only additional error incurred is that associated with the accuracy of the actual phase estimation in the inference process of the eigenvalues of $\tilde{H}_{\mbox{\tiny eff}}(t)$.

We aim to bound $|E_0-E_{\mbox{\tiny eff}(t),0}|$, which is the difference between the ground-state energy $E_0$ of $H$ (which we aim to estimate) and the lowest eigenvalue of $\tilde{H}_{\mbox{\tiny eff}}(t)$ (which we actually estimate). According to Lemma 3 in the supplementary material of Ref.~\cite{reiher2017elucidating}, for any given target error bound $\epsilon$, $\|H-H_{\mbox{\tiny eff}}(t)\|\le \epsilon$ also implies $|E_0-E_{\mbox{\tiny eff}(t),0}|\le \epsilon$. Moreover, according to Lemma 4 in Ref.~\cite{reiher2017elucidating}, the assumption that $\left\| \exp\left(-itH\right)-\exp\left(-itH_{\mbox{\tiny eff}}(t)\right)\right\|\le \gamma(t)\,t$ is true for some nondecreasing continuous function $\gamma(t)$ on $[0,\infty)$ implies $\|H-H_{\mbox{\tiny eff}}(t)\|\le \gamma(t)$. We can use these implications to deduce a relation between the error in energy estimation and the errors associated with the Trotter--Suzuki approximation and the circuit synthesis as follows. Using the triangle inequality multiple times and the analytic bounds given in \Cref{eq:bound-PF-order1,eq:bound-PF-order2}, we may infer the following:

\begin{align}
&\left\|\exp\left(-itH\right)-\left[\mathcal{S}_p\left(\tau\right)\right]^r\right\|= \nonumber\\
&\quad\quad\quad=\left\|\left[\exp\left(-i(t/r)H\right)\right]^r-\left[\mathcal{S}_p\left(t/r\right)\right]^r\right\|
\nonumber\\&\quad\quad\quad\le r
\left\|\exp\left(-i(t/r)H\right)-\mathcal{S}_p\left(t/r\right)\right\|
\nonumber\\&\quad\quad\quad= r
\left\|\exp\left(-iH\tau \right)-\mathcal{S}_p\left(\tau\right)\right\|
\nonumber \\&\quad\quad\quad
=:  \Delta E_{\mbox{\scriptsize TS}[p]}(t)\,t\,,
\end{align}
where
\begin{align}
\label{eq:E_TS-order1}
\Delta E_{\mbox{\scriptsize TS}[1]}(t)
   &:=\tau\sum_{l=1}^L\sum_{j=l+1}^L C_{lj}|\gamma_l\gamma_j|\,,\\
   \Delta E_{\mbox{\scriptsize TS}[2]}(t)
   &:=\frac{\tau^2}{3}\sum_{l_1=1}^L\sum_{l_3=l_1+1}^L \sum_{l_2=l_1+1}^L C_{l_1l_2l_3}|\gamma_{l_1}\gamma_{l_2}\gamma_{l_3}|\nonumber\\&\quad +\frac{\tau^2}{8}\sum_{l_1=1}^L\sum_{l_2=l_1+1}^LC_{l_1l_2}|\gamma_{l_1}^2\gamma_{l_2}|.
   \label{eq:E_TS-order2}
\end{align}
Moreover, by repeated use of the triangle inequality, we can prove by induction that
\begin{equation}
\left\|\left[\mathcal{S}_p\left(\tau\right)\right]^r-\left[\tilde{\mathcal{S}}_p\left(\tau\right)\right]^r\right\|\le
\begin{cases}
        rL\Delta_{\mbox{\tiny synth}}\quad & \mbox{for $p=1$},
        \\
        r(2L-1)\Delta_{\mbox{\tiny synth}}\quad & \mbox{for $p=2$}.
        \end{cases}
\end{equation}
Hence, using the triangle inequality, we may infer the following:
\begin{widetext}
\begin{eqnarray}
\left\|\exp\left(-itH\right)-\left[\tilde{\mathcal{S}}_p\left(\tau\right)\right]^r\right\|
&\le & \left\|\exp\left(-itH\right)-\left[\mathcal{S}_p\left(\tau\right)\right]^r\right\|
+\left\|\left[\mathcal{S}_p\left(\tau\right)\right]^r-\left[\tilde{\mathcal{S}}_p\left(\tau\right)\right]^r\right\|\nonumber\\
&\le &
\begin{cases}
        \Delta E_{\mbox{\scriptsize TS}[1]}(t)\,t+rL\Delta_{\mbox{\tiny synth}}\quad & \mbox{for $p=1$},
        \\
       \Delta E_{\mbox{\scriptsize TS}[2]}(t)\,t+ r(2L-1)\Delta_{\mbox{\tiny synth}}\quad & \mbox{for $p=2$}.
        \end{cases}
\end{eqnarray}
\end{widetext}
Thus, according to Lemmas 3 and 4 and Theorem 1 in the supplementary material of Ref.~\cite{reiher2017elucidating}, we can conclude that the
error in the ground-state energy that results from such a simulation is at most
\begin{widetext}
\begin{equation}
|E_0-E_{\mbox{\tiny eff}(t),0}|
\le
\begin{cases}
        \Delta E_{\mbox{\scriptsize TS}[1]}(t)+rL\Delta_{\mbox{\tiny synth}}/t\quad & \mbox{for order-1 PF},
        \\
       \Delta E_{\mbox{\scriptsize TS}[2]}(t)+ r(2L-1)\Delta_{\mbox{\tiny synth}}/t\quad & \mbox{for order-2 PF}.
        \end{cases}
\end{equation}
\end{widetext}
Similar to the approach used in Ref.~\cite{reiher2017elucidating}, we define $\epsilon_1:=\Delta E_{\mbox{\scriptsize TS}[p]}$, $\epsilon_2:=rL\Delta_{\mbox{\tiny synth}}/t$ or
$\epsilon_2:=r(2L-1)\Delta_{\mbox{\tiny synth}}/t$ depending on whether we use the first-order or second-order PF, and  $\epsilon_3$ to be the error in phase estimation. For chemical significance, the total overall target error $\epsilon:=\epsilon_1+\epsilon_2+\epsilon_3$ should be at most 0.1 millihartrees, that is, our ideal overall error budget is $\epsilon=10^{-4}$ hartrees.
The split of the total error budget into three parts is non-trivial; in our QRE analysis, we have treated $\epsilon_1,\,\epsilon_2$, and $\epsilon_3$  as parameters and optimized the error budget allocation to these three parts so as to minimize the expected $T$-gate count.

Finally, to determine an appropriate evolution time for the unitary $ \exp\left(-itH\right)$, we require that the phase that we estimate using QPE, $\theta:= E_0t$ (where $E_0$ is the ground-state energy), is within $[0,2\pi]$. Since we do not have knowledge of the eigenvalues of $H$, we require that $\|H\|t\le 2\pi$. As we do not have knowledge of the spectral norm of the Hamiltonian (which is equal to the largest eigenvalue) either, we use $\Gamma:= \sum_l |\gamma_l|\ge \|H\|$ and choose
   $t= 2\pi/\Gamma\le  2\pi/\|H\|$. This choice of $t$
   implies that, when using the first-order Lie--Trotter formula, the number of Trotter slices is given by
\begin{equation}
r=\max\left\{1, \left\lceil\frac{\pi\sum_{\ell=1}^L\sum_{j=\ell+1}^L C_{\ell j}|\gamma_\ell\gamma_j|}{\epsilon_1\sum_{\ell=1}^L |\gamma_{\ell=1}|}\right\rceil\right\},
\end{equation}
which directly follows from \Cref{eq:E_TS-order1}. A similar expression for $r$ can be derived when using the second-order Trotter--Suzuki formula by using \Cref{eq:E_TS-order2}. The parameters $r$ and $L$ determine the size of the logical quantum circuits. From the knowledge of  $r$ and $L$, we can also infer the error budget per gate for the SK algorithm as defined in \Cref{eq:delta-synth}, that is,
\begin{equation}
\Delta_{\mbox{\tiny synth}}\le\delta:=
\begin{cases}
        2\pi\epsilon_2 /(rL\Gamma)\quad & \mbox{for order-1 PF},
        \\
        2\pi\epsilon_2 /[r(2L-1)\Gamma]\quad & \mbox{for order-2 PF}.
        \end{cases}
\end{equation}

Estimations of Trotter errors via rigorous analytic upper bounds can be loose, which can result in overestimating the number of Trotter slices by many orders of magnitude. For this reason, several recent studies instead have attempted to predict the number of Trotter slices practically required using various heuristics, such as those based on Monte Carlo sampling. Following this trend, we have conducted an additional empirical QRE analysis based on more-realistic Trotter numbers that we inferred through extrapolation. More concretely, we empirically computed the Trotter error $\left\|\exp\left(-itH\right)-\left[\mathcal{S}_2\left(\tau\right)\right]^r\right\|$
via full numerical computations for $p$-benzyne Hamiltonians pertaining to small active spaces HL$\pm n$ for $n=0,\,1,\,2$. We inferred the corresponding required evolution time $\tau$ for a single Trotter slice 
and the associated Trotter number $\beta:=1/\tau$ to satisfy an error budget associated with a constant accuracy in the energy estimation. Based on the obtained data, we then inferred the approximate scaling of $\beta$ as a function of the number of active spin orbitals. We found the inferred scaling to be consistent with the results of a prior empirical study based on Monte Carlo sampling~\cite{poulin2015trotter}. Based on the deduced scaling, we have estimated $\beta$ values for larger active spaces via regression. 

\subsection{Propagation of errors in qubitization}
\label{sec:error-analysis-qubitization}

In this section, we discuss the various errors incurred in the process of generating the logical quantum circuits for the double-factorized (DF) qubitization algorithm, and how we can guarantee electronic-structure quantum computations to estimate the Hamiltonian eigenvalues to 
a given target accuracy by satisfying certain error bounds within predetermined error budgets. In QPE based on qubitization, the spectrum of the molecular Hamiltonian $H$ is sampled by measuring the phase that is accumulated on an initial input quantum state through multiple controlled executions of the quantum walk operator $e^{i\arccos(H/\lambda)}$ (or $e^{i\arcsin(H/\lambda)}$) in place of the time-evolution
operator $\exp(-iHt)$, see Refs.~\cite{berry2018improved,poulin2018quantum}. In the DF qubitization algorithm, there are four main sources of error (see Refs.~\cite{vonBurg2021quantum,lee2021even}): 
\begin{enumerate}
     \item Error associated with the truncation in the double-factorization procedure;
     \item Error associated with the binary approximation of the Hamiltonian coefficients when implementing the \textsc{prepare} oracle;
    \item Error associated with the approximation of the individual Givens rotations in implementing the basis rotations; and
    \item Error associated with the measurement readout in phase estimation.
\end{enumerate}
The first source of error arises from truncating the small eigenvalues of the double-factorized Hamiltonian, but by keeping track of the values of the truncated eigenvalues it is possible to bound the 2-norm of the difference between the original Hamiltonian and the truncated Hamiltonian. 
The second and third sources of error arise from approximations in the implementation of the LCU oracles associated with the qubitization operator,  
namely, binary approximations of the coefficients involved in the \textsc{prepare} operation as well as binary approximations of the rotation angles in the diagonalization operations involved in the innermost decomposition of the double-factorized Hamiltonian (with finite bits of precision). 
The final contribution to the error is the imprecision in the measurement readout of QPE; this error and the requisite number of logical ancillary qubits can be bounded using the total number of repetitions of the qubitization operator. For chemical significance, the total overall target error in estimating the ground-state energy should be at most 1.6 millihartrees, that is, the overall error budget for estimating the lowest eigenvalue of the molecular Hamiltonian is $\epsilon=1.6$ mHa (or the much lower target error $\epsilon = 0.1\text{ mHa}$ for quantitative accuracy).

At the logical level, it is possible to break down the error in QPE of the qubitization approach as in \mbox{Equation (23)} of Ref.~\cite{babbush2018encoding}, namely, the output energy of the QPE is within 
\begin{equation}
    \Delta E \leq \lambda \sqrt{ \left(\frac{\pi}{2^m}\right)^2 + (\epsilon_H + \pi \epsilon_{\text{\tiny QFT}})^2}\label{eq:QPE-error}
\end{equation}
of the Hamiltonian used in the qubitization approach. Here, $\lambda$ is the 1-norm of the Hamiltonian (computed as the sum of the absolute values of the Hamiltonian weightings), $m$ is the number of bits in the QPE, $\epsilon_{\text{\tiny QFT}}$ is related to the error in implementing the quantum Fourier transform and is usually negligible, and $\epsilon_H$ is the systematic error in the phase resulting from the gate synthesis in implementing the qubitization operator of the double-factorized Hamiltonian.  
According to \Cref{eq:QPE-error}, if we aim 
to estimate spectra to within error $\Delta E$, we can achieve this by choosing (see \cite{babbush2018encoding}):
\begin{equation}
m=\left\lceil\log_2 \left(\frac{\pi \lambda}{\sqrt{2}\Delta E}\right)\right\rceil,\;
\epsilon_H \le \frac{\sqrt{2}\Delta E}{4\lambda},\; \epsilon_{\text{\tiny QFT}} \le \frac{\sqrt{2}\Delta E}{4\pi \lambda}.\label{eq:QPE-Error-Bounds}
\end{equation}

Following the analysis outlined in the Appendix of  Ref.~\cite{babbush2018encoding}, we can bound the error contributions  arising from finite-bit precision approximations of the quantum circuit encoding of the double-factorized Hamiltonian. Let $H=\sum_{\ell=0}^{L-1}w_\ell H_\ell$ denote the 
double-factorized Hamiltonian (after truncation of the terms pertaining to small
eigenvalues), and let $\widetilde{H}=\sum_{\ell=0}^{L-1}\widetilde{w}_\ell \widetilde{H}_\ell$ denote the corresponding approximate encoding of $H$ resulting from finite-bit precision approximations of 
Hamiltonian coefficients and Givens rotations. Then, the  associated propagated error into the eigenphase obeys: 
\begin{align}
    \epsilon_H &\leq \| e^{i \arccos(H/\lambda)} - e^{i \arccos(\widetilde{H}/\lambda)} \| \nonumber\\
    &\leq \| \arccos(H/\lambda) - \arccos(\widetilde{H}/\lambda) \| \nonumber\\
    &\leq \sum_{p = 0}^\infty \frac{(2p-1)!!}{\lambda^{2p+1}(2p+1)(2p)!!}\| H^{2p+1} - \widetilde{H}^{2p+1} \|.
\end{align}
Let $\chi:= \| H - \widetilde{H}\|$. Then the term $\| H^{2p+1} - \widetilde{H}^{2p+1} \|$ can be bounded from above by $(2p+1)(\| H \|+\chi)^{2p}\chi$, which yields
\begin{align}
    \epsilon_H &\leq \sum_{p = 0}^\infty \frac{(2p-1)!!}{\lambda^{2p+1}(2p)!!} (\| H \|+\chi)^{2p}\chi\nonumber\\
    &\le \frac{\chi}{\lambda} \left[ 1 - \left(\frac{ \| H \| + \chi }{\lambda}\right)^2 \,\right]^{-1/2}.
\end{align} 
Thus, in order for the error contributions arising from finite-bit precision approximations in implementing the qubitization quantum circuit to satisfy the bound specified in \Cref{eq:QPE-Error-Bounds}, we must choose the bit precisions to be large enough such that 
\begin{eqnarray}
    \chi 
    &\leq \frac{\sqrt{2} \Delta E}{4 \left( 1 + \frac{\Delta E^2}{8\lambda^2}\right)}\left( 1 - \|H\|^2/\lambda^2\right).
\end{eqnarray}
This inequality is used to determine the number of bits required for the binary approximations of the Hamiltonian coefficients $w_\ell$ and the binary approximations of the rotation angles associated with the diagonalizing fermionic basis transformations that yield the double-factorized form.  

\section{Quantum resource estimation\\ using TopQAD}
\label{sec:resource_estimation}

Once a target logical quantum circuit has been generated, we construct a fault-tolerant architecture that can implement this circuit to conduct a QRE analysis using the TopQAD toolkit~\cite{1qbit2024topqad}. The software's approach to creating a fault-tolerant architecture is described in Ref.~\cite{silva2024optimizing}. For a given quantum circuit and success probability, we generate an architecture that would feasibly run the computation at the requisite precision. This architecture allows us to estimate the resources required for a specific quantum circuit using hardware that can implement a rotated surface code layout. By abstracting the hardware away, we are able to focus on the layout, and from there construct an architecture that can be used to implement those operations required for FTQC.

\subsection{The compilation process}
\label{sec:topqad-compiler}

The main idea behind this construction is to transform the given circuit into an optimized sequence of $\pi/8$ Pauli rotations, and then to process these rotations using multi-qubit lattice surgery to connect distant qubits~\citep{litinski2019game, beverland2022assessing, silva2024lattice}. These $\pi/8$ Pauli rotations can then be implemented on the underlying architecture, where magic states are distilled and consumed through specific applications of lattice surgery.  This procedure results in a nontrivial simultaneity condition, as the bus qubits used in the lattice surgeries can only be used for a single rotation at a given point in time. Additionally, there are several conditions for how the bus qubits can interact with qubits storing data for the circuit, leading to complications in the underlying architecture. However, once these conditions have been taken into account, various classical scheduling processes can be used to generate the necessary schedule of operations that we then implement via lattice surgery.

The pipeline followed to generate the QREs for a given quantum circuit and a target success probability starts by transforming the circuit into one in which only Clifford and $T$ gates are used. This requires some algorithm to decompose an arbitrary gate into known elements. The most well-known of such procedures is an implementation of the SK theorem, which, while efficient in a complexity theoretic sense, is actually quite costly in practice. A different procedure with slightly less applicability is the RS algorithm, which results in significantly shorter circuits. Additionally, such implementations quickly become a bottleneck in terms of the reachable error rates, as the per-gate error budgets for the SK algorithm quickly approach machine precision. 

Given that the architecture considered requires that quantum operations are represented by Pauli rotations, the next step is to {\em transpile} the circuit consisting of Clifford and $T$ gates into a circuit consisting of Pauli rotations. The circuit described in the Clifford+$T$ gate set is first converted to a sequence of $\pi/4$ (Clifford) and $\pi/8$ (non-Clifford) Pauli rotations according to the conversion rules described in Ref.~\cite{litinski2019game}. After conversion, a procedure is run to remove the Clifford operations from the circuit using commutation rules, leaving only $\pi/8$ rotations.  This procedure can be run efficiently using the symplectic representation of Clifford gates~\cite{silva2024lattice}. Additionally, since commutable $\pi/8$ rotations can be reordered such that adjacent $\pi/8$ rotations with same axis of rotation can be combined, this allows us to perform operations corresponding to multiple rotation commutations in a single step effectively reducing the $T$ count. As discussed in Ref.~\cite{silva2024lattice}, this transpilation procedure drastically decreases the overall running time of the circuit even if the resulting circuit makes the operations less parallelizable due to its increased density.

\subsection{The assembly process}
\label{sec:topqad-assembly}

At this point, we have constructed a logic circuit tailored to an implementation on a surface code encoding logical qubits. The next step in the pipeline to generate the QREs is the assemble of the structures required for FTQC that will allow the scheduling of the $\pi/8$ Pauli rotations in the circuit. As illustrated in \Cref{fig:full_layout}, the architecture considered for the scheduling of the logical operations features a core processor, comprising a memory fabric with two-tile two-qubit patches of data qubits and an auto-correcting buffer, which is connected to the MSF using bus qubits, mirroring the configuration used in Ref.~\cite{litinski2019game}.

Central to our approach is the utilization of a multi-level MSF for magic state distillation, where the fidelity of magic states undergoes iterative enhancement across successive distillation levels. Low-fidelity magic states are created from operations on physical qubits at magic state preparation units following a magic state preparation protocol \cite{Gidney2023cleaner, Gidney2024magic} as described in~\Cref{sec:other-FTQC-protocols}. Then, at each distillation level, the MSF used the lower-fidelity magic states to create higher-fidelity magic states that are dispatched to a dedicated area where magic states can be enlarged to the required code distance that interfaces with the next round of distillation. A 15:1 distillation protocol is assumed to be used by the distillation units at all levels due to its capacity to improve magic state fidelity in $O(P_T^3)$, where $P_T$ is the logical error rate for input magic states~\cite{beverland2022assessing}. The uppermost level of the MSF connects to the memory fabric via a buffer space that allows magic states to be temporarily stored before being consumed within the memory fabric. This space is designed to incorporate auto-correcting buffers, named for their capability to execute corrective measures concurrently with magic state consumption, notably enabling the auto-correcting of $\pi/8$ operations. The auto-correcting buffers and the memory fabric comprise the core processor of the device.

The scheduling methodology employed in the studied topological architecture is presented in Ref.~\cite{silva2024lattice} and addresses the sequencing of operations and the allocation of logical resources required for establishing connections between distant qubits in the core processor as needed. Due to the reduced parallelization potential of the $\pi/8$ operations, we assume a serial scheduling is employed in the QREs presented here. If nonrestrictive availability of magic states is ensured, and considering that the expected time to execute a $\pi/8$ rotation is equal to one logical cycle because of the auto-correcting buffers, the minimum number of logical cycles required to execute the entire circuit in a serial scheduling is equal to its $T$ count, ignoring the warm-up time. Each logical cycle requires performing $d_{\text{core}}$ parity checks, where $d_{\text{core}}$ is the code distance of the logical qubits in the core processor, each taking a time $T_M + 4T_2 + 2T_1 + T_M + t_R$, considering the measurement time $t_M$, the reset time $t_R$, the single-qubit gate time $t_1$, and the two-qubit gate time $t_2$. Therefore, while it is easy to generate time estimates for circuits scheduled in serial, generating the space estimates require creating an MSF with enough distillation units capable of distilling magic states quickly enough to keep the core processor constantly busy.

The assembler of the quantum architecture described requires minimizing the space (i.e., the physical qubits required) under a given error budget. The decisions to be made are related to sizing the components of the architecture, i.e., the core processor and the MSF, while ensuring fault tolerance. The given error budget is distributed between errors that arise in the execution of quantum operations in the core processor, $E_{\text{core}}$, and in the distillation of magic states in the MSF, $E_{\text{MSF}}$. Therefore, 
\begin{equation}\label{eq:total_error_budget}
    E_{\text{core}} + E_{\text{msf}} \leq E.
\end{equation}
The errors of the core processor and the MSF are modelled and predicted following the pipeline described in Ref.~\cite{silva2024optimizing}. Since we provide QREs for the case with a never idling core processor, the accumulated errors in the core are only resulting from the Clifford operations required for the multi-qubit lattice surgeries performed and the protection of the idling data qubits while lattice surgeries involving other data qubits are occurring in the core. The accumulated errors in the core processor is approximated as
\begin{equation}\label{eq:E_core}
    E_{\text{core}} \approx (2Q + \sqrt{8Q} + 29) T e_{\text{mem},\text{core}},
\end{equation}
which assumes that all $(2Q + \sqrt{8Q} + 29)$ logical qubits in the core processor (approximated size of the memory fabric and buffer) following the design presented in Figure \ref{fig:full_layout}, where $Q$ is the number of data qubits in the circuit, are susceptible to result in an error with probability $e_{\text{mem},\text{core}}$ during all the $T$ logical cycles required to run the circuit. The error rate $e_{\text{mem},\text{core}}$ is derived from emulations of the FTQC protocol for quantum memory. These emulations establish the correlation between code distance and logical error rates based on a given choice of physical parameters, resulting in a predictive model obtained by regression from numerical simulations at low code distances, using efficient stabilizer circuit simulators~\cite{gidney2021stim} following the descriptions in \Cref{sec:hardware-noise-modeling}. In the MSF, the error rate of output magic states is resulting from the preparation, distillation and expansion procedures. Following Ref.~\cite{silva2024optimizing}, considering $e_{\text{prep}}$ as the error rate for the magic states prepared from physical qubits and that the Clifford and growth accumulated errors can be approximated to the memory errors, i.e., $e_{\text{cliff}} = e_{\text{grow}} = e_{\text{mem}}$, the magic state error rates of the entire MSF using 15:1 distillation units can be calculated recursively as follows:
\begin{itemize}
    \item Input to level 1: $e_{\text{in},1} = e_{\text{prep}}$;
    \item Output from level $l$ for all $l \in \{1,\ldots, L\}$:  $$e_{\text{out},l} = 35e_{\text{in},l}^3 + 7.1e_{\text{mem},l};$$
    \item Input to level $l+1$ for all $l \in \{1,\ldots, L\}$: $$e_{\text{in},l+1} = 1 - (1-e_{\text{out},l})(1-e_{\text{mem},l}).$$
\end{itemize}
Therefore, the error rate of the magic state input to the core processor is $e_{\text{core}} = e_{\text{in},L+1}$ for an MSF with $L$ distillation levels. Given that $T$ magic states needs to be distilled for the entire execution of the quantum program, we have
\begin{equation}\label{eq:E_msf}
    E_{\text{msf}} = e_{\text{core}}T.
\end{equation}
The choice of hardware parameters to determine the required number of distillation levels $L$ and code distances $d_l, \forall l \in {1, \ldots, L+1}$, which includes the core processor as $l = L+1$, is such that it must reach the target logical error rates based on~\Crefrange{eq:total_error_budget}{eq:E_msf}. The process followed to make these decisions is described in Ref.~\cite{silva2024optimizing}. In summary, the core processor's code distance $d_{L+1}$ is minimized, assuming $E_{\text{msf}} = 0$. Then, it sets the first level code distance $d_1$ considering the residual error budget left after the core level logical encoding is decided, that is, $E_{\text{msf}} \leq E - E_{\text{core}}$. Next, it determines the number of distillation levels $L$ required to meet the magic state error rate requirement $e_{\text{core}}$ derived from \Cref{eq:E_msf} for the residual error budget. Finally, if $L > 1$, it calculates the code distances $d_l$ for all levels that can meet the error budget using the minimum number of physical qubits across the whole architecture. Once these decisions have been made, the assembler determines the number of distillation units required for a steady flow of magic states to the core processor such that the distillation rate of magic states output from the MSF matches the consumption rate of magic states in the core processor.

While it is possible for each distillation level to contain only a single distillation unit, such a configuration introduces significant idling time, thereby prolonging the expected runtime of executing quantum circuits in our proposed architecture. In addition, although having fewer units implies that there are fewer logical qubits, this solution potentially increases physical space requirements due to the larger code distances resulting from the additional overhead incurred from logical operations executed on data qubits to mitigate decoherence during idling time~\cite{silva2024optimizing}.

\subsection{Comparison with the AzureQRE toolkit}
\label{sec:azure_qre}

\begin{table*}[t!]
    \centering
    {\scriptsize
    \begin{tabular*}{\textwidth}{|c|c|c|@{\extracolsep{\fill}}ccc@{\extracolsep{\fill}}ccc@{\extracolsep{\fill}}ccc}
        \hline\hline
       \multicolumn{3}{c}{}&&&&&&&&\\
       \multicolumn{3}{c}{} & \multicolumn{3}{c}{\bf Baseline Parameter Set } & \multicolumn{3}{c}{\bf Target Parameter Set } & \multicolumn{3}{c}{\bf  Desired Parameter Set } \\
        \cmidrule{4-6} \cmidrule{7-9} \cmidrule{10-12}
       \multicolumn{1}{c}{} & \multicolumn{1}{c}{\multirow{2}{*}{\shortstack[c]{Target \\ error $\epsilon$}}} & \multicolumn{1}{c}{\multirow{2}{*}{\shortstack[c]{$N_{\text{orb}}$}}} & \multirow{2}{*}{\shortstack[c]{\# Phys.\\ qubits}} & \multirow{2}{*}{\shortstack[c]{Phys. \\ time}}& \multirow{2}{*}{\shortstack[c]{QEC code\\ distances}} &  \multirow{2}{*}{\shortstack[c]{\# Phys.\\ qubits}} & \multirow{2}{*}{\shortstack[c]{Phys. \\ time}} & \multirow{2}{*}{\shortstack[c]{QEC code\\ distances}} & \multirow{2}{*}{\shortstack[c]{\# Phys.\\ qubits}} & \multirow{2}{*}{\shortstack[c]{Phys. \\ time}} & \multirow{2}{*}{\shortstack[c]{QEC code\\ distances}}\\
        \multicolumn{3}{c}{}&&&&&&&&\\
        \hline\hline
          \parbox[t]{4mm}{\multirow{11}{*}{\rotatebox[origin=c]{90}{{AzureQRE}}}}&
          &&&&&&&&&\\ 
        & \multirow{4}{*}{\rotatebox[origin=c]{90}{$1.6$ mHa}} & 6 &  - & - & - & 1.1$\times$10\textsuperscript{7} & 17.4 minutes & 75 & 3.1$\times$10\textsuperscript{6} & 10.0 minutes & 43 \\
        & & 18 & - & - & - & 2.3$\times$10\textsuperscript{7} & 16.2 hours & 89 & 7.6$\times$10\textsuperscript{6} & 9.3 hours & 51 \\
        & & 26 & - & - & - & 3.1$\times$10\textsuperscript{7} & 2.8 days & 93 & 9.9$\times$10\textsuperscript{6} & 1.6 days & 53 \\
        & & 76 & - & - & - & 9.7$\times$10\textsuperscript{7} & 286.3 days & 109 & 3.0$\times$10\textsuperscript{7} & 160.2 days & 61 \\
         &&&&&&&&&\\
        \cline{2-12}
         &&&&&&&&&\\
        & \multirow{4}{*}{\rotatebox[origin=c]{90}{$0.1$ mHa}} & 6 &  - & - & - & 1.6$\times$10\textsuperscript{7} & 6.1 hours & 85 & 5.0$\times$10\textsuperscript{6} & 3.4 hours & 47 \\
        & & 18 & - & - & - & 3.0$\times$10\textsuperscript{7} & 15.0 days & 97 & 9.4$\times$10\textsuperscript{6} & 8.5 days & 55 \\
        & & 26 & - & - & - & 4.1$\times$10\textsuperscript{7} & 62.1 days & 103 & 1.3$\times$10\textsuperscript{7} & 34.4 days & 57 \\
        & & 76 & - & - & - & 1.3$\times$10\textsuperscript{8} & 21.0 years & 119 & 4.0$\times$10\textsuperscript{7} & 11.8 years & 67 \\
         &&&&&&&&&\\
        \hline
        \parbox[t]{4mm}{\multirow{11}{*}{\rotatebox[origin=c]{90}{{TopQAD}}}}&
          &&&&&&&&&\\
        & \multirow{4}{*}{\rotatebox[origin=c]{90}{$1.6$ mHa}} & 6 & 1.4$\times$10\textsuperscript{7} & 3.3 minutes & 25, 59 | 73 & 1.3$\times$10\textsuperscript{6} & 1.2 minutes & 23 | 27 & 8.2$\times$10\textsuperscript{5} & 35.8 seconds & 17 | 21 \\
        & & 18 & 4.8$\times$10\textsuperscript{7} & 1.4 days & 15, 31, 75 | 93 & 3.6$\times$10\textsuperscript{6} & 7.3 hours & 11, 27 | 33 & 1.7$\times$10\textsuperscript{6} & 5.6 hours & 21 | 25 \\
        & & 26 & 5.5$\times$10\textsuperscript{7} & 5.4 days & 15, 33, 79 | 95 & 4.7$\times$10\textsuperscript{6} & 1.3 days & 11, 29 | 35 & 2.6$\times$10\textsuperscript{6} & 23.6 hours & 23 | 27 \\
        & & 76 & 1.2$\times$10\textsuperscript{8} & 1.4 years & 17, 37, 89 | 107 & 1.3$\times$10\textsuperscript{7} & 121.8 days & 13, 33 | 39 & 7.7$\times$10\textsuperscript{6} & 96.8 days & 25 | 31 \\
          &&&&&&&&&\\ 
          \cline{2-12}
         &&&&&&&&&\\
        & \multirow{4}{*}{\rotatebox[origin=c]{90}{$0.1$ mHa}} & 6 & 2.3$\times$10\textsuperscript{7} & 12.6 hours & 29, 77 | 91 & 3.0$\times$10\textsuperscript{6} & 2.7 hours & 11, 27 | 31 & 1.4$\times$10\textsuperscript{6} & 2.2 hours & 21 | 25 \\
        & & 18 & 5.8$\times$10\textsuperscript{7} & 30.8 days & 15, 35, 91 | 105 & 5.1$\times$10\textsuperscript{6} & 6.5 days & 13, 31 | 35 & 2.6$\times$10\textsuperscript{6} & 5.4 days & 23 | 29 \\
        & & 26 & 7.9$\times$10\textsuperscript{7} & 132.1 days & 19, 35, 85 | 115 & 6.8$\times$10\textsuperscript{6} & 28.5 days & 13, 31 | 39 & 3.4$\times$10\textsuperscript{6} & 21.2 days & 25 | 29 \\
        & & 76 & 1.5$\times$10\textsuperscript{8} & 28.5 years & 17, 41, 99 | 119 & 1.8$\times$10\textsuperscript{7} & 6.5 years & 15, 35 | 43 & 9.8$\times$10\textsuperscript{6} & 5.0 years & 29 | 33 \\
         &&&&&&&&&\\
        \hline\hline
    \end{tabular*}
    }
    \caption{Physical resource estimates generated by TopQAD~\cite{1qbit2024topqad} and AzureQRE~\cite{microsoft2022azureQRE} for implementing 
    the QPE algorithm on electronic-structure quantum circuits associated with the \mbox{$p$-benzyne} and FeMoco molecules, for two precisions in energy estimation: qualitatively accurate computation within a target error \mbox{1.6 mHa}, and quantitatively accurate computation within a target error \mbox{0.1 mHa}, respectively, using   
    a circuit-level error budget of 0.01 using the double-factorized qubitization algorithm. 
    We report estimates for the physical wall-clock time and the number of physical qubits required for fault-tolerant implementations of the QPE algorithm for electronic spectra associated with various molecular active spaces with sizes specified by the number of orbitals $N_{\text{orb}}$. The data for $N_{\text{orb}}=6, 18, 26$ correspond to active space selections HL$\pm 2, 8, 12$ (using HL$\pm n$ to denote ``HOMO$-n$ and LUMO$+n$''; see  \Cref{sec:logical_circuit} for an explanation of these terms) for \mbox{$p$-benzyne} using the \texttt{6-31G} basis to represent the fermionic orbitals; the data for $N_{\text{orb}} =76$ pertains to the active-space model for FeMoco proposed in Ref.~\cite{li2019the}. In addition, we also report the QEC code distances that are required for running the corresponding circuits fault-tolerantly. 
    Here, physical resources are reported only for the quantum circuits based on running the  DF qubitization algorithm.
    The associated resource requirements are reported for three hardware specifications, namely, baseline, target, and desired  hardware, as summarized in \Cref{tab:physical_params}. Note that the symbol ``-'' represents that AzureQRE estimates that the baseline parameter set is above the QEC threshold. }
    \label{tab:resource_estimates_physical_azure}
\end{table*}

To provide additional logical and physical resource estimates, we use the Azure Quantum Resource Estimator (AzureQRE)~\cite{van2023using,microsoft2022azureQRE}. We used the surface code option within AzureQRE and the hardware parameters of \Cref{tab:physical_params} and estimate the resources required only for the double-factorized qubitization algorithm. We refer the reader to Refs.~\cite{van2023using} and~\cite{beverland2022assessing} for full details about the architectural assumptions, but briefly highlight AzureQRE assumes a 2D nearest-neighbor layout which has the ability to perform parallel operations and utilizes 15-to-1 magic state distillation. We show the results for physical resource estimates in \Cref{tab:resource_estimates_physical_azure}, where we have also included the results from TopQAD's resource estimates for the DF qubitization algorithm in \Cref{tab:resource_estimates_physical_trotter} for easy comparison. Compared with the estimates based on TopQAD presented in the main text (see \Cref{sec:resource-estimates-pbenzyne}), AzureQRE finds that the baseline parameter set is not below the surface code threshold, specifically because of the measurement error rate. The TopQAD suite's QRE pipeline includes more-advanced FTQC protocols, including magic state factories and space--time trade-offs~\cite{silva2024optimizing} compared with those implemented in AzureQRE. For the target and desired hardware parameter sets, TopQAD's estimates are about an order of magnitude lower in terms of the number of physical qubits and by around a factor of 3 lower in terms of the runtime than AzureQRE, due to its use of more advanced FTQC protocols.

\section{Double-factorized quantum chemistry}
\label{sec:Double-factorized_quantum-chemistry}

The standard quantum chemistry Hamiltonian is
\begin{equation}\label{eq:ham_qchem}
    H = \sum_{ij,\sigma} h_{ij} a^\dagger_{i\sigma} a _{i\sigma} 
 + \frac{1}{2}\sum_{ijkl,\sigma\rho} h_{ijkl} a^\dagger_{i\sigma} a^\dagger_{j\rho}a_{k\rho} a_{l\sigma},
\end{equation}
where $h_{ij}$ and $h_{ijkl}$ are the one- and two-electron integrals, $\sigma$ and $\rho$ index spin, and $a_{p\sigma}$ are fermionic creation and annihilation operators. To implement the time evolution of this Hamiltonian on a quantum computer, double-factorization can be used as a resource-efficient alternative to Trotterization~\cite{von2021quantum}.

The fourth-order Coulomb tensor $ h_{ijkl} $ can be written as a $ N_o^2/4 \times N_o^2/4 $ electronic repulsion integral (ERI) matrix, $A$, where $N_o$ is the number of \textit{spin orbitals}. $A$ is positive semi-definite and generally has a rank $ L = \mathcal{O}(N) $ for chemical systems. We can diagonalize $A$, leading to a decomposition in terms of an auxiliary tensor $\mathcal{L}$ such that~\cite{Motta2021}:
\begin{eqnarray}
A = \sum_{\ell=0}^{L-1} (\mathcal{L}^{(\ell)})^2 = \sum_{\ell=0}^{L-1} \sum_{ijkl=0}^{N/2-1} \mathcal{L}_{ik}^{(\ell)} \mathcal{L}_{jl}^{(\ell)} a^\dagger_{i} a_{k}a^\dagger_{j} a_{l} .
\end{eqnarray}
Each matrix $\mathcal{L}^{(\ell)}$ can then be be further decomposed, giving a set of eigenvalues $\{\lambda^{(\ell)}_m \}$ and a diagonalizing unitary $U^{(\ell)} $. This leads to the double-factorized form of the Hamiltonian $H_{\text{DF}}$:

\begin{eqnarray}
    \label{eq:df-ham}
H_{\text{DF}} &=& \sum_{ij,\sigma} \tilde{h}_{ij} a^\dagger_{i\sigma} a_{j\sigma}+\nonumber\\ &&+\frac{1}{2} \sum_{\ell=0}^{L-1} \left(\sum_{ij, \sigma} \sum_m \lambda_m^{(\ell)}  U_{m,i}^{(\ell)} U_{m,j}^{(\ell)} a^\dagger_{i\sigma} a_{j\sigma}\right)^2, \nonumber\\
\end{eqnarray}
where $\tilde{h}_{ij} \equiv h_{ij} - \frac{1}{2}\sum_{l}h_{illj}$ comes from the reordering of the creation and annihilation operators. By truncating some of the eigenvalues, a low--rank approximation can be obtained. There exist efficient walk operators, $W$, which implement this Hamiltonian as a quantum circuit, as described in Ref.~\cite{von2021quantum}.

\section{Runtime of classical algorithms for quantum chemistry}
\label{sec:classical_re}
To provide realistic estimates of the classical resources required for various classical quantum chemistry algorithms, we extrapolate the results of recent publications that use full configuration interaction (FCI)~\cite{gao2024distributed} and density matrix renormalization group (DMRG)~\cite{bellonzi2024feasibility}. For the FCI calculations, we note that Ref.~\cite{gao2024distributed} reports running their largest system of C$_3$H$_8$ in an \texttt{STO-3G} basis, which has 26 electrons in 23 orbitals, a calculation involving 1.3 trillion determinants, took 113.6 hours using 512 processes, a total of around 58k CPU hours. Assuming quadratic scaling with number of determinants ($\mathcal{O}(N_{\text{det}}^2)$) scaling for the FCI algorithm, we use this single data point to compute a realistic prefactor for the computational time scaling. Note that the number of determinants scales exponentially with number of orbitals. We then take the worst-case number of determinants for each number of orbitals, where the number of electrons ($N_e$) is equal to the number of orbitals ($N_o$), and calculate the total number of determinants as $N_{\text{det}} = \big(N_o!/(N_o-N_e)!N_e!\big)^2$ and, assuming a factor of 1000 in parallelism, compute the time for various numbers of orbitals, consistent with the 512 CPUs used in Ref.~\cite{gao2024distributed}.

For the DMRG calculations, we assume cubic scaling with bond dimension ($\mathcal{O}(\chi^3)$). Note that there is no definitive scaling of bond dimension $\chi$ with number of orbitals for generic quantum chemistry problems, but it is generally expected to scale exponentially for strongly correlated systems. Estimates of the necessary bond dimension for various homogeneous catalysts are reported in Ref.~\cite{bellonzi2024feasibility}, as well as runtimes for smaller bond dimension DMRG calculations. Using the data in Table 3 of Ref.~\cite{bellonzi2024feasibility}, specifically the data which was run on a computer cluster, we fit parameters $a$ and $b$ in the scaling function $f(\chi) = a \chi^3 + b$ and then use those coefficients to predict the runtime necessary for the reported bond dimensions necessary to reach chemical accuracy, assuming a factor of 100 parallelism, consistent with the 40 CPUs used in the Ref.~\cite{bellonzi2024feasibility}.

\section{Circuit-level noise model}
\label{sec:Circuit-Level-Noise-Model}

We employ a circuit-level depolarizing noise model for the benchmarking and sensitivity analysis simulations in~\Cref{sec:hardware-noise-modeling}.  It consists of three types of errors: gate errors, idling errors, and state preparation and measurement (SPAM) errors, where the strength of each type is determined from the values of hardware noise parameters, such as those provided  in \Cref{tab:physical_params}.

Imperfect gates are modeled by adding a depolarizing noise channel at rate $p$ to the gate qubits after the application of each gate. For one-qubit gates, the noise channel randomly applies one of $X$, $Y$, or $Z$, each with probability $p/3$. Similarly, for two-qubit gates, the noise channel applies one of the 15 two-qubit non-identity Pauli gates, each with probability $p/15$. The rate $p$ is determined by utilizing the formula for the depolarizing channel's average gate fidelity,
\begin{equation}
    F_{\text{dep}, n} = 1- \frac{(2^n-1)2^n}{2^{2n}-1} p,
\end{equation}
where $n$ is the number of gate qubits. 

Idle qubits are susceptible to errors dependent on both the decoherence time $T_1$ of the qubit and the time $t$ it takes to apply the gate(s) to the active qubits. The error is modeled with a single-qubit depolarizing noise channel with rate equal to
\begin{equation}
    p = \frac{3}{4}\left[1-\exp\left(-\frac{t}{T_1}\right)\right].
\end{equation}

The errors in state preparation and reset are captured by assuming that with rate $p$ the orthogonal state is produced, that is, $|0\rangle$ is prepared instead of $|1\rangle$ and vice versa. Similarly, measurements in the $Z$ basis are flipped at rate $p$. In all three cases the fidelity of the operation is
\begin{equation}
    F_{\textsc{spam}} = \frac{P(0|0) + P(1|1)}{2},
\end{equation}
from which we directly determine the rate $p = 1 - F_{\textsc{spam}}$.

\section{High-performance real-time control and decoding with NVQlink}
\label{sec:dgx}

Nvidia NVQlink~\cite{dgx_quantum, NVQlink2025}, developed in collaboration with Quantum Machines, is a groundbreaking architecture that tightly integrates advanced classical and quantum computational capabilities. It combines Nvidia Grace Hopper superchips with the Quantum Machines OPX1000 control system, achieving microsecond-scale latency between GPU and QPUs. This connection enables latency-critical work like QEC to be accelerated with GPUs. The system is designed to scale with both quantum and classical computing demands; as quantum computers advance, additional NVQlink nodes can be incorporated to interface with more QPUs. Moreover, these nodes can be connected by both classical and quantum interconnects to scale up to large accelerated quantum supercomputing systems, as shown in Figure \ref{Figure_HPCquantum}. NVQlink is also QPU agnostic, enabling all qubit modalities to be integrated within the architecture.  

\begin{figure}[h]

\centering
\includegraphics[width=0.71\linewidth]{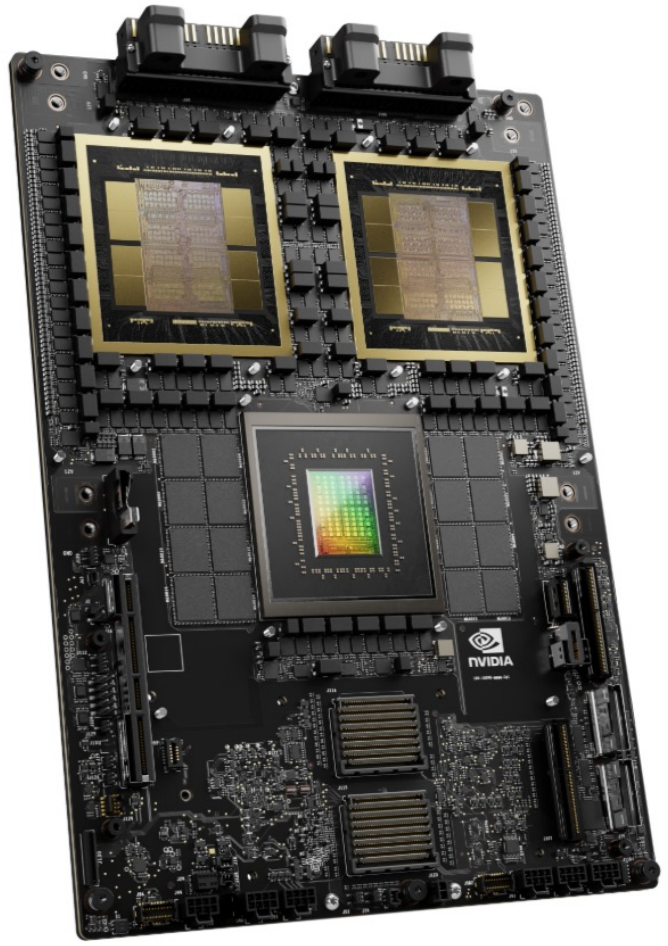}\\
\caption{Picture of the Nvidia Grace-Blackwell GB200 superchip. The GB200 contains two Nvidia Blackwell B200 Tensor Core GPUs with an Nvidia Grace CPU, interconnected via a 900 GB/s NVLink-C2C interface, creating a unified memory architecture. This configuration provides a total of 896 GB of memory, combining 384 GB of HBM3e memory on the GPUs and 512 GB of LPDDR5 memory on the CPU.}
\label{fig:gb200}
\end{figure}

The Nvidia GH200 superchip is a key part of the DGX Quantum integrated system a reference architecture for NVQlink, delivering exceptional performance for hybrid quantum--classical workloads through its heterogeneous architecture.  This architecture eliminates bottlenecks in data movement and ensures optimal performance for latency-sensitive quantum--classical interactions, as required in QEC and hybrid algorithm execution. Advanced features of the Hopper GPU, including asynchronous execution engines and thread block reconfiguration, further enhance computational efficiency, ensuring that all resources are fully utilized in the DGX Quantum integrated system. Future DGX Quantum  systems may use Grace Blackwell GB200 superchips  (see Figure \ref{fig:gb200}). A comparison of Grace Hopper and Blackwell superchips is shown in Table \ref{tab:grace_comparison}, which demonstrates the GB200 superchip's significantly improved specifications over those of the GH200. It is an important example of improvement because to calculate the future true costs of quantum computing, one needs to factor in the decreasing cost of error correction due to improvement in classical computing infrastructure.

\begin{table*}
\centering
\begin{tabular}{|l|l|l|}
\hline
\textbf{} & \textbf{GH200 Grace Hopper Superchip} & \textbf{GB200 Grace Blackwell Superchip} \\ \hline
\textbf{Configuration} & 1 Grace CPU : 1 Hopper GPU & 1 Grace CPU : 2 Blackwell GPUs \\ \hline
\textbf{FP4 Tensor Core} & 20 PFLOPS & 40 PFLOPS \\ \hline
\textbf{FP8/FP6 Tensor Core} & 4 PFLOPS & 20 PFLOPS \\ \hline
\textbf{INT8 Tensor Core} & 4 POPS & 20 POPS \\ \hline
\textbf{FP16/BF16 Tensor Core} & 2 PFLOPS & 10 PFLOPS \\ \hline
\textbf{TF32 Tensor Core} & 1 PFLOPS & 5 PFLOPS \\ \hline
\textbf{FP32} & 67 TFLOPS & 180 TFLOPS \\ \hline
\textbf{FP64} & 34 TFLOPS & 90 TFLOPS \\ \hline
\textbf{FP64 Tensor Core} & 67 TFLOPS & 90 TFLOPS \\ \hline
\textbf{GPU Memory | Bandwidth} & 96 GB HBM3 | 4 TB/s & Up to 384 GB HBM3e | 16 TB/s \\ \hline
\textbf{NVLink Bandwidth} & 0.9 TB/s & 3.6 TB/s \\ \hline
\textbf{CPU Core Count} & 72 Arm Neoverse V2 cores & 72 Arm Neoverse V2 cores \\ \hline
\textbf{CPU Memory | Bandwidth} & Up to 480GB LPDDR5X | Up to 512 GB/s & Up to 480GB LPDDR5X | Up to 512 GB/s \\ \hline
\end{tabular}
\caption{Grace Hopper vs. Grace Blackwell superchip specifications.}
\label{tab:grace_comparison}
\end{table*}

DGX Quantum also enables many other control mechanisms, such as error mitigation, dynamic circuit programming, and adaptive circuit knitting. By directly writing quantum control data into the GH200's local memory, latency overhead is eliminated, ensuring that the system can respond to quantum hardware feedback in real time. This feature is particularly valuable for iterative classical--quantum applications. For less latency-critical scenarios, users can queue quantum programs locally and remotely, enabling resource flexibility while maintaining access to DGX Quantum's high-performance computational capabilities.

\section{CUDA-Q: A software platform for heterogeneous quantum--classical computing}
\label{sec:cudaq}

DGX Quantum integrated systems are built to run Nvidia CUDA-Q~\cite{cuda_q}, which is an open source programming framework and compiler toolchain designed as an efficient and unified platform for programming hybrid quantum--classical systems. CUDA-Q supports both Python and C++ programming interfaces. Its kernel-based programming capabilities are specifically tailored for hybrid quantum--classical systems. CUDA-Q programs use the concept of a quantum kernel to differentiate between host and quantum device code, with kernels specifying a target for compilation and execution. The CUDA-Q platform also includes the NVQ++ compiler, which supports split compilation by lowering quantum kernels to a multi-level intermediate representation (MLIR) and a quantum intermediate representation (QIR). This ensures the seamless integration of classical and quantum resources needed for accelerating applications in large-scale quantum computing.

CUDA-Q includes specialized libraries designed to streamline the development of quantum algorithms and support advanced research. The CUDA-Q QEC library provides optimized implementations of key  quantum error correction primitives (such as GPU-accelerated decoders) and the CUDA-Q Solvers library includes application development primitives (such as prebuilt optimized kernels for the VQE, ADAPT-VQE, and QAOA methods).
CUDA-Q also offers GPU-accelerated quantum dynamics simulations, enabling accurate modeling of open quantum systems---a critical capability for the design, characterization, optimization, and scaling of QPUs.

Another key advantage of CUDA-Q is its ability to scale the performance of quantum simulations. Leveraging Nvidia's multi-GPU acceleration, CUDA-Q can achieve a massive speedup over traditional CPU-based simulations~\cite{cuda_q}, enabling researchers to efficiently simulate large quantum circuits. CUDA-Q supports multiple simulation technique backends such as state vector and tensor networks, allowing users to select from among various simulation methods for their specific problem.

Additionally, CUDA-Q is interoperable with AI software and the CUDA software ecosystem. This  ecosystem is a comprehensive suite of tools, libraries, and frameworks designed to harness the computational power of GPUs for diverse applications, ranging from AI and data science to HPC and quantum simulation. It includes components that provide developers with optimized building blocks, scalable programming models, and integration capabilities for hybrid quantum--classical workflows.

\end{appendices}


\begin{center}
\noindent\rule[0.5ex]{0.1\linewidth}{0.3pt}\rule[0.5ex]{0.1\linewidth}{0.8pt}\rule[0.5ex]{0.3\linewidth}{1.5pt}\rule[0.5ex]{0.1\linewidth}{0.8pt}\rule[0.5ex]{0.1\linewidth}{0.3pt}
\end{center}
\end{document}